\documentclass[11pt,fleqn]{book} 
\usepackage{pdfpages}

\usepackage[top=3cm,bottom=3cm,left=3cm,right=3cm,headsep=10pt,a4paper]{geometry} 
\usepackage{bm}
\usepackage{graphicx} 
\graphicspath{{Pictures/}} 

\usepackage{lipsum} 
\usepackage{tabularx}
\usepackage{tikz} 
\usepackage{subfiles}
\usepackage{xr}
\usepackage[english]{babel} 
\usepackage{dcolumn}
\usepackage{enumitem} 
\setlist{nolistsep} 

\usepackage{booktabs} 

\usepackage{xcolor} 
\definecolor{ocre}{RGB}{243,102,25} 
\usepackage{tipa}

\usepackage{amsthm}
\usepackage{algorithm}
\usepackage[noend]{algpseudocode}

\usepackage{avant} 
\usepackage{mathptmx} 

\usepackage{microtype} 
\usepackage[utf8]{inputenc} 
\usepackage[T1]{fontenc} 

\usepackage{calc} 
\usepackage{makeidx} 
\makeindex 

\usepackage{titletoc} 

\contentsmargin{0cm} 

\titlecontents{part}[0cm]
{\addvspace{20pt}\centering\large\bfseries}
{}
{}
{}

\titlecontents{chapter}[1.25cm] 
{\addvspace{12pt}\large\sffamily\bfseries} 
{\color{cyan!60}\contentslabel[\Large\thecontentslabel]{1.25cm}\color{cyan}} 
{\color{cyan}}  
{\color{cyan!60}\normalsize\;\titlerule*[.5pc]{.}\;\thecontentspage} 

\titlecontents{section}[1.25cm] 
{\addvspace{3pt}\sffamily\bfseries} 
{\contentslabel[\thecontentslabel]{1.25cm}} 
{}
{\hfill\color{black}\thecontentspage} 
[]

\titlecontents{subsection}[1.25cm] 
{\addvspace{1pt}\sffamily\small} 
{\contentslabel[\thecontentslabel]{1.25cm}} 
{}
{\ \titlerule*[.5pc]{.}\;\thecontentspage} 
[]

\titlecontents{figure}[0em]
{\addvspace{-5pt}\sffamily}
{\thecontentslabel\hspace*{1em}}
{}
{\ \titlerule*[.5pc]{.}\;\thecontentspage}
[]

\titlecontents{table}[0em]
{\addvspace{-5pt}\sffamily}
{\thecontentslabel\hspace*{1em}}
{}
{\ \titlerule*[.5pc]{.}\;\thecontentspage}
[]

\titlecontents{lchapter}[0em] 
{\addvspace{15pt}\large\sffamily\bfseries} 
{\color{cyan}\contentslabel[\Large\thecontentslabel]{1.25cm}\color{cyan}} 
{}  
{\color{cyan}\normalsize\sffamily\bfseries\;\titlerule*[.5pc]{.}\;\thecontentspage} 

\titlecontents{lsection}[0em] 
{\sffamily\small} 
{\contentslabel[\thecontentslabel]{1.25cm}} 
{}
{}

\titlecontents{lsubsection}[.5em] 
{\normalfont\footnotesize\sffamily} 
{}
{}
{}

\usepackage{fancyhdr} 

\fancypagestyle{plain}{\fancyhead{}} 

\makeatletter
\renewcommand{\cleardoublepage}{
\clearpage\ifodd\c@page\else
\hbox{}
\vspace*{\fill}
\thispagestyle{empty}
\newpage
\fi}

\usepackage{amsmath,amsfonts,amssymb,amsthm} 

\newtheoremstyle{cyannumbox}
{0pt}
{0pt}
{\normalfont}
{}
{\small\bf\sffamily\color{cyan}}
{\;}
{0.25em}
{\small\sffamily\color{cyan}\thmname{#1}\nobreakspace\thmnumber{\@ifnotempty{#1}{}\@upn{#2}}
\thmnote{\nobreakspace\the\thm@notefont\sffamily\bfseries\color{black}---\nobreakspace#3.}} 

\newtheoremstyle{blacknumex}
{5pt}
{5pt}
{\normalfont}
{} 
{\small\bf\sffamily}
{\;}
{0.25em}
{\small\sffamily{\tiny\ensuremath{\blacksquare}}\nobreakspace\thmname{#1}\nobreakspace\thmnumber{\@ifnotempty{#1}{}\@upn{#2}}
\thmnote{\nobreakspace\the\thm@notefont\sffamily\bfseries---\nobreakspace#3.}}

\newtheoremstyle{blacknumbox} 
{0pt}
{0pt}
{\normalfont}
{}
{\small\bf\sffamily}
{\;}
{0.25em}
{\small\sffamily\thmname{#1}\nobreakspace\thmnumber{\@ifnotempty{#1}{}\@upn{#2}}
\thmnote{\nobreakspace\the\thm@notefont\sffamily\bfseries---\nobreakspace#3.}}

\newtheoremstyle{cyannum}
{5pt}
{5pt}
{\normalfont}
{}
{\small\bf\sffamily\color{cyan}}
{\;}
{0.25em}
{\small\sffamily\color{cyan}\thmname{#1}\nobreakspace\thmnumber{\@ifnotempty{#1}{}\@upn{#2}}
\thmnote{\nobreakspace\the\thm@notefont\sffamily\bfseries\color{black}---\nobreakspace#3.}} 
\makeatother

\newcounter{dummy} 
\numberwithin{dummy}{section}
\theoremstyle{cyannumbox}
\newtheorem{theoremeT}[dummy]{Theorem}

\newtheorem{exerciseT}{Exercise}[chapter]
\theoremstyle{blacknumex}
\newtheorem{exampleT}{Example}[chapter]
\theoremstyle{blacknumbox}

\newtheorem{definitionT}{Definition}[section]
\newtheorem{corollaryT}[dummy]{Corollary}
\theoremstyle{cyannum}

\RequirePackage[framemethod=default]{mdframed} 

\newmdenv[skipabove=7pt,
skipbelow=7pt,
backgroundcolor=black!5,
linecolor=cyan,
innerleftmargin=5pt,
innerrightmargin=5pt,
innertopmargin=5pt,
leftmargin=0cm,
rightmargin=0cm,
innerbottommargin=5pt]{tBox}

\newmdenv[skipabove=7pt,
skipbelow=7pt,
rightline=false,
leftline=true,
topline=false,
bottomline=false,
backgroundcolor=cyan!10,
linecolor=cyan,
innerleftmargin=5pt,
innerrightmargin=5pt,
innertopmargin=5pt,
innerbottommargin=5pt,
leftmargin=0cm,
rightmargin=0cm,
linewidth=4pt]{eBox}	

\newmdenv[skipabove=7pt,
skipbelow=7pt,
rightline=false,
leftline=true,
topline=false,
bottomline=false,
linecolor=cyan,
innerleftmargin=5pt,
innerrightmargin=5pt,
innertopmargin=0pt,
leftmargin=0cm,
rightmargin=0cm,
linewidth=4pt,
innerbottommargin=0pt]{dBox}	

\newmdenv[skipabove=7pt,
skipbelow=7pt,
rightline=false,
leftline=true,
topline=false,
bottomline=false,
linecolor=gray,
backgroundcolor=black!5,
innerleftmargin=5pt,
innerrightmargin=5pt,
innertopmargin=5pt,
leftmargin=0cm,
rightmargin=0cm,
linewidth=4pt,
innerbottommargin=5pt]{cBox}

\newenvironment{theorem}{\begin{tBox}\begin{theoremeT}}{\end{theoremeT}\end{tBox}}

\makeatletter
\renewcommand{\@seccntformat}[1]{\llap{\textcolor{cyan}{\csname the#1\endcsname}\hspace{1em}}}                    
\renewcommand{\section}{\@startsection{section}{1}{\z@}
{-4ex \@plus -1ex \@minus -.4ex}
{1ex \@plus.2ex }
{\normalfont\large\sffamily\bfseries}}
\renewcommand{\subsection}{\@startsection {subsection}{2}{\z@}
{-3ex \@plus -0.1ex \@minus -.4ex}
{0.5ex \@plus.2ex }
{\normalfont\sffamily\bfseries}}
\renewcommand{\subsubsection}{\@startsection {subsubsection}{3}{\z@}
{-2ex \@plus -0.1ex \@minus -.2ex}
{.2ex \@plus.2ex }
{\normalfont\small\sffamily\bfseries}}                        
\renewcommand\paragraph{\@startsection{paragraph}{4}{\z@}
{-2ex \@plus-.2ex \@minus .2ex}
{.1ex}
{\normalfont\small\sffamily\bfseries}}

\newcommand{\@mypartnumtocformat}[2]{%
\setlength\fboxsep{0pt}%
\noindent\colorbox{cyan!20}{\strut\parbox[c][.7cm]{\ecart}{\color{cyan!70}\Large\sffamily\bfseries\centering#1}}\hskip\esp\colorbox{cyan!40}{\strut\parbox[c][.7cm]{\linewidth-\ecart-\esp}{\Large\sffamily\centering#2}}}%
\newcommand{\@myparttocformat}[1]{%
\setlength\fboxsep{0pt}%
\noindent\colorbox{cyan!40}{\strut\parbox[c][.7cm]{\linewidth}{\Large\sffamily\centering#1}}}%
\newlength\esp
\newlength\ecart
\def\@part[#1]#2{%
\ifnum \c@secnumdepth >-2\relax%
\refstepcounter{part}%
\addcontentsline{toc}{part}{\texorpdfstring{\protect\@mypartnumtocformat{\thepart}{#1}}{\partname~\thepart\ ---\ #1}}
\else%
\addcontentsline{toc}{part}{\texorpdfstring{\protect\@myparttocformat{#1}}{#1}}%
\fi%
\startcontents%
\markboth{}{}%
{\thispagestyle{empty}%
\begin{tikzpicture}[remember picture,overlay]%
\node at (current page.north west){\begin{tikzpicture}[remember picture,overlay]%
\fill[cyan!20](0cm,0cm) rectangle (\paperwidth,-\paperheight);
\node[anchor=north] at (4cm,-3.25cm){\color{cyan!40}\fontsize{220}{100}\sffamily\bfseries\thepart}; 
\node[anchor=south east] at (\paperwidth-1cm,-\paperheight+1cm){\parbox[t][][t]{8.5cm}{
\printcontents{l}{0}{\setcounter{tocdepth}{1}}%
}};
\node[anchor=north east] at (\paperwidth-1.5cm,-3.25cm){\parbox[t][][t]{15cm}{\strut\raggedleft\color{white}\fontsize{30}{30}\sffamily\bfseries#2}};
\end{tikzpicture}};
\end{tikzpicture}}%
\@endpart}
\def\@spart#1{%
\startcontents%
\phantomsection
{\thispagestyle{empty}%
\begin{tikzpicture}[remember picture,overlay]%
\node at (current page.north west){\begin{tikzpicture}[remember picture,overlay]%
\fill[cyan!20](0cm,0cm) rectangle (\paperwidth,-\paperheight);
\node[anchor=north east] at (\paperwidth-1.5cm,-3.25cm){\parbox[t][][t]{15cm}{\strut\raggedleft\color{white}\fontsize{30}{30}\sffamily\bfseries#1}};
\end{tikzpicture}};
\end{tikzpicture}}
\addcontentsline{toc}{part}{\texorpdfstring{%
\setlength\fboxsep{0pt}%
\noindent\protect\colorbox{cyan!40}{\strut\protect\parbox[c][.7cm]{\linewidth}{\Large\sffamily\protect\centering #1\quad\mbox{}}}}{#1}}%
\@endpart}
\def\@endpart{\vfil\newpage
\if@twoside
\if@openright
\null
\thispagestyle{empty}%
\newpage
\fi
\fi
\if@tempswa
\twocolumn
\fi}

\newif\ifusechapterimage
\usechapterimagetrue
\newcommand{\thechapterimage}{}%
\newcommand{\chapterimage}[1]{\ifusechapterimage\renewcommand{\thechapterimage}{#1}\fi}%
\newcommand{\autodot}{.}
\def\@makechapterhead#1{%
{\parindent \z@ \raggedright \normalfont
\ifnum \c@secnumdepth >\m@ne
\if@mainmatter
\begin{tikzpicture}[remember picture,overlay]
\node at (current page.north west)
{\begin{tikzpicture}[remember picture,overlay]
\node[anchor=north west,inner sep=0pt] at (0,0) {\ifusechapterimage\includegraphics[width=\paperwidth]{\thechapterimage}\fi};
\draw[anchor=west] (\Gm@lmargin,-9cm) node [line width=2pt,rounded corners=15pt,draw=cyan,fill=white,fill opacity=0.5,inner sep=15pt]{\strut\makebox[22cm]{}};
\draw[anchor=west] (\Gm@lmargin+.3cm,-9cm) node {\huge\sffamily\bfseries\color{black}\thechapter\autodot~#1\strut};
\end{tikzpicture}};
\end{tikzpicture}
\else
\begin{tikzpicture}[remember picture,overlay]
\node at (current page.north west)
{\begin{tikzpicture}[remember picture,overlay]
\node[anchor=north west,inner sep=0pt] at (0,0) {\ifusechapterimage\includegraphics[width=\paperwidth]{\thechapterimage}\fi};
\draw[anchor=west] (\Gm@lmargin,-9cm) node [line width=2pt,rounded corners=15pt,draw=cyan,fill=white,fill opacity=0.5,inner sep=15pt]{\strut\makebox[22cm]{}};
\draw[anchor=west] (\Gm@lmargin+.3cm,-9cm) node {\huge\sffamily\bfseries\color{black}#1\strut};
\end{tikzpicture}};
\end{tikzpicture}
\fi\fi\par\vspace*{270\p@}}}

\def\@makeschapterhead#1{%
\begin{tikzpicture}[remember picture,overlay]
\node at (current page.north west)
{\begin{tikzpicture}[remember picture,overlay]
\node[anchor=north west,inner sep=0pt] at (0,0) {\ifusechapterimage\includegraphics[width=\paperwidth]{\thechapterimage}\fi};
\draw[anchor=west] (\Gm@lmargin,-9cm) node [line width=2pt,rounded corners=15pt,draw=cyan,fill=white,fill opacity=0.5,inner sep=15pt]{\strut\makebox[22cm]{}};
\draw[anchor=west] (\Gm@lmargin+.3cm,-9cm) node {\huge\sffamily\bfseries\color{black}#1\strut};
\end{tikzpicture}};
\end{tikzpicture}
\par\vspace*{270\p@}}
\makeatother

\usepackage{hyperref}
\hypersetup{hidelinks,backref=true,pagebackref=true,hyperindex=true,colorlinks=false,breaklinks=true,urlcolor= cyan,bookmarks=true,bookmarksopen=false,pdftitle={Title},pdfauthor={Author}}
\usepackage{bookmark}
\bookmarksetup{
open,
numbered,
addtohook={%
\ifnum\bookmarkget{level}=0 
\bookmarksetup{bold}%
\fi
\ifnum\bookmarkget{level}=-1 
\bookmarksetup{color=cyan,bold}%
\fi
}
}
{

\def\biblio{\bibliographystyle{amsalpha}\bibliography{bibgraf}}  

\begin{document}
\def\biblio{}
\includepdf[pages=-]{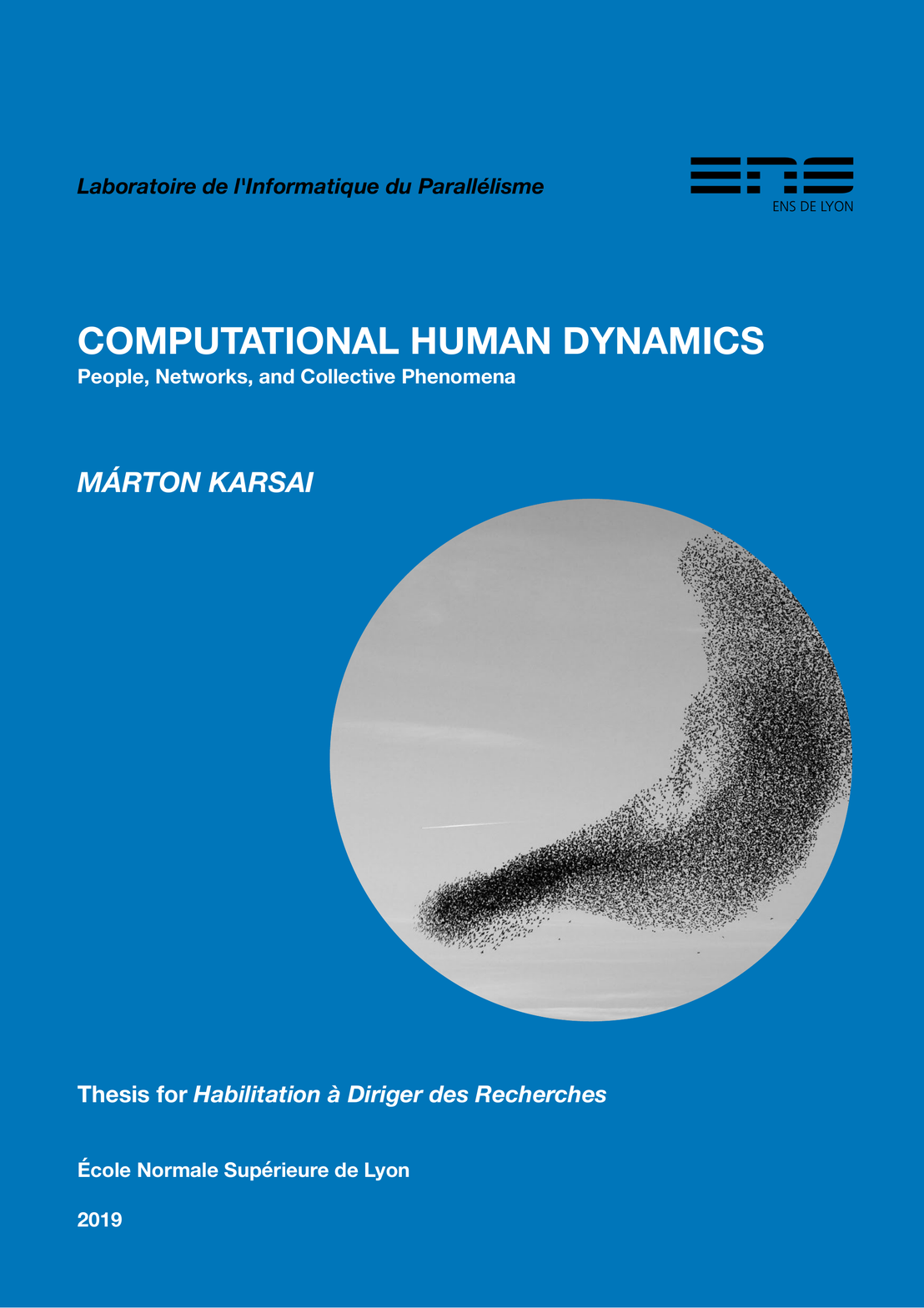}
\frontmatter

\begingroup
\thispagestyle{empty}

\begin{center}
{\Large \textsc{Acad\'emie de Lyon}}\\[10pt]
{\LARGE \textsc{\'Ecole Normale Sup\'erieure de Lyon}}\\[10pt]
{\LARGE \textsc{Laboratoire de l’Informatique du Parall\'elisme}}\\[40pt]
{\huge \textsc{Habilitation \`a Diriger des Recherches}} \\
\vspace{60pt}

{\Huge \textbf{\textsc{Computational Human Dynamics}}}\\[20pt]
{\LARGE People, Networks, and Collective Phenomena}\\[20pt]
{\large presented by}\\[20pt]
{\LARGE \textbf{\textsc{M\'arton Karsai}}}\\
\vspace{20pt}

{\large to obtain the Habilitation \`a Diriger des Recherches in Computer Science}\\[5pt]
{\large from \'Ecole Normale Sup\'erieure de Lyon} \\[20pt]
{\large \textsc{Section CNU: 27 - Informatique}} \\[20pt]
{\large Defence date: 23rd April 2019} \\[20pt]
{\large Defense committee:}\\[20pt]
\end{center}
{\large \textbf{President:}}\\[5pt]
\hspace*{.2in}{\large \textsc{Isabelle Gu\'erin Lassous} - Professor, LIP, Universit\'e Claude Bernard Lyon 1}\\[5pt]
{\large \textbf{Reviewers:}}\\[5pt]
\hspace*{.2in}{\large \textsc{Matthieu Latapy} - Research Director, LIP6, CNRS}\\[5pt]
\hspace*{.2in}{\large \textsc{Alain Barrat} - Research Director, CPT-Aix-Marseille Universit\'e, CNRS}\\[5pt]
\hspace*{.2in}{\large \textsc{Tina Eliassi-Rad} - Associate Professor, Khoury College of Computer and}\\
\hspace*{1.7in}{\large Information Sciences, Northeastern University}\\[5pt]
{\large \textbf{Members:}} \\[5pt]
\hspace*{.2in}{\large \textsc{James Gleeson} - Professor, MACSI, University of Limerick}\\[5pt]
\hspace*{.2in}{\large \textsc{Markus Strohmaier} - Professor, HumTec Center, Aachen University}\\[5pt]


\endgroup


\newpage
~\vfill
\thispagestyle{empty}

\noindent Copyright \copyright\ 2018 M\'arton Karsai\\ 



\noindent Licensed under the Creative Commons Attribution-NonCommercial 3.0 Unported License (the ``License''). You may not use this file except in compliance with the License. You may obtain a copy of the License at \url{http://creativecommons.org/licenses/by-nc/3.0}. Unless required by applicable law or agreed to in writing, software distributed under the License is distributed on an \textsc{``as is'' basis, without warranties or conditions of any kind}, either express or implied. See the License for the specific language governing permissions and limitations under the License.\\ 

\noindent \textit{First printing, December 2018} 


\usechapterimagefalse 

\chapterimage{net_5581081_cutn.pdf} 

\pagestyle{empty} 

\tableofcontents 

\cleardoublepage 

\pagestyle{fancy} 






\chapter*{Acknowledgement}
\addcontentsline{toc}{chapter}{Acknowledgement} 
\label{ch:ackn}

Over the years I had the opportunity to work with several exceptional people who truly shaped my opinion, sophisticated my scientific approach, and set high standards to follow in scientific and personal manner. First of all I would like to acknowledge my mentors who helped me to develop, and to reach opportunities since the beginning of my scientific career. Pr. Ferenc Igl\'oi and Dr. Jean-Christian Angl\`es d'Auriac were my PhD supervisors who set me on the path towards my scientific journey. Pr. Kimmo Kaski, Pr. Jari Saram\"aki and Pr. J\'anos Kert\'esz were my mentors at Aalto University who introduced me to new scientific challenges, which are still in the core of my research. They opened me the door for new opportunities, which entirely changed my professional motivations and goals. Their continuous encouragement, patient, stimulating suggestions and healthy criticism helped me greatly to develop my scientific identity. I am especially beholden to Pr. Alessandro Vespignani at Northeastern University, who first of all provided me the unconditioned opportunity to explore my own scientific path, and who set me an example of such high professional and personal standards, that I can use as a compass to follow during the rest of my career. Finally, I am truly grateful to Pr. Eric Fleury, who welcomed and integrated me at ENS Lyon. His continuous support helped me to grow in every sense. He trained me on how to propose and manage projects, how to supervise students and a team, while not loosing connection with the core of scientific problems and with the fun they provide in life.

I regard science as a social effort and thus I have built several collaborations with excellent researchers. Exploring problems together with them helped me to develop ideas without compromises, while in turn many of them became my friends. They are J. I. Alvarez-Hamelin, A.-L. Barab\'asi, A. Barrat, S. Bernhardsson, G. Bianconi, V. Blondel, H. Bouchet, R. Burioni, J-P. Chevrot, J-P. Cointet, C. Cherifi, H. Cherifi, V. Colizza, W. Du, A. Flamini, L. Gauvin, M. G\'enois, P. Jensen, H.-H. Jo, J. Karikoski, R. Kikas, M. Kivelä, L. Kovanen, G. Krings, S. Liu, J-P. Magu\'e, F. Menzer, P. Merckl\'e, M. Minnoni, D. Mocanu, M. Musolesi, A. Nardy, R. K. Pan, N. Perra, W. Quattrociocchi, L. Rossi, Z. Ruan, C. Sarraute, F. Schweitzer, K. Sun, T. Takaguchi, C. J. Tessone , G. Tibely, M. Tizzoni, M. V. Tomasello, E. Ubaldi, E. Valdano, C. L. Vestergaard, A. Vezzani, L. Weng, Q. Zhang, and K. Zhao.

Transition from being a student to supervise students is a journey I try to make every day. That is why I am very grateful for all students and postdocs I have had the chance to work with, for their open minded approach and their susceptibility and commitment to explore new ideas. They are L. Alessandretti, G. Laurent, J. Cambe, M. Morini, Y. Leo, S. Unicomb, J. Levy Abitbol, S. Dai, H. Hours, Y. Liao, and S. Lerique.

I also would like to acknowledge the support and friendly environment created by the DANTE research team, the IXXI and LIP laboratories, and the Computer Science Department at ENS Lyon. Namely, I would like to mention P. Gon\c{c}alves, P. Nain, A. Busson, T. Begin, C. Crespelle, I. Gu\'erin Lassous, T. Venturini, L. Lecot, N. Trotignon, P. Borgnat, S. Thomasse, and D. Stehl\'e, with all of whom I enjoy to interact on the daily base. I am especially thankful for P. Gon\c{c}alves, A. Busson and E. Fleury who helped and advised me during the preparation of this thesis.

I am indebted to my parents and overly thankful for my brother, my sister and their families, who support and encourage me until today despite the large geographic distance between us.

Finally and most importantly, I would like to express my deepest gratitude to my little family. My wife Sophie and my children Lilla and \'Abel are the most important things in my life. Their unconditional love and trust provide me an invaluable support; any of these achievement would have been simply impossible without them.

\biblio


\chapter*{R\'esum\'e des travaux}
\addcontentsline{toc}{chapter}{R\'esum\'e des travaux} 
\label{ch:resume}

Cette th\`ese regroupe mes contributions scientifiques dans le domaine des science des r\'eseaux et des dynamiques humaines. Le document repose sur mes travaux publi\'es depuis l’obtention de mon doctorat en 2009. Mes contributions \'emargent sur plusieurs disciplines et mes sujets d’\'etude font appel \`a des m\'ethodes emprunt\'ees aux sciences informatique, \`a la physique, aux statistiques, aux math\'ematiques appliqu\'ees et aux sciences sociales computationnelles. Cette h\'et\'erog\'en\'eit\'e peut rendre difficile la lecture du document si celui-ci est abord\'e sous l’angle d’une seule discipline. Aussi, pour guider le lecteur, je me suis efforc\'e d’\'etablir un fil conducteur qui partant des aspects les plus techniques li\'es aux m\'ethodologies aboutit \`a la pr\'esentation de r\'esultats tr\`es concrets et donc accessibles \`a un plus large public. Apr\`es une mise en contexte de mes travaux dans l’introduction, la pr\'esentation de mes resultats s’organise en trois chapitres: dynamiques humaines temporelles h\'et\'erog\`enes, r\'eseaux temporels et ph\'enom\`enes collectifs. Pour chacune de ces contributions, je d\'ecris des travaux empiriques qui m’ont conduit \`a d\'evelopper des m\'ethodologies innovantes et \`a \'etablir des r\'esultats th\'eoriques originaux. Au del\`a d’une pr\'esentation synth\'etique de mes contributions scientifiques les plus marquantes, l’objectif de ce manuscrit est aussi de pr\'esenter ma vision et mes interrogations sur mon domaine de recherche et sur ses perspectives. 

Je commence mon manuscrit avec l’introduction de mon domaine de recherche et resitue dans ce contexte, mes propres contributions. Je pr\'esente un point de vue personnel sur les d\'efis actuels que posent les donn\'ees, qu’il s’agisse de leur collecte, leur observation et analyse, leur mod\'elisation, leur v\'erification mais aussi des  applications qui en d\'ecoulent. Cette introduction a \'egalement pour but de fixer la terminologie utilis\'ee tout au long de la dissertation, de poser les concepts et caract\'eristiques des r\'eseaux complexes, ainsi que de la th\'eorie des r\'eseaux sociaux. Enfin, afin de limiter les r\'ep\'etitions et les r\'ef\'erences crois\'ees r\'eflexives, je pr\'esente les jeux de donn\'ees qui sont \'et\'e utilis\'es et analys\'es \`a de nombreuses reprises dans le manuscrit. Le chapitre se termine sur un  r\'esum\'e de mes contributions.

Dans un premier temps, je me concentre sur les dynamiques humaines dites en rafales (« bursty »), propres au caract\`ere temporels naturellement h\'et\'erog\`ene des actions et interactions humaines. Ayant r\'ecemment \'ecrit un livre sur le sujet,  il m’est relativement ais\'e de brosser une vue d’ensemble du domaine et d’y situer mes contributions. Apr\`es l’introduction, je discute de syst\`emes de r\'ef\'erences et de mesures \'el\'ementaires, puis je d\'eveloppe mon travail sur la caract\'erisation des corr\'elations temporelles locales, responsables des rafales que l’on observe dans diff\'erentes dynamiques de communication. Je synth\'etise \'egalement mes travaux sur l’impact des effets circadiens sur les mod\`eles bursty dans les interactions humaines. Par la suite, les observations dynamiques des interactions dyadiques sont discut\'ees. Je termine cette partie en pr\'esentant un mod\`ele \'etendu, capable de reproduire simultan\'ement les rafales et les d\'erives des balance des communications, souvent observ\'ees dans les services de communication mobiles.

La chapitre suivant traite des r\'eseaux temporels. Il synth\'etise mon travail de ces derni\`eres ann\'ees sur diff\'erentes m\'ethodes de repr\'esentation, de caract\'erisation et de mod\'elisation des structures variables dans le temps. Cette partie commence par une discussion sur la notion centrale d’\'echelle de temps, puis souligne l’importance de la repr\'esentation des r\'eseaux statiques et temporels, et enfin insiste sur le poids de la mesure des r\'eseaux dynamiques. Plus pr\'ecis\'ement, les probl\`emes abord\'es portent sur le niveau de caract\'erisation des r\'eseaux temporels: le r\^ole de la fen\^etre d’agr\'egation temporelle dans l’\'etude des r\'eseaux de communication mobiles; une nouvelle mesure d’entropie sur les r\'eseaux dynamiques sociaux; des tests d’hypoth\`ese bas\'es sur des mod\`eles al\'eatoires de r\'ef\'erence. Sur ce dernier point, je propose une nouvelle technique pour identifier les corr\'elations temporelles et structurelles dans des syst\`emes variants dans le temps.  Je discute \'egalement d’une m\'ethode algorithmique pour l’identifications de motifs temporels, et d\'eveloppe une nouvelle approche pour la repr\'esentation d’ordre sup\'erieur des r\'eseaux temporels sous forme de graphiques acycliques dirig\'es pond\'er\'es statiques. Dans une derni\`ere partie, je r\'esume mes travaux sur les mod\`eles g\'en\'eratifs de r\'eseaux temporels. 

Enfin, le troisi\`eme volet pr\'esente mes travaux sur les observations bas\'ees sur les donn\'ees et la mod\'elisation de ph\'enom\`enes sociaux collectifs. Je r\'esume ici des \'etudes sur les observations statiques de mod\`eles \'emergents d’in\'egalit\'es socio\'economiques et leurs corr\'elations avec les r\'eseaux de communication sociale et des mod\`eles linguistiques. Je pr\'esente aussi des observations dynamiques des processus de contagion sociale. Dans la deuxi\`eme partie du chapitre, je m’int\'eresse \`a la mod\'elisation des processus de propagation avec une discussion sur les diff\'erences entre la contagion simple et complexe. Dans le cas de la propagation simple, je pr\'esente un travail th\'eorique sur le contr\^ole de la contagion \'evoluant sur des r\'eseaux variant dans le temps et sur une nouvelle mesure de centralit\'e bas\'ee sur un processus de diffusion d'informations. Pour la contagion complexe, je pr\'esente une s\'erie d’\'etudes sur les mod\`eles de processus bas\'es sur les seuils et leurs solutions utilisant des techniques de champ moyen approximatif, et des simulations num\'eriques bas\'ees sur des donn\'ees. 

Ma th\`ese se termine sur les objectifs \`a court et \`a moyen terme de mes recherches. Une synth\`ese des perspectives \`a long terme de mes th\`emes de recherche est \'egalement pr\'esent\'ee. Enfin, je dresse une br\`eve r\'etrospection sur l’ensemble de ms contributions, sur le travail effectu\'e, et sur la finalit\'e de ce manuscrit.

\biblio


%
%

\chapter*{List of symbols}
\addcontentsline{toc}{chapter}{List of symbols} 

This is a non-complete list of acronyms, which only collects the most frequently used notations in the text. Some notation specific to a given study will be introduced at the place of discussion, this way allowing the multiple usage of the same symbols.
\vspace{.2in}

\begin{enumerate}
\setlength{\itemsep}{2pt}
\item[$G$]{graph}
\item[$V$]{set of nodes of a graph}
\item[$E$]{set of links of a graph; bursty train size}
\item[$e_{u,v}$]{link between nodes $u$ and $v$}
\item[$N$]{size, i.e., number of nodes in a network}
\item[$M$]{number of links in a network}
\item[$w_{u,v}$; $w$]{weight of the link $e_{u,v}$; link weight in general}
\item[$G_v$]{egocentric network of node $v$}
\item[$V_v$]{neighbour set of node $v$}
\item[$E_v$]{set of links in $G_v$}
\item[$G_t$]{temporal network}
\item[$E_t$]{set of events (temporal events)}
\item[$G_{[0,t]}$]{aggregated temporal network}
\item[$D$]{event graph}
\item[$E_D$]{set of links in $D$}
\item[$k_v$; $k$]{degree of node $v$; node degree in general}
\item[$k_{in}$; $k_{out}$]{in and out-degree of a node}
\item[$\langle k \rangle; z$]{average degree}
\item[$s_v$; $s$]{strength of node $v$; node strength in general}
\item[$C_v$]{clustering coefficient of a node $v$}
\item[$C$]{average local (or global) clustering coefficient}
\item[$O_{u,v}$; $O$]{overlap of link $e_{u,v}$; link overlap in general}
\item[$n_{u,v}$]{number of common neighbours of nodes $u$ and $v$}
\item[$\rho$]{network density}
\item[$P(x)$]{Probability density function of a variable $x$}
\item[$\langle x \rangle$]{average value of variable $x$}
\item[$\gamma$]{degree distribution exponent}
\item[$d(u,v)$]{graph distance between nodes $u$ and $v$}
\item[$BC(u)$]{betweenness centrality of node $u$}
\item[$T$]{observation time period}
\item[$A^{ev,ac}_i$]{An attribute $i$ of an event $ev$ or action $ac$}
\item[$A_j^v$]{A meta-data attribute of an individual $v$.}
\item[$t$; $t_i$]{time; time of the $i$th event.}
\item[$x(t)$; $ev(t_i)$]{binary event sequence; sequence of event timings}
\item[$\tau$]{inter-event time}
\item[$\tau_r$]{residual time}
\item[$\alpha$]{inter-event time distribution exponent}
\item[$A(t_d)$; $t_d$]{autocorrelation function, delay time}
\item[$\gamma$]{autocorrelation exponent}
\item[$\beta$]{train size distribution exponent}
\item[$\omega$]{event frequency}
\item[$B$]{bursty parameter}
\item[$p(n)$]{memory function}
\item[$\mathbf{k}$, $\mathbf{w}$]{sequence of degrees, weights, etc.}
\item[$\langle \ell \rangle$]{average path length}
\item[$\Delta t$]{maximum time between causal events}
\item[$a_i$]{activity of node $i$}
\item[$\varphi$]{individual adoption threshold}
\item[$\Phi$]{integer adoption threshold}
\item[$\phi$]{average adoption threshold}
\item[$r$]{fraction of blocked nodes for adoption}
\end{enumerate}



\mainmatter

\chapter{Introduction, positioning and terminology}

\section{Introduction}

The goal of this thesis is to summarise my scientific achievements since my PhD in 2009 and to position these contributions on the scientific landscape. My scientific motivations have always been \emph{curiosity driven} and focusing on answers to the question ``\emph{why?}'' (rather than to develop novel methods or technologies). This attitude may be rooted in my interdisciplinary training but certainly resulted in a very heterogeneous scientific portfolio with projects and publications ranging between and combining knowledge from several domains. However, these results all have a common ground to address some aspects of human dynamics using computational methods. That is why the title of my thesis is \emph{computational human dynamics}, which refers to studies focusing on the \emph{data-driven observations and computational modelling of individual, social, and collective human dynamical phenomena}. It builds on the conventional results and recent developments of several fields and as such it is truly interdisciplinary, which makes it difficult to fit into the traditional categories of academic disciplines. However, I judge this scientific multi-pluralism as an advantage, as it provides a ground for the recombination of knowledge of foreign disciplines to solve unconventional problems.

\subsection{Scientific landscape}

Computational human dynamics is naturally related to \emph{complex system science}, which studies the emergence of collective phenomena that arise from the interactions of entities in many-agent multivariate systems. The emergence of collective behaviour has been historically studied in the framework of statistical physics~\cite{stanley1987introduction} of phase transitions and critical phenomena. My PhD landed within these fields as it is mostly about phase-transitions of cooperative behaviour emerging in modelled physical systems like in spin models, diffusion, or percolation phenomena~\cite{karsai2006nonequilibrium,karsai2007rounding,karsai2008density,karsai2010interface,karsai2009nonequilibrium}. Meanwhile, collective phenomena occur not only within the realm of the physics of matter, but also in many other areas, including biological, social, and economic systems. The global spreading of a pandemic, the emergence of social movements, the collective migration of animals, the growth of tumours, and the coordinated firing of neurons are all examples of interdependence and collective behaviour~\cite{smelser2011theory}. However, advances in these areas have been hampered by difficulties in collecting the vast amounts of detailed data necessary for validating theories and developing quantitative approaches. Especially in the social systems arena, the community has confronted the insurmountable obstacle of a lack of data on human behaviour at multiple scales. Consequently, in many ways, it is currently easier to observe tiny bacteria or galaxies light years away than our fellow humans~\cite{vespignani2009predicting}.

These foundational limitations are now being obliterated by the \emph{digital data revolution}~\cite{mayer2012big,chen2014big}. Remarkably, every 1.2 years, more human-driven socioeconomic data are produced than during all preceding years of human history combined. Finally, we are in the position to follow the evolution of large real-world systems and detect the emergence of collective social behaviour ``in vivo''. This boom in data collection was induced by widely adopted new technologies, such as mobile phones or online services, and was escorted by the development of new computational designs. Recent availability of \emph{high performance computational resources}, the advances in \emph{social computation methods}, and the advent of \emph{advanced machine learning and statistical data analysis} all contributed to the success of data-driven research of human behaviour. However, at the same time, such studies highlight our limitations in knowledge about complexity itself and challenge us on the foundational aspects (such as conceptual, theoretical, and modelling issues) that form the basis of our understanding of complex systems.

In this context, \emph{network science} has been assigned an increasingly relevant role in defining a conceptual framework for the analysis of complex systems (as discussed in details in Section~\ref{sec:cn}). Network science is concerned with graphs that map real entities and their interactions to graph nodes and links~\cite{watts1998collective,barabasi1999emergence,albert2002statistical,newman2003structure}. For a long time, this mathematical abstraction has contributed to the understanding of real-world systems in physics, biology, chemistry, social sciences, and economics. Recently, however, the enormous amounts of detailed data, electronically collected and meticulously catalogued, which finally become available for scientific analysis and study, led us to the discovery that most networks describing real world systems show the presence of complex properties and heterogeneities, which cannot be neglected in their topological and dynamical description. This has called for a major effort in developing the methodology to characterise complex networks~\cite{newman2010networks,holme2012temporal,dorogovtsev2002evolution,zhao2011entropy,krings2012effects}, to describe the observed structural and temporal heterogeneities~\cite{albert2002statistical,holme2012temporal,barabasi2005origin,karsai2018bursty}, to capture the multilayer nature of connectedness~\cite{kivela2014multilayer}, to detect and measure emerging community structure~\cite{fortunato2010community}, to identify the effects of spatial embeddedness~\cite{barthelemy2011spatial}, and higher-order structural~\cite{yan2007graph,goyal2017graph} and temporal~\cite{karsai2011small,kivela2012multiscale,miritello2011dynamical,kovanen2011temporal,kivela2017mapping} correlations determining the emerging network structure, etc. All these efforts have brought us to a point where the science of complex networks has become advanced enough to help us to disclose the deeper roles of complexity and gain understanding about the behaviour of very complicated systems like global epidemic, transportation systems, the brain, or society.

Finally, we are able to tackle challenges, which were not addressable earlier due to lack of data, but now they are possible through the quantitative observations of the social behaviour of individuals and groups in global settings. These advancements called for the emergence of the new field of \emph{computational social science} (CSS)~\cite{lazer2009life} with the aim to develop the methodology for the quantitative description and modelling of social systems. Beyond topical relevance, any question of CSS translates to computational problems, which are challenging to the state-of-the-art of \emph{computer science} and opens possible new interesting directions in this field. Moreover, CSS is based on methodologies borrowed from conventional social sciences, cognitive and behavioural sciences, psychology, statistics, physics, computer science, and network science. Direct applications of methodologies from this broad set of disciplines were proofed to be successful at the outset, but at the same time they inherited concepts, which potentially mislead the description of the systems in focus. This way, for the quantitative description of human behaviour, the development of entirely new concepts were called for, leading to the emergence of domains like computational economics, human dynamics, social simulations, or computational linguistic, etc. Potentials of these advancements are not only to enhance data-driven reasoning in explaining social phenomena~\cite{vespignani2009predicting} but to fuel a paradigm shift to introduce social sciences as a more quantitative field.

\emph{Dynamical processes} like information or infection spreading, the adoption dynamics of innovations, memes, fads, or opinions can be effectively studied through the adoption of the network picture. Central question here is how the structure of interactions effects the critical behaviour and phase transitions, commonly characterising such phenomena~\cite{barrat2008dynamical}. The theoretical description and synthetic modelling of these systems have been earlier studied with tools borrowed from statistical physics, critical phenomena and computational modelling, while \emph{data-driven modelling}~\cite{rudas2017understanding,karsai2011small} and \emph{statistical learning}~\cite{foster2016big} are recent promising directions to bring predictive modelling of dynamical processes closer to real observations. It has been shown that several systems are crucially influenced by the heterogenous number and strength of interactions as well as by the time-varying nature, multiplexity, and structural and temporal correlations of them. Computational social science is naturally concerned with these problematics by quantitatively studying any collective social phenomena and in a broader sense dynamical processes, which evolves on social networks and is driven by information (e.g. influence, rumours, or memes) transmitted via social interactions.

\emph{Human dynamics}~\cite{skalka2009understanding}, on one hand, concentrates on individuals in terms of their actions, autonomous behaviour, mobility and migration, opinion formation, or spatio-temporal dynamics in geographic or mental spaces, etc. However, even at this level of description the ego\footnote{Note that naming a person an individual or an ego are deemed equivalent in this thesis, as they may be conventional in different terminologies, but they refer to the same, a person.} in focus is always embedded in a socio-economic environment, which cannot be ignored. On the other hand, human dynamics is concerned also with dyadic and egocentric interaction dynamics to unveil temporal patterns of interactions, mesoscopic group formation, inter and intra-community dynamics in social networks. On the system level, it addresses the dynamical emergence of the social network and any global human dynamical phenomena, like patterns of collective motion, emergence of collective movements, spreading of opinion, products or information, etc. This way, it can be identified as a sub-field of computational social science even its general focus is somewhat broader than to address sole social behaviour. It is also strongly entangled with network science, data science, computational science and other fields mentioned above.

Human dynamics as a field originates from direct observations of individual human behaviour~\cite{barabasi2005origin,gonzalez2008understanding}. In this thesis, I propose a broader interpretation of the field, by considering not only directly studying dynamical human behaviour, but also its consequences~\cite{karsai2018bursty}. This way, I include problems under this umbrella, which are conventionally associated to other fields, but their understanding is crucially depending on some aspects of human dynamics.

\subsection{Methodological challenges}

The methodological steps of research in human dynamics basically follow the conventional epistemological structure common in natural sciences, but with innovative, topic relevant techniques applied at each stages. Next, I give a short and certainly non-complete summary about the  methods, techniques and paradigms applied at each phases, to obtain generalisable and verifiable knowledge from data-driven observations of human behaviour.

\paragraph{Data collection}
Digital behavioural data has the advantage to record human behaviour in its own settings, in fine details, potentially accompanied with spatial, temporal and demographic details, for millions of individuals, without the common observational biases~\cite{chen2014big}. This way, it may provide statistically more significant, variant, and better generalisable observations about human behaviour. This is opposed to census data where individual details are aggregated, or to earlier designs of experimental social studies where observations were typically made in artificial settings involving a handful of participants. Digital behavioural and especially relational (network) data can be collected from various sources. \emph{Life-logs} and \emph{individual tracking}~\cite{eagle2006reality,jo2012spatiotemporal,stopczynski2014measuring} relies on recent technologies like mobile phones or radio-frequency IDs (RFID)~\cite{barrat2008high,lucet2012electronic} to follow people with their consent. Such projects provide temporally detailed records about a limited number of participants potentially including their whereabouts, communication (email, mobile call, short messages, online social networks,...), physical proximity, application usage, sleeping times, circadian patterns, just to mention a few examples. Other great data sources are provided by the automatically collected communication records for archiving or billing purposes, like \emph{emails}~\cite{eckmann2004entropy} or records of \emph{mobile phone communications}~\cite{blondel2015survey}. Such data usually record temporally detailed lists of dyadic communication actions of millions of anonymised users (e.g. customers of the same mobile provider), potentially with their actual location and length of interaction, but certainly without any information about the content of their communication. Typically they are accompanied with meta-data sets including demographic and socio-economic details about the users. The recent popularity of \emph{online social services} contributed enormously to this digital data endeavour. Services like Facebook, Google, Twitter, Spotify or gaming companies all record and store detailed personal and service usage informations about their customers and use them for development, marketing and advertisement purposes. At the same time, they share data via open APIs or data challenges, which in turn are extremely useful for external developers and scientists~\cite{weller2014twitter,ngai2015social}. Finally, \emph{online collaborative platforms}~\cite{kalliamvakou2014promises} and \emph{archiving services}~\cite{arxiv} record relational data about professional interactions like scientific collaborations or project developments. In addition, there are myriad other ways to collect and access digital data valuable for research, and this list will ever expand due to the emergence of new technologies and online services.

Some major disadvantages are also rooted in the data-driven approach to human behaviour. First of all, such digital datasets are more like field experiments, they are not collected in a controlled experimental setting and this way they are \emph{not reproducible} at the finest level of details. They are somewhat similar to astronomic observations, where despite recurrent patterns and regularities, the observation of the Universe is unique at any point in time. Possible solutions to these shortcomings are provided by the recent and more-and-more frequently used designs of online social experiments~\cite{reips2007methodology}, which are controllable, scalable, and relatively un-expensive. Depending on the problem setting, they may employ unique experimental designs~\cite{centola2010spread} or use online survey platforms like Amazon Mechanical Turk~\cite{mechanicalturk,buhrmester2011amazon} or SurveyMonkey~\cite{surveymonkey}. There are also some \emph{demographic biases} commonly characterising digital behavioural datasets. The digital data collection about individuals is conditional to the usage of digital devices (like mobile phones or computers) and communication services (email, online social networks, etc.). This way some demographic groups like elderly people, ones living at rural areas, or people with smaller income may appear under-represented in the data. However, due to the development of the telecommunication infrastructure, the increasing availability of smart devices, and the ever seen popularity of online services, these biases are vanishing even in developing countries, yet they are necessary to be noted.  Another disadvantage of the actual digital data practice has been raised on the ethical side. The automatic collection of \emph{highly sensitive personal data} and the potential tracking and identification of people without their consent have been identified as major concerns. Customers using free online services, unintentionally pay with their data, which in turn is used and re-sold for marketing and other purposes. Recent developments in data ethic and personal information treatment has been enforced by policy makers to put the rights to own data to the hand of the customers, this way closing the ethical gap between practice and privacy~\cite{eudatalaw}. However, this also indicates challenges in research and development, to design appropriate standards of data privacy and methods for data collection, which aggregate individual details on a necessary level to avoid potential re-identification~\cite{de2013unique}, but yet contains valuable information for scientific investigations.

\paragraph{Observation and analysis}

Hundreds years old methodologies provided by \emph{statistics} were developed for the analysis of conventional social and economic data, where the usual obstacle was the small sample size. Due to this limitation advanced statistical models, measures of variance and confidence were developed to be able to state something meaningful from small data and to identify biases~\cite{stevens2012applied}. These techniques were inherited to the contemporary methods of \emph{statistical data analysis}~\cite{ott2015introduction}, where the sample size is usually not a bottleneck anymore, but the heterogeneity of the observed quantities raises new challenges. To tackle large digital datasets, the new field of \emph{data mining}~\cite{witten2016data} grew out on the basis of \emph{statistics}, \emph{machine learning}, and \emph{data bases} to identify patterns from large datasets and transform them into understandable knowledge. The methodology of this field circles around data retrieval, pre-processing, transformation, data mining, and interpretation and aims to solve problems like anomaly detection, association learning, clustering and classification, regression and inference. 

\emph{Network analysis} provides other tools, which were found useful to learn about the dynamics of humans, embedded in a social, economical, geographical or abstract spaces~\cite{wasserman1994social}. Building on tools originated from \emph{graph theory} and \emph{statistical physics}, network analysis operates with measures characterising nodes and links, local and mesoscopic structures and the entire connected network to identify locally and globally important vertices and relationships, assortative correlations, communities, core-peripheral structures, and heterogeneous statistical characteristic of the structure, and many others~\cite{newman2010networks}. On the other hand, network analysis provides several ways to represent relational data as network structures with increasing level of complexity. Beyond the simple static, weighted and directed description of interactions, recently it grounded the methodology to take into account the evolution of networks, the time-varying nature of interactions~\cite{holme2012temporal}, the multiplexity of links associated with several types~\cite{kivela2014multilayer,boccaletti2014structure}, the spatial embeddedness of nodes~\cite{barthelemy2011spatial}, or several attributes assigned to nodes and links. In addition, \emph{randomised reference models}~\cite{karsai2011small,gauvin2018randomized} provide further analysis tools to identify significant structural and temporal patterns of interactions via the controlled shuffling of the temporal and structural informations encoded in the network.

\paragraph{Modelling}
\label{sec:modelling}

Modelling systems of human dynamics borrowing concepts from complex systems, machine learning, and computational social science, with the aim of better understanding and predicting human behaviour. One can identify three main branches in this endeavour:

\emph{Statistical models} are common in main stream empirical social science to detect statistically significant correlations between a variable describing a social phenomenon and a variable thought to explain it~\cite{holme2015mechanistic,wasserman1994social}. This approach is originated in advanced statistical methods, however recently, it applies technics borrowed from machine learning, data mining, and Bayesian learning. The main advantages of this approach are to account for individual differences to identify groups of egos with similar behaviour and to make statistical predictions even on the single ego level. Moreover, these methods allow the detection of causal inferences in data~\cite{friedman2001elements}, while crucially they do not address mechanistic processes driving human social behaviour. In this way they may help to build hypothesis of social behaviour but they have limited capacities in proof-of-concept and scenario-testing type of modelling.

\emph{Mechanistic models} are based on concepts borrowed from physics, where egos are identified as interacting entities with decisions driven by presumed mechanisms~\cite{holme2015mechanistic,hedstrom2010causal}. They typically model emergent behaviour whereby larger complex phenomena arise through interactions among smaller or simpler entities such that the emergent system exhibits properties the smaller entities do not. These generative processes are sometimes formalised into coupled dynamical differential equations with analytical or numerical solutions, and emulated in the realm of social simulations~\cite{gilbert2005simulation} as large-scale computer simulations where the stochastic nature of human behaviour is considered by probabilities of possible decisions at the individual level. School examples are generative network models~\cite{borge2011structural,karsai2014time} employing the minimal but sufficient set of microscopic mechanisms, e.g. preferential attachment, memory, or social reinforcement, which lead to some emerging property of the global network structure, like heterogeneous degrees or community structure. These models can consider social mechanisms influencing the decisions of egos, they intrinsically take into account causal relationships in reasoning, and provide ways to understand emerging phenomena in social systems~\cite{castellano2009statistical}. On the other hand, they usually assume that egos are identical and statistically equivalent entities (above the cognitive level) and can be characterised by their ``average'' behaviour. This approach has been proofed to be successful to simulate system level behaviour, and it provides an ideal test bed for hypothesis and scenario-testing. On the other hand, as it completely neglects individual heterogeneities, it has limitation for a fine-grained description of a social system.

\emph{Data-driven models} is a new branch in modelling social systems and co-evolving dynamical processes~\cite{tizzoni2012real,karsai2011small}. They integrate real-world data with synthetic models to concentrate on the effect of selected mechanisms, while keeping the rest of the system as close to reality as possible. Data can enter a synthetic model in several ways, e.g. by parameters determined from empirical systems~\cite{tizzoni2012real}, by replacing some parts or the entire system with real data and only model the effects of selected mechanisms~\cite{karsai2014complex}, or even by replacing real entities in an observed population by model agents to observe their simulated behaviour in a real setting~\cite{gautrais2009analyzing,calovi2015collective}. In addition, to better understand simulated dynamical processes evolving on real networks , the effects of structural and temporal correlations can be testified by using \emph{randomised reference models}~\cite{gauvin2018randomized,karsai2011small,kivela2012multiscale}.

Although these three modelling paradigms have relevance on their own, they provide complementary advantages for better understanding. To understand, model, and predict social systems, after observation it is crucial to ask the question ``why?''. Using mechanistic and data-driven models one can build and demonstrate hypothesis(es) by identifying the minimal set of causal mechanisms leading to the emergent behaviour in focus. This knowledge may provide relevant information to select features to train statistical models for prediction. Applying models alone from these three paradigms would leave us (a) with self-serving mechanistic models with multiple hypothesises explaining the same phenomena, and (b) with simple rules-of-thumb like black-box experiments using statistical models.

\paragraph{Verification and validation}

Verification and validation of models of human dynamics are difficult as social systems are far from being deterministic and as such they do not provide any 'ground-truth' (unlike systems in physics) for formal evaluation of modelling results~\cite{hahn2013conundrum}. Nevertheless, various techniques exist to verify results provided by models of one of the three approaches we discussed above. In case of statistical learning datasets are commonly divided for training and validation sets. After training a model on the former one, verification standards are defined via the \emph{sensitivity} (true positive) and \emph{specificity} (true negative) recalls from an independent validation set~\cite{ott2015introduction,witten2016data}. These measures are able to quantify the proportion of correctly recovered positive and negative attempts of the predicted features even on the individual level and can help to identify deterministic independent features for prediction. However, since they are simply considering the predicted output of a black-box experiment they are unable to proof dependencies beyond the phenomenological level. 

Verification of generative models is impossible on the level of single entities but on the level of the emergent phenomena. This can be done by directly comparing the solution of the model's formal description to its simulated dynamics or stationary state~\cite{barrat2008dynamical}, or by verifying the emergence of some expected characters on the system level~\cite{newman2010networks}. Data-driven models are typically compared to real world observations, whether directly or by feeding the modelled process with real parameters and compare predictions to empirical outcomes~\cite{tizzoni2012real}. Moreover, biases induced by complex mechanisms can be verified by data-driven simulations by comparing their outcome to corresponding simplified synthetic processes.

\paragraph{Applications}

Beyond scientific merit and understanding, human dynamics has far-reaching applications in several domains. First of all results of statistical modelling provides functional knowledge for predictions and inference of e.g. individual communications, mobility, or interactions patterns or online and offline behavioural dynamics. More concrete applications and algorithmic solutions of data mining results on human dynamics are usually disseminated on KDD conferences~\cite{kdd}. In terms of generative and data-driven modelling applications are also manifold. The prediction and scenario testing of crowd dynamics, cooperative behaviour, traffic congestion, public transportation, adoption of products and services, or the forecast of global pandemic and diffusion of information, memes, trends and opinion are only a few examples, which rely on data, observations, and modelling of human dynamical systems.

This way, human dynamics as a field has an un-questionable and increasingly relevant role in understanding human behaviour embedded in modern techno-social societies~\cite{vespignani2009predicting}. It contributes to the transformation of social sciences to become a more quantitative field, it pushes the frontier of computer science in terms of data science and computation, and it fuels the paradigm shift introduced by digital data in science, technology and several application domains.

\subsection{Positioning and outline}

As I explained in the introduction my research is very heterogeneous and builds on several disciplines. If I would need to summarise in short, I am a network scientist who is using methods from computer science, statistics, and physics to answer questions about emergent phenomena in general, and about dynamical human behaviour more specifically. My contributions include theoretical works, methodological contributions, data analysis and modelling studies to push forward the fields of network science, computer science, data science and computational social science.

\paragraph{Positioning on the scientific landscape}
To better situate my contributions on the general scientific landscape, I assign here their positions on the axes of various dimensions qualifying scientific works.

\vspace{.05in}

\emph{Fundamental vs. applied:} My personal scientific challenge has always been to understand causal dependencies in systems to see \emph{why} they behave in a way we observe them. This is opposed to the approach where the central question is to gain phenomenological understanding via correlations about \emph{what} systems do. While the two approaches are indeed not independent and complementary, without detecting the causal laws and mechanisms driving complex behaviour, we undermine our scientific understanding and turn scientific cognition into an engineering problem. This way, I judge the vast majority of my scientific works as contributions to fundamental sciences. This is moreover true as none of them aim to develop direct applications.

\vspace{.05in}

\emph{Interdisciplinary vs. disciplinary:} Disciplines in science has been provided a structural division between fields, which historically evolve parallel around well distinguishable fields. However, in modern science the seemingly rigid borders between disciplines are softening up as the combination of knowledge, the common mathematical background, and borrowed methodologies and technologies engage fields appearing independent at the outset. Even apparently independent systems e.g. from physics, biology, or sociology has been found to share common characters suggesting some universality and common underlying laws in their evolutions. Best examples are complex networks, which as a field builds on a single concept, yet disclose the similarity of systems in several domains. My education, which is built on computer science and physics as well my research, which combines knowledge from these fields to address dynamics on/of networks of various systems, are strongly interdisciplinary from the ground level.

\vspace{.05in}

\emph{Quantitative vs. qualitative:} Network science, which builds on the fundaments of graph theory, is a quantitative field in terms of its theory, observations, and modelling. Although it cannot be claimed to be an entirely exact domain, its ground concepts and basic models are provable exactly or approximately. Computational social science and human dynamics are also quantitative fields, although far being exact. They aim to develop quantitative approaches to problems in social sciences where qualitative reasoning has long lasting traditions. Most of my studies are going back and forth between these domains thus naturally they are all quantitative. 

\vspace{.05in}

\emph{Hypothesis vs. discovery driven:} Positioning my works along this axe is not as straightforward as in the cases above as I followed both approaches in different contributions. Some of my works are data-mining studies and as such they are entirely discovery driven. Some of them goes around the complete methodological circle, starting from discoveries via data-analysis, hypothesis building, modelling and verification. Finally, several of them are modelling works relying on some well established theories with the aim to extend, or starting from some pre-assumed hypothesis with the aim to demonstrate and to explore their consequences.

\paragraph{Relation to computer science}

My research is embedded in the field of computer science in several ways. My main expert field, \emph{network science}, strongly relies on the fundamental results of graph theory, which is an active field of theoretical computer science. Problems here are related to the theoretical foundation and algorithmic solutions of the representation, metrology and navigability of very large and sparse networks, which may be weighted, directed, signed, coloured, multilayer, temporal, spatial, hyperbolic, featured, or exhibit other properties. On the application side, network science contributes to the design of technological, infrastructure and virtual systems, which are also in the focus of computer science and telecommunication. Infrastructure systems like the Internet, mobile communication networks, ad-hoc networks, sensor networks, p2p networks, and the internet of things (at home, in buildings, in cities), or virtual systems like the world wide web, online social systems, communication systems like emailing or blog spheres are all examples of applied computer science which benefits from the theoretical and applied results of network science.

My research is also related to \emph{data science}, a sub-field of computer science, which deals with the collection, storage, and analysis of enormous digital datasets commonly recording informations about millions of individuals. Topics like data management, data representation, data analysis, data mining, statistical learning via deep and shallow neural structures, personal data protection, data encryption are all topics, which are central both in data science and computer science.

Finally, a large portion of my research lands within the field of \emph{computational science}. I have several works to develop generative models of interacting many-body systems, which are based on stochastic computational modelling using Monte Carlo techniques, and commonly implemented as large-scale computational simulations built on well optimised distributed algorithms and executed on high performance computational systems.






\paragraph{Outline of the thesis}
Although my works are at the intersection of various domains they all address emergent behaviour in systems of human dynamics. Most of them can contribute to three main topics, which also defines the overall structure of this thesis.

After this short introduction, in the remaining of the first Chapter I am going to introduce the general definitions and concepts in complex networks, together with some theories and general properties of social networks, which will be used throughout the whole thesis. After that I give short summaries of various datasets what I used during the last years, and which are recurrent in several studies discussed later. In Chapter~\ref{ch:burst}, I discuss bursty dynamical systems, which is the first main topic of my contributions. This Chapter builds on a set of  studies, in which I proposed novel methodologies to characterise and model real-world heterogeneous temporal behaviour of individuals and once they are connected in a network structure. The second main topic of my research is discussed in Chapter~\ref{ch:tnet}. There I summarise my theoretical and modelling contributions to the foundation of temporal networks. These works circle around new theoretical methods to interpret and characterise temporal networks at the micro-, meso-, and macroscopic scales. In addition, I propose the observations of various mechanisms and two modelling frameworks to explain the emergent properties of temporal structures. 
In Chapter~\ref{ch:dynpr} I address my works on dynamical processes evolving on static or temporal networks. These studies investigate the structural and temporal correlations, which control emergent collective phenomena like spreading processes. I will address simple contagion models on real and modelled temporal networks, and complex contagion processes, especially the empirical observation and modelling of social spreading phenomena. Finally, in Chapter~\ref{ch:disc} I will summarise my contributions and will draw my motivations for the future.














\section{General concepts and terminology}

In this Section, I briefly introduce some general definitions and concepts, which will be used throughout the whole thesis. All the definitions are related to complex networks, first introducing general characteristics, while second some specific concepts which concerns social networks. This list of definitions, however, is far from being complete to give a thorough introduction to the field, while some specific definitions will be introduced at the related Sections in the following Chapters.

\subsection{Complex networks}
\label{sec:cn}

Networks provide a way to map the architecture of complex systems by identifying interactions between their interconnected components. A network is commonly interpreted as a graph defined by a set $G=(V,E)$. Here $V$ defines the set of $v\in V$ nodes representing components, and $E \subseteq V \times V$ is the set of $e_{u,v}=(u,v)\in E$ links, which indicate if two components $u,v\in V$ are interacting in the system\footnote{Note that several naming conventions of the same objects are accepted in different disciplines. Networks are also called structures or simply graphs, nodes are equivalently named as sites, components or vertices, while links can be called as edges or ties. In this thesis I treat equivalent the corresponding names.}. The cardinality of these two sets assign the size $N=|V|$ and the number of links $M=|E|$ of the network. In their first approximation networks are defined as simple graphs, thus they are undirected (do not distinguish between links $(u,v)$ and $(v,u)$), while assume no self-loops $(v,v)$, or multiple connections between the same pairs of nodes. On the other hand several extension of this representation is possible. For example, if we assign an orientation to each link, i.e., we distinguish between $(u,v)$ and $(v,u)$, we arrive to the definition of \emph{directed networks}, while \emph{weighted networks}, defined as a triplet $G=(V,E,w)$, assume a weight function $w: (u,v) \rightarrow \mathbb{R}$, associating a number (or another quality) to each link to capture, e.g., the interaction strength between connected entities.

Networks can be characterised on various topological scales. To describe them on the local scale, i.e., at the level of nodes and links, we consider the $G_v=(V_v,E_v)$ \emph{egocentric network} of each node $v$, which includes the central node, its first neighbours, and all links among them. At this level the most important character of a network is the \emph{node degree}. The degree $k_v$ of a node $v\in V$ is the number of links it is incident to, or in other words its number of neighbours\footnote{Note that although $k_v$ is a specific character of any node in $V$, later we may use the simpler notation $k$, which assigns degree in general of a node in the network. Similar convention will be used for any other node property.}. Formally it can be defined as $k_v=\sum_{u\in V \setminus \{ v\}}\delta(u,v)$, where $k_v\in [0,N-1]$ and $\delta(u,v)=1$ if $(u,v) \in E$ and $0$ otherwise. Similarly, we can define the $k_{in}$ in-degree and $k_{out}$ out-degree of a node in directed networks, counting the number of its incoming and outgoing links. In case of weighted networks, the equivalent quantity is called the weighted degree or \emph{node strength}, defined as $s_v=\sum_{u\in V \setminus \{ v\}}w(u,v)$, i.e. the sum of weights of adjacent links to a node $v$. Another important node characteristic is the \emph{local clustering coefficient}. It quantifies the local connectedness via counting the fraction of closed triangles (i.e., three nodes connected by three links) in the egocentric network of a node $v$. Formally it is defined as $C_v=\frac{2|\{ e_{v,w} : v,w \in V_v, e_{v,w} \in E\}|}{k_v(k_v-1)}$, where the denominator $(k_v(k_v-1))/2$ counts the maximum number of connections between the neighbours of a node with degree $k_v$. The local clustering coefficient is conventionally used in social network research~\cite{watts1998collective} where its network average value, $C=\sum_{v\in V} C_v/N$, characterises the average local connectedness of the structure. A similar quantity can be defined for links, called the \emph{link overlap}~\cite{granovetter1973strength,onnela2007structure}, which measures the fraction of common neighbours of connected nodes. It is formally introduced as $O_{u,v}=\frac{n_{u,v}}{(k_u-1)+(k_v-1)-n_{u,v}}$, where $n_{u,v}=|\{V_v \cap V_u\}|$ is the number of common neighbours of nodes $u$ and $v$.

On the global scale, networks are commonly characterised by the distributions and statistical means of different local quantities. In the following we define some of them for simple networks (undirected with no loops and multiple links), but all measures can be generalised for more complicated representations. The simplest global measure is \emph{network density}, $\rho=\frac{2|E|}{N(N-1)}$, which is the fraction of present links as compared to the possible number of links in the network. A more informative measure is the probability density function of node degrees, also called the \emph{degree distribution}, defined as $P(k)=\tfrac{1}{N}|\{ v\in V, k_v=k\}|$, and its mean $\langle k \rangle=\tfrac{1}{N}\sum_{v\in V} k_v$ called the \emph{average degree}. This distribution determines the probability that a randomly selected node in the network has degree $k$, while its mean gives the average number of neighbours of a node. In real-world networks, $P(k)$ usually appears as a broad distribution with a long tail ranging over several orders of magnitude~\cite{albert2002statistical}, indicating present high degree heterogeneities. Such distribution are commonly approximated with a power-law distribution scaling as $P(k)\sim k^{-\gamma}$, with a characteristic exponent $\gamma$. Although this approximation has been proven to be useful, the $P(k)$ of several real-world networks may be better fitted by other broad statistical distributions, like lognormal or stretched exponential, as it is summarised in~\cite{clauset2009power}. Note that several other characters of real world networks are distributed heterogeneously, like link weights or node strengths, and can be similarly characterised. \emph{Global clustering coefficient} is another global scale measure, which quantifies the fraction of closed triangles on the network level. It is defined as $C=\frac{3\times |Triangles|}{|Triplets|}$, where triplets are subgraphs of three nodes connected by two links\footnote{I do not distinguish in notation between the average local clustering coefficient and the global clustering coefficient. As default, in this thesis $C$ will refer to the former one if not noted otherwise.}. Note that the weighted average of the local clustering coefficient is equivalent to the value of the global clustering coefficient.

\emph{Paths} in a network are very important as they determine the possible communication routes between nodes and the global connectedness of the structure. Two nodes, $u$ and $v$, in a network are connected by a path it there exist a sequence of links, $(u,a),(a,b),\dots,(x,y),(y,v)$, which starts from $u$, ends at $v$, and each consecutive links share one common node. The \emph{length of a path} is the number of links in the corresponding sequence, while the \emph{distance} $d(u,v)$ between two nodes is given by the length of the shortest path, i.e, the shortest sequence of links which form a path between them. Shortest paths can be used to define sensitive measures of centrality and overall connectedness of a network. First of all, the \emph{average distance}, $\langle d \rangle=\tfrac{1}{N(N-1)}\sum_{u,v\in V}d(u,v)$, captures the average number of steps required to pass information between any pairs of nodes. Another measure is the \emph{network diameter}, defined as $\max_{u,v\in V} d(u,v)$, gives the maximum of any distances between any pairs of nodes. \emph{Betweenness centrality} is a path based measure to identify central nodes in the network, which lays on several shortest paths thus keeping the structure connected, or controlling effectively information diffusing on the network. It is defined as $BC(v)=\sum_{v\neq s \neq v in V}\frac{\sigma_{u,v}(s)}{\sigma_{u,v}}$, where $\sigma_{u,v}$ is the number of shortest paths between nodes $u$ and $v$, while $\sigma_{u,v}(s)$ is the number of those which run through node $s$. Existing paths also determine the connected components of a network. A connected component contains a set of nodes, which are all available from each other via paths in the network. The largest of them, called the \emph{giant connected component} (GCC) or the \emph{largest connected component} (LCC), plays a special role as its size assigns the overall connectedness of the network and determines the maximum size of any macroscopic phenomena emergent in the system. The size of the \emph{second largest component} (or the average size of components other than LCC) is also important in some cases. By varying the number of links in the network, its connectedness can be interpreted as a \emph{percolation process}~\cite{aharony2003introduction} with a phase transition point, where LCC vanishes and the size of the second LCC shows a singularity (in the thermodynamical limit).


\subsection{Social networks}
\label{sec:socialnets}

As most of the works summarised in this thesis concentrate on social systems, it is useful to introduce a non exhaustive list of some commonly accepted concepts and general characters of social networks to help the reader. In social networks nodes are associated to individuals, while edges represent social ties between them. Social networks are inherently \emph{dynamical} on multiple temporal scales, as people enter and leave the system and create and break social relationships, and maintain them via rapid communication events. Social ties can be manifold by distinguishing between family, friend, intimate, professional or service types of relationships. Such variety of interactions can be represented via \emph{multilayer} structures with layers associated to social ties of different types. Social networks are also \emph{embedded} in geographical space due to the whereabouts of people, but some networks are better represented in a \emph{virtual space} (like online) with abstract metrics (like similarity in opinion or interest) defining alternative distance measures between egos.

One of the most important questions in social networks concerns social tie creation. The propensity of two people to become acquaintances or friends may depend on several factors such as co-habitation, common interest, socioeconomic status, age, gender, education level, or occupation, etc. It has been argued in social theory that people who are more similar to each other tend to create new relationships with a higher probability~\cite{mcpherson2001birds}. This selection mechanism~\cite{barash2011dynamics}, called in general \emph{homophily}, has been observed in several studies~\cite{centola2011experimental,aral2009distinguishing,leo2016socioeconomic} and has far-reaching consequences in the emergent network structure. The effects of homophily, however, are hardly distinguishable from the consequences of another process called \emph{influence}. This is a force mechanism where connected people become more similar to each other via social or interpersonal influence~\cite{barash2011dynamics}. This mechanism is arguably behind several macroscopic social phenomena like the spreading of information and memes~\cite{bakshy2012role,weng2012competition}, innovations and services~\cite{karsai2014complex,aral2009distinguishing} or the emergence of collective social movements~\cite{borge2011structural,granovetter1978threshold}. In some cases it is difficult to decide from simple data-driven observations, whether similar people are connected due to homophilic preferences, or became similar via social influence after building the relationship. This question sets one of the most important challenges of computational social science as, depending on the dominant mechanism, competing hypothesis has been set to explain social diffusion processes~\cite{aral2009distinguishing} (as we will discuss later in Section~\ref{sec:ccpoi} and~\ref{sec:ccpmodels}).

Along homophilic preferences, a mechanism called \emph{triadic closure} has been proposed~\cite{simmel19921908,rapoport1953spread} to be a determinant force behind link creation and the emergence of \emph{community structure} in social networks. Triadic closure is a property among three individuals such that if we observe already two ties between them, with a high probability there is or going to be a third tie closing the open triad to become a triangle. This property induces strongly connected local structures built up from several triangles among people usually sharing some common property. Communities in general are connected subgraphs of a network with nodes, which are denser connected with other nodes of their own community than to the rest of the network. Communities can be mutually exclusive partitions, or overlapping sets of nodes, they can be static or dynamic, uni-layer or multilayer, etc. Their detection is an ongoing challenge with several methods and models proposed~\cite{fortunato2010community}.

The \emph{strength of a social tie} is a linear combination of the amount of time, emotional intensity, intimacy, and reciprocal services, which characterise a social relationship. The strength of social ties are commonly assumed whether being weak or strong and to be strongly correlated with the network position of the actual tie, as argued by Granovetter~\cite{granovetter1973strength,granovetter1983strength}. In his seminal paper~\cite{granovetter1973strength} he suggests that \emph{weak ties} are maintained via sparse interactions, bridging between tightly connected communities keeping the network connected~\cite{onnela2007structure}, and play important roles in disseminating information globally~\cite{centola2007complex,centola2010spread}. On the other hand \emph{strong ties}, sustained by frequent communications, are crucial in shaping the local connectivity of social networks. Due to triadic closure mechanisms they are responsible for the emerging clustered topology~\cite{kossinets2006empirical,kumpula2007emergence}, and they exert to keep information locally~\cite{karsai2011small,miritello2011dynamical}.

However, even differentiating ties by strength one cannot simultaneously maintain a large number of social relationships in a meaningful way. This has been argued by Dunbar who proposed the \emph{social brain hypothesis} based on an anthropological correlation between the social group size and neocortex size of primates~\cite{dunbar1998social}. He argues that due to cognitive limitations of people, the size of the egocentric network of an individual is also limited to be around $150$. Although one could assume that modern communication services and social platforms have helped to overcome this constrain called the \emph{Dunbar's Number}, recent studies have observed a similar limit in number of online friends, which after one cannot commit equally to every acquaintances but needs to share attention~\cite{gonccalves2011modeling}. Dunbar further developed his theory by introducing \emph{intimacy circles} by categorising acquaintances by their social tie strength, intimacy and importance to obtain a finer grained structure of one's egocentric network~\cite{dunbar2008cognitive, palchykov2012sex}.




\section{Datasets}
\label{sec:datasets}

The central block of my research concerns the preparation and analysis of longitudinal datasets recording the actions and interactions of large number of individuals. Most of these datasets are temporally detailed, coupled with geographical informations and demographic details about the anonymised participants. In the following Section I give short descriptions of some important large datasets, which I used recurrently in several studies to make various observations which will be discussed later during the course of the thesis.  Note that all unique IDs in any of the following datasets were anonymised in advance by the providers without the involvement of the researchers, thus individual re-identification of any participants were impossible from the data.

\subsection{Data representation}

Datasets on \emph{individual action dynamics} usually come in a format of action sequences represented in general as a set of events $AC \subset T \times V (\times \prod_i A_i^{ac} \times LOC)$ where $T$ assigns the set of discrete time stamps bounded by the observed period and $V$ is the set of individuals. Usually this tuple is extended by $A_i^{ac}$ sets of different attributes $a_i\in A_i^{ac}$ (like type of action, duration, cost, etc.) of the actual activity, and some $LOC$ location information. This way, e.g., one activity record $ac(t, v, a_1, a_2, \dots, loc_{v})$ indicates that at time $t\in T$ the user $v\in V$ were active in doing $a_1\in A_1^{ac}$ for $a_2\in A_2^{ac}$ duration at $loc_v\in LOC$. Such datasets commonly arrive with some accompanying meta-data sets in a general form of $A_j^v \subset \prod_j A_j^v$ describing the attributes $j$ of each user in $v\in V$ regarding different features $A_j^v$, such as age, gender, home location, active period, etc.

\emph{Interaction sequences} record temporal networks~\cite{holme2012temporal,latapy2017stream} and usually come in an event sequence format defined as $EV\subset T\times V \times V ( \times \prod_i A_i^{ev} \times LOC)$. This representation is very similar to action sequences but it describes the interaction between $(u,v)\in V\times V$ ordered (unordered) pairs of entities performing a directed (undirected) interactions. E.g., in case of mobile call communication events, which are typically called Call Detail Records (CDRs), events are like $ev(t,v_{caller}, v_{callee}, duration, cost, loc_{caller})$ and come as long sequences describing all interactions of every users of a mobile phone provider over several months with usually one-second resolution. Using an event sequence and the corresponding meta-data set one can actually define a temporal network (see Chapter~\ref{ch:tnet}), where directed interactions can be followed as function of time and space. In addition, accompanying meta-data of users can be defined as a matching feature vector, which can take more complicated format as explained above, or can be even dynamical, as we will see later. This interpretation is useful to follow the recurrent communication patterns of egos, and the formation of their social ties as a function of time, space, and personal attributes.

Note that several datasets, dynamical or not, could be used to map the \emph{static network structure} of social interactions. I commonly used the standard adjacency list representation~\cite{newman2010networks} of such large networks, or streaming algorithms to process edge list representation of relational data.

\subsection{Datasets in hand}
\label{sec:datasets}



\paragraph{DS1 - Mobile phone communication data}
The datasets I investigated to most in several studies~\cite{jo2012circadian,karsai2011small,karsai2012correlated,karsai2014time,kivela2012multiscale,kivela2017mapping,kovanen2011temporal,laurent2015calls,liu2014controlling,tibely2011communities,ubaldi2016asymptotic,ubaldi2017burstiness,unicomb2018threshold,zhao2011entropy} record sequences of mobile-phone calls (MPC) and short messages (SMS) of a large set of individuals. These datasets were recorded by a single operator with 20\% market share in an undisclosed European country (ethic statement was issued by the Northeastern University Institutional Review Board). I analysed several variants of these datasets, which typically contained $\sim 633$ million time stamped phone call and $\sim 209$ million SMS events recorded during 182 days with 1 second resolution between $6,2$ million ($4,8$ million for SMS) individuals who were connected via $16,8$ million ($10,3$ million for SMS) social ties. In order to take into account only true social interactions and avoid commercial communication, we used only actions which were executed on links between users, who are at least once mutually connected each other.

\paragraph{DS2 - Mobile communication/Credit record data}

Another mobile phone data I extensively studied over the last years~\cite{liao2017prepaid,leo2016correlations,leo2016socioeconomic,leo2018correlations} record the temporal sequence of $\sim 8$ billion call and SMS CDRs between $\sim 112$ million anonymised mobile phone users for $21$ months in a Latin American country. The dataset was collected by a single operator but other mobile phone users called by the customers of the provider, who were not clients of the actual provider, also appear in the dataset with unique anonymised IDs. The initially constructed static social network contained all users (whether clients or not of the actual provider), while links were drawn between them if they interacted (via call or SMS) at least once during the observation period. In order to filter out call services and other non-human actors from the network, we recursively removed all nodes (and connected links) who appeared with either in-degree $k_{in}=0$ or out-degree $k_{out}=0$. We repeated this procedure recursively until we received a network where each user had $k_{in}, k_{out} > 0$, i.e. made at least one outgoing and received at least one incoming communication events during the nearly 2 years of observation. After construction and filtering the network remained with $\sim 82$ million users connected by $\sim 1$ billion links.

This mobile call dataset was coupled with another data provided by a Bank from the same country. This data records financial details of $\sim 6$ million people assigned with unique anonymised identifiers over 8 months. Data records the time varying customer variables as the amount of debit card purchases, purchase categories of purchased goods, monthly personal loans, and static user attributes such as billing postal codes (zip code), the age and gender of customers. A subset of IDs of the anonymised bank and mobile phone customers were matched\footnote{Note, that the matching, data hashing, and anonymisation procedure was carried out through direct communication between the two providers (bank and mobile provider) without the involvement of the scientific partner.}. This way of combining the datasets allowed us to simultaneously observe the social structure and estimate economic status (for definition see Section ~\ref{sec:esSCL}) of the connected individuals. The combined dataset contained $\sim 1$ million people, all of them assigned with communication events and detailed bank records, connected by $\sim 2$ million links into a single giant component.

\paragraph{DS3 - Twitter data}

In a project we collected a large data corpus from the online news and social networking service, Twitter~\cite{hours2016link,abitbol2018socioeconomic,abitbol2018location}. There, users can post and interact with messages, "tweets", restricted to 140 characters. Tweets may come with several types of metadata including information about the author's profile, the detected language, where and when the tweet was posted, mentions of other users (denoted by the \texttt{@} symbol) and hashtags (denoted by the \texttt{\#} symbol) to assign topics, etc. Specifically, we recorded $170$ million tweets written in French, posted by $2.5$ million users in the timezones GMT and GMT+1 over three years (between July 2014 to May 2017). These tweets were obtained via the Twitter powertrack API feeds provided by Datasift and Gnip with an access rate varying between $15-25\%$. We used this dataset to obtain personal \emph{linguistic data} from the written text of each post and to infer the \emph{social network structure} between users. Tweet messages may be direct interactions between users via mentions by using the symbol (\texttt{@username}). We took direct mutual mentions as proxies of social interactions and used them to identify social ties between pairs of users. This constraint lead to a structure of $500,000$ users and $2$ million undirected links. In addition, about $2\%$ of tweets contained some \emph{location information} regarding either the tweet author's self-provided position or the place from which the tweet was posted. We extensively used this information to combine the Twitter datasets with census informations (see Section ~\ref{sec:TwSESLang} and~\ref{sec:TwSESinf}). Note that in some other studies~\cite{tizzoni2015scaling,ubaldi2017burstiness} I participated, we reported results based on the analysis of another, yet very similar, Twitter dataset.

\paragraph{DS4 - Skype service adoption data}

In a couple of projects I had the chance to access and analyse the social network of one of the  largest online communication services at the time, the Skype network. The centrepiece of this dataset is the contact network, where nodes represent users and edges assign confirmed Skype connections. A user’s contact list is composed of friends of mutually confirmed relationship assigned with two time stamps indicating the moment the contact request was approved, and deleted if it happened before the end of the observation period. In some studies we considered the temporal evolution of this network~\cite{kikas2013bursty}, while in others~\cite{karsai2014complex,karsai2016local} we took its static representation, aggregated for $99$ months between September 2003 and November 2011. The largest connected component of this structure includes roughly $510$ million users and $4.4$ billion edges. In addition, this dataset comes with time stamped purchase records of different Skype services. To study social contagion processes, we followed how users purchase “credits” for calling phones. For each user, the dataset includes the date when a user first adopted the paid product “buy credit” (first credit purchase, for all purposes). We selected this service since its lifetime of $89$ months is considerably long (it was introduced in 2004), and it can be adopted by registered Skype users only. This way the aggregated Skype network provided a complete description of the mediating social structure, which allowed us to calculate the correct network and dynamical properties. To make additional observations, we performed calculations on a second paid service called “subscription”, which was introduced in April 2008, lasts for over 42 months, and could also be adopted by registered Skype users only.



\paragraph{Other datasets}


I addition there are several other datasets which I analysed during my research to make data-driven observations, design data-driven models or to train and verify model predictions. I used various social interaction datasets recording emails~\cite{karsai2011small,karsai2011universal}, Facebook wall posts~\cite{mocanu2015collective}, face-to-face interactions~\cite{zhao2011entropy}, or sexual contacts~\cite{kivela2017mapping}; individual tracking data of communication, mobility and service usage records~\cite{karsai2011small,jo2012spatiotemporal}; collaboration datasets as scientific co-publication~\cite{ubaldi2016asymptotic,ubaldi2017burstiness} or co-citations~\cite{jensen2015detecting}, business alliances~\cite{tomasello2014role}, or the election of wikipedia editors~\cite{unicomb2018threshold}; transportation datasets as air-traffic networks~\cite{kivela2017mapping,jensen2015detecting} or public transportation systems in large cities~\cite{alessandretti2016user}; or even activity dynamics of firing neurons or earthquakes~\cite{karsai2011universal}. All these datasets provided the solid ``observational'' ground for my studies, which will be summarised in the following three Chapters.

\biblio









\chapter{Bursty Human Dynamics}
\label{ch:burst}

\section{Introduction}
\label{sec:brst_intro}
Bursty behaviour is a temporal character of some dynamical systems, which alternate between active periods with high frequency of events and long periods of inactivity. Dynamics characterised by such large temporal fluctuations cannot be explained by the conventional picture using Poisson processes assuming a single temporal scale, but rather can be the result of non-Poissonian dynamics inducing strong temporal heterogeneities on various temporal scales\footnote{Non-Poissonian bursty dynamics is in general characterised by the heterogeneous distribution of inter-event times passing between the consecutive occurrences of a given type of event. In contrast, in a system with Poissonian dynamics, inter-event times are distributed exponentially. However, many empirical inter-event time distributions are broad and follow a log-normal, Weibull, or power-law form, implying that the underlying mechanisms behind them maybe different than a Poisson process.}.

\begin{figure}[!ht]
\centering
  \includegraphics[width=.6\textwidth]{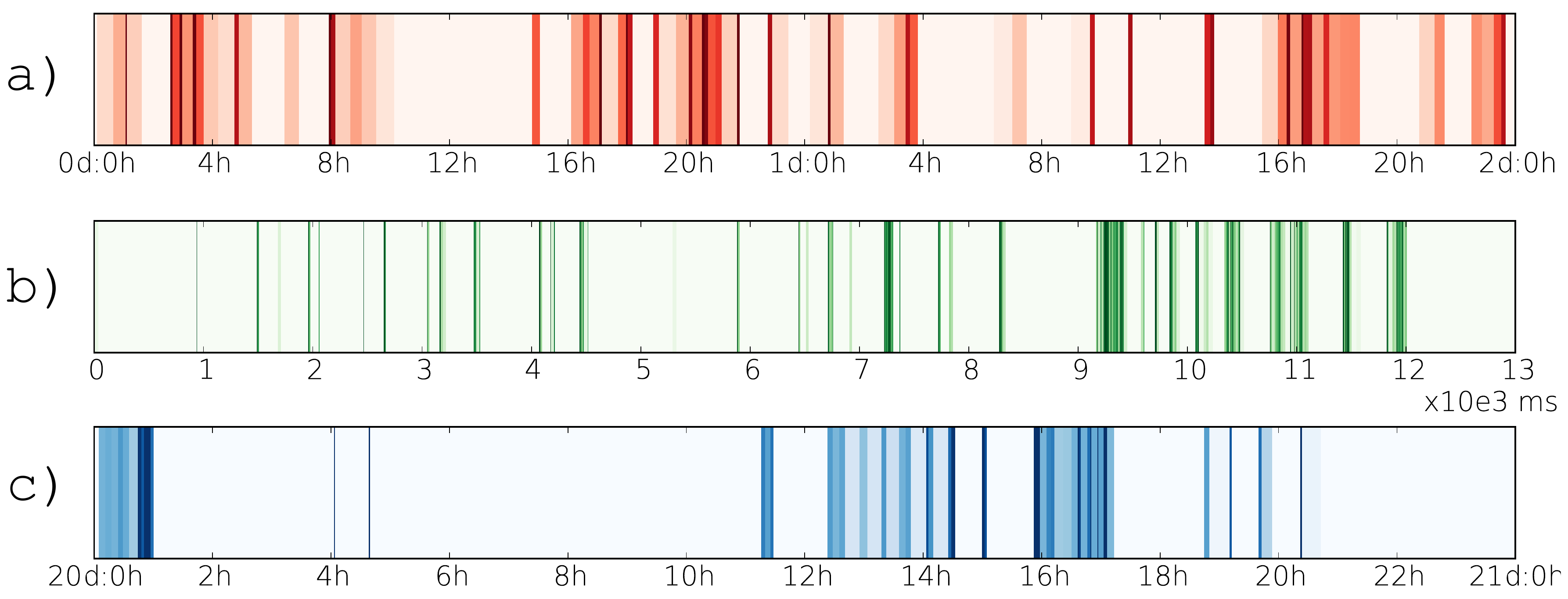}
\caption{\small (a) Sequence of earthquakes with magnitude larger than two at a single location (south of Chishima Island, 8th--9th October 1994). (b) Firing sequence of a single neuron from a rat's hippocampal. (c) Outgoing mobile phone call sequence of an individual. The shorter the time between the consecutive events are, the darker colour is coded. This figure was published in~\cite{karsai2011universal}. 
}
\label{fig:BurstySignals}
\end{figure}

There are a number of systems in Nature that evolve following non-Poissonian dynamics~\cite{karsai2018bursty}. One example are earthquakes~\cite{corral2004longTerm, davidsen2013earthquake,bak2002unified, deArcangelis2006universality,smalley1987a}, in which the times of shocks occurring at a given location show bursty temporal patterns, as illustrated  in Fig.~\ref{fig:BurstySignals}(a). Another example are solar flares with bursty emergence induced by huge and rapid releases of energy~\cite{mcAteer2007the,wheatland1998the}. It has been shown that the stochastic processes underlying these apparently different phenomena show such universal properties that lead to the same distributions of event sizes, inter-event times, and temporal clustering~\cite{deArcangelis2006universality}, which can be arguably modelled by the frame of self-organised criticality (SOC)~\cite{bak1996how}. Burstiness is also seen in the contexts of neuronal systems where the firing of a single neuron (as shown in Fig.~\ref{fig:BurstySignals}(b)) or collective of neurons appears with such dynamics~\cite{kemuriyama2010power}. Moreover, bursty patterns has been observed in ecology and animal dynamics in the context of initiating conflicts \cite{proekt2012scale}, communication, foraging~\cite{sorribes2011origin}, predators waiting in ambush~\cite{wearmouth2014scaling}, or the displacement of monkeys or mice~\cite{boyer2012nonrandom,nakamura2008of}, which form complex self-similar temporal patterns reproduced on multiple time scales very similarly to examples observed in human behaviour. In addition, scale-invariant bursty temporal patterns have also been found in several man-made systems, like written text, in which successive occurrences of the same word display bursty patterns~\cite{altmann2009beyond}. In case of engineering systems perhaps some of the best examples are in the context of package-based traffic and wireless communication signals, which were found to evolve through non-Poissonian dynamics~\cite{chlebus1995is,janevski2003traffic,lee2013mobile,paxson1995wideArea}. Finally, financial markets, in which non-Poissonian dynamics characterises time series of returns of financial assets, stock sales, order books, and other transactions with dynamics studied in the realm of econophysics~\cite{mantegna2007introduction}.

Bursty patterns have been found to characterise human dynamics in the timings of actions, dyadic social interactions, or even in collective social phenomena. 
The first observations were reported by Eckmann~\emph{et al.}~\cite{eckmann2004entropy} and by Barab\'asi~\cite{barabasi2005origin}, who observed broad inter-event time distributions with a power-law tail by analysing datasets of email correspondence. These seminal papers initiated an avalanche of studies to observe, characterise, and model bursty phenomena detected in a number of human activities. Various examples were found, like emails~\cite{eckmann2004entropy,barabasi2005origin}, letter correspondence~\cite{oliveira2005human}, mobile phone calls and short messages~\cite{karsai2011universal} (like in Fig.~\ref{fig:BurstySignals}(c)), web browsing~\cite{dezso2006dynamics}, printing~\cite{harder2006correlated}, library loans~\cite{vazquez2006modeling}, job submission to computers~\cite{kleban2003hierarchical}, and file transfer in computer network~\cite{paxson1995wideArea}, or even in arm movements of human subjects~\cite{coley2008arm}, just to mention a few. In addition, further examples were identified at the group or societal level, such as the emergence of causal temporal motifs~\cite{kovanen2013temporal}, the evolution of mass demonstrations, revolutions, global information cascades, and wars~\cite{bouchaud2013crises,tang2010stretched}.

All these new observations highlighted some shortcomings of earlier methods to characterise human bursty dynamics and called for novel measures and models to gain deeper understanding about the roots of bursty patterns in human behaviour. Several modelling frameworks of bursty human dynamics have been proposed over the last years, which could be roughly classified into three main groups based on the assumed underlying explanatory mechanisms. In his original study, Barab\'asi suggested that bursty activity patterns could be the consequence~\cite{barabasi2005origin,oliveira2005human,vazquez2006modeling} that people do not execute their ``to-dos'' in a random fashion but assign importance to each task at hand. This induces intrinsic correlations between different tasks and results in bursty patterns of completed activities, which can be effectively modelled by \emph{priority queues} with different constraints. Another direction was proposed by Malmgren~\emph{et al.}~\cite{malmgren2008poissonian,malmgren2009universality} who argued that ``human behaviour is primarily driven by external factors such as circadian and weekly cycles, which introduces a set of distinct characteristic time scales, thereby giving rise to heavy tails''. This approach assumes no intrinsic correlations in human activities but models the dynamics as alternating homogeneous and non-homogeneous \emph{Poisson processes}. The third main modelling concept assumes strong correlations between consecutive events and define \emph{non-Markovian dynamics} by using memory functions~\cite{vazquez2006impact, han2008modeling}, self-exciting point processes~\cite{masuda2013selfexciting,jo2015correlated}, or reinforcement mechanisms~\cite{karsai2012correlated,wang2014modeling} in simulating bursty activity patterns. Finally, several other modelling ideas were suggested assuming self-organised criticality~\cite{tang2010stretched}, local structural correlations~\cite{myers2014bursty}, some dynamical process like random walk~\cite{goetz2009modeling}, contact process~\cite{odor2014slow}, or voter model~\cite{fernandezgracia2013timing} to introduce heterogeneous temporal patterns at the individual or system levels.

Based on these advancements more far-reaching scientific questions have been addressed about the effects of non-Poissonian patterns of individuals on collective dynamical processes. First question raised, whether they are ongoing or co-evolving with bursty actions and interactions. An important example is diffusion processes on temporal networks where bursty dyadic interactions may enhance or slow down the speed and/or control the emergence of globally spreading process, like information diffusion, epidemics, or random walk~\cite{holme2012temporal}. Beyond the conventional modelling and simulation techniques of such processes, data-driven models and random reference systems~\cite{karsai2011small,miritello2011dynamical} were recently shown to be very successful in addressing such problematics.

I entered this field during my first postdoctoral period at Aalto University and worked with several colleagues on various topics to observe, measure, and model bursty human behaviour, and to better understand its consequences on the evolution of dynamical processes. On the methodological level I had two main contributions: an entirely new measure to detect bursty temporal correlations in heterogeneous signals~\cite{karsai2011universal}, and a method to account for the effects of circadian fluctuations to identify to what extent they are responsible for the emergence of bursty patterns in human dynamics~\cite{jo2012circadian}. Using these techniques I conducted data-analysis studies~\cite{kikas2013bursty,karsai2012correlated} to observe and characterise bursty social link creation and maintenance dynamics. I also developed various models using reinforcement mechanisms to explain individual and dyadic correlated bursty behaviour~\cite{karsai2011universal,karsai2012correlated}. We were among the firsts to use random reference models to identify the effects of burstiness on spreading processes~\cite{karsai2011small,kivela2012multiscale} and to develop a generative temporal network model with bursty interactions~\cite{ubaldi2017burstiness}. In addition, together with collaborators I recently published a monograph book~\cite{karsai2018bursty} to review the knowledge accumulated during the last ten years in this domain.

In this Chapter, without aiming a complete overview of the field, I give a brief summary of my main contributions to the area. After this introduction, I lay down some general concepts and measures which I will rely on later during the Chapter. Then, I organise my description by first introducing my methodological contributions, then observational studies, and modelling, and finally I will conclude my understanding and contribution to the overall scientific landscape.

\section{Characterisation of bursty phenomena}

Dynamical systems can be described as time series of sequential observations~\cite{box2008time} where timing of an observation, denoted by $t$, can be either continuous or discrete. Since most datasets of human dynamics we analyse have been recorded digitally, we will here focus on the case of discrete timings. In this sense, the time series can be called an event sequence, where each event indicates an observation with a particular character. In this series the $i$th event takes place at time $t_i$ with the result of the observation $z_i$ describing the actual state of the system with a number, set of numbers, etc\footnote{Note that some events could occur in a time interval or with duration, like phone calls between individuals~\cite{holme2012temporal}, what we neglect in our description at the outset unless stated otherwise.}. At the simplest scenario, we assume that the system at a given time can be in two states only, as being active and performing an event, or being inactive. The event sequence with $n$ events can be represented by an ordered list of event timings, i.e., $ev(t_i)=\{t_0,t_1,\cdots,t_{n-1}\}$, where $t_i$ denotes the timing of the $i$th event. Such dynamics (also called point processes) can be described in a form of binary event sequences of $x(t)$ that takes a value of $1$ at time $t=t_i$ of events, or $0$ otherwise. Formally, for discrete timings, one can write the signal as $x(t)=\sum_{i=0}^{n-1}\delta_{t,t_i}$, where $\delta$ denotes the Kronecker delta.

\subsubsection{The Poisson process}

The temporal Poisson process is a stochastic process, which is commonly used to model random processes such as the arrival of customers at a store, or packages at a router. It evolves via completely independent events, thus it can be interpreted as a type of continuous-time Markov process. In a Poisson process, the probability that $n$ events occur within a bounded interval follows a Poisson distribution $P(n)=\frac{\lambda^{n}e^{-\lambda}}{n!}$, where $\lambda$ denotes the average number of events per interval, which is equal to the variance of the distribution in this case. Since these stochastic processes consist of completely independent events, they have served as reference models when studying bursty systems. As we will see later, bursty temporal sequences emerge with fundamentally different dynamics with strong temporal heterogeneities and temporal correlations. Any deviation in their dynamics from the corresponding Poisson model can help us to indicate patterns induced by correlations or other factors like memory effects.

Throughout the thesis we are going to refer to two types of Poisson processes. One is called the \emph{homogeneous Poisson process}, which is characterised by a constant event rate $\lambda$, while the other type, called the \emph{non-homogeneous Poisson process}, defined such that the event rate varies over time, denoted by $\lambda(t)$. For more precise definitions and discussion on the characters of Poisson processes we suggest the reader to study the extended literature addressing this process, e.g., Ref.~\cite{grimmett2009probability}.

\subsubsection{The inter-event time and residual time}

The first and most important measure to characterise bursty temporal sequences is based on the quantity called the \emph{inter-event time}, $\tau_i\equiv t_i-t_{i-1}$, defined as the time interval between two consecutive events at times $t_{i-1}$ and $t_i$ for $i=1,\cdots,n-1$. For an event sequence of $n\geq 2$, we can obtain the sequence of inter-event times, i.e., $iet(\tau_i)=\{\tau_1,\cdots,\tau_{n-1}\}$, and compute their probability density function, i.e., the \emph{inter-event time distribution} $P(\tau)$. For completely regular time series, all inter-event times are the same and equal to the mean inter-event time, denoted by $\langle\tau\rangle$, thus the inter-event time distribution appears as:
\begin{equation}
P(\tau)=\delta(\tau-\langle\tau\rangle),
\end{equation}
where $\delta(\cdot)$ denotes the Dirac delta function. Here the standard deviation of inter-event times, denoted by $\sigma$, is zero. 

For the completely random and homogeneous Poisson process, it is easy to derive~\cite{grimmett2009probability} that the inter-event times are exponentially distributed as follows:
\begin{equation}
    P(\tau)=\frac{1}{\langle\tau\rangle} e^{-\tau/\langle\tau\rangle},
\end{equation}
where $\sigma=\langle\tau\rangle$ and the event rate is $\lambda=1/\langle \tau \rangle$.

In many empirical processes inter-event time distributions have been observed to be broad with heavy tails ranging over several magnitudes. In such bursty time series the fluctuations characterised by $\sigma$ are much larger than $\langle\tau\rangle$, indicating that $P(\tau)$ is rather different from an exponential distribution, as it would derive from Poisson dynamics. Bursty systems evolve through events that are heterogeneously distributed in time and exhibit a broad $P(\tau)$, which can be fitted with either power law, log-normal, or stretched exponential distributions, just to name a few candidates. Most commonly, they can be approximated by a power-law distribution function with an exponential cutoff, defined as
\begin{equation}
    P(\tau)\simeq C\tau^{-\alpha}e^{-\tau/\tau_c},
\end{equation}
where $C$ denotes a normalisation constant, $\alpha$ is the power-law exponent, and $\tau_c$ sets the position of the exponential cutoff. The power-law scaling of $P(\tau)$ indicates the lack of any characteristic time scale, but the presence of strong temporal fluctuations, characterised by the power-law exponent $\alpha$. Power-law distributions are also associated to the concepts of scale-invariance and self-similarity~\cite{newman2005power} and deemed to have an important meaning, especially in terms of universality classes in statistical physics~\cite{plischke2006equilibrium}.

\begin{figure}[!t]
    \center
    \includegraphics[width=.8\textwidth]{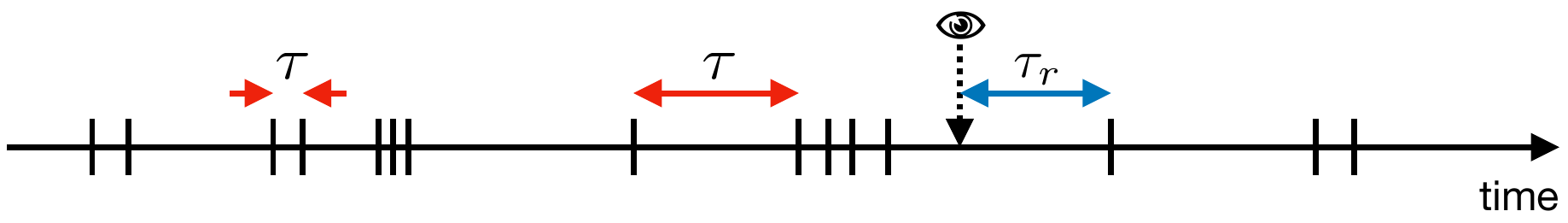}
    \caption{\small Schematic diagram of an event sequence, where each vertical line indicates the timing of an event. The inter-event time $\tau$ is the time interval between two consecutive events. The residual time $\tau_r$ is the time interval from a random moment (e.g., the timing annotated by the vertical arrow) to the next event. This figure was published in~\cite{karsai2018bursty}.}
    \label{fig:scheme1}
\end{figure}

Note that there is another similar quantity, called the \emph{residual time} $\tau_r$ (also called the residual waiting time), what we will use later during our discussion. It's definition considers that the observations of an event sequence always cover a finite period and usually begins at a random moment of time. The time interval between the time of the observation and the first observed event is the residual time $\tau_r$. Its distribution and average can be derived from the corresponding inter-event time distribution as
\begin{equation}
    P(\tau_r)=\frac{1}{\langle \tau\rangle}\int_{\tau_r}^\infty P(\tau)d\tau, \hspace{1in} \langle \tau_r \rangle = \int_0^\infty \tau_r P(\tau_r)d\tau_r =  \frac{\langle \tau^2\rangle}{2\langle \tau\rangle}.
\label{eq:rst_iet}
\end{equation}
This result explains a phenomenon called the \emph{waiting-time paradox}, which has important consequences in dynamical processes evolving on bursty temporal systems as will be discussed later.

\subsubsection{The burstiness parameter}

The heterogeneity of the inter-event times can be quantified by a single measure introduced by Goh and Barab\'asi~\cite{goh2008burstiness}. The burstiness parameter $B$ is defined as the function of the coefficient of variation (CV) of inter-event times $r\equiv \sigma/\langle \tau\rangle$ to measures temporal heterogeneity as follows:
\begin{equation}
    B\equiv \frac{r-1}{r+1}=\frac{\sigma-\langle\tau\rangle}{\sigma+\langle\tau\rangle}.
    \label{eq:burstiness_param}
\end{equation}
Here $B$ takes the value of $-1$ for regular time series with $\sigma=0$, and it is equal to $0$ for random, Poissonian time series where $\sigma=\langle \tau\rangle$. In case when the time series appears with more heterogeneous inter-event times than a Poisson process, the burstiness parameter is positive ($B>0$), while taking the value of $1$ only for extremely bursty cases with $\sigma \rightarrow \infty$. Note that this measure has been recently shown to have some finite size effect, and an alternative measure has been introduced to account for these shortcomings~\cite{kim2016measuring}.

\subsubsection{The autocorrelation function}
\label{sec:autocorr}

The conventional way for detecting correlations in time series is to measure the autocorrelation function. To define we use the representation of event sequences as binary signals $x(t)$ and introduce the delay time $t_d$, which sets a time lag between two observations of the signal $x(t)$. Then the autocorrelation function with delay time $t_d$ is defined as follows:
\begin{equation}
  A(t_d)\equiv \frac{ \langle x(t)x(t+t_d)\rangle_t- \langle x(t)\rangle^2_t}{ \langle x(t)^2\rangle_t- \langle x(t)\rangle^2_t},
\end{equation}
where $\langle \cdot \rangle_t$ denotes the time average over the observation period~\cite{box2008time}. In time series with temporal correlations, $A(t_d)$ typically decays as a power law:
\begin{equation}
A(t_d)\sim t_d^{-\gamma}
\end{equation}
with decaying exponent $\gamma$. In addition, this measure can be related to the power spectrum or spectral density of the signal $x(t)$ as follows:
\begin{equation}
    \label{eq:powerSpectrum}
    P(\omega)=\left|\int x(t)e^{i\omega t}dt\right|^2=\int A(t_d)e^{-i\omega t_d}dt_d,
\end{equation}
which appears as the Fourier transform of autocorrelation function, indicating dominant $\omega$ event  frequencies present in the signal.
\begin{figure}[!ht]
    \center
    \includegraphics[width=\textwidth]{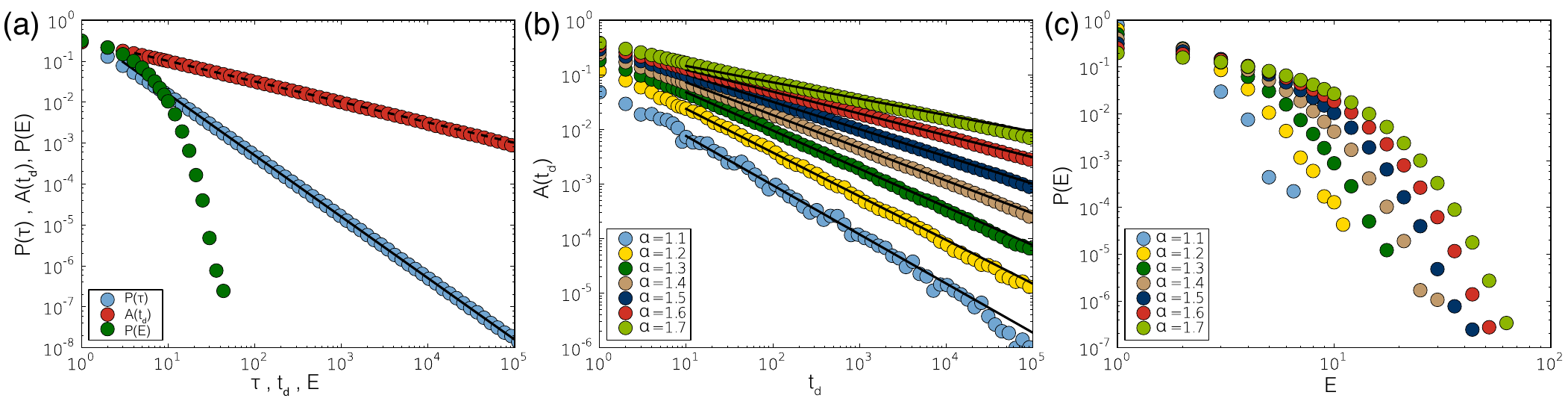}
    \caption{\small The characteristic functions calculated for heterogeneous independent signals. (a) $P(\tau)$, $A(t_d)$ and $P(E)$ functions for $\alpha=1.5$. Solid line is a power-law function with the given $\alpha$ exponent value, while dashed line denotes a a power-law function with an effective $0.5$ exponent value. (b) $A(t_d)$ effective autocorrelation functions for various $\alpha$ exponents. Straight lines are denoting power-law functions with $\alpha$ exponents satisfying the $\alpha+\gamma=2$ relation. (c) Corresponding $P(E)$ distributions for various $\alpha$ exponents. This figure was published in~\cite{karsai2011universal}.}
    \label{fig:pedemo}
\end{figure}

A scaling relation between the $\alpha$ inter-event time and the $\gamma$ autocorrelation exponents has been studied both analytically and numerically~\cite{lowen1993fractal, allegrini2009spontaneous,vajna2013modelling}. It has been shown that they relate as
\begin{eqnarray}
    \label{eq:alpha_gamma}
    \begin{tabular}{ll}
        $\alpha+\gamma=2$ & for $1<\alpha\leq 2$,\\
        $\alpha-\gamma=2$ & for $2<\alpha\leq 3$.
    \end{tabular}
\end{eqnarray}
This indicates that the power-law decaying autocorrelation function could be explained solely by the inhomogeneous inter-event times and not by temporal correlations in the time series. In fact, the observed autocorrelation functions measure not only correlations between events but also between consecutive inter-event times of arbitrary length. Such correlations spuriously appear in independent heterogeneous time series disallowing autocorrelation to be a proper measure of real temporal correlations in bursty sequences. This is demonstrated in Fig.~\ref{fig:pedemo}(a) where we built an independent event sequence by sampling randomly inter-event times from a $P(\tau)\sim \tau^{-\alpha}$ power-law distribution with exponent $\alpha=1.5$ (blue symbols) and measured the $A(t_d)$ autocorrelation function in this uncorrelated signal. Due to the heterogeneity of the inter-event time distribution effective positive correlations are indicated by the autocorrelation, which emerges with a power-law tail (red symbols) even the sequence is independent. Fig.~\ref{fig:pedemo}(b) demonstrates that the scaling exponent $\gamma$ of the emerging autocorrelation function is dependent on the $\alpha$ inter-event time distribution, in full agreement with the relation suggested in Eq.~\ref{eq:alpha_gamma}.

\subsection{The Bursty Train Size Distribution}
\label{sec:brsttrains}

This ambiguity to detect short term temporal correlations in bursty event sequences motivated us to provide a new measure, which can decide evidently the presence of such dependencies, independently from the shape of the inter-event time distribution. More precisely, we made the qualitative observation that bursty events do not usually come in pairs but may form longer trains, where consecutive events may be in a causal relation with each other. However, to detect such bursty clusters in binary event sequence, $x(t)$, first we have to identify those events which we consider to be causally correlated. The smallest temporal scale at which correlations can emerge in the dynamics is between consecutive events. If only $x(t)$ is known, we can assume two consecutive actions at $t_i$ and $t_{i+1}$ to be related if they follow each other within a short time interval, $t_{i+1}-t_i\leq \Delta t$  \cite{wu2010evidence,turnbull2005string}. For events with the duration $d_i$ this condition is slightly modified: $t_{i+1}-(t_i+d_i) \leq \Delta t$.

\begin{figure}[!ht]
    \center
    \includegraphics[width=.8\textwidth]{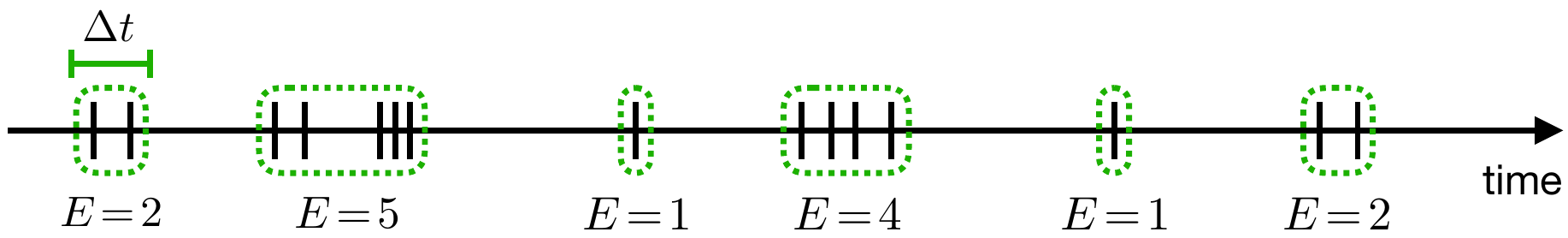}
    \caption{\small Schematic diagram of an event sequence, where each vertical line indicates the timing of the event. For a given time window $\Delta t$, a bursty train is determined by a set of events separated by $\tau\leq \Delta t$, while events in different trains are separated by $\tau>\Delta t$. The number of events in each bursty train, i.e., bursty train size, is denoted by $E$. This figure was published in~\cite{karsai2018bursty}.}
    \label{fig:scheme2}
\end{figure}

This definition allows us to detect bursty periods, defined as a sequence of events where each event follows the previous one within a time interval $\Delta t$, as illustrated in Fig.~\ref{fig:scheme2}. By counting the number of events, $E$, that belong to the same bursty period, we can calculate their distribution $P_{\Delta t}(E)$. For a sequence of independent events, $P_{\Delta t}(E)$ is uniquely determined by $\Delta t$ and the inter-event time distribution $P(\tau)$ as follows:
\begin{equation}
P_{\Delta t}(E)= \left(  \int_0^{\Delta t}P(\tau)d \tau \right)^{E-1} \left( 1-\int_0^{\Delta t}P(\tau)d\tau \right) \hspace{.1in} \approx \hspace{.1in} \dfrac{1}{E_c(\Delta t)}e^{-E/E_c(\Delta t)}
\label{eq:PE}
\end{equation}
for $E>0$. Here the integral $F(\Delta t)=\int_0^{\Delta t}P(\tau)d\tau$ defines the probability to draw an inter-event time $P(\tau)\leq \Delta t$ randomly from an arbitrary distribution $P(\tau)$, $E_c(\Delta t) \equiv \tfrac{1}{-\ln F(\Delta t)}$ is a constant, and approximation appears due to the series expansion of the constant multiplicative term. The first term on the l.h.s. of Eq.\ref{eq:PE} gives the probability that we draw an inter-event times $P(\tau)< \Delta t$ independently $E-1$ consecutive times, while the second term assigns that the $E^{th}$ drawing gives a $P(\tau)> \Delta t$ therefore the evolving train size becomes exactly $E$. If the measured time window is finite ($\Delta t <\infty$), which is always the case here, the integral $\int_0^{\Delta t}P(\tau)d\tau < 1$ is a constant and the asymptotic behaviour appears like in a general exponential form. Consequently, for any finite independent event sequence the $P_{\Delta t}(E)$ distribution decays exponentially even if the inter-event time distribution is fat-tailed. Deviations from this exponential behaviour indicate correlations in the timing of the consecutive events.

\begin{figure}[!ht]
    \center
    \includegraphics[width=\textwidth]{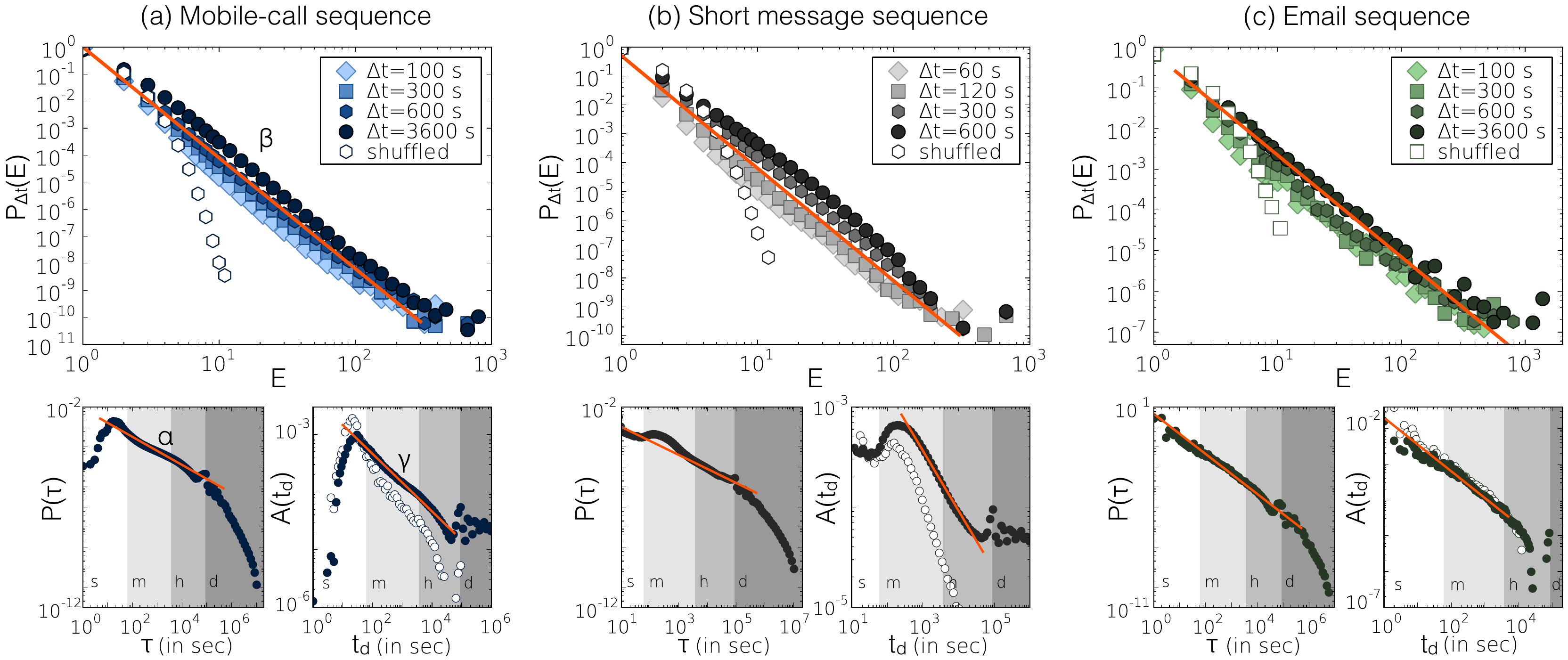}
    \caption{\small The bursty train size distribution $P_{\Delta t}(E)$ with various time windows $\Delta t$ (main panels), the inter-event time distribution $P(\tau)$ (left bottom panels), and autocorrelation functions $A(t_d)$ (right bottom panels) for different human communication datasets such as (a) Mobile phone call dataset: The scale-invariant behaviour was characterised by power-law functions with exponent values $\gamma\simeq 0.5$, $\beta\simeq 4.1$, and  $\alpha\simeq 0.7$ (b) Short message sequences taking values $\gamma\simeq 0.6$, $\beta\simeq 3.9$ and  $\alpha\simeq 0.7$. (c) Email event sequence with exponents $\gamma\simeq 0.75$, $\beta\simeq 2.5$ and $\alpha=1.0$. A gap in the tail of $A(t_d)$ on figure (c) appears due to logarithmic binning and slightly negative correlation values. Empty symbols assign the corresponding calculation results on independent sequences. Vertical stripes called s, m, h and d are denoting seconds, minutes, hours and days, respectively. This figure was published in~\cite{karsai2011universal}. 
}
\label{fig:peemp}
\end{figure}

\subsubsection*{Bursty sequences in human communication}
\label{sec:burstytrainsobs}

To check the scaling behaviour of $P_{\Delta t}(E)$ in real systems we focused on outgoing events of individuals in three selected communication datasets: (a) A mobile-call dataset from a European operator (see DS1 in Section~\ref{sec:datasets}); (b) Text message records from the same dataset (also DS1); and (c) Email communication sequences \cite{eckmann2004entropy}. For each of these event sequences, the distribution of inter-event times measured between outgoing events are shown in Fig.\ref{fig:peemp} (left bottom panels) and the estimated power-law exponent values are summarised in Table \ref{table:PEexp}. The autocorrelation functions, which were averaged over $1,000$ randomly selected users with maximum time lag of $t_d= 10^6$, indicate strong temporal correlation (as seen in Fig.\ref{fig:peemp}.a and b (right bottom panels) with exponents in Table \ref{table:PEexp}). The power-law behaviour in $A(t_d)$ appears after a short period, denoting the reaction time through the corresponding channel, and lasts up to $12$ hours, capturing the natural rhythm of human activities. For emails in Fig.\ref{fig:peemp}.c (right bottom panels) long term correlation are detected up to $8$ hours, which reflects the typical length of office hours (note that the dataset includes internal email communication of a university staff).

\begin{table}[ht!]
\begin{center}
\begin{tabular}{ p{5.3cm} p{1cm} p{1cm} p{1cm} p{1cm}}
  & $\alpha$ & $\beta$ & $\gamma$ & $\nu$\\ \hline
 Mobile-call sequence & $0.7$ & $4.1$ & $0.5$ & $3.0$\\ \hline
 Short message sequence & $0.7$ & $3.9$ & $0.6$ & $2.8$\\ \hline
 Email sequence& $1.0$ & $2.5$ & $0.75$ & $1.3$\\ \hline
 Model & $1.3$ & $3.0$ & $0.7$ & $2.0$\\ \hline
\end{tabular} 
\end{center}
\caption{\small Characteristic exponents of the ($\alpha$) inter-event time distribution, ($\beta$) bursty train size, ($\gamma$) autocorrelation functions and $\nu$ memory functions calculated in different datasets and for the model study (see Section~\ref{sec:datasets}). This table was published in~\cite{karsai2018bursty}}
\label{table:PEexp}
\end{table}

The broad shape of $P(\tau)$ and $A(t_d)$ functions confirm that human communication dynamics is heterogeneous and displays non-trivial correlations up to finite time-scales. However, after destroying event correlations by shuffling inter-event times in the interaction sequence of single individuals (see a how to in Section~\ref{sec:tnet_rrm}) the autocorrelation function still shows slow power-law like decay (empty symbols on bottom right panels) via spurious residual dependencies. This clearly demonstrates the disability of $A(\tau)$ to characterise correlations for heterogeneous signals, just as we have already seen for modelled signals. However as we have discussed earlier, the $P_{\Delta t}(E)$ distribution should indicate evidently the presence of short temporal correlations. Calculating this distribution for various $\Delta t$ windows, we find that it depicts a scale invariant behaviour as
\begin{equation}
P(E)\sim E^{-\beta}
\label{eq:E}
\end{equation} 
for each of the empirical event sequences as shown in the main panels of Fig.\ref{fig:peemp}. $P_{\Delta t}(E)$ evidently indicates that there are strong temporal correlations in the empirical sequences as it is remarkably different from the corresponding distributions calculated for independent sequences which, as predicted by (\ref{eq:PE}), appear with exponential decay (empty symbols on the main panels). 

Exponential behaviour of $P_{\Delta t}(E)$ was also expected from results published in the literature assuming human communication behaviour to be uncorrelated \cite{malmgren2008poissonian,wu2010evidence,anteneodo2010poissonian}. However, the observed scaling behaviour of $P_{\Delta t}(E)$ offers a direct evidence of correlations in human dynamics, which can arguably be responsible for the observed bursty dynamics. These correlations induce long bursty trains in the event sequence rather than short bursts of independent events. In addition, we have found that the scaling of the $P_{\Delta t}(E)$ distribution is quite robust against the choice of $\Delta t$ for an extended regime of time-window sizes, or when it is computed for individuals or group people of similar activity level, or once the effects daily fluctuations are accounted for (for results see~\cite{karsai2011universal}).

Note that we observed long bursty event trains and similar scaling of their size in various natural phenomena, like in earthquake sequences recorded at given locations~\cite{smalley1987a,zhao2010non}, or in the firing patterns of single neurons recorded in rat's hypocampal. Corresponding results are not shown here but reported in~\cite{karsai2011universal}.

\section{Cyclic patterns in human dynamics}\label{subsect:cyclic}

It is evident that humans follow intrinsic periodic patterns of circadian, weekly, and even longer cycles~\cite{malmgren2008poissonian, jo2012circadian, aledavood2015daily}. Such cycles clearly contribute to the inhomogeneities of temporal patterns, and they often result in an exponential cutoff to the inter-event time distributions. Identifying and filtering out such cyclic patterns from a time series can reveal bursty behaviour of different origins than those cycles~\cite{jo2012circadian}. In order to characterise such cyclic patterns, let us consider a time series, i.e., the number of events at time $t$, denoted by $x(t)$, for the entire period of $0\leq t< T$. One may be interested in a specific cycle, like daily or weekly ones, with period denoted by $T_{\circlearrowleft}$. Then, for a given period of $T_{\circlearrowleft}$, the event rate with $0\leq t <T_{\circlearrowleft}$ can be defined as
\begin{equation}
    \rho(t)\equiv \frac{T_{\circlearrowleft}}{X}\sum_{k=0}^{T/T_{\circlearrowleft}} x(t+kT_{\circlearrowleft}), \hspace{.2in} X\equiv \int_0^{T} x(t)dt.
\end{equation}
Such cycles turn out to be also apparent in the inter-event time distributions $P(\tau)$. For example, one finds peaks of $P(\tau)$ corresponding to multiples of one day in mobile phone communication as can be seen in Fig.~\ref{fig:peemp}a and b lower left panels. Note that such periodicities could be characterised by means of a power spectrum analysis in Eq.~(\ref{eq:powerSpectrum}), however here we take another way.

\begin{figure}[!h]
    \center
    \includegraphics[width=\columnwidth]{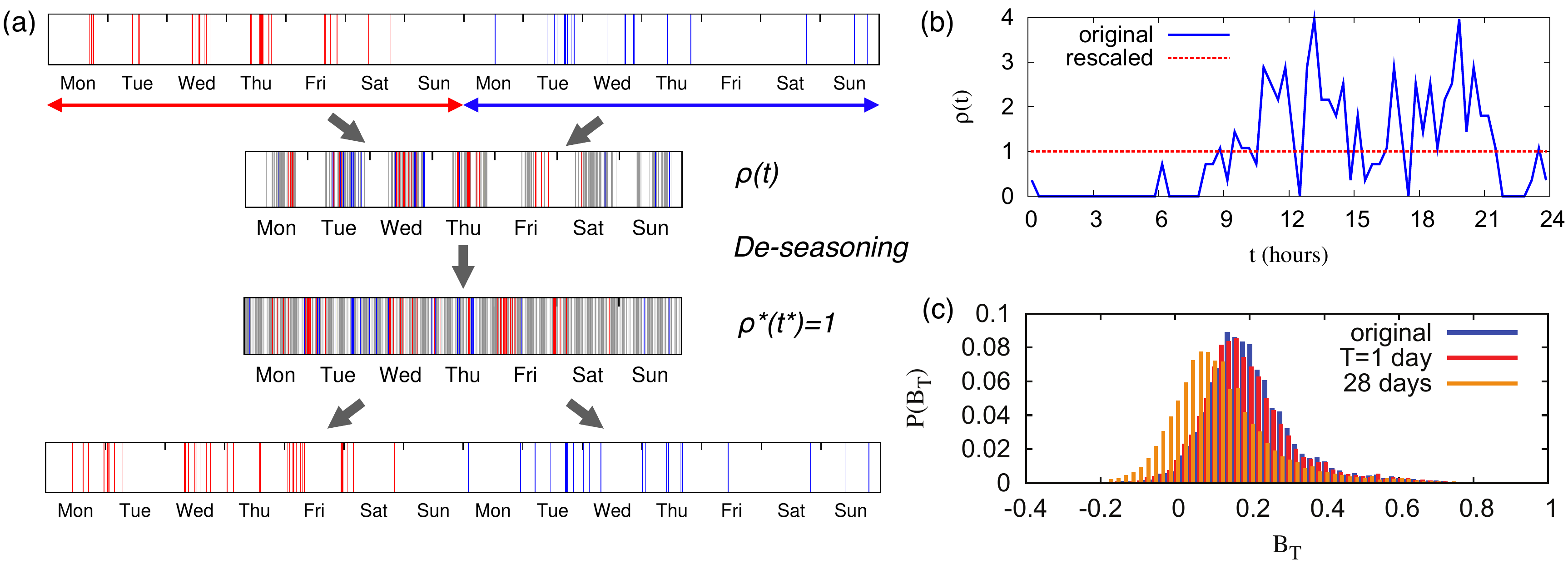}
    \caption{\small (a) An example of the de-seasoning method applied to a mobile call series of a user, with $T_{\circlearrowleft}=1$ week. The top shows the first two weeks of the call series coloured in red (the first week) and blue (the second week). Events for all weeks are collected in one week period to obtain the event rate $\rho(t)$ for $0\leq t< T_{\circlearrowleft}$. After de-seasoning, the events in each week are put back to their original slot. (b) The original (blue) and de-seasoned (red) hourly event rate of communication for individuals with $200$ calls. (c) The distributions of $B_T$ bursty parameters of individual users with the same strength after de-seasoning over $T_{\circlearrowleft}$ period. This figure was prepared by HH. Jo and published in~\cite{jo2012circadian}.
}
\label{fig:circDeseas}
\end{figure}

Once such cycles are identified in terms of the event rate $\rho(t)$, we can filter them by deseasoning the time series~\cite{jo2012circadian}. First, we extend indefinitely the domain of $\rho(t)$ by $\rho(t+kT_{\circlearrowleft})=\rho(t)$ with an arbitrary integer $k$. Then using the identity of $\rho(t)dt=\rho^*(t^*)dt^*$ with the deseasoned event rate of $\rho^*(t^*)=1$, we can get the de-seasoned time $t^*(t)$ as
\begin{equation}
    t^*(t)\equiv \int_0^t \rho(t')dt'.
\end{equation}

For the schematic example of the de-seasoning method, see Fig.~\ref{fig:circDeseas}a. In plain words, the time is dilated (respectively contracted) at the moment of the high (respectively low) event rate resulting an overally constant average event rate as shown in as demonstrated in Fig.~\ref{fig:circDeseas}b. Then the de-seasoned event sequence of $\{t^*(t_i)\}$ is compared to the original event sequence of $\{t_i\}$ to see how strong signature of burstiness or memory effects remained in the de-seasoned sequence. This reveals whether the empirically observed temporal heterogeneities can (or cannot) be explained by the intrinsic cyclic patterns, characterised in terms of the event rate. For example, one can measure the burstiness parameter $B_T$ for both the original and the de-seasoned mobile phone call series as shown in Fig.~\ref{fig:circDeseas}c using DS1 (see Section~\ref{sec:datasets}). Although the $P(B_T)$ distribution slightly changes after de-seasoning over a period $T$, it assigns that the majority of individual sequences remain with positive bursty parameters.

\begin{figure}[!t]
    \center
    \includegraphics[width=\columnwidth]{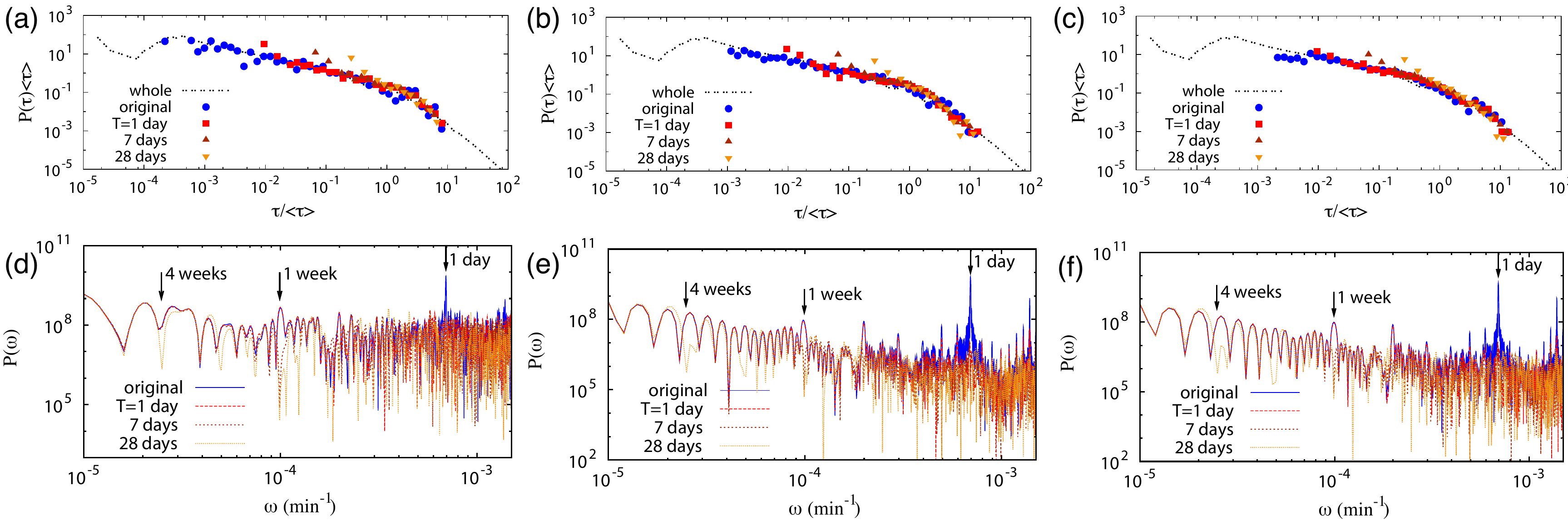}
    \caption{\small (a-c) The original and de-seasoned inter-event times extracted from the individual call sequences of mobile phone users. Dashed line assigns the $P(\tau)$ of the whole population. Rescaling were done for periods of $T=1$, $7$, and $28$ days. (e-f) Corresponding power-spectrum curves as the function $\omega$ event frequencies with unites $1/min$. Individual users with $s = 400$ (a, d), $800$ (b, e), 1600 (c, f) number of calls were analysed. This figure was prepared by HH. Jo and published in~\cite{jo2012circadian}.}
    \label{fig:circIndiv}
\end{figure}

One can also obtain the de-seasoned inter-event time $\tau_i^*$ corresponding to the original inter-event time $\tau_i=t_i-t_{i-1}$ as
\begin{equation}
    \tau_i^* \equiv t^*(t_i)-t^*(t_{i-1})=\int_{t_{i-1}}^{t_i} \rho(t')dt'.
\end{equation}
This way the de-seasoned inter-event time distribution $P(\tau^*)$ can be compared to the original inter-event time distribution $P(\tau)$. As shown in Fig.~\ref{fig:circIndiv}a-c, the inter-event time distributions for the original and de-seasoned event sequences show almost the same shape for various values of $T_{\circlearrowleft}$ and for individuals of various activity level. At the same time, it is evident from Fig.~\ref{fig:circIndiv}d-f, that after de-seasoning the circadian and weekly peaks of $\omega$ event frequencies disappear from the power spectra (for definition see Eq.~\ref{eq:powerSpectrum}), while its overall scaling remains very similar to the original spectrum. All these results together imply that bursty human dynamics cannot be exclusively explained by periodic circadian and weekly fluctuations, but it may have some other intrinsic behavioural origins.

I have another set of works~\cite{karsai2011small,kivela2012multiscale}, which provide new tools to understand the effects and consequences of bursty interactions, through the introduction of random reference models of temporal networks. These works will be addressed in Section~\ref{sec:tnet_rrm} as they provide tools to analyse temporal networks in general, not only bursty phenomena exclusively.


\section{Observation of bursty phenomena}
             
Bursty dynamics characterise human behaviour on the individual level and in turn may determine the evolution of dyadic interactions or the emergence of macroscopic phenomena in the social network. The dynamics of social networks can be discussed in terms of nodes, links, communities, or at the collective level, and can be characterised at different temporal scale, as we will discuss later in Chapter~\ref{ch:tnet} on temporal networks. To address bursty dynamics of interactions, we distinguish between two temporal scales of link dynamics, which can be assigned to rather different types of behaviour. On one hand, we consider the slow dynamics of social link creation and decay, which determines the evolution of the social network. As an example, think about student mates with whom one may maintain a social relationship over years, which typically decay after graduation. On the other hand, we consider temporal interactions appearing with a rapid pace on existing social ties. These are for example calls, messages, or emails, which typically appear recurrently with high frequency and short duration as compared to the lifetime of a social tie. Next, I summarise some observational results we obtained on real social networks, published in~\cite{kikas2013bursty,karsai2012correlated}, to see whether bursty characters appear in the interaction dynamics at these two temporal scales, and if yes, how these bursty interactions are distributed in the egocentric network of an individual.

\subsection{Bursty egocentric network evolution}

First let's concentrate on the evolution of egocentric networks by analysing creation and decay of confirmed contacts in the online social-communication system of Skype~\cite{kikas2013bursty}. As we have already explained in Section~\ref{sec:datasets}, the DS4 dataset contains the time stamps of approval of each Skype contacts (which can be regarded as times of link creation), and deletion of Skype contacts if it happened before the end of the observation period. In addition, we consider time series indicating the number of days in each month when the user connected to the Skype network, and also the adoption time 
(first usage) of each free service [e.g. Skype-to-Skype (S2S) audio calls, video calls, chat, etc.] together with time series indicating the number of days in each month when the user used the given service.

\begin{figure}[!ht]
    \center
    \includegraphics[width=1.\columnwidth]{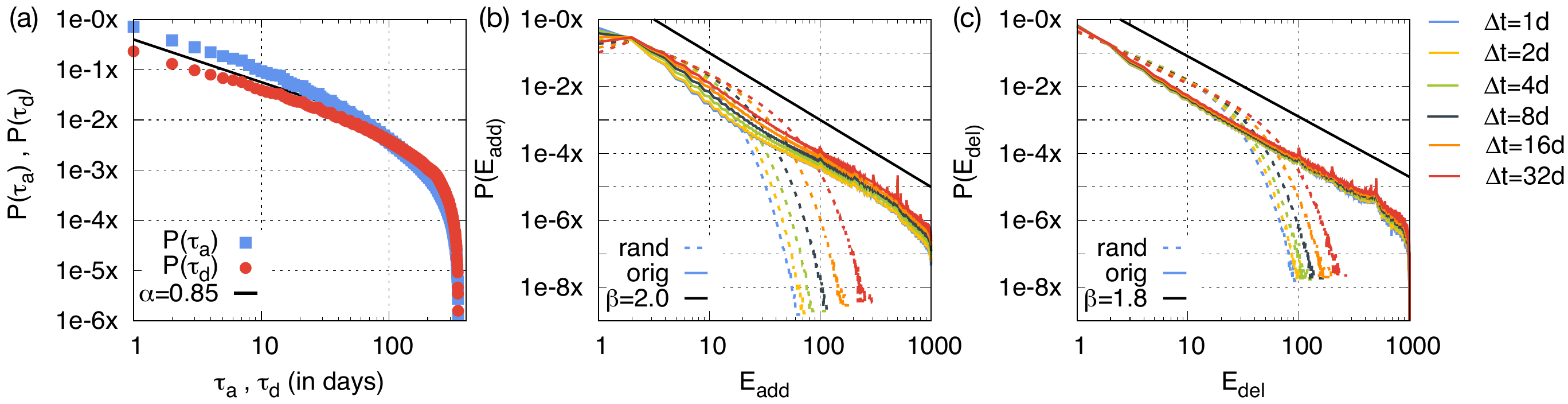}
    \caption{\small (a) Inter-event time distributions of edge addition (blue squares) and deletion (red circles) events of users. The straight line indicates a power-law function with exponent $\alpha=0.85$. (b,c) Distribution of number of events in bursty trains of (b) contact addition and (c) deletion of individuals. Distributions were calculated with time window sizes $\Delta t=1$, $2$, $4$, $8$, $16$ and $32$ days. Distributions for randomly shuffled sequences (dashed line) were calculated with the same $\Delta t$ values. Straight lines indicate power-law functions with exponents (b) $\beta=2.0$ and (c) $\beta=1.8$. This figure was prepared by R. Kikas and published in~\cite{kikas2013bursty}.}
\label{fig:kikasPE}
\end{figure}

To characterise the temporal evolution of contact lists, we first examine the sequences of edge addition and deletion of each individual by calculating the distributions of inter-event times defined as $\tau_{a}=t^{a}_{i+1}-t^{a}_{i}$  (resp. $\tau_{d}=t^{d}_{i+1}-t^{d}_{i}$) between consecutive additions (resp. deletions) events of the same user. As demonstrated in Fig.~\ref{fig:kikasPE}a, these distributions are very heterogeneous both in case of edge additions and edge deletions. They show rather similar scaling, which can be approximated with a power-law function with exponent $\gamma\simeq 0.85$ and an exponential cutoff. This is an interesting observation as one would expect rather different decision mechanisms behind adding and deleting a contact as additions can be assumed to be driven by the desire or need to communicate or to signal a social relation, while deletions are arguably driven by the desire not to be visible or accessible by the deleted contact. 

To identify temporal correlations between consecutive actions of link additions or deletions, we locate bursty event trains and compute their size distributions using the methodology explained in Section~\ref{sec:brsttrains}. More precisely, we analysed the edge modification sequences of each individual and extracted the clusters of events of new edge addition and deletion (the trains) to record their size $E_a$ (resp. $E_d$). The fact that the $P(E_a)$  (resp. $P(E_d)$) distribution in Fig.\ref{fig:kikasPE} spans over orders of magnitude, for several $\Delta t$ values, confirming the presence of correlations between consecutive events of edge additions (resp. deletions). This is even more apparent once we compare the empirical $P(E)$ functions to the equivalent distributions calculated for independent sequences where inter-event times has been randomly shuffled. It puts into evidence that the actions of an individual are not independent and that the evolution of egocentric networks is not only heterogeneous in time but driven by intrinsic correlations. They lead to the presence of high activity bursty periods, where a large number of edges are added or deleted, separated by long low activity intervals.

\begin{figure}[ht!]
  \begin{minipage}[c]{0.70\textwidth}
    \includegraphics[width=\textwidth]{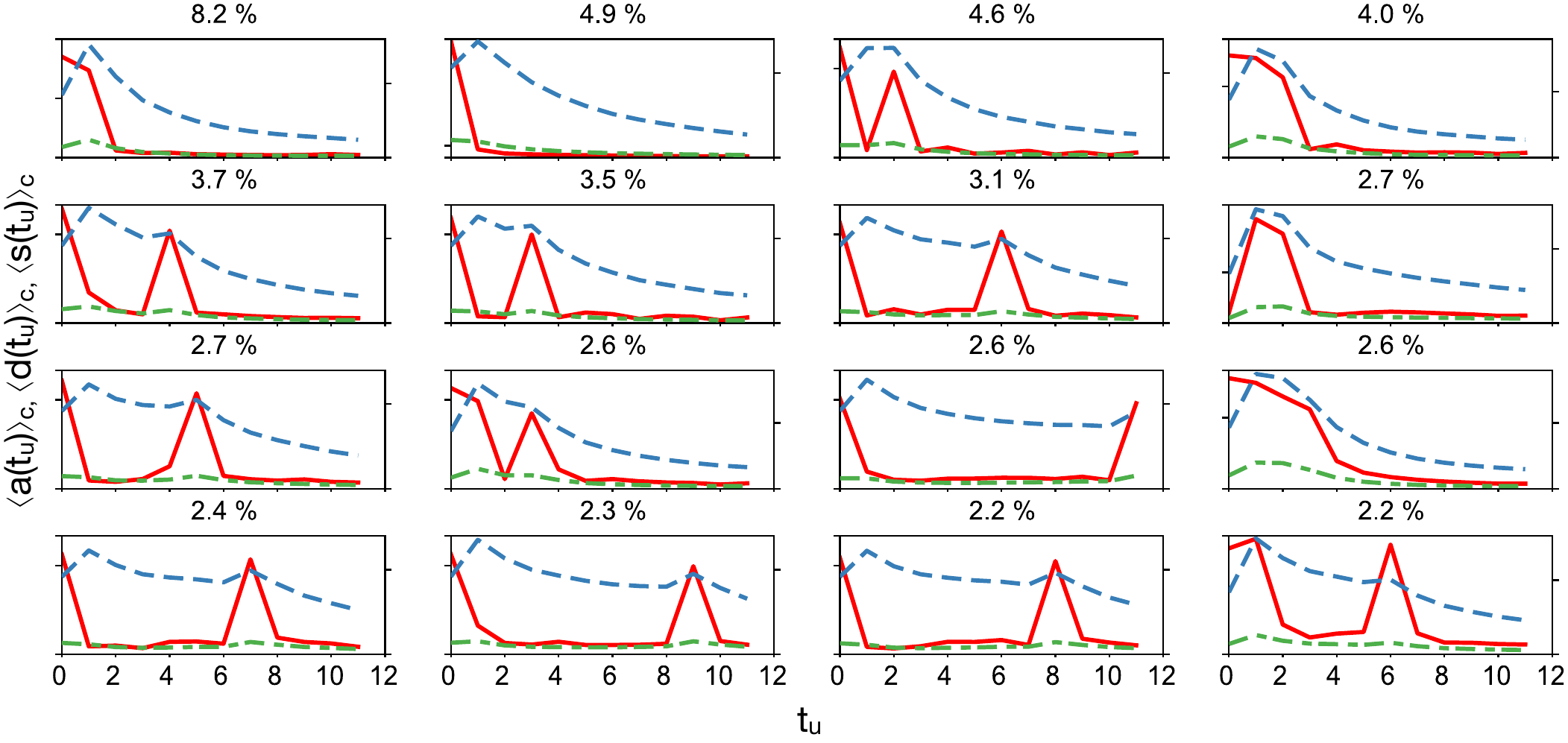}
  \end{minipage}\hfill
  \begin{minipage}[c]{0.29\textwidth}
    \caption{\small Groups of contact addition patterns, where the average number of $\langle a(t_u) \rangle_c$ new contacts added (red), $\langle d(t_u) \rangle_c$ connected days (blue), and $\langle s(t_u) \rangle_c$ days used Skype-to-Skype service at the actual month are shown. Left vertical axe is the added contact number, right axe is in days. This figure was prepared by R. Kikas and published in~\cite{kikas2013bursty}.}
\label{fig:kikasCorrs}
  \end{minipage}
\end{figure}

So far we have observed that edge addition and deletion events of an individual are bursty and clustered in time, yet we know less about when these bursty trains evolve during the lifetime of a user. Do they appear in any time or there are typical activity patterns of edge additions or maybe triggered by other user actions? To answer these questions, we compare the edge addition sequences of individuals during the first year of their $t_u$ user time after their registration~\cite{kikas2013bursty}. We keep track the $a_i$ number of newly added edges of each node $i$ in each month to obtain a discrete $a_i(t_u)$ sequence for each individual with $t_u=1...12$. To be able to compare sequences of users with diverse overall intensity we applied the \textit{Symbolic Aggregate Approximation} (SAX) method~\cite{lin2003symbolic} with alphabet size 10, while to detect groups of users with similar edge addition dynamics, we applied the \textit{k-means clustering}  method on the activity sequences using euclidean distance. We determined the optimal cluster number to be $44$ by using the \emph{Elbow method}. For the most populated clusters, the average link addition activity curves, $\langle a_i(t_u) \rangle_k$, are shown in Fig.~\ref{fig:kikasCorrs} (red line). Looking at the most common patterns it is straightforward that typically people perform their principal (the largest and usually the only one) edge addition burst right after they join the network, a behaviour which is confirmed by other studies~\cite{gaito2012bursty}. This is the time when they explore their social acquaintances who have already joined the system, while later they add contacts just occasionally with lower frequency. This behaviour and its significance has been demonstrated in~\cite{kikas2013bursty} (results not shown here). However, a single bursty peak at early time is not the characteristic of every user. Less common motifs in Fig.\ref{fig:kikasCorrs} show that principal bursty peaks may emerge later or even in multiple times. This observation indicates that events other than user registration (e.g. adoption of different services) may also trigger immediate changes in the egocentric network. Actually, we found that the link addition dynamics is positively correlated (with coefficient $r=0.308137$) with the overall activity of users (shown with blue lines in Fig.\ref{fig:kikasCorrs}), what we measured as the average number of connected days per month. Similarly, we found correlations with the activity of free service usage (with coefficient $r=0.34608$), shown as green lines. More importantly, we showed that the probability of a user performing a link addition burst is strongly conditional to the adoption of free and payed services, which explains the emergent later peaks in the link addition dynamics (for further results see~\cite{kikas2013bursty}).


\subsection{Bursty communication in egocentric networks}
\label{sec:balance}

\begin{figure}[!ht]
    \center
    \includegraphics[width=1.\columnwidth]{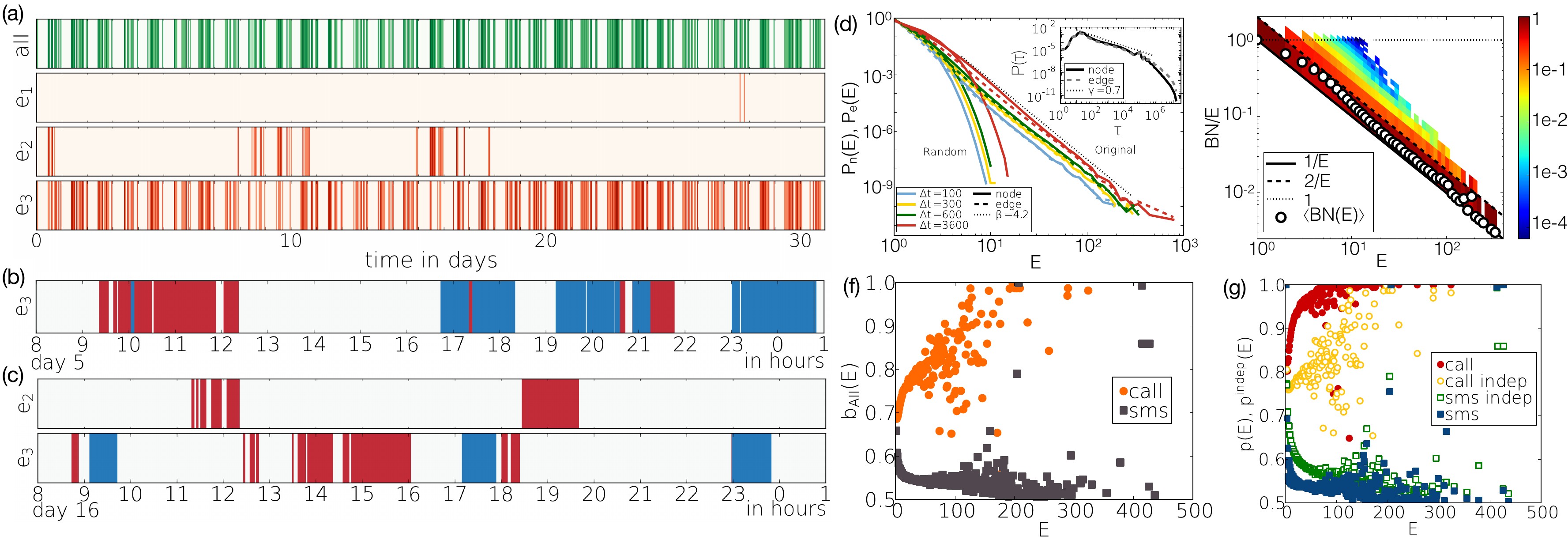}
    \caption{\small (a) Overall activity of the ego (green) and its neighbours ($e_1$, $e_2$, $e_3$). Darker colour scales with call number per hour. (b,c) In- (red) and outgoing (blue) calls with length of the corresponding calls. (d) $P(E)$ distributions of outgoing calls of nodes towards all the neighbours (solid lines) and to single neighbours (dashed lines) for various $\Delta t$ (in second) using the original and inter-event time shuffled (random) sequences. Inset: inter-event time distributions between outgoing events of a user towards all neighbours (solid line) and on a single link (dashed line). (e) The distribution and average of the ratio $BN/E$ for each $E$ train size. Pointed, solid and dashed lines assign limiting cases of $1$, $1/$E, and $2/E$. (f) Average $b_e$ edge balance calculated for trains with the same size for calls (orange circles) and SMS (brown squares). (g) Average $p(E)$ balance values for trains of the same size for calls (red circles) and SMS (blue squares). Corresponding independent event trains are shown with yellow circles for calls and green squares for SMS. This figure was published in~\cite{karsai2012correlated}.}
    \label{fig:DirB_corr}
\end{figure}

In our second empirical study let's move from the level of social tie evolution to the level of time varying interactions in egocentric networks. Analysing interaction dynamics at this finer temporal scale, at which social ties are actually maintained, is a key to the better understanding of the evolution of egocentric networks and the emergent structural correlations in the social network (as we will discuss in Chapter~\ref{ch:tnet}). As a demonstration, Fig.~\ref{fig:DirB_corr}a illustrates mobile phone call patterns in an egocentric network, where the overall activity of the ego (green row) and activities on separated edges with three friends (orange rows) are presented. By looking at this picture we can draw three important observations: (a) the communication dynamics of the central ego (green) is evidently bursty with heterogeneous inter-event times and broad distribution (see Fig.~\ref{fig:DirB_corr}d) with exponent $\alpha=0.7$ (for SMS see~\cite{karsai2012correlated}); (b) The amount of communication efforts is not evenly distributed among ties (orange), but some ties carry the wast majority of interactions, while others are maintained by only a few events. It assigns differences in terms of social tie strengths, arguably associated to various level of intimacy as suggested by Dunbar~\cite{barrett2002human}, and leads to heterogeneous link weight distributions on the network level; Finally, (c) correlated events form trains in bursty periods. The distribution of trains in the egocentric network can be explained by two competing hypothesises. On one hand, correlated event trains of the ego may evolve on single links (as seen on the zoom shown on panel Fig.~\ref{fig:DirB_corr}b), which suggests that bursty periods are actually induced by dyadic interactions. On the other hand, as demonstrated in Fig.~\ref{fig:DirB_corr}c, bursty communication periods of an ego may involve multiple peers, suggesting that bursty patterns are potentially induced by collective behaviour, the effort of a group e.g. to organise an event or to process information. As next~\cite{karsai2012correlated}, our aim will be to empirically decide between these hypothesises by analysing DS1 mobile phone call and SMS datasets with the previously introduced characteristic methods of bursty phenomena.

More precisely, we are going to use the $P(E)$ bursty train size distribution to indicate how trains are distributed in the egocentric topology. At first, let's concentrate on ego-initiated events of outgoing voice calls and SMS’. In case we consider entire event sequences of egos, which combines all of their communications on any links, the $P(E)$ distributions (shown for calls in Fig.~\ref{fig:DirB_corr}b with solid lines) appear approximately as a power-law function with exponent $\beta=4.2$ (for SMS not shown~\cite{karsai2012correlated}). This behaviour is remarkably different from the scaling of the corresponding reference distributions, where inter-event times were randomly shuffled over the whole data (see exponential distributions with solid lines in Fig.~\ref{fig:DirB_corr}b for calls). Based on this node centric view, intuitively one would assume that correlated outgoing communications of an individual may serve the information processing or organisation of a group~\cite{licoppe2005social,engestrom1999center}, resulting events grouped in trains directed towards several neighbours. Assuming this mechanism to be dominant, burstiness would appear as the property of a single node or a group of individuals.

Surprisingly, the generic picture seems to be very different. If we assume that the correlated events in trains are directed toward several neighbours, decoupling event trains on single edges should induce an entirely different, less correlated statistics of bursts. However, this is not the case. The $P(E)$ distributions, detected on single edges (shown for calls in Fig.~\ref{fig:DirB_corr}b with dashed lines), are scaling very similarly and can be characterised by the same exponents as in the node centric case. This suggests that trains of events usually evolve on single links rather among a larger group of individuals. This picture is also supported by the statistics of temporal motifs~\cite{kovanen2011temporal}, where motifs involving two individuals are by far the most common ones. The same conclusion can be drawn by counting the number of neighbours $BN$, whom an individual called in a bursty train of $E$ events. If a user communicates with only one neighbour during a period then the ratio $BN/E=1/E$, or if each call are directed toward different neighbours than $BN/E=1$. The distributions of the $BN/E$ ratios for each $E$ train size (shown in Fig.~\ref{fig:DirB_corr}e for calls, for SMS not shown~\cite{karsai2012correlated}) indicates that on average only one or two people are called in a bursty train independent of its size.


Next let's have a look at the direction of event trains. Are they balanced or contain events dominantly initiated by one partner? Do voice calls and SMS' are different from this point of view? First, let's calculate the overall communication balance over the entire observation period for each edge $e$ as $b_e=max(N_A,N_B)/(N_A + N_B)$, where $N_A$ ($N_B$) are the total number of calls from $A$ to $B$ ($B$ to $A$). Hence $b_e$ can vary between $1/2$ (completely balanced) and $1$ (completely imbalanced, dominated by one of the participants). We use this measure to compute the weighted average \emph{communication balance of trains on single edges} with size $E$ evolving as:
\begin{equation}
b_{All}(E)=\langle b_e\rangle_{E}=\dfrac{\sum_e n_e(E)b_e}{\sum_e n_e(E)}
\label{eq:b_All}
\end{equation}
where $n_e(E)$ is the number of trains of length $E$ on edge $e$. For SMS (brown squares in Fig.\ref{fig:DirB_corr}.f) $b_{All}$ is converging to $1/2$ for larger $E$ sizes thus longer trains are more and more balanced in this case. This can be explained by the uni-directed information flow in case of SMS forcing mutual discussions to be more balanced, as it has been also argued by Wu et al. \cite{wu2010evidence}. However, this constrain does not apply for the mobile calls (orange points in Fig.\ref{fig:DirB_corr}.f) where information can flow in both directions during a call. Here $b_{All}$ reflects strongly unbalanced communication as it increases towards $1$ for trains with larger $E$.

However, the $b_{All}(E)$ average overall balance does not reflect evidently the communication balance evolving in single trains. Hence we define the \emph{communication balance within a train} on an edge $e$ as $p_{e}(E_m)=max(n_A,n_B)/(n_A + n_B)$. Here $p_{e}(E_m)$ is the balance of the $m-$th train of length $E_m$ on edge $e$ connecting $A$ and $B$, and $n_A$ ($n_B$) denotes the number of events initiated by $A$ (resp. $B$) towards $B$ (resp. $A$) in that train; $E_m=n_A+n_B$. Averaging over trains of the same size $p(E)=\langle p_e(E_m)\rangle_{E}$ gives an estimate for the average communication balance in trains of size $E$ (note that in this case  different $p_e(E_m)$ values can evolve even for trains on the same edge $e$). This has to be compared to the reference case, where trains of a given size follow the overall balance $b_e$ of the actual link. This can be calculated as
\begin{equation}
p_e^{indep}(E_m=E)=\dfrac{1}{E}\sum_{i=0}^{E}\left( \left| \frac{E}{2} - i \right|+\frac{E}{2} \right)b_e^i(1-b_e)^{E-i}\binom{E}{i},
\label{eq:prand}
\end{equation}
where the first term after the summation weights the binomial distribution taking into account that the imbalance can evolve in both directions, i.e., parallel or antiparallel to the imbalance of the edge. As $p_e^{indep}(E_m=E)$ depends only on $b_e$ and $E$, the average can be taken as $p^{indep}(E)=\langle p^{indep}_e(E_m) \rangle_{E{\it \ trains}}$ to get a reference estimate of and independent process.

Fig. \ref{fig:DirB_corr}.g shows $p(E)$ and $p^{indep}(E)$  for both voice calls and SMS messages. Interestingly, large difference is observed between $p(E)$ and the corresponding $p^{indep}(E)$ measures. It suggests that call trains (red points) are more unbalanced than one would expect from the overall communication balance of the link, caputured by the independent processes (yellow circles). At the same time for SMS the contrary is true as trains (blue squares) are much more balanced than one would derive from independent processes (green squares) reflecting the $b_e$ balance of the corresponding edges. This demonstrates real correlations between events of the same train and suggests different correlated mechanisms behind call and SMS dynamics.




\section{Models of bursty human phenomena}

As we have already explained in the introduction of this Chapter (Section~\ref{sec:brst_intro}), there are three main modelling frameworks, which have been proposed to explain the origin of bursty patterns in human dynamics. One framework provides a variety of models using \emph{priority queues}~\cite{barabasi2005origin,oliveira2005human,vazquez2006modeling}; a second one is based on the assumption that consecutive actions of individuals are independent and can be modelled by \emph{Poisson processes} with alternating time scales~\cite{malmgren2008poissonian,malmgren2009universality}; while the third direction assumes strong local correlations modelled by \emph{memory processes}~\cite{vazquez2006impact, han2008modeling}, \emph{self-excited point processes}~\cite{masuda2013selfexciting,jo2015correlated}, or \emph{reinforcement mechanisms}~\cite{karsai2012correlated,wang2014modeling}, etc. In the following, we are going to discuss two models~\cite{karsai2012correlated,karsai2011universal}, proposed by me and colleagues, which belongs to the third modelling framework and employs reinforcement processes to model simultaneously temporal heterogeneities, bursty trains, or communication balance. While these models aim to explain phenomena observed on the individual or dyadic level, later in Section~\ref{sec:tnet_rrm} we will discuss other models~\cite{karsai2011small,kivela2012multiscale}, which address the emergence and effects of bursty phenomena on the collective level.

\subsection{Model of individual bursty dynamics with event trains}

According to the Decision Field Theory of psychology \cite{busemeyer1993decision}, each decision to perform a task can be interpreted as a threshold phenomenon, as the stimulus of the task has to reach a given threshold level to be chosen from the enormously large number of possible actions. This theory lays behind one possible interpretation of bursty behaviour, where the dynamics of an individual performing events of one certain task, like writing emails or printing, goes through active and inactive periods. An active state is initiated once the importance of the actual task overreach a certain level, which after the person performs a bursty cascade of events, while otherwise doing something else which in turn appears as long inactive periods in the actual observation. However, in active periods events are not independent from each other but form long bursty trains as we have already demonstrated in Section~\ref{sec:burstytrainsobs}. 
Such dynamics can be explained by memory driven processes and modelled by reinforcement mechanisms as we explain next~\cite{karsai2011universal}.

\subsubsection*{Memory process}




\begin{figure}[!ht]
    \center
    \includegraphics[width=.9\columnwidth]{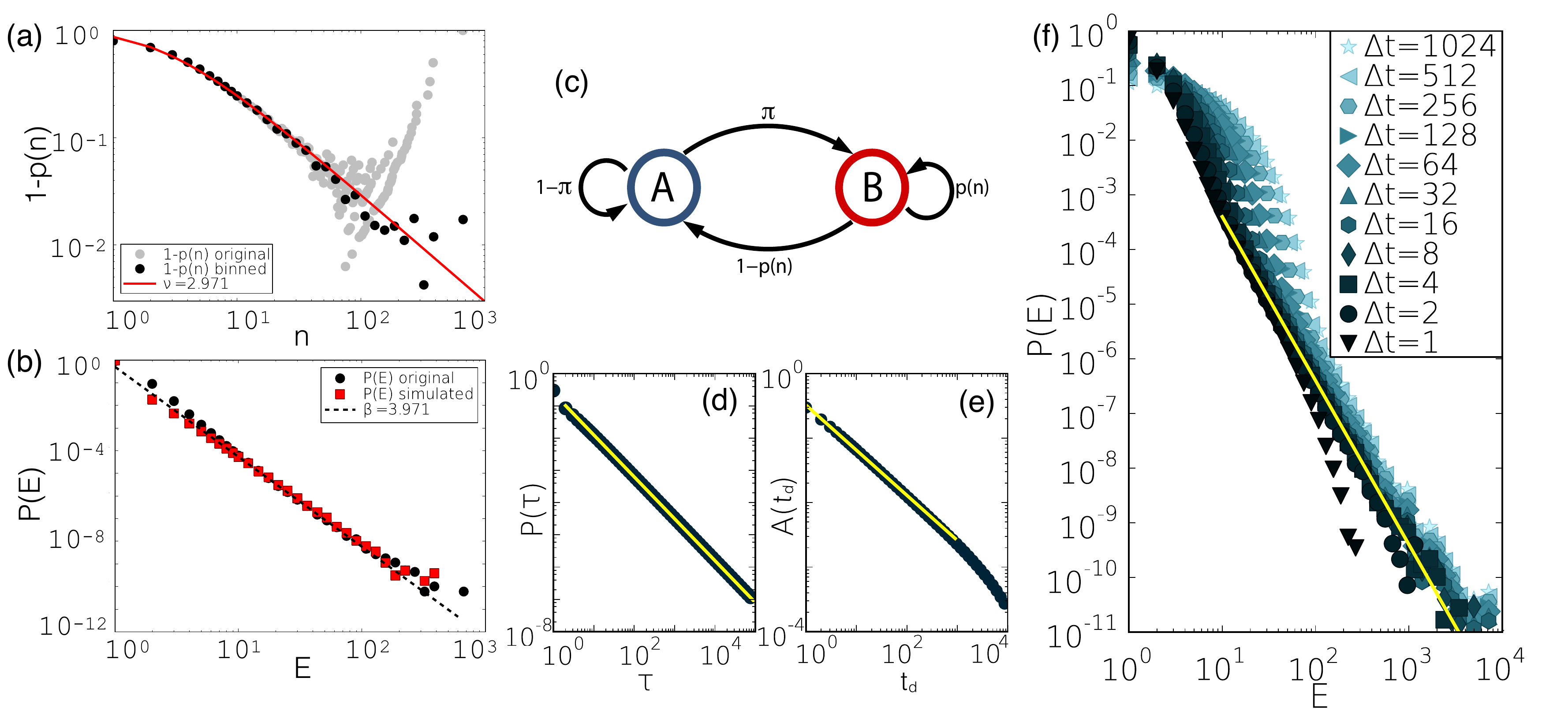}
    \caption{\small (a) The $1-p(n)$ complement of the memory function measured from the mobile call sequence with $\Delta t=600$ second and fitted with the analytical curve defined in Eq.~\ref{eq:pnbrst} with $\nu=2.971$. Grey symbols are the original points, while black symbols denotes the same function after logarithmic binning. (b) The $P(E)$ distributions measured in real and in modelled event sequences together with a power-law function with exponent derived from Eq.~\ref{eq:pnbrst}. (c) Transition probabilities of the reinforcement model with memory. (d) Logarithmic binned $P(\tau)$ inter-event time distribution of the simulated process with a fitted exponent $\gamma=1.3$.  (e) The average logarithmic binned autocorrelation function with a maximum lag $t_d^{max}=10^4$ and fitted exponent $\alpha=0.7$. (f) Logarithmic binned $P(E)$ distributions of the synthetic sequence with window sizes $\Delta t=1...1024$ and fitted exponent $\beta=3.0$.  Simulation were averaged over $1000$ independent realisations with parameters $\mu_A=0.3$, $\mu_B=5.0$, $\nu=2.0$, $\pi=0.1$ and $T=10^9$. For the calculation we chose the maximum inter-event time $\tau^{max}=10^5$. This figure was published in~\cite{karsai2011universal}.}
    \label{fig:BrstModel1}
\end{figure}

The correlations taking place between consecutive bursty events can be interpreted as a memory process, allowing us to calculate the $p(n)$ \emph{probability that the individual will perform one more event within a $\Delta t$ time frame after it executed $n$ events previously} in the actual train. This probability can be written as $p(n)=\frac{\sum_{E=n+1}^{\infty}P(E)}{\sum_{E=n}^{\infty}P(E)}$, thus its functional form is entirely coded in the train size distribution. If we assume that the actual train size distribution scales as $P(E)\sim E^{-\beta}$ (as already discussed in Eq.\ref{eq:E}) the memory function appears as
\begin{equation}
p(n) = \left( \frac{n}{n+1}\right) ^\nu  \hspace{.5in} \mbox{where} \hspace{.5in} \beta=\nu +1
\label{eq:pnbrst}
\end{equation}
scaling relation is expect to hold as derived in~\cite{karsai2011universal}. To demonstrate that Eq.~\ref{eq:pnbrst} holds for real systems, we measured the $P(E)$ distribution for mobile calls in $DS1$ with $\Delta t=600$ seconds and derived the corresponding $p(n)$ function. In Fig.\ref{fig:BrstModel1}.a we show the $1-p(n)$ complement of the empirical memory function, with strong finite size effects (grey dots) and the same function after logarithmic binning (black dots) on which we fit the theoretical memory function defined in Eq~\ref{eq:pnbrst} using a non-linear least-squares method with only one free parameter, $\nu$. As seen in Fig.\ref{fig:BrstModel1}.a, we find that the best fit (red line) offers an excellent agreement with the empirical data with $\nu=2.971\pm 0.072$. Using the above mentioned exponent relation, this way we can estimate $\beta\simeq 3.971$, which is close to the empirical value $\beta\simeq 4.1$ already reported in Table~\ref{table:PEexp} and Fig~\ref{fig:peemp}a. To validate this approximation we generated bursty trains of $10^8$ events by using the theoretical memory function $p(n)$ (defined in Eq.~\ref{eq:pnbrst}) with exponent $\nu=2.971$ and compared the scaling of the generated $P(E)$ distribution to the corresponding empirical result. Results in Fig.\ref{fig:BrstModel1}.b evidently demonstrate a good match between the simulated and empirical results, thus in turn they validate the chosen analytical form for the memory function (for more results see ~\cite{karsai2011universal}).

\subsubsection{Reinforcement model of bursty dynamics}

Based on the above observations, we assume that the activity of an individual performing a task can be described with a two-state model, where a person can be in a normal state $A$, executing independent events with longer inter-event times, or in an excited (bursty) state $B$, performing correlated events with higher frequency. In this model we assume that inter-event times are determined by a reinforcement process, which dictates that the longer the system waits after an event, the larger the probability that it will keep waiting~\cite{stehle2010dynamical,zhao2011social}. Thus our two-state model is strongly non-Markovian as the timing of its events depends on the current and past states of the system. More precisely, given that our model system performed its last event $\tau$ time ago, the probability that it will wait one time unit longer without performing the next event is given as
\begin{equation}
 f_{A,B}(\tau)=\left( \dfrac{\tau}{\tau+1} \right)^{\mu_{A,B}}
 \label{eq:rfunc1}
\end{equation}
where $\mu_A$ and $\mu_B$ control the reinforcement dynamics and the characteristic inter-event times in state $A$ and $B$, respectively.

Finally, the model is defined as follows (for schematic demonstration see Fig.\ref{fig:BrstModel1}.c, while an algorithmic description see Alg.~\ref{alg:pememorymodel}): first the system performs an event in a randomly chosen initial state (line 2 in Alg.~\ref{alg:pememorymodel}). If the last event was in the normal state $A$, it waits for a time induced by $f_A(\tau)$ (line 7), after which it switches to state $B$ with probability $\pi$ and performs an excited event (line 9 and 10); or with probability $1-\pi$ stays in the normal state $A$ and executes a new normal event. In the excited state the inter-event time for the actual event comes from $f_B(\tau)$ after which the system executes one more excited event with a probability $p(n)$ (see Eq.~\ref{eq:pnbrst} and line 14 in Alg.~\ref{alg:pememorymodel}) that depends on the $n$ number of excited events since the last normal event; otherwise it switches back to a normal state with probability $1-p(n)$ (line 16).

\begin{algorithm}[h!]
  \caption{\small \label{alg:pememorymodel} Algorithmic description of the reinforcement model of bursty activity trains.}
\begin{algorithmic}[1]
\Statex
\Function{Bursty reinforcement model}{}
  	\State $\sigma \gets rand(A,B)$
	\State $n=1$
	\State $time=0$
  	\While{$time < T$}
		\If{$\sigma==A$}
			\State $time+=f_{A}(\tau)$
			\If{$rand()<\pi$}
				\State $\sigma \gets B$
				\State $n \gets 1$
			\EndIf
			\State $Out: (time,\sigma)$
		\ElsIf{$\sigma==B$}
			\State $time+=f_{B}(\tau)$
			\If{$rand()<p(n)$}
				\State $n \gets n+1$
			\Else
				\State $\sigma \gets A$
			\EndIf		
			\State $Out: (time,\sigma)$
		\EndIf
	\EndWhile
\EndFunction
\end{algorithmic}
\end{algorithm}

The results of the simulated model process, summarised in Fig.~\ref{fig:BrstModel1} and Table~\ref{table:PEexp}, indicates that the emergent inter-event time distribution appears with strong inhomogeneities (see Fig.\ref{fig:BrstModel1}.d). It can be approximated by a scale-free function with exponent $\alpha=1.3$, which satisfies the expected exponent relation $\alpha=\mu_A+1$, similar to the one derived in Eq.~\ref{eq:pnbrst}. Beyond temporal heterogeneities we detect emergent long temporal correlations reflected by the power-law tail of the autocorrelation function (see Fig.\ref{fig:BrstModel1}.e). It can be characterised by an exponent $\alpha=0.7$, which also satisfies the relation $\alpha+\gamma=2$ (as discussed in Section~\ref{sec:autocorr} and~\cite{karsai2011universal,vajna2013modelling}). Finally, the $P(E)$ distribution appears with a fat-tail for each investigated window size ranging from $\Delta t=1$ to $2^{10}$ (see Fig.\ref{fig:BrstModel1}.f), which can be characterised with an exponent $\beta=3.0$ in agreement with the expected relation in (Eq.~\ref{eq:pnbrst}). Note that the weak $\Delta t$ dependency of the simulated $P(E)$ can be explained by the merge of correlated long bursty trains and uncorrelated single events which is more common for larger $\Delta t$. Finally, once we fix the value of $\alpha$ and $\beta$, the emergent $\gamma$ exponent satisfies the inequality $\gamma < \alpha < \beta$,  observed in empirical data (see Table \ref{table:PEexp}).

\subsection{Model of communication balance of dyadic event trains}

\begin{figure}[!ht]
    \center
    \includegraphics[width=.9\columnwidth]{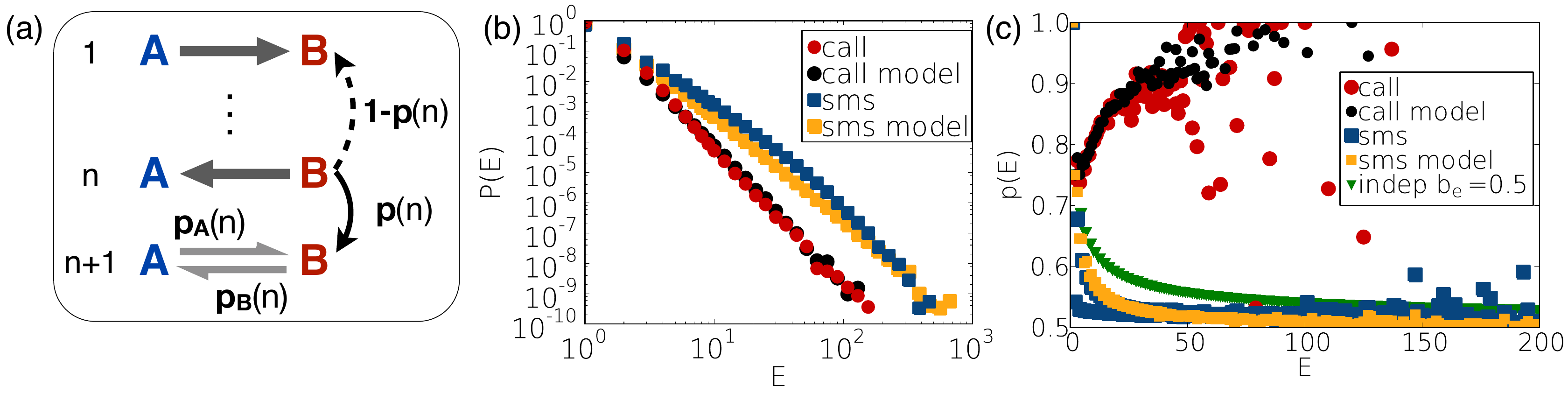}
    \caption{\small (a) Illustrative definition of the communication balance model, simulating events between two nodes $A$ and $B$. The dynamics and direction of the events are determined by probabilities $p(n)$, $p_A(n)$ and $p_B(n)$ defined in the text. (b) $P(E)$ distributions of empirical call trains (red circles) with $\Delta t=600s$ on edges with $0.5 \leq b_e < 0.55$ and in corresponding model trains (black circles). The same functions are shown for SMS trains (blue and yellow squares accordingly). (c) Balance values calculated for empirical call (red circles) and SMS (blue squares) trains and in corresponding model processes (red and yellow symbols). Balance values of independent trains are also shown (green triangles) calculated by using Eq.\ref{eq:prand} with $b_e=1/2$. This figure was published in~\cite{karsai2012correlated}.}
    \label{fig:BrstModel2}
\end{figure}

In an extended model definition, introduced in~\cite{karsai2012correlated}, we further used the $p(n)$ memory function (in Eq.~\ref{eq:pnbrst}) to model the observed cases of communication balance in call and SMS sequences as reported in Section~\ref{sec:balance}. To introduce this new model let us first concentrate on voice calls. One correlation we observed in Fig.~\ref{fig:DirB_corr}.f was that longer event trains tend to be more unbalanced, meaning that they are more dominated by events initiated by one of the callers. Keeping in mind that mobile calls enable bidirectional information change, we assume that the observed unbalanced communication in interaction trains reflects the difference in motivation between the communicating partners. If there is a task to solve, which is more important for one party, it gives motivation for him/her to repeate calls until the issue gets settled. This mechanisms can be incorporated into the reinforcement process demonstrated in Fig.\ref{fig:BrstModel2}.a, and can be summarised in the following way (note that its algorithmic solution is somewhat similar to Alg.~\ref{alg:pememorymodel} thus we do not present it here): We simulate bursty trains, which evolve on a link between a pair of individuals $A$ and $B$. To initiate a train with a probability equal to $b_e$ we randomly select $A$ or $B$ who then perform one event towards the other agent and we set the actual train size to $n=1$. The decision about the next event is carried out in two steps. First we decide with the probability in Eq.\ref{eq:pnbrst} whether to perform one more event in the train or initiate a new train otherwise. If the train should be continued the probabilities that the next event is initiated by $A$ or $B$ are
\begin{equation}
p_{\sigma}(n|\sigma_1)=\dfrac{n}{n+1} \mbox{\hspace{.2in} or \hspace{.2in}} p_{\sigma}(n|\neg\sigma_1)=1-\dfrac{n}{n+1}
\label{eq:pAcall}
\end{equation}
where $\sigma\in \{A,B\}$. Here $p_{\sigma}(n|\sigma_1)$ denotes the probability that the $n$th event of the actual train is performed by the same user who initiated the train at $n=1$, while $p_{\sigma}(n|\neg\sigma_1)$ is the probability that the other agent initiates the event. Consequently, the longer a train evolves, the larger is the probability that the agent, who initiated the actual train, will initiate the next event. Eq.\ref{eq:pnbrst} and Eq.\ref{eq:pAcall} may capture the entangled mechanisms of reinforced motivation of an individual, which is induced by the effort already invested in the actual series of calls to successfully solve a task with the other partner.

In case of SMS sequences the mechanism for developing strong balance in bursty trains is different. There, in single events information can pass only one way and consecutive events in a train usually have reversed direction, possibly forming strongly balanced conversations. To simulate this behaviour we use the above defined generative process but we select the direction of the actual event differently. Here we assume that the direction of an event conditioned only on the orientation of the previous event, and can be determined by the conditional probabilities
\begin{equation}
p_{\sigma}(n|\neg\sigma_{n-1})=\dfrac{n}{n+1} \mbox{, \hspace{.1in}} p_{\sigma}(n|\sigma_{n-1})=1-\dfrac{n}{n+1}
\label{eq:pAsms}
\end{equation}
where $\sigma \in \{ A,B \}$ and $p_{\sigma}(n|\neg\sigma_{n-1})$ denotes the probability to choose the opposite direction for the $n$th event compared to the one in the $(n-1)$th step. Accordingly $p_{\sigma}(n|\sigma_{n-1})$ denotes the probability of choosing the same direction as the previous event. This way, the longer a train evolves, the larger is the probability to revert the direction of consecutive events and consequently to generate more balanced train.

The emergence of enhanced balance/imbalance in trains can be evidently checked on links where the overall communication is completely balanced and the communication balance of trains comes only from actual behavioural differences. To do so we set $b_e=1/2$ and compare the modelled results to averages calculated for similar empirical trains. To analyse real trains we select edges from the mobile call network with overall balance $0.5 \leq b_e < 0.55$ (there are $115,277,534$ calls and $69,288,504$ SMS on such links) and after detecting trains we calculate the corresponding $P(E)$ distribution and $p(E)$ function (defined in Section~\ref{sec:balance}). As expected and shown in Fig.\ref{fig:BrstModel2}.b, the size of call trains (red circles) detected with $\Delta t=600 s$ and SMS trains (blue squares) with $\Delta t=300 s$ are distributed broadly with characteristic exponents $\beta=4.6$ and $\beta=3.5$, accordingly. Using these exponents as parameters we modelled event sequences to simulate calls and SMS trains with the same number of events and corresponding $\nu$ exponents deduced from $\beta$ according to Eq.\ref{eq:pnbrst}. The $P(E)$ size distributions of model call trains (black circles) and model SMS trains (yellow squares) collapse surprisingly well on the corresponding empirical functions, as shown in Fig.\ref{fig:BrstModel2}.b.

At the same time, in Fig.~\ref{fig:BrstModel2}.c the $p(E)$ balance functions calculated for the limited event sets on fully balanced links show surprisingly similar behaviour to the overall averages (seen in Fig.~\ref{fig:DirB_corr}.g). This demonstrates that even an overall balanced link, strong communication imbalances appear due to local correlations within one bursty train. This is even more striking if we compare the empirical P(E) curves to the corresponding independent one (green triangles in Fig.\ref{fig:BrstModel2}.c) which was generated using Eq.\ref{eq:prand} with $b_e=1/2$. Moreover, in Fig.~\ref{fig:BrstModel2}, the average $p(E)$ balance functions emerging without parameters for model trains are in surprisingly good agreement with the empirical observations. Consequently, the assumed mechanisms defined in Eq.~\ref{eq:pAcall} and~\ref{eq:pAsms} are capturing rather accurately the salient features of the dynamics of directed human communication through phone calls and SMS. The only discrepancy is for the $p(E)$ values of short SMS trains, where the empirical data show an even-odd effect, which is not reproduced in the model. This indicates that possibly other mechanism may be present in the communication sequence what are not considered in this modelling study.



\section{Conclusions}

In this Chapter, I summarised my most important results in one of my main research domains on bursty human dynamics. After a brief overview of the field, first I introduced some basic characteristic measures, some of them defined by me and colleagues, which are commonly used to quantify heterogeneous patterns in human dynamics. Subsequently, I systematically walked through a series of studies I published over the last years for the advanced characterisation, observation, and modelling of bursty human dynamics.

Due to the diverse experiences, broad overview, and devoted interest towards this field, together with Dr. Hang-Hyun Jo and Pr. Kimmo Kaski, we recently took the timely opportunity to write a monograph book, entitled as ''Bursty Human Dynamics``, to review all relevant knowledge on the field cumulated over the last decade. This book has been published by Springer in January 2018.

Finally note that some of my studies published on the system-level observations, modelling, and effects of bursty dynamical patterns are not mentioned in this Chapter as they land close to field of temporal networks, which is the topic of the coming Chapter.

\biblio





\chapter{Temporal Networks}
\label{ch:tnet}

\section{Introduction}
\label{sec:tnet_intro}

The success of Networks Science is built on the operational representation of complex systems as graphs, which in turn can be quantified, observed, and modelled for the better understanding of emerging phenomena~\cite{newman2010networks}. From the early years of the field a common simplifying assumption has been taken to neglect that real networks may evolve in time, and may consist of nodes and links of different types. Although these approximations were obviously vague in various cases, yet the static/monolayer network approach was extremely successful to obtain crucial knowledge about several empirical systems. However, after a decade the field became advanced enough to overcome these limitations and to develop the contemporary domains of \emph{temporal networks} to consider the dynamic nature of interactions~\cite{holme2012temporal,masuda2016guide}, and \emph{multilayer networks} to take into account the multiplicity of interaction types~\cite{kivela2014multilayer,boccaletti2014structure}. These recent developments were fuelled by the ever increasing network data available, which more-and-more commonly comes in the form of time-stamped interaction sequences and/or with metadata recording details on nodes and links. Advancements in data collection and data sharing together with the novel theoretical foundation of these domains help us to break the glass ceiling of the static/monolayer network picture and push us to think about networks as dynamical systems with various types of interactions and agents. In this Chapter we concentrate on temporal networks as I mainly contributed to the theoretical, methodological, and empirical foundation of this domain. First we lay down some general thoughts on temporal networks, which after we will introduce a set of representations and characteristic measures which are necessary to understand the body of this Chapter. Subsequently, we will discuss four main directions of my contributions on to the system level characteristic, random reference models, higher order representations and generative modelling of temporal networks.

\begin{figure}[!ht]
\centering
  \includegraphics[width=1.\textwidth]{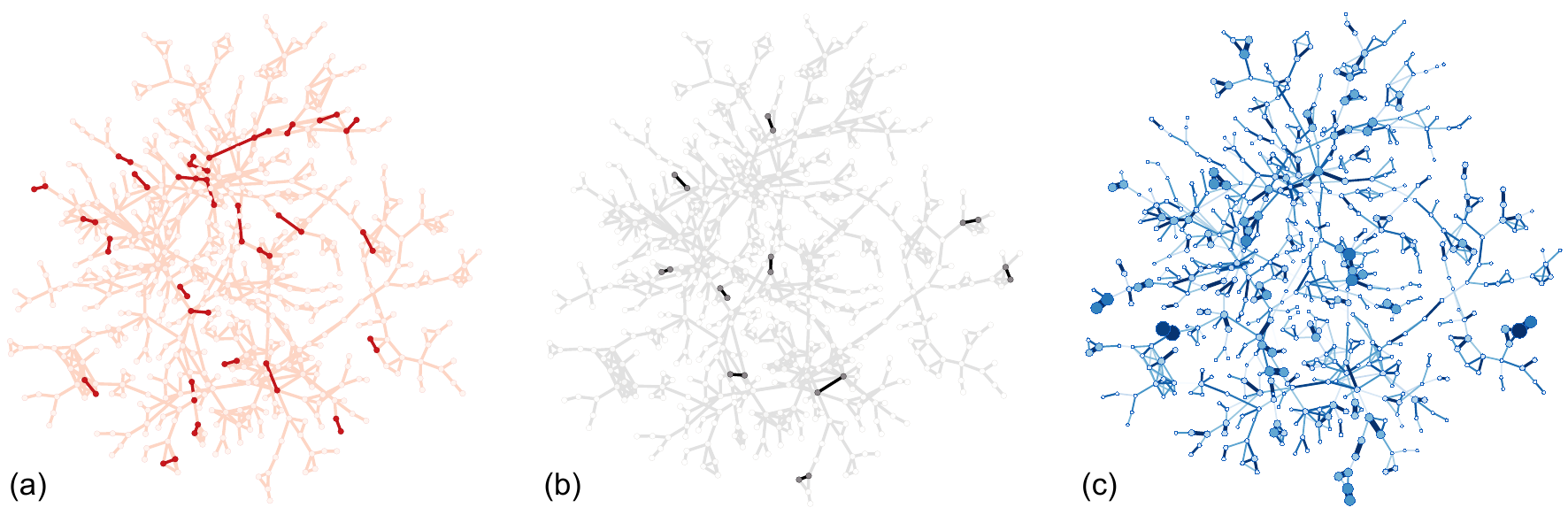}
\caption{\small Dynamics of a mobile call network. Panels (a), and (b) show calls over $3$ hours between people in the same town at two different time stamps. Panel (c) presents the backgrounding weighted social network structure, which was recorded by aggregating interactions evolved between people during $6$ months. Node size and colours describe the activity of users, while link width and colour represent weight. This figure was published in~\cite{karsai2014time}.}
\label{fig:tnet_schem}
\end{figure}

\subsection{Static vs. Temporal Networks}

Most network structures are the results of some emergent phenomena, consequently they continuously evolve in time, or dynamical evolution was necessary at some point of their existence. This way, by only looking at the static description of their aggregated final or actual state we may obtain limited understanding about their "morphogenesis" and emergent properties. A good example is the commonly observed degree heterogeneity, which has been argued to emerge due to some reinforcement mechanisms (preferential attachment, Matthew process, etc) driving the interaction dynamics in a network~\cite{barabasi1999emergence}. Other examples are degree correlations, weight heterogeneities, network communities, etc., which all can be observed in a static network as emergent properties, due to mechanisms driving the underlying network dynamics at the first place. Temporal networks contributes to their understanding by studying the network dynamics at the level of rapid recurrent interactions between nodes~\cite{holme2012temporal}. As we will see later, this level of description can be used to identify important mechanisms and correlations between single events of interactions~\cite{karsai2011small,laurent2015calls,karsai2014time}, which leads to emergent heterogeneous properties on the aggregated scale (as demonstrated in Fig.~\ref{fig:tnet_schem}). Taking interactions as static links between nodes subsequently entails that information can flow between the connected peers at any time~\cite{karsai2014time}. This is evidently not true in many cases, like in human communication networks~\cite{karsai2011small,kivela2017mapping}, where information can be passed between nodes only at the time of interactions and only between the interacting nodes. This way, information flow, and as a consequence the emergence of any macroscopic phenomena on networks, are limited by the emerging time respecting paths determined by the the timing, direction, and ordering of events~\cite{kivela2017mapping}. This puts network analysis in a rather different perspective, as all characteristic network properties, centrality measures, or structural properties become time dependent~\cite{zhao2011entropy,krings2012effects}. On the other hand, better understanding of time respecting paths opens the door to more realistic simulations of dynamical processes, like epidemic or information spreading, diffusion processes, opinion dynamics, or synchronisation, which are actually crucially altered by the time-varying nature of interactions (for further discussion on this matter see Section~\ref{sec:ADNmemory}).


\subsection{Time-scales of network dynamics}
\label{sec:ts}

\begin{figure}[!ht]
\centering
  \includegraphics[width=1.\textwidth]{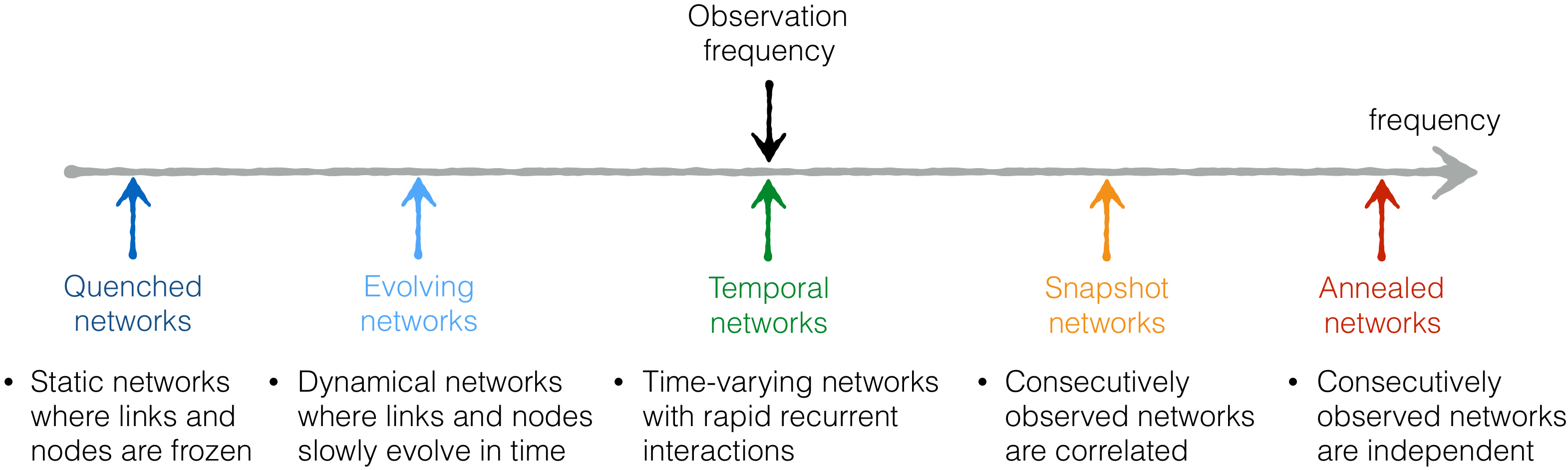}
\caption{\small Relative times-scales of network dynamics as compared to the time-scale of observations.}
\label{fig:tnet_scales}
\end{figure}

Taking an observer point of view, a network can dynamically evolve in various temporal scales relative to the observation frequency~\cite{ribeiro2013quantifying}. The typical cases are demonstrated schematically in Fig.~\ref{fig:tnet_scales}, where we fix the observation frequency and vary the relative time-scales between two extremes. On one end of this scale we find the so called \emph{quenched networks}, which are invariant in time or evolve on such a slow pace that appears to be invariant from the observer point of view. Static networks are belonging in this category with frozen nodes and links, which are always present in the structure. Once we consider that nodes and links may be created and deleted in the network, but yet on a slower temporal scale as the observation, we arrive to the \emph{evolving network} picture~\cite{albert2000topology}. Best examples for this representation are social networks where individuals may born and die and social ties are created and broken but all being present for longer periods lasting for several observations. Other examples are the internet, or other infrastructure networks. Once the observation frequency closely matches the temporal scale on which the network is evolving we arrive to \emph{temporal networks}~\cite{holme2012temporal}. Here we consider rapid and potentially recurrent time-varying interactions between nodes like emails or phone calls in social communication networks, or sexual contact networks. Note that temporal networks contain all information provided by the other network pictures evolving slower than the observation frequency. One can take first and last observations of interactions between nodes to reconstruct the evolving network representation, while aggregating interactions over time would give us the static network description of a time-varying network~\cite{holme2012temporal}. However, we loose this advantage once the network is evolving faster than we can observe. In this case observations provide \emph{snapshot networks}, which aggregate the structure of the time-varying network over short consecutive periods. If the time-scale difference is still not too large, consecutive snapshots are not entirely independent from each other and can be used as an approximate method to represent temporal networks. For a good example see Fig.~\ref{fig:tnet_schem}.a and b where two snapshots of a mobile phone call network is shown. On the other hand, once the network is evolving on a way faster time-scale than we can observe we arrive to the \emph{annealed network} picture in which case consecutive observations of the network are weakly correlated or entirely independent~\cite{noh2009critical,ferreira2011quasistationary}. Neural networks  in the brain are good example for this case, where imaging technologies have not reached yet the temporal (neither spatial) resolution necessary to observe the precise electric signals running between neurons. Also note that the so called \emph{mean-field approximation} becomes valid in this extreme.

\subsection{Representations of temporal networks}
\label{sec:tempnetrepr}

In order to study temporal networks we need to introduce a representation, which can be further used to define general measures and which maps between a mathematical description and the common data collection format of time-varying interactions. The most straightforward description, which we will consistently use later on, is the event list (also called event-based) representation. To introduce, let us consider a temporal network as a set $G_t=(V,E_t,T)$ defined as sets of vertices $v\in V$, and edges defined as a set of events $E_t \subset V \times V \times [0,T]$ over a time period $T$. This way a temporal network is described as a sequence of events, which in their simplest form appear as triplets $(u, v, t)$ indicating an interaction between nodes $u$ and $v$ at time $t \in [0,T]$. Note that an event can be directed or undirected depending on whether we consider $V\times V$ as an ordered or unordered set. Furthermore, this way of representation allows for both continuous and discrete time description of temporal networks. As data commonly comes in a discrete time format, we chose this as our default assumption and we consider events undirected if not mentioned otherwise. This representation can be further extended by defining events as $E_t \subset V \times V \times [0,T] \times \prod_i A_i^e$ where $A_i^e$ is a set of meta informations like duration, cost, or the location of participating nodes, which may extend the description of each event depending on the actual data.

Several other representations of temporal networks has been provided over the last couple of years~\cite{holme2012temporal,masuda2016guide}. One represents temporal networks as a discrete time sequence of snapshot networks captured by adjacency matrices $\mathcal{A}=\{ A(t) \}$. Each matrix $A(t)=|V|\times |V|$ aggregates all events, which are present between nodes at time $t$. However, since the ordering of events within one snapshot is not determined, this representation does not contain all information about the original temporal structure, and the information loss depends on the relative time-scales of the snapshot time window and the evolving network. Furthermore, the notion of duration in this representation is not straightforward. If events have shorter duration than the sampling frequency we may loose information again. Or in the contrary, if events have long duration, they appear as multiple events in several consecutive time windows. Another recently proposed representation is so called \emph{link streams} of a temporal network~\cite{latapy2017stream}. This description aims to provide the most general representation of dynamic networks by using stream graphs defined as a $S=(V, W, E, T)$, where $W$ being the set of temporal nodes $W\subset V\times [0,T]$. In case we assume that $W$ is time invariant, i.e., all nodes are present always, we receive the definition of link streams which provides a similar description with the even list representation discussed above. As follows we are going to mostly apply the even list representation, while in Section~\ref{sec:wDAGs} we will propose a new lossless representation of temporal networks as static weighted directed acyclic graphs.

\subsection{Some characteristic measures of temporal networks}

There are several ways to extend the definition of characteristic measures of static networks (e.g. ones mentioned Section~\ref{sec:cn}) for temporal structures. Here, without aiming a complete review, we will summarise some concepts and metrics, which will be frequently used in the forthcoming Chapters for the analysis of time-varying networks. For a more complete list of definitions we refer the reader to recent review articles and books like~\cite{holme2012temporal,masuda2016guide,latapy2017stream}.

\subsubsection{Dynamical measures of characteristic metrics}

The first thing to notice in case of temporal networks is that characterising metrics, like degree, strength, weight, clustering coefficient, or centrality measures may vary in time. However, despite their timely variance, their distributions or average values may reach stationary values. The most straightforward way to capture their dynamics is via assuming a time aggregation process. There, initially at $t=0$, we have an empty graph with $N$ nodes and take each event in a timely order to build an aggregated static network $G_{[0,t]}=(V,E_{[0,t]})$. At time $t$ this static structure will have links induced by interactions appeared between $[0,t]$ and can have link weights defined as the number of interactions, the sum of event durations etc. Measuring the average or the distribution of characteristic network metrics in the aggregated network $G_{[0,t]}$ as $t\rightarrow T$ give us some information about the evolution of the network and the emergence of its stationary properties~\cite{krings2012effects}. Note that in the limit $t=T$ the representations $G_{[0,T]}\equiv G$ are equivalent. There are several other ways to measure general characteristics, like via the analysis of networks observed in consecutive discrete aggregated snapshots. However, these methodologies are not discussed here as they are not necessary for the understanding of the rest of the thesis.

\subsubsection{Path-based measures}
\label{sec:paths}

Some of the most important characters of temporal networks build on the concept of temporal paths (also called time-respecting paths). Paths in temporal networks have rather different properties as compared to static networks. Most importantly, they fully determine the outcome of any collective phenomena as they constrain possible information flow between nodes. However, before we ground the definition of temporal paths we need to introduce some other concepts related to event adjacency and temporal walks. In the following we are going to use the event based representation of an undirected temporal networks, but all definitions are generalisable for other representations as well.

\emph{Event adjacency} defines a directed relationships between events which are performed on neighbouring links. More precisely, taking two events $e_1=(v_i,v_j,t_1)$ and $e_2=(v_j,v_k,t_2)$ from a temporal network (with $v_i,v_j,v_k \in V$ and $t_1,t_2\in T$), the event $e_1$ is adjacent to event $e_2$ if they share at least one common node and $t_1<t_2$. Next, let's imagine a walker on the temporal network, which can pass between nodes but only at the time of their interactions while respecting the ordering of events, i.e., it can use only events from the future. Then, a temporal walk between two nodes $v_0$ and $v_n$ is defined as a sequence of $\{ (v_0,v_1,t_1),(v_1,v_2,t_2) \dots (v_{n-1},v_n,t_n) \}$ of adjacent events, where $t_i<t_{i+1}$ ($0 \leq i \leq n-1$). In general, a temporal walk depends on the time $t_1$ when it starts, it can visit a node multiple times, and its length is defined as $t_n-t_1$. Furthermore, a temporal walk is non-symmetric (even in an undirected temporal network), non-transitive, and cannot be infinitely long (for further explanation of these properties see~\cite{masuda2016guide}). Subsequently, we can define the concept of \emph{reachability}, i.e., a node $v_j$ is reachable from a node $v_i$ if there exists a temporal walk from $v_i$ to $v_j$. We call a temporal walk between two nodes to be a \emph{temporal path} if the walker visits one node maximum once (thus no node is visited multiple times during the walk).


Finally, we introduce the concept of \emph{connected components} in temporal networks, which is different from the corresponding static network definition due to the non-symmetric property of temporal paths. A connected component in a temporal network is a maximal set of nodes in which each pair of nodes is temporally connected. Following this definition one can define weakly and strongly connected components, but one needs to keep in mind that a node can participate in multiple connected components depending on the temporal paths in which it participates. This way connected components do not provide a well defined node-partitioning in temporal networks. In terms of computation, the complexity to find all connected components in a temporal network is an $\mathcal{O}(N^2)$ problem compared to the static network solution with $\mathcal{O}(M)$ complexity. Finding all temporal paths in a temporal network has been an $NP$-complete problem. However, in a recent paper~\cite{kivela2017mapping} we provided some better algorithmic solution what will discuss in Section~\ref{sec:wDAGs}, where we will introduce a new, higher-order representation of temporal networks together with new definitions of connected components.


\section{System level characterisation}
\label{sec:tnet_sys}

\subsection{Aggregation time for social communication networks}

After this brief introduction to temporal networks first we will discuss some results characterising time-varying networks on the system level. As we have discussed earlier, temporal aggregation of time-varying interactions provides a static, weighted network representation, which contains no temporal informations, but codes the underlying structure. This aggregated structure may have various heterogeneous properties and may emerge with communities and other consequences of structural correlations. However, when taking a sequence of time-varying interactions it is not clear how long one needs to aggregate in order to capture the most important static network properties.

Next, we address this question by monitoring and analysing the features of network structure emerging from aggregation over different time intervals for an empirical data set of mobile phone communication~\cite{krings2012effects}. We are interested in the effects of the aggregation window on the structural features of human communication networks that are known to display dynamics on multiple overlapping time scales. The data comes in the form of a time-stamped sequence of mobile telephone calls between anonymised customers of a Belgian mobile operator for a period of 6 months. Note that although it has been collected in another country, it is very similar to DS1 in Section~\ref{sec:datasets} (for more about the data see~\cite{krings2012effects}).

The aggregated network possesses all structural heterogeneities typically characterising social communication networks. It appears with a broad degree, weight and node strength distribution as reported in~\cite{krings2012effects}. As a first approximation, let's characterise its dynamics by simply measuring the evolution of basic network characteristics. As shown in~\cite{krings2012effects}, the number of nodes or links, the average node degree or link weight all increase monotonously in time without indicating any characteristic time scale. Moreover, the interaction dynamics clearly displays the usual daily and weekly pattern, and it appears bursty, indicated by the long tail of the distribution of inter-even times measured between consecutive calls on individual edges. 

\begin{figure}[!ht]
\centering
  \includegraphics[width=1.\textwidth]{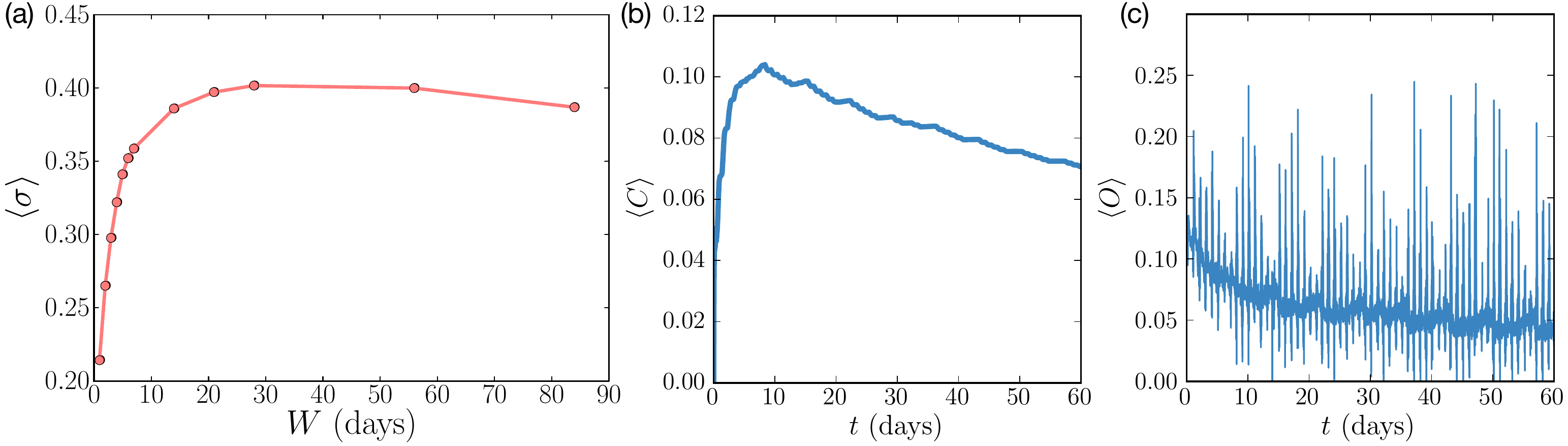}
\caption{\small (a) The average fraction of links $f$ common to consecutive aggregation windows of duration $W$. (b) Global clustering coefficient and (c) average final overlap of added edges as a function of aggregation time, for the first 2 months. This figure was prepared by G. Krings and published in~\cite{krings2012effects}.}
\label{fig:tnet_krings1}
\end{figure}

Further observations suggest that links in a social network appear with various time span. While some of them are stable and remain active for prolonged periods of time, others exist or can be detected only within limited time periods. In reflection, the aggregation window length should obtain a representative, “backbone” networks that capture the stablest connections in the system? Thus, we compare the similarity of networks aggregated for different periods of time when the observation period is divided into multiple consecutive aggregation windows. We then calculate the similarity $\sigma$ of two consecutive networks $G_1=(V_1,E_1)$ and $G_2=(V_2,E_2)$ as
\begin{equation} \label{eq:sim}
\sigma(G_1,G_2) = \frac{|E_1 \cap E_2|}{|E_1 \cup E_2|},
\end{equation}
such that $\sigma=1$ if the networks are the same, and $\sigma=0$ if they share no links. In Fig.~\ref{fig:tnet_krings1}a, the average similarity $\sigma$, computed for different durations $W$, indicates small similarity if the windows are very short, as the networks are very sparse with only a few common links. Then, the similarity increases with increasing window duration, reaching a maximum at $\sim 30$ days; subsequently, the similarity begins to decrease slowly as the aggregation process captures more and more of spuriously appearing very weak ties or random links. 

To better understand the network evolution it is important to learn about the characteristics of links that emerge early on in the process. As we have already discussed in Section~\ref{sec:socialnets}, the Granovetter hypothesis suggests that link weights correlate with the network topology such that high-weight links are associated with dense network neighbourhoods, whereas low-weight links connect such neighbourhoods. This is directly related to the presence of community structure~\cite{fortunato2010community} in social networks; links within communities are stronger and have higher-than-average weights~\cite{tibely2011communities}. For the network aggregation, this means that clusters and communities are likely to appear early on in the process. In order to investigate this effect, we measure the evolution of the global clustering coefficient $C(t)$ and the average final overlap values $\langle O \rangle (t)$ of added links (both defined in Section~\ref{sec:cn}) as the function of aggregation time. As seen in Fig.~\ref{fig:tnet_krings1}b, the clustering coefficient does indeed increase initially rapidly, and then decreases after a peak at around $t=9$ days. This decrease can be attributed to the weak links observed later in the process: those links contribute less frequently to triangles. For the case of the average overlap in Fig.~\ref{fig:tnet_krings1}c, the early observed links have on average a higher overlap than those added later; the final overlap is a decreasing function. Hence, even when the aggregation times are short, the networks capture features of the community structure of the final aggregated networks. Interestingly, the overlap also shows a strong circadian and weekly pattern -- its highest peaks correspond to the early morning when the overall call rate is very low. Thus, if calls are made during these hours, they are likely to be targeted towards people in the strongest clusters of friends and family. 

\begin{figure}[!ht]
\centering
  \includegraphics[width=1.\textwidth]{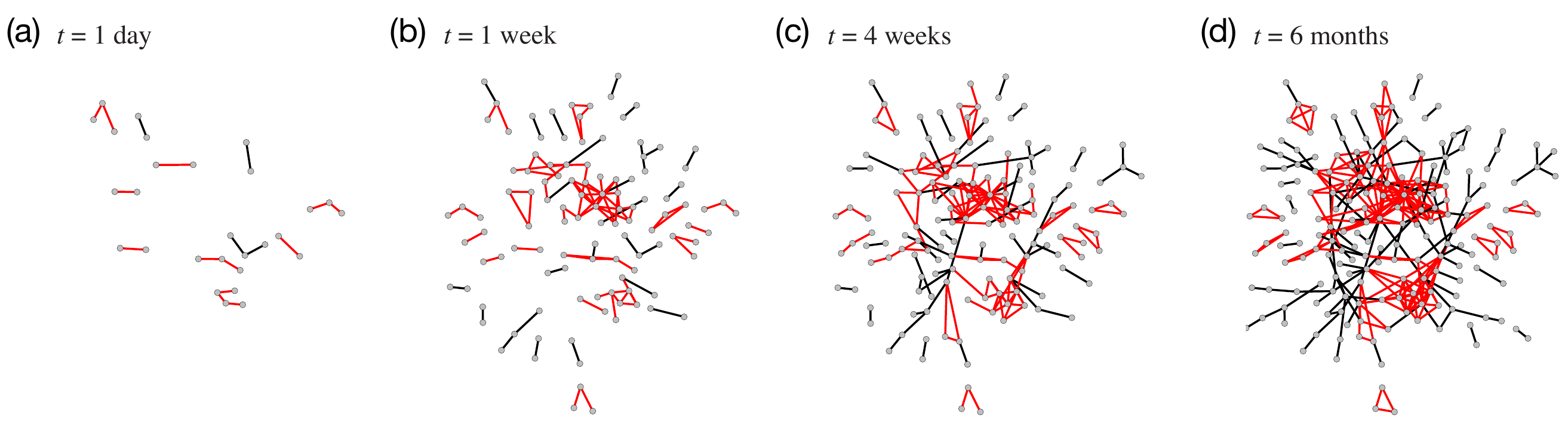}
\caption{\small Series of aggregated networks with a growing aggregation interval. Networks here represent small subnetworks of individuals from a single postal code. Links that participate in triangles in the final 6-month aggregated network are coloured red, while the rest are black. This figure was prepared by G. Krings and published in~\cite{krings2012effects}.}
\label{fig:tnet_krings2}
\end{figure}

In order to illustrate the network growth, we have visualised small subnetworks corresponding to different aggregation times $t$ of people from a single town (see Fig.~\ref{fig:tnet_krings2}). Panels a) to d) of Fig.~\ref{fig:tnet_krings2} show that at the early times of the observation, those edges appear mostly which participate in triangles in the final aggregated network (coloured in red). These edges are the ones forming communities and clusters. On the other hand, not all intra-community edges are discovered early on; rather, those links that appear early are associated with communities with a high probability.

Further investigating the statistical distributions of node degrees and link weights, we observed (not shown here but presented in~\cite{krings2012effects}) that after a short initial transition period, sufficient amount of data has been collected to correctly estimate the stationary degree distribution of the network, and the shape of the distribution remains similar for longer intervals longer than $3$ weeks. The weight distribution displays a slightly slower convergence, and is still slowly changing even for aggregation intervals of around $3$ weeks.

\subsection{Entropy of Dynamical Social Networks}

To characterise temporal networks on the system level we turn to a more complex measure capturing the entropy of time-varying information coded in the interaction dynamics of individuals. Network entropy has been earlier introduced for static networks~\cite{eckmann2004entropy,bianconi2007entropy,anand2009entropy}, but its first definition for temporal networks was proposed by us in~\cite{zhao2011entropy}. We have seen through several earlier examples that human activities are commonly bursty and not Poissonian and modulated by periodic daily (circadian) or weakly patterns. To understand the convoluted effects of these characters, our question here is: How much can humans intentionally change the statistics of social interactions and the level of information encoded in the dynamics of their social networks? To answer this question, through the analysis of a mobile phone-call network (DS1 in Section~\ref{sec:datasets}), we show that the entropy of dynamical networks is able to quantify the information encoded in the dynamics of time-varying interactions.

\subsubsection*{The entropy measure}

\begin{figure}[!ht]
\begin{center}
\includegraphics[width=1.\textwidth]{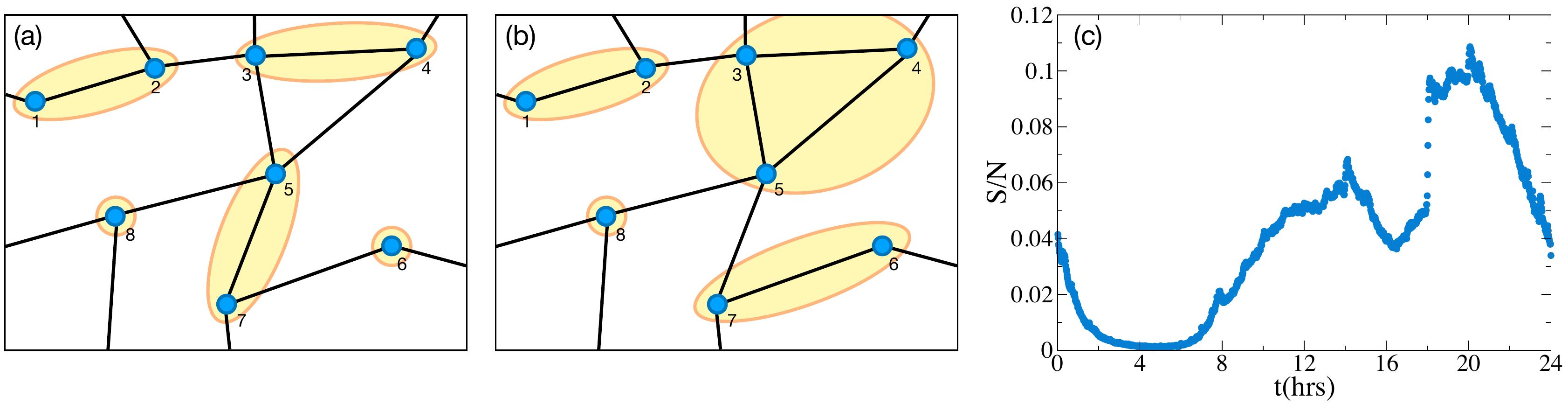}
\end{center}
\caption{\small  The dynamical social networks are composed by different dynamically changing groups of interacting agents. In panel (a) we allow only for groups of size one or two as it typically happens in mobile phone communication. In panel (b) we allow for groups of any size as in face-to-face interactions. (c) Mean-field evaluation of the entropy of the dynamical social networks of phone calls communication in a typical week-day. In the nights the social dynamical network is more predictable. This figure was published in~\cite{zhao2011entropy}.}
\label{fig:entropy}
\end{figure}

Here we introduce the entropy of dynamical social networks as a measure of information encoded in their dynamics~\cite{zhao2011entropy}. We assume to have a quenched network $G$ formed by $N$ agents and we allow a dynamics of interactions on this network. If two agents $i,j$ are linked in the network they interact at each given time giving rise to the time-varying network. In the network $G$, agents $i_1,i_2,\ldots i_n$ can interact in a group of size $n$. Therefore at any given time the static network $G$ will be partitioned in connected components or groups of interacting agents as shown in Fig \ref{fig:entropy}a or b. In order to indicate that an interaction is occurring at time $t$ in the group of agents $i_1,i_2,\ldots , i_n$ and that these agents are not interacting with other agents, we write $g_{i_1,i_2,\ldots , i_n}(t)=1$ otherwise we put $g_{i_1,i_2,\ldots, i_n}(t)=0$. Therefore each agent is interacting with one group of size $n>1$ or non interacting (interacting with a group of size $n=1$). Consequently at any given time, the condition
\begin{equation}
\sum_{{\cal G}=(i,i_2,\ldots,i_n )|i\in {\cal G}}g_{i,i_2,\ldots, i_n}(t)=1.
\end{equation} 
has to be valid, where we  indicate with ${\cal G}$ an arbitrary connected subgraph of $G$. The history ${\cal S}_t$ of the dynamical social network is given by ${\cal S}_t=\{g_{i_1,i_2,\ldots, i_n}(t^{\prime})\, \forall t^{\prime}<t\}$. If we indicated by $p(g_{i_1,i_2,\ldots, i_n}(t)=1|{\cal S}_t)$ the probability that $g_{i_1,i_2,\ldots, i_n}(t)=1$ given the story ${\cal S}_t$, the likelihood that at time $t$ the dynamical networks has a group configuration $g_{i_1,i_2,\ldots,i_n}(t)$ is given by 
\begin{equation}
{\cal L}=\prod_{{\cal G}} p(g_{i_1,i_2,\ldots, i_n}(t)=1|{\cal S}_t)^{g_{i_1,i_2,\ldots, i_{n}}(t)} 
\end{equation}

The entropy $S$ characterises the logarithm of the typical number of different group configurations that can be expected in the dynamical network model at time $t$ and is given by $S=-\langle \log{\cal L}\rangle_{|{\cal S}_t}$ that we can explicitly express as
\begin{equation}
S=- \sum_{{\cal G}}p(g_{i_1,i_2,\ldots, i_n}(t)=1|{\cal S}_t)\log p(g_{i_1,i_2,\ldots, i_n}(t)=1|{\cal S}_t).
\end{equation}
According to the information theory results \cite{cover2006elements}, if the entropy is vanishing, i.e. $S=0$ the network dynamics is regular and perfectly predictable, while if the entropy is larger, the number of future possible configurations is growing and the system is less predictable. In general, we have to allow the possible formation of groups of any size. However, if we model the mobile phone communication,  we need to allow only for pairwise interactions. Therefore, if we define the adjacency matrix of the network $G$ as the matrix $a_{ij}$,  the log likelihood takes the very simple expression given by 
\begin{equation}
{\cal L}=\prod_i p(g_i(t)=1|{\cal S}_t)^{g_i(t)}\prod_{ij|a_{ij}=1} p(g_{ij}(t)=1|{\cal S}_t)^{g_{ij}(t)} \hspace{.2in} \mbox{with} \hspace{.2in} g_i(t)+\sum_{j} a_{ij} g_{ij}(t)=1,
\end{equation}
for every time $t$. The entropy is then given by 
\begin{equation}
S=- \sum_i p(g_{i}(t)=1|{\cal S}_t)\log p(g_{i}(t)=1|{\cal S}_t)\nonumber -\sum_{ij}a_{ij} p(g_{ij}(t)=1|{\cal S}_t)\log p(g_{ij}(t)=1|{\cal S}_t).
\end{equation}

\subsubsection*{Social dynamics and entropy of phone call interactions}

For demonstration we have analysed the mobile phone-call sequences recorded in DS1. For the entropy calculation we selected $562,337$ users who executed at least one call per a day during a week period and we have studied how the entropy of this dynamical network is affected by circadian rhythms. We assigned to each agent $i=1,2$ a number $n_i=1,2$ indicating the size of the group where she belongs. If an agent $i$
had coordination number $n_i=1$ she is isolated, and if $n_i=2$ she was interacting with a group of $n=2$ agents. We also assigned to each agent $i$ the variable $t_i$ indicating the last time at which the coordination number $n_i$ has changed. If we neglect the feature of the nodes, the most simple transition probabilities that includes for some memory effects present in the data, is given by a probability $p_n=p_n(\tau,t)$ for an agent in state $n$ at time $t$  to change her state given that she has been in her current state for a duration  $\tau=t-t_i$.

We have estimated the probability $p_n(\tau,t)$ in a typical week-day. Using the data on the probabilities $p_n(\tau,t)$  we have calculated the entropy, estimated by a mean-field evaluation (for more on this see~\cite{zhao2011entropy}) of the dynamical network as a function of time in a typical week-day. The entropy of the dynamical social network is reported in Fig. \ref{fig:entropy}c. It significantly changes during the day describing the fact that  the predictability of the phone-call networks change as a function of time. In fact, as if the  entropy of the dynamical network is smaller and the network is in a more predictable state.

\section{Random Reference Models}
\label{sec:tnet_rrm}

Human actions and interactions are driven by various intrinsic decision mechanisms and influenced by several environmental factors. As the consequence of these convoluted processes various correlations appear at the phenomenological level, which in turn characterise the structure and dynamics of the emerging temporal network. Differentiation of these correlations is not only important to identify and understand the mechanisms underlying the network evolution, but because they have important consequences on the ongoing dynamics processes like in case of spreading phenomena. Spreading processes are relevant for a number of fields and applications ranging from epidemiology of biological viruses to the dynamics of social processes, such as opinion dynamics and information transmission~\cite{barrat2008dynamical}. While certain static characteristics of complex networks work to enhance spreading, such as the small-world or the scale-free properties, it has been shown that the temporal characteristics of links may slow spreading down~\cite{vazquez2007impact,karsai2011small,kivela2017mapping,miritello2011dynamical}. These results indicate that dynamical processes cannot necessarily take advantage of topologically characters~\cite{pan2011path} but they are determined by the combination of structural and temporal characters of the underpinning temporal network.

For static networks, a common way to assess the significance of chosen topological features is to compare their abundance or characteristics against some reference model where the network is randomised. This approach has also been applied in assessing the importance of such features for dynamical processes. The most widely applied reference model is the \emph{configuration network model}~\cite{molloy1995critical}, where the links of the original network are rewired pairwise randomly. This reference model preserves the original degree sequence but yields networks that are as random as the degree sequence allows. Then, one can assess the significance of topological characteristics of the empirical graph, e.g. by measuring the extent to which the dynamics of some processes differ when they take place on the original networks or the reference ensemble.

Our aim in this Section is to introduce \emph{random reference models} (RRMs) for temporal networks. In this case, the original event sequences are randomised or randomly reshuffled to remove time-domain structure and correlations~\cite{karsai2011small,kivela2012multiscale,miritello2011dynamical,holme2005network,gauvin2018randomized}. Thus a reference model for an empirical event sequence is a maximally random (micro-canonical) ensemble of event sequences, for which some predefined set of properties are the same as for the empirical sequence. There are various kinds of temporal correlations such like burstiness, causal events, or circadian fluctuations, etc., and thus no single, general-purpose reference model (a ‘temporal configuration model') can be designed. Rather, by applying appropriate reference models, one may switch off correlations of selected types in order to understand their contribution to the observed properties. Here, we will briefly introduce the methodological concept, a canonical naming convention and a hierarchical organisation of RRMs~\cite{gauvin2018randomized} and will apply some of these models on an empirical temporal network to demonstrate their potential in identifying important structural and temporal correlations controlling spreading dynamics~\cite{karsai2011small,kivela2012multiscale}.

\subsection{Naming and and organisation of RRMs}

One can think about random reference models as constrained shuffling methods of interaction sequences, where some property of the original temporal network is retained, otherwise the shuffled sequence is maximally random. Equivalently, such methods sample uniformly from the micro-canonical ensemble of all networks, which have a given set of features constrained to the same value as that of the original data, but have maximum entropy otherwise~\cite{kivela2012multiscale}\footnote{Note, that it is possible to define grand canonical ensembles of temporal networks, but as most of the published methods (including the ones discussed in this Chapter) fall within the micro-canonical definition, we limit our discussion to these cases.}. Consequently, a RRM is determined by the temporal network and a set of selected constrains, which can be related to temporal, structural, or environmental properties, or can be specific to the actual representation of the network. The sequence $\mathbf{t}$ of interaction times, the sequence of $\mathbf{k}$ node degrees, or the $p(\mathbf{w})$ distribution of number of interactions on single links are all good examples for such constraints, which in turn can be used for naming RRMs. For example, the simple random reference model, $P[ E ]$, constrains only the total number of events $E$ in the network, and permutes all the instantaneous events at random otherwise. On the other hand RRMs can be defined by constraining on multiple characters, or via the intersection of multiple ways of shuffling. The $P[ \mathbf{w},\mathbf{t}]$ 'time-shuffling' RRM, randomly permutes the timestamps $t$ of all events, while keeping the participating nodes of the events fixed. Completely equivalently, we may define the timestamp shuffling by constraining the timestamps $t$ and permuting the pairs of interacting nodes among events. Due to the indistinguishability of networks obtained through permutation of event indices or times, both are equivalent to conserving number of interactions on each link $\mathbf{w}$ and the global sequence of interaction times $\mathbf{t}$. This framework enables us to build a taxonomy of existing RRMs, which lists their effects on important temporal network features and (partially) orders them by the amount of features they constrain. This hierarchy allows to apply RRMs so that the fixed features of the original data are systematically reduced. Without going into details, we refer the interested reader to our recent article~\cite{gauvin2018randomized}, where we categorically list all identified constrains, and formally introduce the naming convention and a hierarchical organisation of simple and combined RRMs.

\subsection{Modelling concept and demonstration of RRMs}

As it has been discussed in Section~\ref{sec:modelling}, random reference models are school examples of data-driven modelling, where real-world data and synthetic models are combined for a realistic simulation of a given phenomenon. In this case, taken a real temporal network and its randomly shuffled variates, synthetic dynamical processes are simulated on the top of them to understand how much structural and temporal correlations, present in the network, influence the unfolding of the synthetic dynamical process. To answer this question, we assume that taken some initial conditions for the process, information (infection, influence, etc.) in the temporal network can pass between nodes only at the times of their interactions and in a direction respecting the orientation of the actual event (if not noted otherwise). This way the initial conditions and the order of interactions determine the possible time respecting paths, which along the information can be transmitted, and in turn the final outcome of the dynamical process. To carry out such data-driven model experiments we follow general algorithmic recipe:
\begin{enumerate}
\item Take a temporal network.
\item Carry out a large number of simulations of a dynamical process on the temporal network with given initial conditions until it reaches a pre-defined equilibrium state.
\item Compute average quantities characterising the dynamics and final outcome of the dynamical process.
\item Apply a selected RRM on the temporal network to eliminate the effects of selected correlations and repeat points 2-3.
\item Compare the average outcome of the dynamical processes simulated on the original network and its RRMs to draw conclusions.
\end{enumerate}

For demonstration~\cite{karsai2011small,kivela2012multiscale}, as a temporal network we use here a six months long sequence of mobile-phone call interactions (DS1 in Section~\ref{sec:datasets}), and as a synthetic process we use one of the simplest dynamical model, the \emph{Susceptible-Infected} (SI) process, which is to simulate information or infection spreading on networks~\cite{barrat2008dynamical}. This model assigns one of two mutually exclusive states, susceptible (S) or infected (I), to each node in the network and assumes that initially every node are susceptible expect one infected seed, which is chosen randomly at the beginning of the process. During the process, a susceptible node can become infected with a rate $\beta$ (or probability per unit time step) at the time when it interacts with an infected node\footnote{Note, that we are going to discuss in details this and other model processes in Section~\ref{sec:obssp}, where we focus on the unfolding and critical behaviour of various dynamical processes on static and temporal network.}. In the coming simulations we use a deterministic SI process with $\beta=1$ and neglect the direction of events, which means that a susceptible node necessarily becomes infected once interacting with an infected node. We initiate the process by setting infected a single randomly selected node at a randomly selected time, and simulate the process until every node get infected, i.e, when the rate of infected nodes $i(t)=I(t)/N$ reaches $1$. Since the seeding time of the SI process is randomly chosen, it is possible that the process reaches the last event in the event sequence before every node gets infected. In this case we apply a \emph{periodic temporal boundary condition}, or in other words, we continue the process from the beginning of the event sequence. This method evidently introduces some invalid time-respecting paths~\cite{pan2011path} connecting events from the end of the sequence to ones in the beginning, but as we demonstrated, this change marginally the overall dynamics of the process~\cite{kivela2012multiscale}. To obtain a statistical characterisation of the dynamics we measure the average infection rate $\langle i(t) \rangle$ and the $P(t_f)$ distribution of full prevalence time, which fluctuates due to the random initial conditions of the $10^3$ independent simulations.

\begin{figure}[!ht]
\centering
  \includegraphics[width=.7\textwidth]{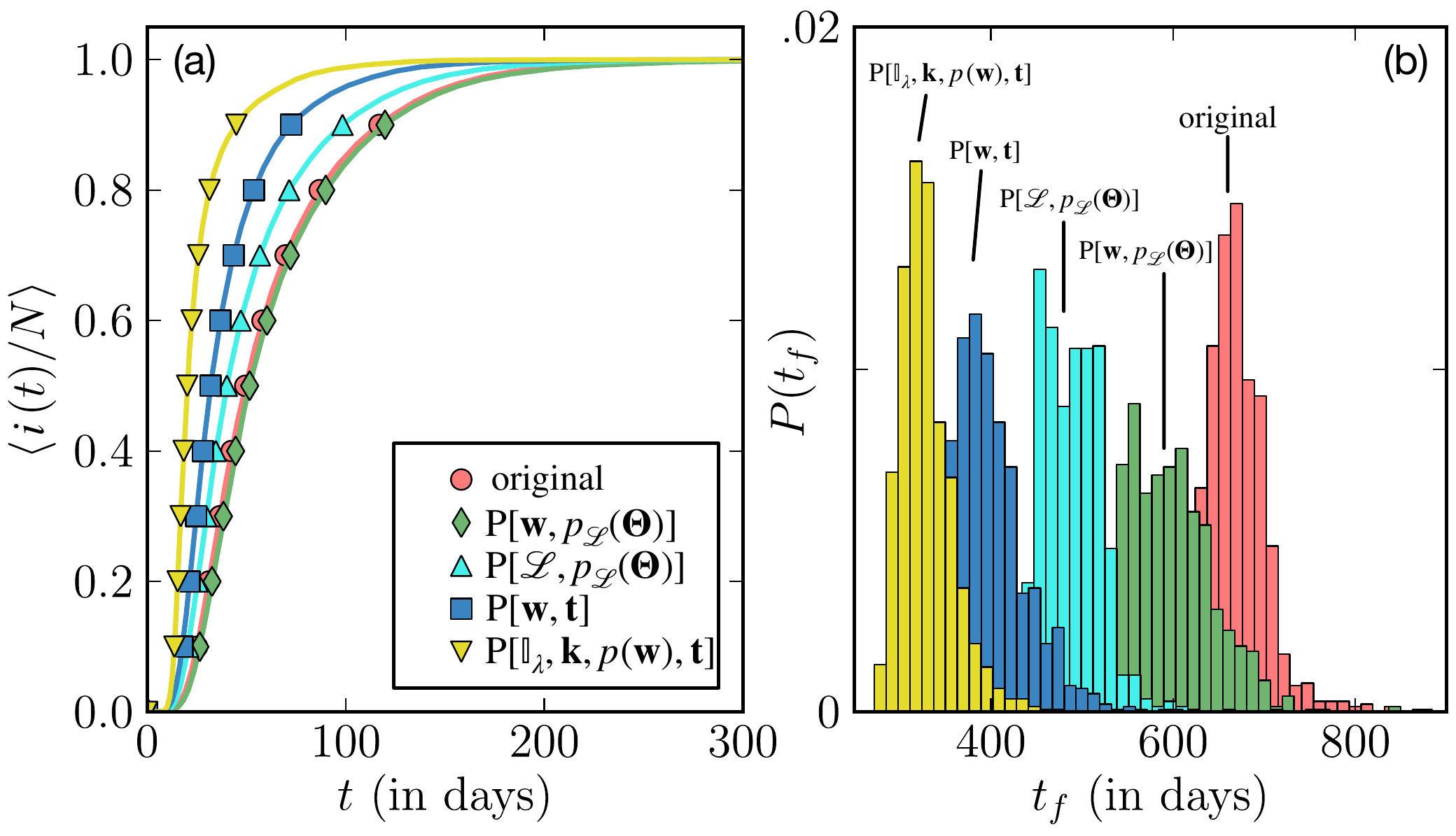}
\caption{\small (a) Fraction of infected nodes $\left<i(t)/N\right>$ as a function of time for the original event sequence and null models. (b) Distribution of full prevalence times $P(t_f)$ due to randomness in initial conditions. This figure was published in~\cite{karsai2011small}.}
\label{fig:SbSW}
\end{figure}

Simulations of the SI process on the original event sequence unfold in a surprisingly slow dynamics and takes on average $\sim 700$ days to reach every node in the network. This is even more puzzling as the underlying static network structure exhibits a small-world property, thus the average shortest paths between the $\sim 4.5M$ nodes is around $\langle \ell \rangle=12.31$. Consequently, the spreading process on the temporal network deviates from the ideal structural paths and follow time-respecting paths determined by the ordering of the time-varying interactions. Such slow spreading dynamics is not un-precedential in real life. Spreading of pandemics, electronic viruses, and information, follow their own pathways, which are not necessarily topologically efficient and, could be surprisingly slow, e.g., new infections are reported years after the emergence of a new computer virus or the introduction of an antivirus. The observed slow dynamics may be caused by the combined effects of several structural and temporal characters. First of all, static topological characteristics such as prominent community structure (C) have been shown to give rise to considerable decelerating effects on spreading speed~\cite{pastor2007evolution,eames2008modelling}, while a fat-tailed degree distribution has been shown to be an accelerating property~\cite{toivonen2009broad}. Second, in weighted networks, the relationship between weights and topology (W) provides an additional source of possible influence on the spreading dynamics. In particular for social networks it is known that links within communities are strong, while links between them are weaker~\cite{mucha2010community} - such Granovetterian structure enhances the trapping effect by the communities, leading to further slowing down of spreading~\cite{pastor2007evolution,mucha2010community}. Third, temporal characters like the daily activity cycle (D), the bursty character (B) of interactions and the causal correlations between adjacent interactions (E) (for definition see Section~\ref{sec:paths}) of the same person may give rise to important temporal inhomogeneities, which in turn influence the speed of spreading dynamics. Next, we introduce some RRMs which remove these effects in a controlled fashion to identify which of them cause the observed slow dynamics.

\paragraph{$\mathrm{P}[\mathbf{w},p_{\mathcal{L}}(\bm{\Theta})]$ - Equal-weight link-sequence shuffled model} Whole event sequences with time stamps (timelines) are randomly exchanged between links that have the same weights, i.e. numbers of events. Timing correlations between adjacent links are destroyed. While temporal characteristics of link event sequences are retained, any correlations between them and the topology are lost. All other temporal and structural correlations are retained. In other words, we constrain our shuffling on the $\mathbf{w}$ number of events (weights) on links, and the $p_{\mathcal{L}}(\bm{\Theta})$ distribution of timelines in the static structure $\mathcal{L}$.

\paragraph{$\mathrm{P}[\mathcal{L},p_{\mathcal{L}}(\bm{\Theta})]$ - Link-sequence shuffled model} As above, but sequences are exchanged between links of any weight. Thus, weight-topology correlations are destroyed, while keeping the $\mathcal{L}$ static structure and the $p_{\mathcal{L}}(\bm{\Theta})$ distribution of timelines.

\paragraph{$\mathrm{P}[\mathbf{w},\mathbf{t}]$ - Time shuffled model} The time stamps of the whole event sequence are randomly reshuffled. Thus all temporal correlations with the exception of network-level frequency envelope (for calls, the daily pattern) are destroyed, while the $\mathbf{t}$ event times and topological features like the $\mathbf{w}$ weight sequence are retained.

\paragraph{$\mathrm{P}[\mathbb{I}_{\lambda},\mathbf{k},p(\mathbf{w}),\mathbf{t}]$ - Configuration network model} The original aggregated network is rewired according to the configuration network model, where the degree distribution $\mathbf{k}$ of the nodes and connectedness $\mathbb{I}_{\lambda}$ are maintained but the topology is uncorrelated. Then, original single-link event sequences are randomly placed on the links thus the overall weight distribution $p(\mathbf{w})$ remains unchanged, and time shuffling as above is performed keeping the event times $\mathbf{t}$ unaltered. All other temporal correlations, except seasonalities like the daily cycle, are destroyed.

\begin{table}[h!]
\centering
\begin{tabular}{|l|l||c|c|c|c|c||c|}\hline
Name & Description & D & C & W & B & E & $\langle t_{20\%} \rangle$ \\ \hline\hline
Original & Empirical event sequence&  \checkmark & \checkmark  & \checkmark & \checkmark & \checkmark & $38.24 \pm .74$ \\ \hline
$\mathrm{P}[\mathbf{w},p_{\mathcal{L}}(\bm{\Theta})]$ & Equal-weight link-sequence shuffling  & \checkmark &  \checkmark & \checkmark & \checkmark & & $41.20 \pm .83$ \\ \hline
$\mathrm{P}[\mathcal{L},p_{\mathcal{L}}(\bm{\Theta})]$ & Link-sequence shuffling& \checkmark &  \checkmark & & \checkmark &  & $30.67 \pm .55$ \\ \hline
$\mathrm{P}[\mathbf{w},\mathbf{t}]$ & Time shuffling & \checkmark & \checkmark  & \checkmark & & & $26.48 \pm .52$ \\ \hline
$\mathrm{P}[\mathbb{I}_{\lambda},\mathbf{k},p(\mathbf{w}),\mathbf{t}]$ & Configuration network model & \checkmark  & & & & & $17.23 \pm .22$ \\ \hline
\end{tabular}
\caption{\small Correlations retained in different null models. D: daily pattern, C: community structure, W: weight-topology correlations, B: bursty single-edge dynamics, E: correlations between adjacent events. Average times to reach $20\%$ prevalence with the error of the mean values. This table was published in~\cite{karsai2011small}.}
\label{table:SbSW}
\end{table}

In our analysis of the results, first we focus on the two temporal reference models that remove temporal correlations while leaving the static network intact (equal-weight link-sequence shuffled and time shuffled). Fig.\ref{fig:SbSW}a shows that the spreading dynamics for those event sequences that contain bursts (the original and equal-weight link-sequence shuffled sequences) are slower than those for the reference model from which burstiness has been removed (sequences from the time shuffled model). This clearly indicates that burstiness of the event sequences slows down the spreading process significantly. Further, the dynamics for the original sequence and equal-weight link-sequence shuffled model closely resemble each other. This similarity means that event-event correlations have only a small influence on the speed of spreading and it causes even a small acceleration for the process in early times. This can be explained by the emergence of rapid chains of causal events, which helps information spreading locally. However, when looking at the distributions of the full prevalence times $t_f$, shown in Fig.\ref{fig:SbSW}b, it is seen that for long times, event-event correlations somewhat slow down the process.

Next, we turn to those reference models that modify structural features of the static aggregated network. We find that when the network topology is retained but weight-topology correlations are removed with the link-sequence shuffled reference model, the spreading significantly speeds up compared to the original. This is because the reference model removes the known Granovetterian weight-topology correlations where weak links connecting dense communities of nodes act as bottlenecks. In addition, if topological correlations such as the community structure are removed with the configuration model, the dynamics of spreading becomes even faster.

Finally, we cross-compare the relative importance for the structural and temporal correlations in the call sequence on the spreading speed. As seen in Fig.\ref{fig:SbSW}a, the spreading dynamics for the time-shuffled model where weight-topology correlations are retained but the bursts are destroyed is faster compared to that for the link-sequence shuffled model, where bursts are retained but weight-topology correlations are destroyed. Consequently, burstiness of events on individual links plays a more important role than weight-topology correlations in slowing down the spreading dynamics.

The same conclusions can be drawn once we quantify the slowing down effects by measuring the average times $\langle t_{20\%}\rangle$ to reach $20\%$ prevalence (see Table~\ref{table:SbSW}). The difference between the original and the fastest model is $\sim 21$ days, i.e., a factor 2. Similarly for the $100\%$ prevalence this factor also 2 ($\sim 342$ days), showing that the effects of correlations are consistent for the duration of the whole process and for individual runs. As for the effect of the random initial conditions, the small error of mean values in Table~\ref{table:SbSW} show that the mean curves in Fig.\ref{fig:SbSW}a characterise the overall behaviour well. The effect of initial conditions are demonstrated in Fig.\ref{fig:SbSW}b, where the distributions are clearly separable at full prevalence.

As a conclusion we can draw that the bursty non-Poissonian dynamics and the Granovetterian weight-topology correlations are the dominating characters which slow down the spreading dynamics, while other correlations like the community structure or causal event correlations play less important roles. Further, in the related studies~\cite{karsai2011small,kivela2012multiscale} we showed that non-Poissonian bursts, evidenced by fat tailed $P(\tau)$ inter-event time distributions, are consistently characterising all activity groups of people. Actually we derived that the decelerating effects of burstiness is coupled with the second moment of $P(\tau)$, leading to the extremely slow dynamics in strongly bursty systems like mobile-phone communication sequences. In addition we found that, interestingly, circadian patterns have negligible effects in influencing the speed of information spreading.

\vspace{.1in}

In this Section we briefly introduced the concept and a taxonomy of random reference models and through a data-driven modelling study we demonstrated their potential in characterising temporal networks and ongoing dynamical processes at the system level. As a next challenge we turn our focus on causal correlations between events to see how they lead to the emergence of mezoscopic temporal motifs and to long time-respecting paths in the temporal network.

\section{Higher order representations}
\label{sec:tnet_hons}

So far we looked at temporal networks as a sequence (or snapshots) of events, which connect two nodes in the network at a given time. Such first order representations have been proven to be useful to characterise temporal structures and ongoing processes but they make it difficult to study the non-Markovian character of temporal networks where events are not necessarily independent from each other. Such correlations, like causal relationships between adjacent events, potentially explain phenomena like bursty interaction patterns~\cite{scholtes2014causality}, temporal motifs~\cite{kivela2017mapping} and the emergence of long time-respecting paths~\cite{kivela2017mapping}, but can be explanative for the emergence of homophilic motifs of correlated interactions of similar individuals~\cite{kovanen2013temporal}. They can be effectively studied by higher-order representations of temporal structures, where conceptually we identify events (or set of events) as nodes and connect them with a directed link respecting the timely order of the connected events, if they are adjacent at the given order of representation. Such way of description is not only important to study social interactions, but they are central in transportation networks~\cite{xu2016representing}, in predicting human mobility~\cite{peixoto2017modelling}, or potentially in gene-regulation, or neural networks, just to mention a few examples.

In the following we are going to discuss two methods proposed by us over the last years for higher-order representation of temporal networks. One provides a mesoscopic level description to identify recurrent isomorphic temporal motifs~\cite{kovanen2011temporal,kovanen2013temporal}, while the other provides a representation at macroscopic scale of the whole temporal structure by mapping all temporal paths simultaneously at once~\cite{kivela2017mapping}. For simplicity, we assume that nodes in the network cannot participate simultaneously in multiple events, and that events have no duration (although both methods can be easily generalised in this sense).  Also note that the investigations of non-dyadic interactions may lead to the recent fields of simplicial complexes~\cite{wu2015emergent,petri2013topological} and hyper-graphs~\cite{johnson2011hypernetworks}, which are out of the scope of the present thesis.

\subsection{Temporal motifs}

Static motifs are classes of connected isomorphic subgraphs, which appear with a significantly higher frequency in a real structure rather than by chance in a random reference model. Such motifs can be identified as the mesoscopic building blocks of a static network, and more importantly, they can be assigned to special functionality~\cite{milo2002network}. Analogous definition of motifs in temporal network is a non-trivial problem as it requires the extension of the definition of connectedness, which in case should intuitively include time, and isomorphism, which should respect structural and order equivalence between sets of adjacent events~\cite{kovanen2011temporal,kovanen2013temporal}. As an example, in a social communication network one might detect an event sequence where Alice calls Bob, who then calls Carol and Dave. A similar sequence might be frequently observed to take place between the same people, or between other sets of four individuals. All these sequences are members of the same class, which we call a \emph{temporal motif}. Other example can be found in genetic regulation data, where the event sequence would correspond to regulatory interactions switching on and off as the intercellular system performs its function. Beyond providing insight into the operation of the system under study, temporal motifs allow the study of similarities and differences of temporal networks, as originally proposed for static motifs in~\cite{milo2002network}. In addition they may help in building models of network evolution~\cite{kumpula2007emergence}.

\subsubsection{Definition of temporal motifs}

Following the logic of static network motifs~\cite{milo2002network}, temporal motifs~\cite{kovanen2011temporal,kovanen2013temporal} are defined as temporally connected isomorphic temporal subgraphs of causally correlated events. Their definition relies solely on a sequence of directed interactions $e=(u,v,t)$ to identify causally correlated sets of adjacent events (already introduced in Section~\ref{sec:paths}), while neglecting any information on the nature/type/content of interactions (which is usually not available anyway). Consequently, their definition is limited to structural and temporal informations to decide about two adjacent events to be causal. One realistic assumption suggests that two adjacent events happening within a $\Delta t$ time window can be considered causal if the selected time window is sufficiently small. This time window may represent the memory length of a person to remember some information, or the infectious period of an infected agent, or the time one is willing to wait for a connection at an airport. Taking this assumption, we consider two events \emph{$\Delta t$-adjacent} if they have at least one node in common and the time difference between the end of the first event and the beginning of the second event is no longer than $\Delta t$. Equivalently, two events are \emph{$\Delta t$-connected} if there exists a sequence of events $e_i=e_{k_0} e_{k_1} \dots e_{k_n} = e_j$ such that all pairs of consecutive events are $\Delta t$-adjacent.

\begin{figure}[!ht]
\centering
  \includegraphics[width=1.\textwidth]{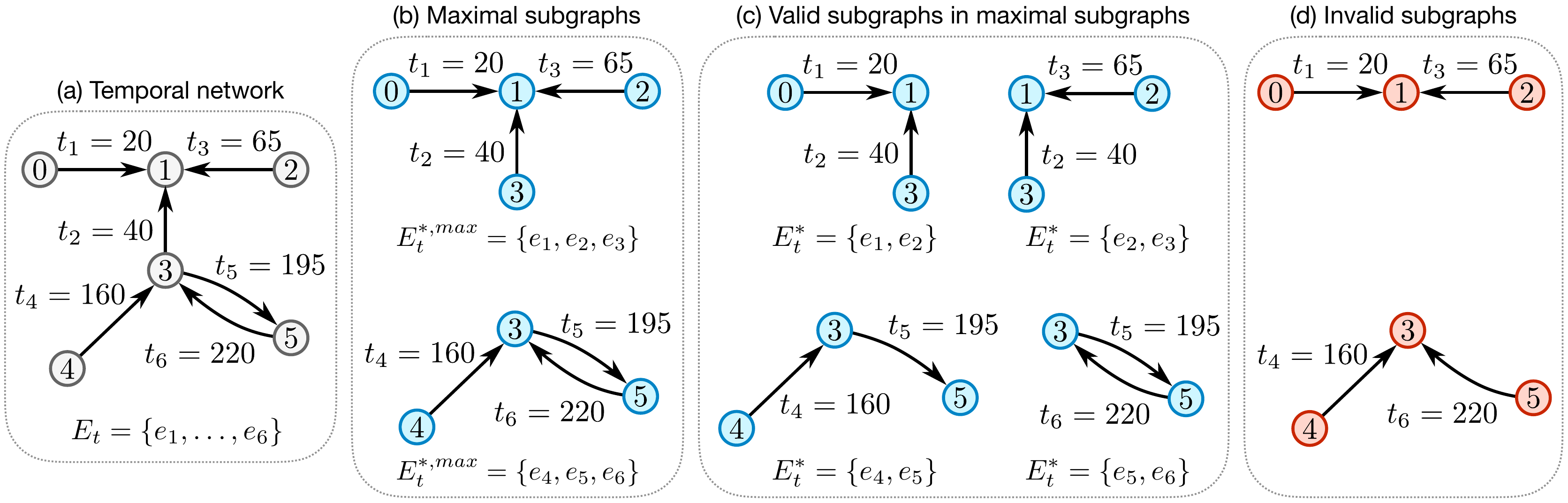}
\caption{\small (a) An example temporal network with event list $E_t$ of six events. With $\Delta t$=40 there are two maximal subgraphs, shown in (b). (c) Valid subgraphs contained in the maximal subgraph in (b). In addition to these the maximal subgraph itself and all unit subgraphs are valid subgraphs. (c) Event sets that are contained in (b) but are not valid subgraphs: the upper one because it does not include all consecutive $\Delta t$-connected events, and the lower one because it is not $\Delta t$-connected. This paper was published in~\cite{kovanen2013temporal} and partially prepared by L. Kovanen.}
\label{fig:motif1}
\end{figure}

Using these definitions, a \emph{connected temporal subgraph} consists of a set of events such that all pairs of events in it are $\Delta t$-connected. This ensures that subgraphs are connected both topologically and temporally. While this definition could already be used as a basis for temporal motifs, it suffers from the same shortcoming as its static cousin: in some simple cases the number of connected subgraphs explodes. For example an $n$-star, where all events take place within $\Delta t$, contains $\binom{n}{k}$ connected temporal subgraphs with $k$ events, which would make the resulting motif statistics difficult to interpret in any intuitive fashion.

One alternative to resolve this shortcoming is to consider only those connected subgraphs where all $\Delta t$-connected events of each node are consecutive. This not only solves the problem with the $n$-star (we now get $n-k+1$ subgraphs with k events) but also offers an intuitive interpretation: each subgraph takes into account all relevant events for each node within the time span covered for that node, in the sense that no events can be skipped (as demonstrated in Fig.~\ref{fig:motif1}c). We call connected subgraphs that satisfy this requirement \emph{valid temporal subgraphs} and denote them by $E_t^*$. For every event $e_i$ there is a unique \emph{maximal subgraph} $E_t^{*,max}$ that contains $e_i$ and in which all event pairs are still $\Delta t$-connected (as shown in Fig.~\ref{fig:motif1}b). Note that a maximal subgraph is always a valid subgraph.

\emph{Temporal motifs} are now defined as classes of isomorphic valid subgraphs, where the isomorphism also includes the similarity of the temporal order of events. Accordingly, two temporal subgraphs are isomorphic if they are topologically equivalent and the order of their events is identical. If a temporal motif is based on a maximal subgraph, we call it a \emph{maximal motif}.


\subsubsection{Identification of temporal motifs}
\label{sec:idtempmot}

Because maximal subgraphs are temporally separated from all other events by at least time $\Delta t$, all subgraphs are fully contained in some maximal subgraph. Based on this observation, the process to identify all temporal motifs in a given event set $E_t$ can be separated into three steps:
\begin{enumerate}
\item Find all maximal connected subgraphs $E_t^{*,max}$.
\item Find all valid subgraphs $E_t^* \subseteq E_t^{*,max}$.
\item Identify the motif corresponding to $E_t^*$.
\end{enumerate}

In \emph{step 1}, to find the maximal subgraph which the event $e_i$ belongs to, we start from $e_i$ and iterate forward and backward in time to find all $\Delta t$-adjacent events; this process is then repeated recursively with all new events encountered. Assuming the $\Delta t$-adjacent events can be found in constant time, the time complexity of this step is $\mathcal{O}(|E_t^{*,max}|)$. Since the same maximal set is discovered starting from any event in it, the total time complexity of this part is $\mathcal{O}(|E_t|)$.

\begin{figure}[!ht]
\centering
  \includegraphics[width=0.8\textwidth]{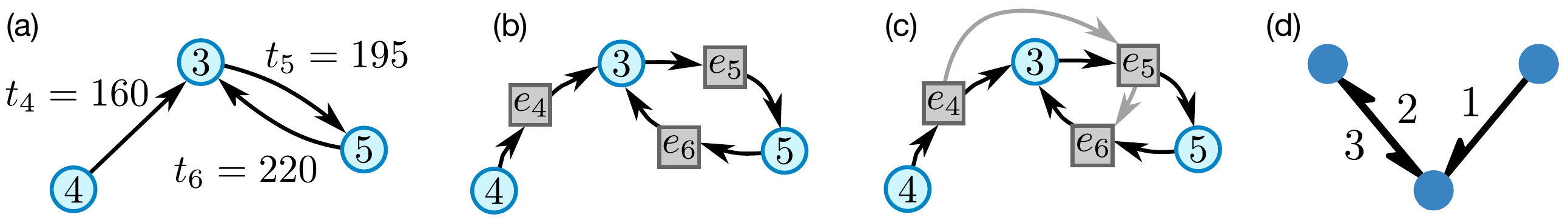}
\caption{\small Illustration of the algorithm for identifying temporal motifs. (a) A valid subgraph $E_t^*$ with three events. (b) A vertex is created for each event and edges are added to connect them to the corresponding nodes. Colours are used to distinguish between the two types of vertices; the labels of the event vertices are arbitrary. (c) Directed edges are created between event vertices to denote their order. A canonical labelling is then calculated for this graph; all temporal subgraphs that are isomorphic at this stage will yield the same canonical labelling. (d) A concise presentation for the temporal motif. The numbers next to edges denote the order of the events. Note that the numbers are always on the side of the arrow heads. This paper was published in~\cite{kovanen2013temporal} and partially prepared by L. Kovanen}
\label{fig:motif2}
\end{figure}

\vspace{.1in}

To solve \emph{step 2}, consider the following theorem~\cite{kovanen2011temporal,kovanen2013temporal}:

\begin{theorem}
Let $G(E_t^{*,max})$ be an undirected graph that has a vertex for each event in $E_t^{*,max}$ and every vertex is connected to the next and previous $\Delta t$-adjacent event of both nodes in that event (there are at most four such events). Then every valid subgraph contained in $E_t^{*,max}$ corresponds to a connected subgraph of $G(E_t^{*,max})$.
\end{theorem}

\begin{proof}
Consider a valid subgraph $E_t^{*} \subseteq E_t^{*,max}$ and the corresponding vertex set in $G$. Because all event pairs in $E_t^{*}$ are $\Delta t$-connected and the events of every node are consecutive, there is at least one path between all vertex pairs in this set. Therefore there is at least one connected subgraph of $G$ that corresponds to $E_t^{*}$.
\end{proof}

Now each valid subgraph contained in $E_t^{*,max}$ corresponds to some connected vertex set of $G$, and the problem of finding all valid temporal subgraphs reduces to identifying all induced subgraphs of $G$ and checking that the events of each node are consecutive. The pseudo-code for identifying vertex sets of induced subgraphs is given in Algorithm \ref{alg:identify_subgraphs} (the code assumes that nodes are labeled with integers from 1 to $|V|$). In function \textsc{FindInducedSubgraphs} we first start $|V|$ search trees so that the tree initialised with node $i$ will include all sets where $i$ is the smallest node. The nodes in the set $V_{-}$ are excluded from that search tree; initially this set contains all nodes smaller than $i$. The set $V_{+}$ includes the nodes where the search can progress, initially all neighbours larger than $i$. Because each search tree finds only sets where $i$ is the smallest node, they are necessarily distinct. Now we need to make sure they are complete.

The function \textsc{SubFind} first adds the current set to be returned (line 10) and grows then sets recursively. The set $V_{-}$ is updated to exclude nodes smaller than $i$; thus each subtree has a different smallest node (from those in $V_{+}$) and the subtrees are again distinct. The set $V_{+}$ contains nodes where the search may progress: previously allowed nodes larger than $i$, or neighbours of $i$ not yet excluded.

\begin{algorithm}[h!]
  \caption{\small \label{alg:identify_subgraphs} Find the vertex sets of all induced, connected subgraphs of a given graph. The parameter $n_{\max}$ can be used to limit the size of the vertex sets returned. We use $N(i)$ to denote the neighbours of node $i$.}
\begin{algorithmic}[1]
  \Require $G=(V,L)$ is an undirected graph.
\Statex
\Function{FindInducedSubgraphs}{$G$, $n_{\max}$}
  \State $S_{\textrm{all}} \gets \emptyset$
  \For{$i$ in $V$}
    \State $S \gets \{i \}$
    \State $V_{-} \gets \{j \in V\, | \, j \leq i \}$
    \State $V_{+} \gets \{j \in N(i) \, | \, j > i \}$
    \State \Call{SubFind}{$G$, $n_{\max}$, $S_{\textrm{all}}$, $S$, $V_{-}$, $V_{+}$}
  \EndFor
  \State \Return{$S_{\textrm{all}}$}
\EndFunction
\Statex
\Function{SubFind}{$G$, $n_{\max}$, $S_{\textrm{all}}$, $S_{\textrm{curr}}$, $V_{-}$, $V_{+}$}
  \State $S_{\textrm{all}} \gets S_{\textrm{all}} \cup S$
  \If{$|S| = n_{\max}$}
    \Return
  \EndIf
  \For{$i$ in $V_{+}$}
    \State $S^{*} \gets S \cup \{i \}$
    \State $V_{-}^{*} \gets V_{-} \cup \{j \in V_{+}\,|\,j \leq i \}$
    \State $V_{+}^{*} \gets \{j \in V_{+}\,|\, j > i \} \backslash \{j \in N(i)\, |\, j \not \in V_{-}^{*} \}$

     \State \Call{SubFind}{$G$, $n_{\max}$, $S_{\textrm{all}}$, $S^{*}$, $V_{-}^{*}$, $V_{+}^{*}$}
  \EndFor
\EndFunction

\end{algorithmic}
\end{algorithm}

Because the subtrees are distinct at each step, the algorithm will return each set at most once. To see that it returns all possible connected set, consider how we could arrive at an arbitrary connected set $S$. The search path is rooted at $i_1 = \min S$. Let $S_k$, $k \leq |S|$, be the set of elements added at depth $k$. Because $S$ is connected, there is at least one node in $S \backslash S_k$ that is a neighbour of some node in $S_k$. The only way the construction can fail is if for some $k$ there is a node $i^* \in S \backslash S_k$ that has already been excluded, i.e. it is in $V_{-}$. It is not possible that $i^*$ was excluded in the beginning---the tree was rooted at $i_1 \leq i^*$ and only nodes smaller than $i_1$ were excluded---so it must have happened during the search. But if $i^*$ was added to $V_{-}$ it means that it was in $V_{+}$ but some larger node of $S$ was added instead, which is a contradiction---in the subtree leading to $S$ we would have added $i^*$. Hence no node $i^*$ can exist and the construction can always proceed until $S$ is obtained.

\vspace{.1in}

Finally for \emph{step 3}, identifying the motif for subgraph $E_t^*$ requires solving the isomorphism problem such that we also include information about the order of the events. One can do this by mapping all relevant information into a directed and coloured graph as illustrated in Fig~\ref{fig:motif2}, for which the isomorphism can be readily solved with existing algorithms. In practice we calculate for this graph its canonical form, a labelling of vertices that is identical for all isomorphic graphs, so that we can easily tell if two valid subgraphs correspond to the same motif. Finding the canonical form is a non-trivial task, but many efficient algorithms have been developed for this purpose; one is called bliss and is described in~\cite{junttila2007engineering}.

\vspace{.1in}

As a final step, to make temporal motifs more accessible we convert the information about the order of events back into plain integers. The above described converting procedure is visually demonstrated in Fig.~\ref{fig:motif2} where panel Fig.~\ref{fig:motif2}d shows a concise presentation of the motif corresponding to the original temporal subgraph in panel Fig.~\ref{fig:motif2}a.

\subsubsection{Temporal motifs in communication networks}

To demonstrate our method on a large-scale temporal network, we identify temporal motifs in a mobile-phone call network (a variant of DS1 described in Section~\ref{sec:datasets}), which consists of $320$ million mobile phone calls between nearly $9$ million customers. We chose the time window to be $\Delta t=10$ minutes, thus keeping the $35\%$ of events which are $\Delta t$-adjacent to at least one other event. As a reference system, to quantify the significance of the observed motifs, we used the time-shuffled RRM ($\mathrm{P}[\mathbf{w},\mathbf{t}]$ in Section~\ref{sec:tnet_rrm}). Measures in the reference system have been averaged over five independent runs.

\begin{figure}[!ht]
\centering
  \includegraphics[width=1.0\textwidth]{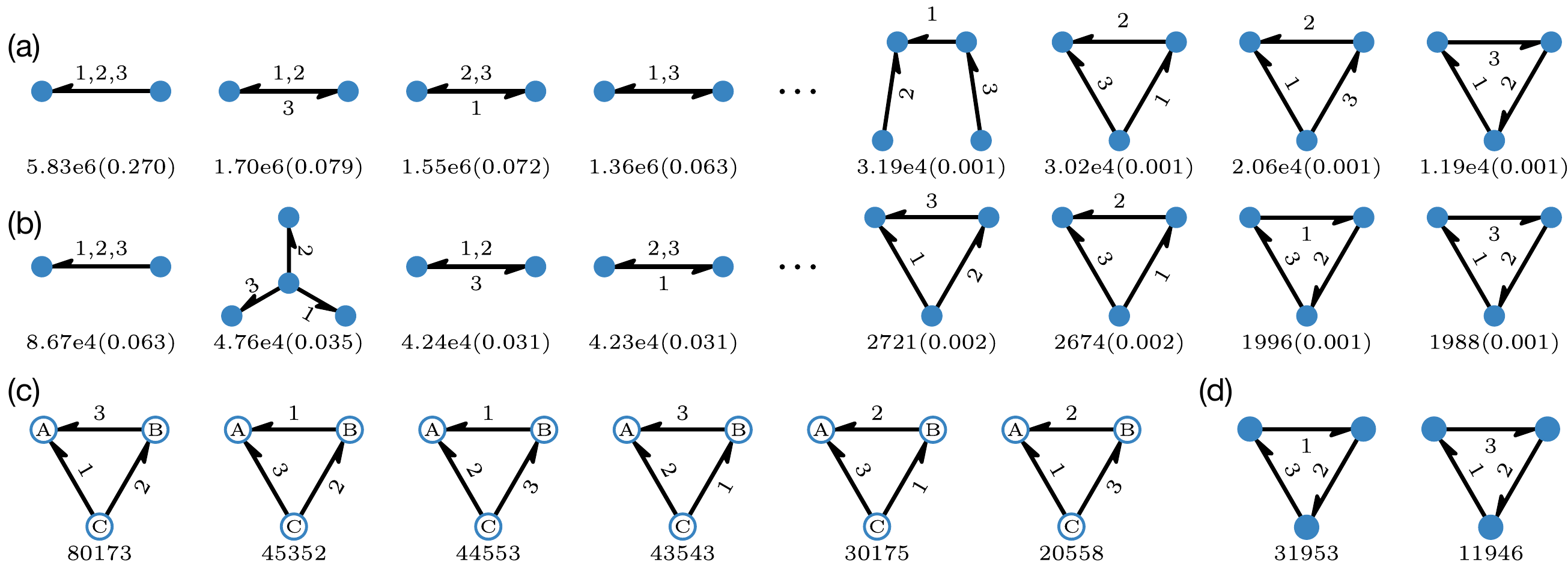}
\caption{\small (a,b)The four most common (on the left) and least common (on the right) motifs in (a) the empirical data, and (b) unbiased time-shuffled random reference model. The values below each motif denote the total count and, in parenthesis, the fraction out of all motifs with three events. (c,d) The two different kinds of directed triangle motifs with three events. Both groups have been ordered by count in the empirical data that are also shown below the motifs. All motifs in (c), as well as those two in (d), differ only in the order of events. This paper was published in~\cite{kovanen2011temporal} and partially prepared by L. Kovanen}
\label{fig:motif3}
\end{figure}

Fig.~\ref{fig:motif3}a shows the four most and least common three-event temporal motifs (there are 68 three-event motifs in total) in the data, and Fig.~\ref{fig:motif3}b depicts the same in the time shuffled reference. Unsurprisingly, the number of non-trivial motifs in the reference is lower (only $8.6\%$ of events are $\Delta t$-adjacent to some other event) but the two cases still appear qualitatively similar. The most common motifs illustrate the bursty nature of the mobile phone data, while the least common motifs are triangles even though triangles are often considered to be the building blocks of social networks. The distribution of different motifs is more balanced in the reference: in the empirical data the most common three-event motif makes up $27\%$ of all three-event motifs, but only $6.3\%$ in the time-shuffled reference.

As a further example clarifying this point, we present in Fig.~\ref{fig:motif3}c and d all motifs based on the different directed triangles with three events. The six motifs in Fig.~\ref{fig:motif3}c would be equally common in the time-shuffled reference, but in the empirical data we observe a four-fold difference between the most and least common triangles. There are two factors that explain this: burstiness and causality. Burstiness appears in the fact that in the four most common motifs the two calls made by $C$ are consecutive; in the two least common motifs they are not. Causality is most apparent when comparing the most and the least common motifs. In the most common motif the caller of the second call ($C$) knows about the first call (because he made it herself), and the caller of the third call ($B$) could know about both previous calls. In the least common motif the caller of the second call ($B$) cannot know about the first one, and the caller of the third call ($C$) cannot know about the call made by $B$. The most common motif is both bursty and causal, while the least common is neither. Causality is also an obvious explanation for the counts in Fig.~\ref{fig:motif3}d: the triangle where events could cause one another is three times as common as the one where events are independent.

\vspace{.1in}

Temporal motifs rely on the higher-order representation of temporal networks as their definition is built on $\Delta t$-adjacent events. Although this mesoscopic level characterisation of temporal networks is very informative and useful to identify frequently appearing patterns in the tissue of temporal networks, it is far from providing a complete description of the higher-order structure of the network. In the coming Section we aim to progress in this direction by using adjacent events for a complete and information lossless representation of temporal networks.

\subsection{Weighted event graphs}
\label{sec:wDAGs}

Temporal paths in temporal networks are outmost important as they determine the evolving structure of a temporal network and the way any collective phenomena can unfold on its fabric. As we discussed in Section~\ref{sec:paths}, finding all shortest paths in a temporal network between any nodes at any time is an $NP$-complete problem by using conventional path detection algorithms~\cite{cormen2009introduction}. A faster way would be to compute paths for a range of values, taking use of redundancy, while even better methods can be designed by using higher-order representations of temporal networks. In this Section, using the concept of event-adjacency, we introduce an information lossless representation of temporal networks. This way of description transforms temporal structures to weighted static graphs, which encodes all temporal and structural informations at once and maps simultaneously all time-respecting paths, with the advantage to be analysed as a static network. Beyond its most general definition, it is capable to constrain on the detection of all $\Delta t$-connected temporal paths, which are outmost important in case of limited waiting time process.

Processes with limited waiting times at nodes are particularly sensitive to broad distributions of inter-contact times; the longest inter-contact times may stop the process. Such processes include variants of epidemiological models with recovery mechanism, like Susceptible-Infectious-Recovered (SIR) and Susceptible-Infectious-Susceptible (SIS) models (for definition see Section~\ref{sec:scp})\cite{iribarren2009impact,miritello2011dynamical,rocha2011simulated,holme2014birth}, where nodes only remain infectious for finite periods. Other examples include social contagion \cite{daley2001epidemic,castellano2009statistical}, ad-hoc message passing by mobile agents \cite{tripp2016special}, and passenger routing~\cite{nassir2016utility}. In these processes, the spreading agent must be transmitted onward from a node within some time $\Delta t$ or the process stops. This waiting time limit can be directly incorporated into time-respecting paths by requiring that their successive contacts are separated by no more than $\Delta t$ units of time.

\emph{Weighted event graphs}~\cite{kivela2017mapping} are static, weighted, and directed acyclic graphs (DAGs) that encapsulate the complete set of $\Delta t$-constrained time-respecting paths for all values of $\Delta t$ simultaneously. The subset of paths corresponding to a specific value of $\Delta t$ can be quickly extracted from the weighted event graph by simply thresholding it. Weighted event graphs can be viewed as a temporal-network extension of the line-graph representation of static networks. There is some similarity with the approach of \emph{Scholtes et al.}~\cite{scholtes2014causality} that maps two-event sequences onto aggregated second-order networks, and with that of \emph{Mellor}~\cite{mellor2017temporal} where an unweighted event graph is constructed from pairs of temporally closest events. Our approach~\cite{kivela2017mapping} builds on concepts introduced in~\cite{backlund2014effects,ayala2015temporal}.

\subsubsection{Definition of weighted event graphs}

\begin{figure}[!ht]
\centering
\includegraphics[width=.6\linewidth]{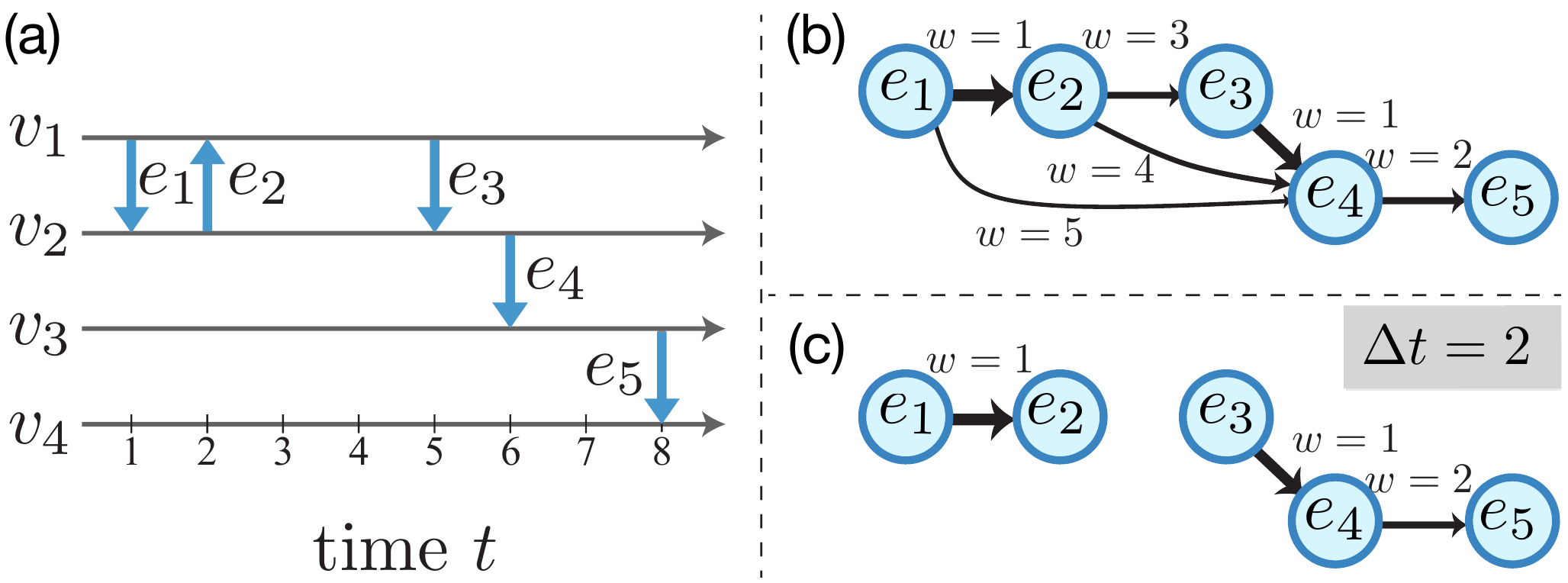}
\caption{\small Constructing and thresholding the weighted event graph. a) The time line of a temporal network with four nodes $v_1-v_4$ and five events $e_1-e_5$. b) The weighted event graph representation of the temporal network. c) The thresholded event graph, containing only pairs of events with a maximum time difference of $\Delta t=2$. This figure was published in~\cite{kivela2017mapping} and partially prepared by J. Saram\"aki.}
\label{fig:wdag}
\end{figure}

Let us consider a temporal network $G_t=(V,E_t,T)$ as we defined in Section~\ref{sec:tempnetrepr} (and demonstrated in Fig.~\ref{fig:wdag}a) with no self-edges or simultaneous events. 
The weighted event graph representation of $G_t$ is defined as the graph $D=(E_t,E_{D},w)$ where the set of nodes $E_t$ is the set of events in $G_t$ and the edges $e_D \in E_{D}$ represent the adjacency of the events $e_{D}=e\rightarrow e'$ with weights defined as temporal distances $w(e_{D})=t'-t$ (see Fig.~\ref{fig:wdag}b). That is, $D$ is a directed acyclic graph with links weighted with temporal distances, contains all time-respecting paths in $G_t$. For paths with a waiting time limit $\Delta t$, we get the subgraph $D_{\Delta t}$ by thresholding $D$ so that only links with $w\leq\Delta t$ are retained (see Fig.~\ref{fig:wdag}c).

Constructing the weighted event graph representation $D=(E_t, E_D, w)$ of a temporal network can be done efficiently by noting that the edges in $E_D$ can be listed by inspecting the sequence of events around each node $v \in V$ separately (as described in Algorithm~\ref{alg:wdag}). For some data sets the full weighted event graph $D$ might be large, and it is convenient to construct $D_{\Delta t_{\max}}$ that can, for example, be later used to sweep through all values $\Delta t < \Delta t_{\max}$.

For each node in the temporal network $v \in V$ we can build a time-ordered sequence of events $\{ e_1, \dots, e_k \}$ in which $v$ participates. In the case where there are no durations we can then simply iterate over each event $e_i$, and for each of them search forward in the ordered event sequence until we find an event $e_j$ for which $t_j - t_i \leq \Delta t_{\max}$. We then add a link $e_i \rightarrow e_j$ at each step of this process until the event $e_j$ that is too far from the starting event $e_i$ is found. (Note that some $\Delta t$ adjacencies are found twice.) Creating the event sequences and sorting them can be done in $\mathcal{O}(|E| \log |E|)$ time, and as each step of the algorithm produces a single link (with possibility of some links being visited twice) the algorithm runs in total $\mathcal{O}(|E| \log |E| + |E_D|)$ time. Including the durations of events only requires a small adjustment to this algorithm.

\begin{algorithm}[h!]
\caption{\small \label{alg:wdag}Weighted event graph edges for a node~\cite{kivela2017mapping}.}
\begin{algorithmic}[1]
\Function{$D$}{$v,\{ e_1, \dots, e_k \}$}
\For{$i \gets 1 \textrm{ to } k$}
 \State{$j \gets i + 1$}
 \While{$t_j - t_i \leq \Delta t_{\max}$ and $j \leq k$}
   \State{Output: $e_i \rightarrow e_j$}
   \State{$j \gets j + 1$}
 \EndWhile
\EndFor
\EndFunction
\end{algorithmic}
\end{algorithm}

The $\Delta t$-thresholded event graph $D_{\Delta t}$ is a superposition of the time-respecting paths that a $\Delta t$-limited spreading process can follow. Therefore, its structure tells if the process can percolate the network. A closer look at the problem reveals that here, the concept of percolation is more complex than for static networks. The components of $D_{\Delta t}$ are directed, (even if the events of $G$ are undirected). There are only weakly connected components--there are no strongly connected components because $D_{\Delta t}$ is by definition acyclic. Each event graph node has an in-component and out-component that contain events on up- and downstream temporal paths; these components may overlap for different event graph nodes~\cite{nicosia2012components}. In the following, we will limit our analysis to weakly connected components because of their uniqueness in $D_{\Delta t}$.

\subsubsection{Temporal network percolation}

In percolation analysis, the relative size of the largest connected component is defined as the order parameter. Here, there are three ways of measuring the size of a component of $D_{\Delta t}$. (1) One can count the number of \textit{event graph nodes} $S_{E_t}(E')=|E'|$ in a connected component $E' \subseteq E_t$ of $D_{\Delta t}$. This gives an upper bound for the number of events on the time-respecting paths that a spreading process can follow if it includes an event from that component. (2) One can count the number of \textit{temporal-network nodes} $S_{G}(E')=|\bigcup_{(u,v,t)\in E'}(u \cup v)|$ that are covered by the event graph component $E'$. This is an upper bound for the number of temporal-network nodes that any spreading process can reach via the component's time-respecting paths. Note that a temporal-network node can belong to multiple event-graph components; this can result in multiple giant components that cover most nodes but are separated in time. (3) One can measure the \textit{lifetime} of the event graph component $S_{LT}(E')=(\max_{(u,v,t)\in E'} t-\min_{(u,v,t)\in E'} t)$. This is an upper bound for the lifetime of any spreading process on the component. Note that there can be many co-existing components with long (or infinite) lifetimes; frequent and sustained contacts between a small number of nodes can already induce such components.

With these measures, we can define the order parameter as the relative size of the largest connected component, 
\begin{equation}
\rho_{*}(D_{\Delta t}) = \dfrac{1}{N_{*}} \max_{n_{S_*}\neq 0} S_*,
\label{eq:rho}
\end{equation}
where $n_{S_*}$ is the number of components of size $S_*$ for the chosen definition of size $* \in \{E_t,G,LT\}$, and $N_*$ is the maximum possible value that $S_*$ can get as a single component, i.e., $N_E=|E_t|$, $N_G=|V|$, and $N_{LT}=T$. In conventional percolation analysis, the average size of the other connected components is a quantity of interest that is equivalent to magnetic susceptibility. It can be introduced for the $S_{*}(E')$ event graph components in $D_{\Delta t}$ as
\begin{equation}
\chi_{*}(D_{\Delta t}) = \dfrac{1}{N_{*}} \sum_{{S_*}<\max{S_{*}}} n_{S_*} S_*^2.
\label{eq:chi}
\end{equation}
One would expect this quantity to have a maximum at the critical $\Delta t_c$, where the percolating connected component emerges in the event graph; in the thermodynamic limit this maximum would become a singularity. 
However, this quantity might behave differently for $S_{G}(E')$ and $S_{LT}(E')$ components due to $\sum n_{S_*} $ not being a conserved quantity, and because of the possible multiplicity of giant components in these representations.

Note the link to \emph{directed percolation} ~\cite{hinrichsen2000non}, where there are two correlation lengths, temporal and spatial, characterising correlations parallel and perpendicular to the directed lattice. In our case, the arrow of time defines the direction, $\rho_{E}$ gives the probability that a randomly selected event in $D_{\delta t}$ belongs to a structurally percolating infinite cluster, while $\rho_{LT}$ is the typical temporal correlation length for a given $\delta t$. However, these correspond to two different order parameters, as the largest and most long-lived components might not be the same.

\subsection*{Weighted event graphs of modelled temporal networks}

To explore how $\Delta t$ controls temporal-network connectivity, we introduce a simple toy model. We define an ensemble of temporal networks $\mathcal{G}_{p,r}(n,k,\lambda)$ where the topology is that of an Erd\H{o}s-R\'enyi (E-R) random graph with $n$ nodes and average degree $k$, and events are generated on each link by a Poisson process with $\lambda$ events per link on average.  We set the observation period $T$ long enough so that  $\Delta t \ll T$ and $\lambda \ll T$.

In this model, there is a transition from the disconnected to the connected phase when the independent Poissonian events become $\Delta t$-adjacent and form a giant weakly connected component in $D_{\Delta t}$. In terms of degree, a lower bound for this critical point can be estimated as the point where the average out-degree of the event graph becomes $\langle k^{out}_{_{D_{\Delta t}}}\rangle=1$. In the underlying E-R network, each edge is adjacent to  $2(k-1)+1$ edges (including the edge itself), and therefore the average out-degree of $D_{\Delta t}$ is $\langle k^{out}_{D_{\Delta t}} \rangle=\lambda \Delta t \left[2(k-1)+1\right]$. The condition for the critical point  can then be written as
\begin{equation}
k_c= \frac{(\lambda \Delta t)^{-1} -1}{2}+1  \hspace {.2in}\mbox{and}\hspace{.2in} \Delta t_c= \frac{1}{\lambda(2k-1)}.
\label{eq:line}
\end{equation}

\begin{figure}[ht!]
\centering
\includegraphics[width=.7\linewidth]{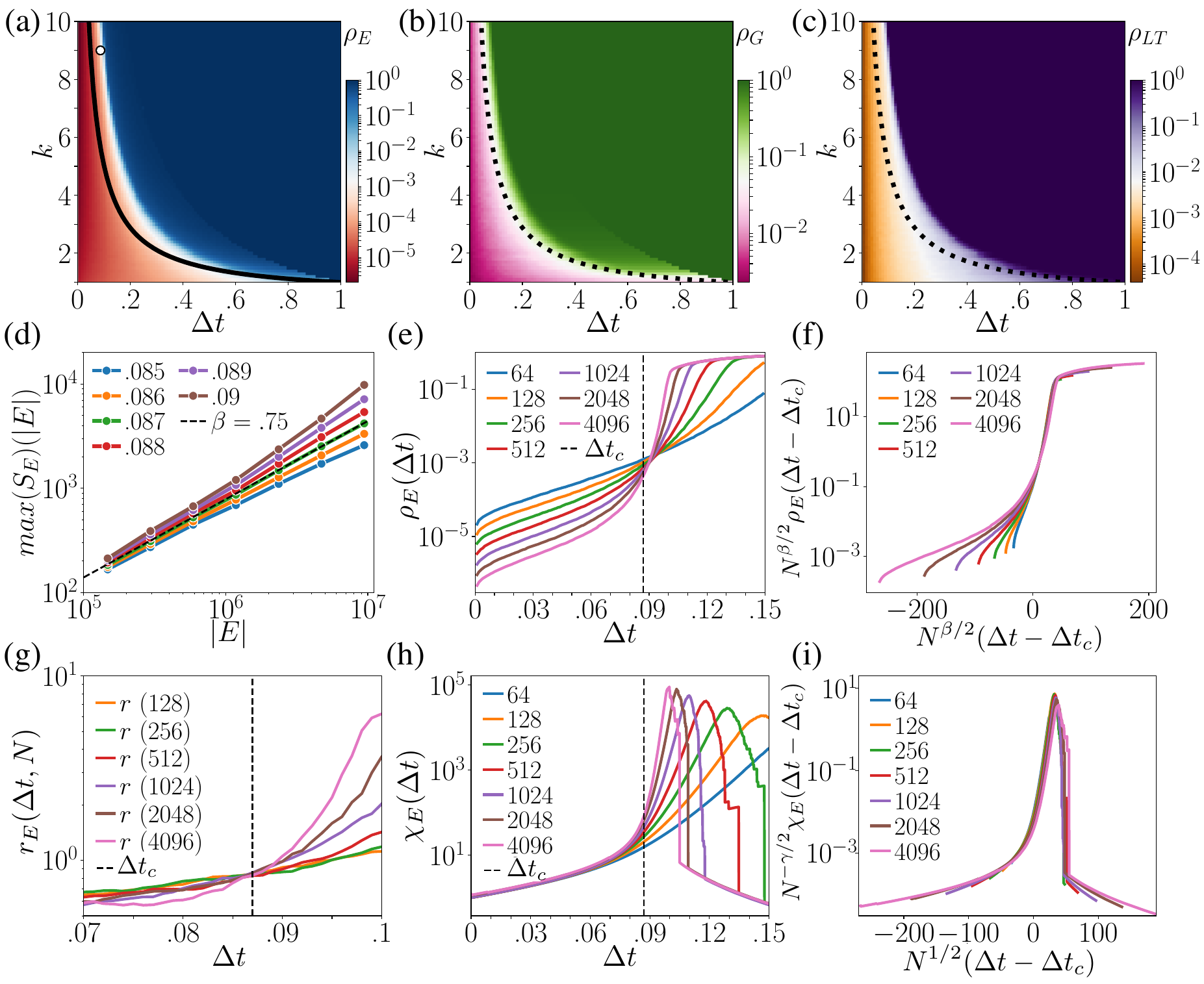}
\caption{\small Phase diagrams for the random temporal network model as a function of the average network degree $k$ and the maximum waiting time between events, $\Delta t$. The colour maps show the (ensemble-averaged) relative size $\rho_{*}(k,\Delta t)$ of the giant weakly connected components, measured as (a) the number of events in the event graph components $S_E$, (b) the number of temporal-network nodes that the largest event graph component covers, and (c) the lifetime of the event graph component $S_{LT}$. The solid line in (a) (dashed lines in (b) and (c)) is the analytic estimate of Eq.\ref{eq:line}. The circle in the upper left corner shows the critical point for $k=9$ determined as explained in the text. (d) Scaling of $\max(S_E)$, the size of the largest weakly connected component in $D_{\Delta t}$, with the size of $D_{\Delta t}$ measured in number of event-nodes $|D_{\Delta t}|=|E|$, for different $\Delta t$. The dashed line assigns the critical $\Delta t_c=0.87$. (e) The order parameter $\rho_E(\Delta t)$ for different network sizes $N=|V|$. (f) Same as (e) after finite-size scaling using the function defined in Eq.\ref{eq:rhofs}. (g) The ratios $r(\Delta t,N)$ crossing at $\Delta t_c$. (h) Susceptibility curves $\chi_{E}(\Delta t)$ for different sizes. (i) Same as (h) after finite-size scaling using the function defined in Eq.\ref{eq:chifs}. Dashed lines in labels (i), (g), and (h) show the critical point determined in (d). Computations for (a-c) are with network size $|V|=2048$ over for $T=512$ time units with an event rate of $\lambda=1$ averaged over $10$ realisations. Results for (d-i) have the same parameters but are averaged over $100$ realisations and may differ in size. This figure was published in~\cite{kivela2017mapping}.}
\label{fig:wdagphase}
\end{figure}

This theoretical line $\Delta t_c(k)$ is shown together with simulated results in Fig.\ref{fig:wdagphase}a, with the number of events determining the relative size of the largest component, $\rho_E$. $\Delta t_c(k)$ separates the simulated percolating and non-percolating regimes well. Fig.\ref{fig:wdagphase}.b and c show the relative largest component sizes in terms of temporal-network nodes ($\rho_G$) and component lifetime ($\rho_{LT}$); a percolation transition appears to take place near the theoretical line $\Delta t_c(k)$ for the number of events from Eq.~\ref{eq:line}. Note generally, the phase transition lines for events, nodes, and lifetime can be different. 

Let us investigate the model's critical behaviour in detail, fixing the average degree to $k=9$. This makes the thresholded event graph $D_{\Delta t}$ dense enough for the mean-field (MF) approach; the MF approximation works well for regular lattices above the critical dimension $d_c=5$~\cite{hinrichsen2000non}. We locate the critical point with two methods. First, when the system reaches a stationary state where the order parameter becomes time-invariant beyond fluctuations, the scaling relation $max(S_E) \sim |D_{\Delta t}|^{\beta}$ is expected to hold around the critical point $\Delta t_c$, where $|D_{\Delta t}|$ is the size of the thresholded event graph in events, and $\beta$ is the critical exponent of the order parameter. We measured this relation for several system sizes and values of $\Delta t$ and found a power-law scaling of $S_E(|D_{\Delta t}|)$ around $\Delta t_c\simeq 0.087$ with the exponent $\beta\simeq 0.75$ (see Fig.\ref{fig:wdagphase}d). This point is shown as a circle in Fig.\ref{fig:wdagphase}a; it is above the analytical estimate, which provides the lower bound for the critical point. Note that for the directed-percolation university class, the MF solution suggests $\beta_{MF}=1$.

The second way of determining the critical point is to calculate the ratios
\begin{equation}
r(\Delta t, N)=\rho_E(\Delta t,N)/\rho_E(\Delta t,N/2)
\end{equation}
for varying $N$~\cite{karsai2006nonequilibrium}. These curves should cross around the critical point $\Delta t_c$ where $r(\Delta t_c,N)=2^{-x}$, and $x$ is related to the finite-size scaling exponent. In Fig.\ref{fig:wdagphase}g, they indeed cross close to $\Delta t_c\simeq 0.087$ with $r(\Delta t)\simeq 0.82$ suggesting an exponent $x\simeq 0.2863$, should be compared to $\beta/2\simeq 0.375$.

Finite-size scaling in networks is naturally related to the network volume $N$ (number of nodes) instead of a linear size scale $\ell$, which usually cannot be defined. Assuming that $N \leftrightarrow \ell^d$, one can derive finite-size scaling functions, which are expected to hold in the conventional mean-field regime $d>d_c$, or for dense networks. This leads to a finite-size scaling function of the order parameter:
\begin{equation}
\rho_{E}(\Delta t,N)\sim N^{-\beta / d \nu^*}\widetilde{\rho}_{E}(N^{1/ d \nu^*}(\Delta t-\Delta t_c)),
\label{eq:rhofs}
\end{equation}
where $\nu^*=2/d$ is the finite-size scaling exponent (of linear size), which depends on the dimension $d$. If $d<d_c$ it is the spatial correlation length exponent, and above the critical dimension $d_c=5$ it takes the value $\nu^*=2/d_c$~\cite{karsai2006nonequilibrium}. At the same time a similar scaling function is expected to hold for susceptibility:
\begin{equation}
\chi_{E}(\Delta t,N)\sim N^{\gamma / d \nu^*}\widetilde{\chi}_{E}(N^{1/ d \nu^*}(\Delta t-\Delta t_c)),
\label{eq:chifs}
\end{equation}
where $\gamma$ is the mean cluster-size exponent. From the definition of $\chi_{E}$ (in Eq.\ref{eq:chi}) and the scaling of $\rho(\Delta t,N)$ at $\Delta t_c$ we can derive the simple exponent relation $\gamma/(d \nu^*)=1-\beta/(d \nu^*)$, where $\nu*=2/d$, $d=d_c=4$ and $\beta\simeq 0.75$, which gives us a value $\gamma\simeq 1.25$ (which is slightly different from the directed-percolation MF value of $\gamma_{MF}=1.0$).

To check whether the predicted finite-size scaling behaviour holds around the critical point, we took the simulated $\rho_{E}(\Delta t,N)$ and $\chi_{E}(\Delta t,N)$ measured for various $N$ (see Fig.\ref{fig:wdagphase}e and h respectively). Using the scaling functions in Eq.\ref{eq:rhofs} and Eq.\ref{eq:chifs} with the determined exponents, we scaled the order parameter and susceptibility as a function of $(\Delta t-\Delta t_c)$. The expected scaling behaviour appears for both quantities close to the critical point (Fig.\ref{fig:wdagphase}.f and i).

\vspace{.1in}

Finally, we carried out a percolation analysis on three empirical temporal networks, a mobile-phone call networks (DS1 in Section~\ref{sec:datasets}), in a sexual contact network~\cite{rocha2011simulated}, and an air transportation network~\cite{USAirline} (detailed results are not reported here but in~\cite{kivela2017mapping}). This analysis showed that these empirical systems goes through similar percolation transitions as we demonstrated for the toy-model. The percolation point $\Delta t_c$ is identifiable in each cases, with $\delta t_c\sim 4$ h $20$ min for the calls, $\delta t_c\sim 7$ d for the sexual-interaction network, and $\sim 20$ minutes for the transportation network. A spreading agent has to survive at least this long at a node to percolate the network, or in case of transportation, this time-scale corresponds the minimum time required to change flights at an airport. In addition, scale-invariant distributions of structurally and temporally connected components verify the located critical points in each system.


\section{Generative models of temporal networks}
\label{sec:tnet_adn}

Since the birth of network science, generative modelling is the most important and frequently used modelling technique to understand the emergent properties of complex networks. Such models are overall useful to identify and to test underlying mechanisms, which lead to features characterising real network features. School examples are preferential attachment, which lead to degree heterogeneities, or triadic closure, which can explain the emergence of community structure in networks. However, although simulations of generative models are necessary dynamical, usually the final emergent aggregated structure has been taken as a result and used further as a static structure. Only recently, some new mechanistic modelling techniques have been proposed to simulate temporal networks, where beyond the emergent structural properties, the dynamical features of interactions are also taken into account. It has been recognised, that this finest level of description of networks is necessary for the deeper understanding of emergent global structural and temporal properties, as after all they are the consequences of individual decision mechanisms driving single interactions between people.

In the following we are going to study one promising direction for the generative modelling of temporal networks. We will introduce the basic concept of \emph{activity-driven network models} and will discuss various ways to extend this framework with decision mechanisms, which lead to more and more realistic models of temporal networks.

\subsection{Activity-driven network model}

The modelling framework of \emph{activity-driven networks} (ADN) has been first proposed by Perra et al. for the agent-based simulation of time-varying networks~\cite{perra2012activity}. This model builds on a single assumption that people are not active with the same pace but there are individual differences in the number of interactions one participates due to differences in personality, age, sociability, etc. Such variation has been found in social systems, where communication frequencies were shown to vary from people-to-people over several orders of magnitudes in a larger population. In the ADN framework these variations are introduced a-priory through a quantity called the \emph{activity potential} $x_i$ associated to each $N$ number of nodes in the network with values sampled from an arbitrary distribution $F(x_i)$. Note that $x_i \in [ \epsilon, 1 ]$ is defined with lower bound assuming a minimum activity level $\epsilon$. The activity potential is a time-invariant function characterising the activity level of agents by determining their $a_i=\eta x_i$ probability to participate in an interaction per a unit time. Here $\eta$ is a time rescaling factor assuring the average number of active nodes per unit time to be $\eta \langle x \rangle N$. Built on these definitions, the ADN model is introduced as an iterative process evolving through global time-steps of $\Delta t$ length, in which on average each node is updated once. More precisely, the generative network process is defined by taking $N$ disconnected vertices at each discrete time step $t$ and activate each vertex with probability $a_i\Delta t$. Once a node is active it generates $m$ links (temporal events) that are connected to randomly selected vertices. Finally, in the end of each iteration step, all links are deleted and the next iteration starts again with a disconnected set of nodes. This algorithm is summarised in Alg.\ref{alg:adn}, where we only save the generated events instead of updating the actual network.

\begin{algorithm}[h!]
\caption{\small \label{alg:adn}Activity-driven network}
\begin{algorithmic}[1]
\Function{$ADN$}{$G=(V,E=\emptyset,a(i)),T,\Delta t$}
\For{$t \gets 1 \textrm{ to } T$}
\For{$N$ randomly selected nodes $i \in V$}
\If{$rand() \leq a_i \Delta t$}
\For{$m$ randomly selected nodes $j \in V \setminus i$}
\State{Output: $event(t,i,j)$}
\EndFor
\EndIf
\EndFor
\EndFor
\EndFunction
\end{algorithmic}
\end{algorithm}

As Perra et al. explains~\cite{perra2012activity}, one can easily realise that the ADN model generates a sequence of random structures. Moreover, if we aggregate these structures over time, the nodes' degree in the aggregated network structure will intuitively depend on their activity potential. Actually there are two mechanisms which can increase the aggregated degree of a node. On one hand, it can be increased by new neighbours contacted by the node. Since at each time step $t\in T$, a node with activity potential $a_i$ selects $m$ other nodes \emph{randomly} for interaction, this out-degree can be computed by making an analogy to the Polya urns problem: it will be equal to the number of different balls extracted from a urn with $N$ balls, performing $Tma_i$ extractions~\cite{mahmoud2008polya}. On the other hand, the degree of a node can be increased by incoming interactions of other active nodes. After some simple calculations (for details see~\cite{perra2012activity}), these two terms can be written as:
\begin{equation}
k_T(i)=k^{out}_T(i)+k^{in}_T(i)=N(1-e^{-Tma_i/N})+Tm\eta \langle x \rangle e^{-Tma_i/N}\sim N(1-e^{-Tm\eta x_i/N})
\end{equation}
in the limit of large $N$ and small $T/N$. Based on this calculation and by using some further approximation we can show that the emerging degree distributions for early times, where $k/N$ is small, appears as
\begin{equation}
P_T(k)\sim \frac{1}{Tm\eta} F\left[ \frac{k}{Tm\eta} \right],
\end{equation}
which indicates that the functional form of the aggregated degree distribution will emerge with the same scaling form as the activity distribution.

This model is very simple and far from being realistic as it produces a sequence of random structures, with an emerging degree distribution trivially the consequence of a parameter of the model. In its simplest form, it does not reproduce common emerging properties of real social networks like link weight heterogeneities, community structure, weight-topological correlations, or burstiness. On the other hand it has the potential to serve as the reference model and the starting point for more realistic network models with integrated microscopic decision mechanisms inducing more realistic network or dynamical characters. As we will see later, there are miriad ways to extend the general ADN model and to use it for hypothesis testing on emerging network features or their effects on dynamical processes. At the same time, ADNs model temporal networks, thus they provide an ideal way to study the importance of multiple temporal scales~\cite{ribeiro2013quantifying}, and dynamical or structural features of interactions on ongoing diffusion, epidemics, opinion formation, complex contagion, etc. processes, just to mention a few examples. I had contributions to extend the general ADN models in two directions, whether by addressing various structural mechanisms~\cite{karsai2014time,ubaldi2016asymptotic,laurent2015calls,tomasello2014role}, or by mechanisms driving the dynamics of dyadic interactions~\cite{ubaldi2017burstiness}. In the following Section, I will summarise some of these studies to familiarise the reader with the concept and potential of this modelling framework.

\subsection{Memory processes in egocentric network formation}
\label{sec:ADNmemory}

In the activity-driven framework we have considered only memoryless generative processes so far. At each time step, nodes select their partners with a uniform probability. The model thus neglects the heterogeneous nature of individuals' social interactions, which is a common character of real social systems where egos may have strong and weak ties. The heterogeneity of social ties is a key ingredient of the social structure and plays a crucial role on diffusion processes~\cite{onnela2007structure}. Our goal here is to understand the mechanism driving their formation, explicitly considering the network's time-varying nature.

\begin{figure*}
\centering
\includegraphics[width=1.\textwidth,angle=0]{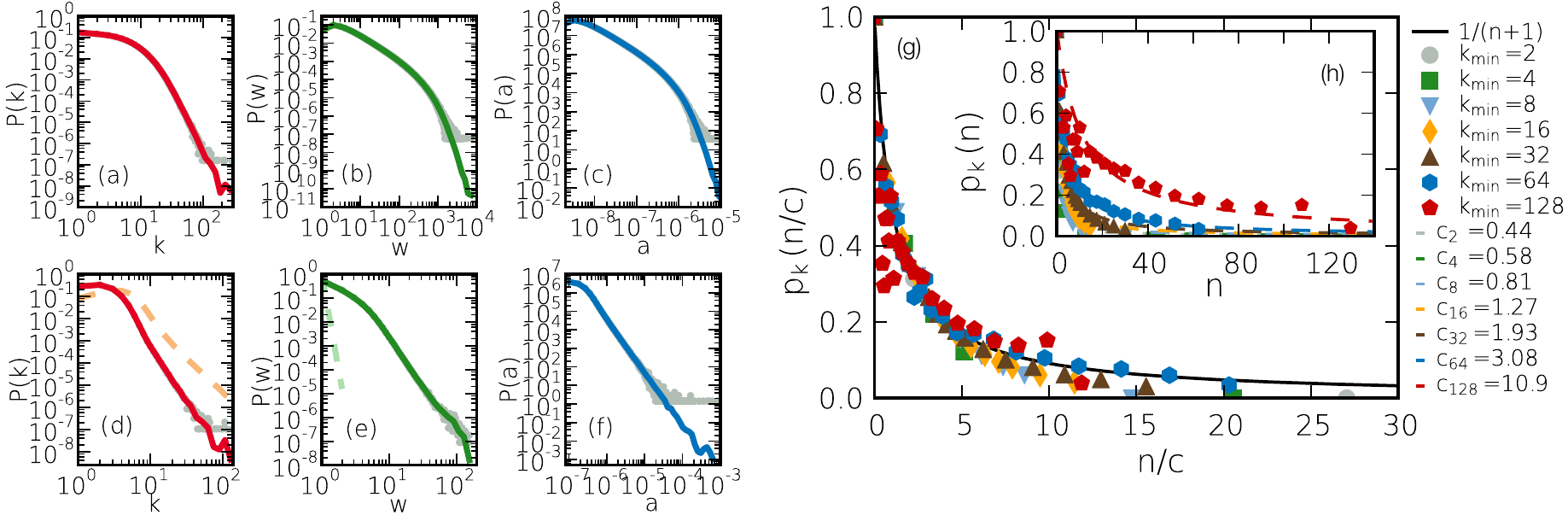}
\caption{\small Panels (a), and (d) plot the degree distributions; panels (b), and (e) plot the weight distributions; and panels (c), and (f)  plot the activity distributions of the empirical and modelled networks (respectively). Grey symbols are the original while coloured symbols are the corresponding logarithmic binned distributions. In panels (d-f) solid lines assign distributions induced by the reinforced process, while dashed lines are due to the memoryless process. Model calculations were parameterised as $N=10^6$, $\epsilon=10^{-4}$ and $T=10^4$. Panel (g) plots the $p_k(n)$ probability functions calculated for different degree groups in the MPC network. In the inset (h), symbols show the averaged $p_k(n)$ for groups of nodes with degrees between the corresponding $k_{min}...k_{min}^2-1$ values. Continuous lines are the fitted functions of Eq.\ref{eq:pn} with $c$ parameter values showed in the legend. The main panel depicts the same functions after rescaling them using Eq.\ref{eq:pnsc}. The continuous line describes the analytical curve of Eq.\ref{eq:pnsc}. This figure was published in~\cite{karsai2014time}.}
\label{fig:ADNMemchar}
\end{figure*}

In our analysis~\cite{karsai2014time}, we focus on a prototypical large-scale mobile phone call network (DS1 in Section~\ref{sec:datasets}), which structure is characterised by heavy-tailed degree $k$, link weight $w$ - defined here as the number of call between people, and activity rate $a$ distributions - defined as the probability of any given node to be involved in an interaction at each unit time (see Fig.~\ref{fig:ADNMemchar}a-c respectively). Our goal first is to identify the mechanisms driving the dynamics of single interactions of the egocentric networks (egonets) and which is responsible for the emergent heterogeneities. We will demonstrate that memory processes characterising each agent plays a crucial role here, which can be incorporated in the ADN model via a simple non-Markovian reinforcing mechanism, that allows to reproduce with great accuracy the empirical data.   

\subsubsection{Egocentric network dynamics.}

As we briefly discussed in Section~\ref{sec:socialnets}, social networks are characterised by two types of links. The first class describes strong ties that identify time repeated and frequent interactions among specific couples of agents. The second class characterises weak ties among agents that are activated only occasionally. It is natural to assume that when observing interactions, strong ties are the first to appear in the system, while weak ties are incrementally added to the egonet of each agent, as it has been demonstrated in~\cite{krings2012effects}. To quantify this hypothesis in observations, we observe the interaction (call) dynamics of egos and measure the probability, $p(n)$, that their next communication event will establish a new $(n+1)^{th}$ link, or will be a repeated interaction on one of their $n$ already observed social ties\footnote{Note, that here and later in the Section, $n$ is equivalent with the degree $k$ of a node at a given time $t$. We distinguish between these notation to avoid confusion between the static degree $k$ (here when $t=T$) and the evolving degree $n$ (at time $t$) of a node.}. We calculate these probabilities in the MPC dataset averaging them for users with the same degree $k$ at the end of the observation time. We therefore  measure the quantity $p_k(n)$ for the egonets with the same degree $k$ and  $n\leq k$.  The empirical $p_k(n)$ functions for different degree groups are shown in Fig.\ref{fig:ADNMemchar}g inset (coloured symbols). Interestingly, the probabilities are decreasing with $n$ for each degree class indicating a slowing down in the egocentric network evolution. The larger the egocentric network, the smaller the probability that the next communication will be with someone who was not contacted before. In other words, agents have memory to remember their social ties and they tend to repeat interactions on them.

\begin{figure*}
\centering
\includegraphics[width=1.0\textwidth,angle=0]{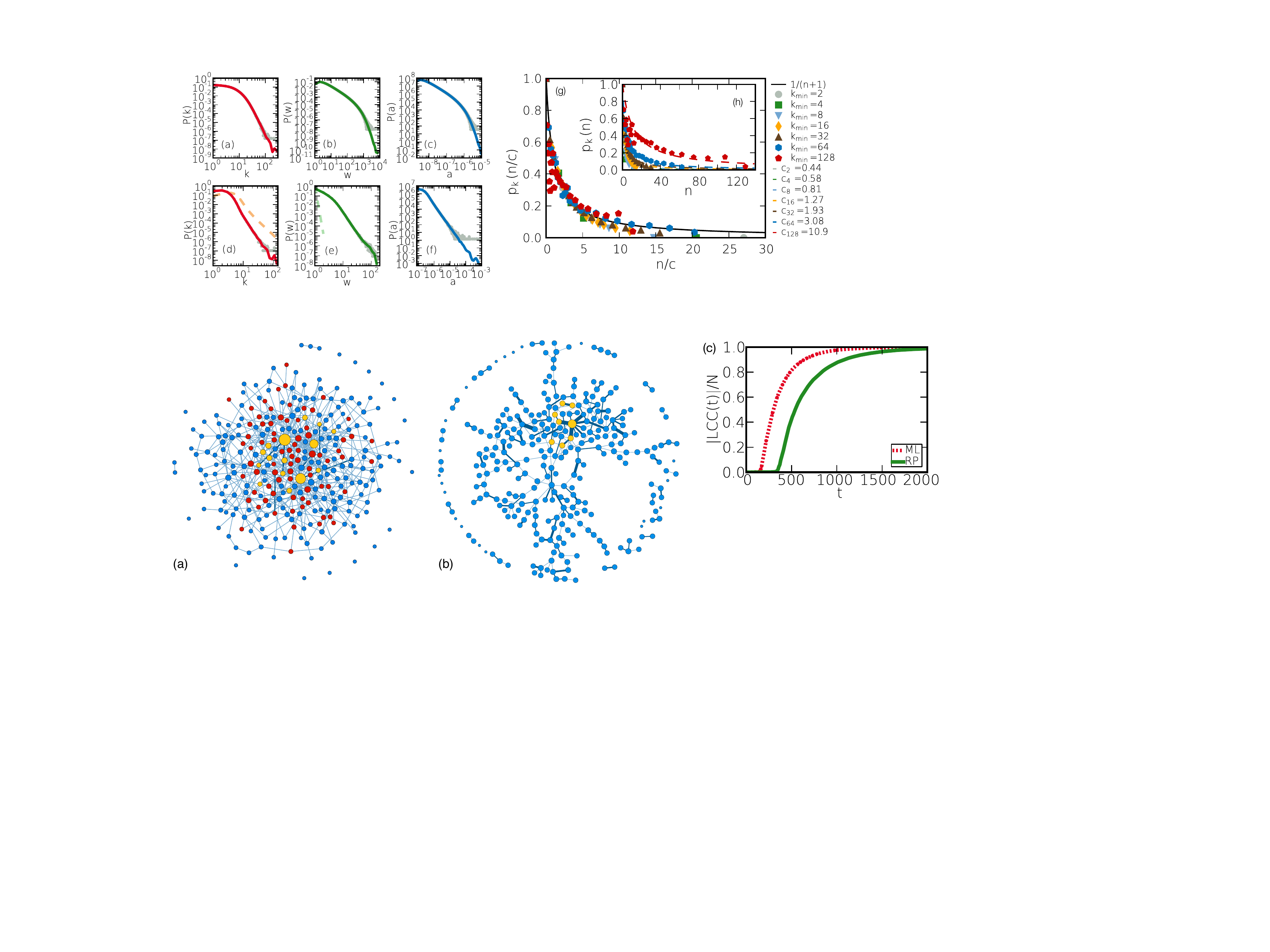}
\caption{\small Emergent structure and rumour spreading processes in (a) ML and (b) RP activity-driven networks. Node colours describe their states as ignorant (blue), spreader (red) and stifler (yellow). Node sizes, colour, and width of edges represent the corresponding degrees and weights. The parameters of the simulations are the same for the two processes: $N=300$, $T=900$, $\lambda=1.0$, and $\alpha=0.6$. The processes were initiated from a single seed with maximum strength. In panel (c) we show the sizes of the largest connected components (LCC) as a function of time for time aggregated ML and RP networks. Simulations were run with the same parameters considering $N=10^5$ nodes. This figure was published in~\cite{karsai2014time}.}
\label{fig:ADNMemDemo}
\end{figure*}

The empirical growth of the egonet  can be captured by a simple mechanism. We find that the probability that a node, characterised by a social circle of size $n$, will establish a new tie is well fitted by the expression :
\begin{equation}
 p(n)=1-\frac{n}{n+c}=\frac{c}{n+c}.
\label{eq:pn}
\end{equation}
Analogously, the probability of having an interaction with someone who is already in the egocentric network is $n/(n+c)$. Here $c$ is an offset constant depending on the degree class considered. By fitting the function in Eq.\ref{eq:pn} on the empirical data (solid lines in Fig.\ref{fig:ADNMemDemo}g inset) we can determine the corresponding constant $c$ for each degree group and using them rescale the empirical $p_k(n)$ functions as
\begin{equation}
 p_k(n/c)=1/(n/c+1).
\label{eq:pnsc}
\end{equation}
This rescaling collapse the data points of different degree groups on a single curve  (see Fig.\ref{fig:ADNMemchar}g main panel), which suggests that the same mechanism is driving the evolution of the egonets of all individuals independently of their final number of connections.

Next question, how can we integrate such a mechanism into the ADN model to test its effect on the emerging structure. Inspired by the observations in the MCP dataset, we impose a reinforcement mechanism in the ADN model definition by extending it in two ways. First, we assume that nodes have memory, in other words they remember the set of other nodes they have been in contact during the course of the simulation. Second, we assume that when a node, with $n$ previously established social ties, becomes active in an iteration step it can make two kinds of actions: (a) it will contact randomly a new node with probability $p(n)=c/(n+c)$, or (b) with probability $1-p(n)=n/(n+c)$ it will interact with a node already contacted, thus reinforcing earlier established social ties. In this case, the selection is done randomly among the $n$ actual neighbours. This model, that in the following we will denote as RP (reinforcement process), is non-Markovian as memory is explicitly introduced in the ego network dynamics. For the moment we fix $c=1$ for all the nodes while we will generalise our model later~\cite{ubaldi2016asymptotic}.

A side by side comparison of the time-aggregated representations of networks generated by the memoryless (ML) and reinforced process (RP) models (using the same parameters) is shown in Fig.\ref{fig:ADNMemDemo}-a and b (where colours should be disregarded for the moment). In both cases we assumed that the distribution of the activity potential followed a power-law $F(x)\sim x^{-\nu}$, with an exponent value $\nu=2.8$ matching the observed MPC value. As a consequence, in case of the ML dynamics (Fig.\ref{fig:ADNMemDemo}a), we obtained an aggregated network with a degree distribution $P(k)\propto k^{-\gamma}$ where $\gamma=\nu$ and a weight distribution decaying exponentially~\cite{perra2012activity,starnini2013topological}. This is also confirmed by large scale simulation results reported in Fig.\ref{fig:ADNMemchar}d and e (dashed lines). In case of the RP dynamics (Fig.\ref{fig:ADNMemDemo}.b), the  memory process induces a considerably different structure. These effects are quantified in Fig.\ref{fig:ADNMemchar}d, e, and f (solid lines). We observe a degree distribution that is heavy-tailed but more skewed in the RP model than the ML ($\nu_{RP}>\nu_{ML}$). This distribution is qualitatively matching better the corresponding empirical measure in Fig.\ref{fig:ADNMemchar}a. Furthermore, the RP model generates heterogeneous weight distributions (see Fig.\ref{fig:ADNMemchar}e solid line) capturing extremely well real data (see Fig.\ref{fig:ADNMemchar}b). This is not the case in the ML model where the absence of memory induces exponential weight distributions far from reality (see Fig.\ref{fig:ADNMemchar}e dashed line). The RP dynamics not only induces realistic heterogeneities in the network structure, but also controls the evolution of the macroscopic network components. Indeed, due to the reinforcement mechanism, the largest connected component (LCC) in RP networks grows considerably slower than in the case of ML models (for illustration see Fig.\ref{fig:ADNMemDemo}.c). This is an important feature because dynamical process evolving on time-varying networks will progress with a time-scale that cannot be smaller than the LCC growth time-scale. As consequence, any dynamical phenomena taking place on time-varying networks with memory will evolve at a slower rate than in memoryless time-varying networks.

\subsection{Individual heterogeneities in social capacity}
\label{sec:soccap}

In the definition of the memory driven ADN model we took some assumptions for simplicity, which limited our model to account for differences potentially characterising individuals. First of all we assumed that the scaling constant in Eq.~\ref{eq:pn} is a constant value for everyone. This constant determine the intrinsic characteristic limit of an ego to maintain multiple ties, thus it may vary from a person to another. At the same time we assumed that speed of exploration of new friends is entirely coded in the activity potential of people. Recent findings demonstrate, that this ability may vary strongly between people depending whether they are more like \emph{social explorers} or \emph{social keepers}~\cite{miritello2013limited}. Such variance can be easily incorporated in our memory function as:
\begin{equation}
p_b(n)=\left( 1+ \frac{n}{c_b} \right)^{-\beta_b}
\end{equation}
where $\beta_b$ modulates the tendency to explore new connections. Just as earlier, here we assign nodes to degree classes $b$ containing actors with statistically equivalent characteristics, i.e. nodes that engaged a similar number of interactions and that feature a comparable cumulative degree in the observation period. We have shown in~\cite{ubaldi2016asymptotic} that this functional form fits very well on memory functions measured in several empirical temporal networks like Twitter mention interactions (TWT), scientific co-publications (PRA, PRB, PRL), or in case of mobile phone call (MPC) interactions. Moreover, we found that the $\beta_b$ parameter of the $p_b(n)$ functions appears highly invariant over various final degree groups. One exception was the MPC network, where some variance and negative correlations have been found.

By leveraging on this result we can define a variation of the memory driven ADN model by implementing $p_b(n)$ in the decision process. Moreover, for this model it is possible to write explicitly the Master Equation (ME)
describing the evolution of the probability distribution $P_i(n,t)$ that a node $i$ has degree $n$ at time $t$:

{
\begin{align}
    \label{eq:ME_mm}
        &P_i(n, t + 1) = P_i(n - 1, t)  \bigg[ {a_i p_i(n-1)} + \sum_{j\nsim
        i}{ a_j \sum_{n_j}{p_j(n_j)\over (N - n_j)}  P_j(n_j, t) } \bigg] + &\nonumber \\
        &P_i(n_, t)  \bigg[ {a_i [1 - p_i(n)]} + \sum_{j\nsim
        i}{ a_j \sum_{n_j}{\Big( 1 - {p_j(n_j)\over (N - n_j)}\Big) P_j(n_j, t)}
        } \bigg] + P_i(n, t)  \bigg[ 1 - \sum_{j}{a_j} \bigg].&
\end{align}
}

In the above equation the sums in $j\sim i$ and $j\nsim i$ run over the nodes already contacted and not yet contacted by $i$, respectively. $n_j$ describes the degree of each node $j$. Moreover, we work in the $a\ll1$ limit, so that we assume that only one node is active for each evolution step. The first two terms on the right hand side of Eq.\ref{eq:ME_mm} account for the increment of the number of nodes having degree $n-1$. The former occurs when node $i$ having degree $n-1$ gets active and contacts a new node with probability $a_i p_i(n)$. On the other hand the latter one is effective when node $i$ gets contacted by node $j$ of degree $n_j$ (that never got in contact with $i$ before) that activates and attaches to node $i$ with probability $a_j p_j(n_j)/(N-n_j)$. In the latter, the $1/(N-n_j)$ factor accounts for the probability of $j$ to exactly select node $i$ amongst the $N-n_j$ nodes outside of the $j$'s neighbourhood of size $n_j$. Likewise, the third and fourth terms of the r.h.s. of the equation account for the conservation of the number of nodes of degree $n$. This is achieved either when node $i$ gets active and contacts one of its neighbours with probability $a_i(1-p_i(n))$, or when $i$ gets contacted by one of its neighbours. The last term of Eq.~\ref{eq:ME_mm} accounts for the possibility in which no node gets active in the current evolution time step, thus conserving the $P_i(n,t)$. Given the $a\ll1$ approximation this term reads $\prod_j (1-a_i) \simeq 1-\sum_j a_j$.

\begin{figure*}
\centering
\includegraphics[width=1.0\textwidth,angle=0]{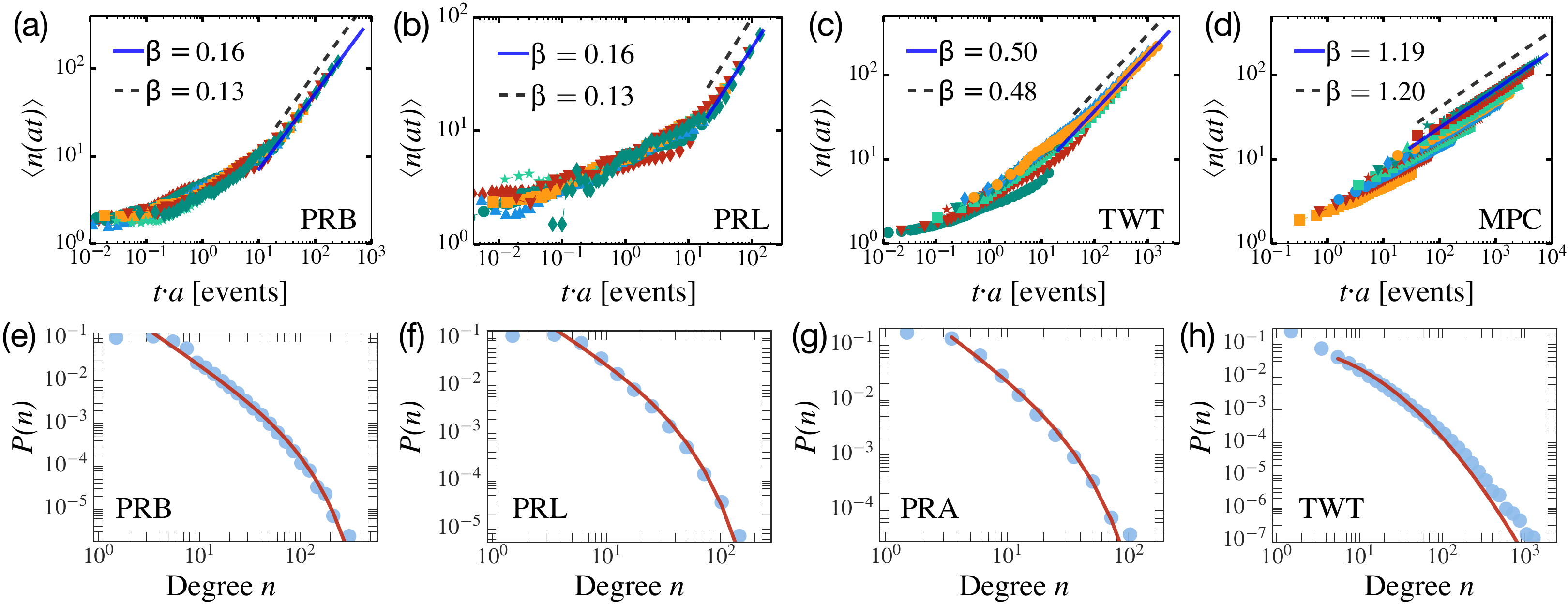}
\caption{\small The rescaled $\langle n(at)\rangle$ curves for selected nodes classes belonging to the (a) PRB , (b) PRL, (c) TWT, and (d) MPN datasets. The time of the original data (symbols) is rescaled with the activity value $t\rightarrow a t$. We also show the fitting curve $\langle n(t)\rangle \propto t^{1/ 1 + \beta}$ (blue solid lines) and the expected asymptotic behaviour (black dashed lines). In the MPN case (d) we fit using $\beta =\beta_{\rm min} = 1.2$. (e-h) The degree distribution $P(n)$ for the (e) PRB, (f) PRL, (g) PRA, and (h) TMN datasets. The predicted functional form of $P(n)$ found in Eq.~(\ref{rhok}) is shown for comparison (red solid lines). We show the data starting from the lower bound of the degree distribution, including in the plot all the statistically significant measures of the probability density function $P(n)$. This figure was prepared by E. Ubaldi and published in~\cite{ubaldi2016asymptotic}.}
\label{fig:ADNDegEv}
\end{figure*}

In case the network is characterised by a single exponent $\beta$, in the large time and degree $1\ll n\ll N$ limits (so that $n$ can be approximated by a continuous variable and $N-k \approx N$), the probability distribution $P_i(n-1,t)$ can be obtained explicitly as
\begin{equation}
    P_i(n, t) = A \exp{\left[ -\frac{\left( n - B(a_i,c_i) t^{\frac{1}{1 + \beta}}
    \right)^2}{C t^{\frac{1}{1 + \beta}}} \right]},
    \label{eq:Pakt}
\end{equation}
where $A$ is a normalisation constant, $C$ a constant and $B(a_i,c_i)$ a multiplicative factor of the $t^{1/(1+\beta)}$ term that depends on the activity $a_i$ and $c_i$ of the considered agent $a_i$ (for further details see~\cite{ubaldi2016asymptotic}). From Eq.~\ref{eq:Pakt} we can determine the evolution in time of the average degree $\langle n(a,t) \rangle$ of nodes belonging to a given activity class as:
\begin{equation}
    \langle n(a, t) \rangle \propto \left( at \right)^{1\over 1 + \beta}.
    \label{evol}
\end{equation}
The growth of the system is thus modulated by the parameter $\beta$ that sets the strength of the reinforcement of ties. The validity of Eq.~\ref{evol} is demonstrate in Fig.~\ref{fig:ADNDegEv}a-d, where the average degree evolution is shown for various activity groups in different empirical datasets, together with analytical predictions. Note that in the limit $\beta=0$ the growth would be linear and in the opposite limit $\beta \rightarrow \infty$ each node would create, and constantly reinforce, just one tie, i.e. the first established. Furthermore, Eq.~(\ref{evol})  connects, at a given time $t$, the actual degree $n$ and the activity $a$ of a given node, as $n \propto a^{1\over 1+\beta}$. Thus, given any specific activity distribution $F(a)$, we can infer the functional form of the degree distribution $P(n)$ by substituting $a\to n^{1+\beta}$, finding:
\begin{equation}
    P(n) dn \propto   F(n^{(1+\beta)}) n^{\beta} dn.
    \label{rhok}
\end{equation}
This prediction is demonstrated on empirical data in Fig.~\ref{fig:ADNDegEv}e-h. It is important stressing that the analytical framework is not limited to a specific functional form of the activity. Indeed, with an arbitrary functional form of $F(a)$, Eq.~(\ref{evol}) gives us the possibility to predict the behaviour and parameters of the corresponding degree distribution (for demonstration see~\cite{ubaldi2016asymptotic}).

In case of different values of $\beta$ in the system complicates the model beyond analytical tractability. Nevertheless, we find that the leading term of the evolving average degree can be described by introducing a simplified model, where the minimum $\beta$ value observed in the ensemble~\cite{ubaldi2016asymptotic}.

\subsection{Dyadic closure and node removal mechanisms}
\label{sec:ADNdyadic}

Building on the model with the simple memory process we discussed in Section~\ref{sec:ADNmemory}, here we further introduce three mechanisms, which are assumed to shape social networks~\cite{laurent2015calls}. They arguably lay behind the emergence of several realistic characters like the community structure, weight-topological correlations, and the stationary evolution of the temporal network structure. 

We consider two \emph{dyadic closure} mechanisms~\cite{laurent2015calls}, one called \emph{cyclic closure}, responsible for triangle formation in social networks, it shapes the social structure at mesoscopic scale, and it leads to the emergence of communities \cite{hebert2015complex}. The other mechanism called \emph{focal closure}, on the other hand, is independent of network structure and represents the formation of ties between individuals with shared attributes or interests. It is driven by the propensity to seek cognitive balance between connected egos~\cite{heider2013psychology,granovetter1973strength} as suggested by earlier theories in sociology~\cite{simmel1964sociology, rapoport1977contribution}. An applicative definition of \textit{cyclic} and \emph{focal closure} in general is given by Kumpula \emph{et al.}~\cite{kumpula2007emergence,kumpula2009model}, who modelled cyclic closure as biased local search, as contrary to \emph{focal closure}, which is modelled as an unbiased global random search. Finally, the third ingredient of the model is a \emph{node removal} process, which ensures the network to reach an equilibrium state where its overall characteristics become invariant of time.

We introduce these mechanisms in the memory-driven ADN model by first, at each iteration step, deleting nodes with probability $p_d$. To keep the network size constant, for each deleted node, we add a new disconnected node to the network in the next iteration step. If a node is not deleted, we let it to follow the memory driven activation process but when it repeats an interaction on an existing link, instead of completely randomly, it selects one of its neighbours $j$ with probability $p^w_{ij}=w_{ij}^t/\sum_{k\in V_t^i} w_{ik}^t$ weighted by the number of their past interactions. The two nodes then interact and increase their link weight $w_{ij}^t$ by $\delta$, a parameter which mimics a \textit{social reinforcement process}. On the other hand, if the node decides to form a new link, it may follow different strategies. In all cases, the new tie will initially have unit weight $w^t=1$. If the degree of the focal node is $0$, it randomly picks another node from the entire network $j$ (\textit{focal closure}) and forms a tie. Otherwise, it attempts to create a new link with a triadic closure mechanism. First, it chooses one of its neighbours $j$ randomly with a weighted probability $p^w_{ij}$. If $j$ has no other neighbours than $i$, node $i$ looks for another random node to interact with (\textit{focal closure}) and forms a link. Otherwise, it looks for a random neighbour $k$ of $j$ ($i\neq k$) with a weighted probability $p^w_{jk}$. If $k$ is not an already existing neighbour of $i$ ($k\notin V_t^i$), the two nodes interact with probability $p_{\Delta}$, and close the triad by forming a link (\emph{cyclic closure}). Otherwise, with probability $1-p_{\Delta}$, node $i$ follows the \textit{focal closure} strategy and instead forms a link with a randomly selected node (other than $j$ and $k$). Finally, if $k$ is already a neighbour of $i$, that is $k\in V_t^i$, the two nodes interact and increase the weight of their existing link by $\delta$ (\textit{reinforcement process}). At the end of each iteration step, all nodes finish their active interactions but remember their already connected neighbours $j\in V_t^{i}$ and the weight $w_{ij}^t$ of interactions with each of them. For a pseudocode of the algorithm see~\cite{laurent2015calls}.

\begin{figure*}[h!]
\centering
\includegraphics[width=.8\textwidth,angle=0]{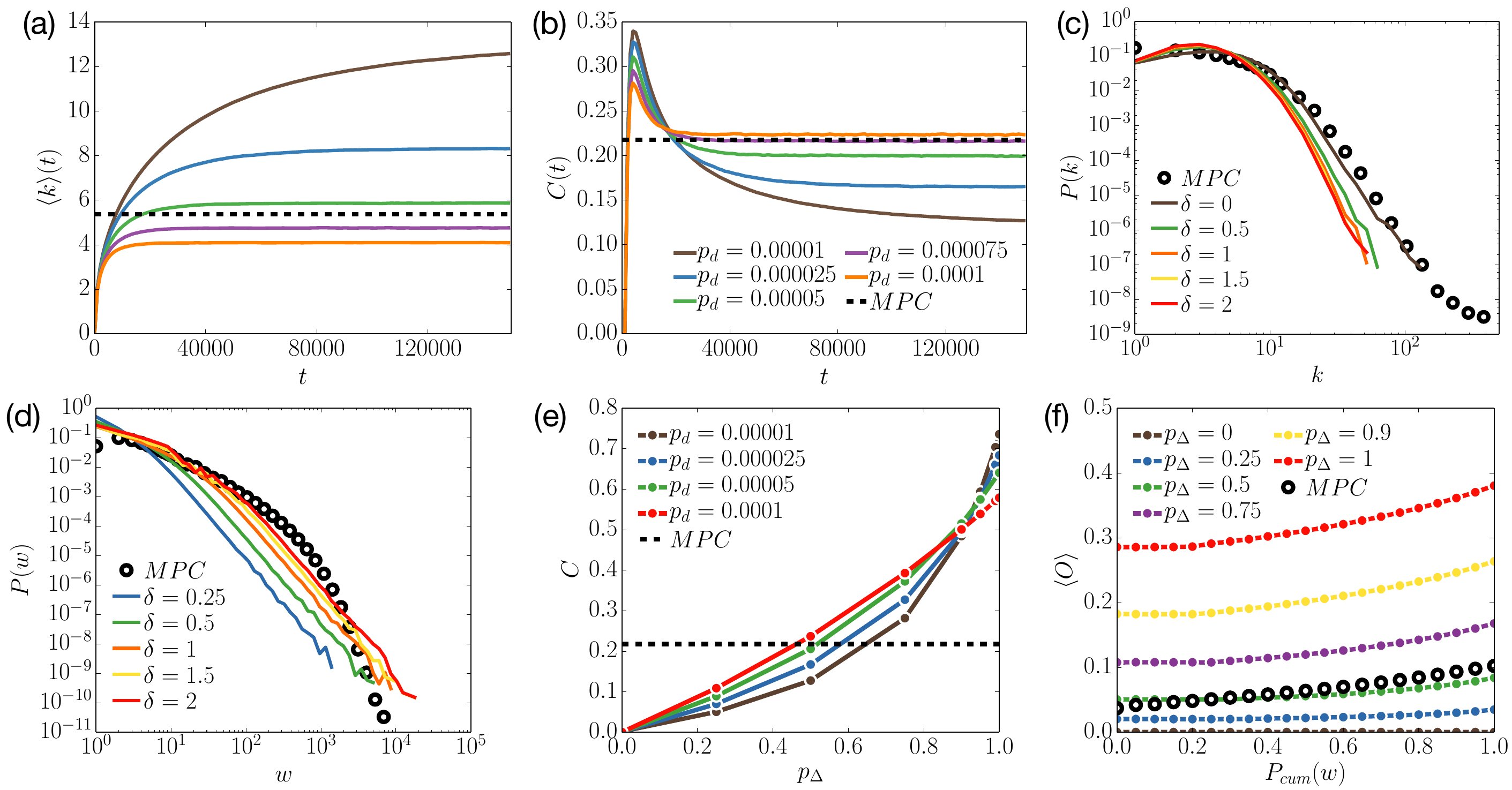}
\caption{\small The evolution of (a) the average degree $\langle k \rangle(t)$ and (b) clustering coefficient $C$ as the function of time and deletion probability  $p_d$ (for exact values, see legend). Panel (c) and (d) depicts the $P(k)$ degree and $P(w)$ weight distributions (respectively) of model networks with varying $\delta$ values. Panel (e) shows the dependence of the average local clustering coefficient $C$ on the parameter $\delta$, while emerging weight-topology correlations are shown in panel (f) as the function of the same parameter measured as the average overlap $\langle O \rangle$ vs the cumulative tie strength $P_{sum}(w)$ in modelled networks. Black circles and dashed lines denote the corresponding empirical MPC measures. This figure was published in ~\cite{laurent2015calls}.}
\label{fig:ADNDyad1}
\end{figure*}

In addition to the activity-driven model parameters whose values are fixed ($\eta=1$, $\epsilon=10^{-3}$, $\Delta t=1$ and $\gamma=2.8$), our model has three intensive parameters, $p_{\Delta}$, $p_d$, and $\delta$. By varying them, one can simulate a rich variety of time-varying networks with several emergent structural properties and correlations. In the following, we explore how the properties of the emerging network structure depend on time and on the intensive parameters, and whether these properties match empirical observations. As a real reference network we use an aggregated representation of a mobile phone communication network (DS1 in Section~\ref{sec:datasets}), with characters shown as black dashed lines or empty circles in Fig.~\ref{fig:ADNDyad1}. In the following, model networks were generated via simulations with $N = 10,000$ nodes (if not noted otherwise), and results were averaged over $100$ independent realisations. We measured network characteristics by considering links that are actually present in the network, i.e. we disregarded links of removed nodes.

The simulated networks are inherently temporal thus they generate time-varying interactions and an evolving network structure. To explore their evolution, we measured their $\langle k \rangle(t)$ average degree and the $C(t)$ average local clustering coefficient as functions of time. Since each process starts from a set of disconnected agents, and hence all measured properties are trivially zero at time t = 0. However, as time goes by and ties are formed via temporal interactions, $\langle k \rangle(t)$ increases until the network reaches a stationary state with constant $\langle k \rangle(t)$. The time which takes to reach this equilibrium state strongly depends on the node deletion probability $p_d$ as shown in Fig.~\ref{fig:ADNDyad1}a. The same is true for the clustering coefficient (see Fig.~\ref{fig:ADNDyad1}b), however before its equilibrium, it increases due to triangles introduced in the beginning of the process, and then it decreases once ties are formed and new interactions appear only as reinforcement or by creating random links. Such ties decrease the relative density of triangles until the clustering coefficient reaches a stationary value.

\begin{figure*}
\centering
\includegraphics[width=1.0\textwidth,angle=0]{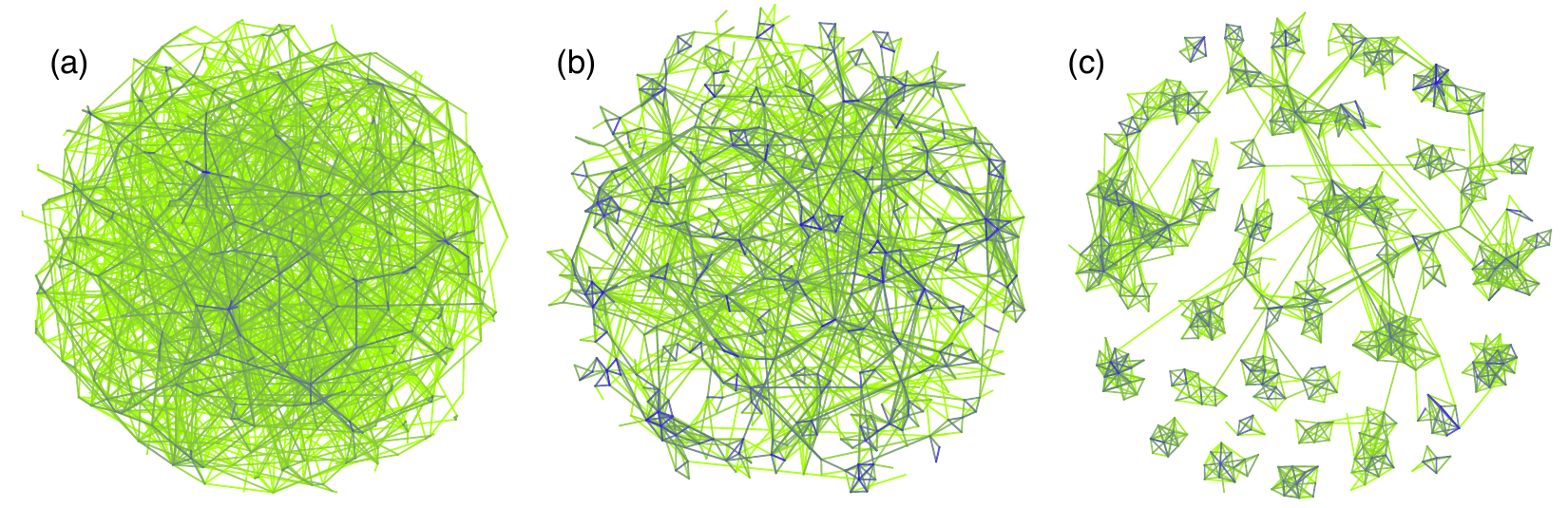}
\caption{\small Demonstration of the emerging structure in the time-varying network model. Panels (a-c) depict simulated networks with fixed $\delta=1$ and varying $p_{\Delta}=0.5$, $0.9$, and $0.995$ ($p_d=4e-5$, $2e-5$, and $1.04e-5$) respectively. Each panel depicts the actual structure of a network with $N=500$, in its stationary state. Links are coloured according to their weight (darker link colour = stronger link weight). This figure was published in ~\cite{laurent2015calls}.}
\label{fig:ADNDyad2}
\end{figure*}

Next we see how the emerging structure depends on the reinforcement parameter $\delta$. We fix $p_{\Delta}=1$ and $p_d= 5\times10^{-5}$ and vary $\delta$ between $0$ and $2$, while measuring network properties in the stationary state. As we see in Fig.~\ref{fig:ADNDyad1}c and d the network emerges with heterogeneous degree and weight distributions, which relatively well approximate the empirical distributions (shown with black empty circles). They appear with a tail robust against $\delta$ but they are shifted, especially $P(w)$, assigning that the average weight naturally depends on the strength of reinforcement.  

At the same time cyclic closure mechanism dominantly control the emergence of triangles and cluster in the network. This is shown in Fig.~\ref{fig:ADNDyad1}e (where we set $\delta=1$), where by increasing $p_{\Delta}$, cyclic closure becomes more dominant, reflected by the steep increase of clustering. Several of the depicted $C(p_{\Delta})$ curves cross the empirical value, indicating that various sets of parameters ($p_{\Delta}$ and $p_d$ as well) are eligible to yield networks with realistic clustering value. In addition, we measured weight-topology correlations to check if the emerging model networks recover the Granovetterian weak-tie structure (for more on this theorem see Section~\ref{sec:socialnets}). More precisely, we computed the average link overlap $\langle O \rangle$ (for definition see Section~\ref{sec:cn}) as the function of the cumulative tie strength $P_{cum}(w)$~\cite{onnela2007structure}. In the MPC network, this quantities are positively correlated (black circles in Fig.\ref{fig:ADNDyad1}f) in accordance with earlier observations~\cite{onnela2007structure}. Interestingly, such correlations spontaneously emerges in the modelled networks for any positive values of $p_{\Delta}$ (see Fig.~\ref{fig:ADNDyad2}f) or $\delta$ (not shown here~\cite{laurent2015calls}). For larger $p_{\Delta}$, this function remains positive and shifted upward as more triangles evolve. More importantly, the combined effects of these two mechanisms, cyclic closure and reinforcement, lead to the emergence of a community like structure in the modelled aggregated network. It is demonstrated in Fig.~\ref{fig:ADNDyad2}a-c where we plot aggregated structures for increasing $p_{\Delta}$ (left to right). When $p_{\Delta}$ is small, the network structure is like a densely connected random structure, but with heterogeneous weights due to the present reinforcement process ($\delta=1$). On the other hand, by increasing $p_{\Delta}$, communities emerge simultaneously with strong ties inside communities, and weak ties between them (darker link colour = stronger link weight). The cyclic closure alone would not be eligible for the emergence of such communities and the Granovetterian picture, as although the clustering of the network wold be large, no strong ties would bias the link creation process and the network would appear rather homogeneous on the mesoscopic scale~\cite{laurent2015calls}.

\begin{center}
  $\ast$~$\ast$~$\ast$
\end{center}

Demonstrated by studies summarised above, activity driven network models have a large potential to define mechanistic models of temporal networks with realistic emerging features. They provide an ideal testbed for identifying the role of microscopic mechanisms in the network formation, and their effects on the final outcome of dynamical processes. This summary already demonstrated the capacities of this framework, thus for the sake of concise description, I decided to excluded two published works of mine from the discussion.

One~\cite{tomasello2014role} addresses the role of endogenous and exogenous link creation mechanisms in cross-country and cross-sectoral R\&D company alliance networks. The extended ADN model is able to reproduce a number of micro-level measures, including the degree distributions, local clustering, path length, component sizes, and the emergence of alliance clusters. Furthermore, by estimating the link probabilities towards newcomers and established firms from the data, we find that endogenous mechanisms are predominant over the exogenous ones in the network formation, thus quantifying the importance of existing structures in selecting partner firms.

The other study~\cite{ubaldi2017burstiness} extends the ADN framework by considering strategies what individuals adopt when selecting between new or old social ties (similar to the model described in Section~\ref{sec:soccap}), and the bursty nature of the social activity setting the pace of these choices. In this paper we studied the non-trivial interplay of these two, structural and dynamical, characters and found that the effects of burstiness might be suppressed when individuals exhibit a strong preference towards previously activated ties. These modelling results provides a principled method to classify the temporal features of real networks, and thus yield new insights to elucidate the effects of social dynamics on spreading processes.

\section{Conclusions}

In this Chapter we focused on my contributions to the theoretical and methodological foundation of the field of temporal networks. Beyond the discussion of my overall view and interpretation of this way of description, we walked through several results addressing the representation, macro-, meso- and microscopic level description of temporal networks, while we also introduced several techniques for their data-driven analysis and modelling, applying random-reference and mechanistic modelling paradigms. On the other hand, we have not addressed so far another important aspects of (static or time-varying) networks: their effects on ongoing dynamical processes. Such questions will be central in the coming Chapter where we will study collective social phenomena and aim to understand how structural and dynamical characters of networks influence the unfolding of processes modelling social or epidemic spreading.












\biblio





\chapter{Collective phenomena on networks}
\label{ch:dynpr}

\section{Introduction}
\label{sec:coll_intro}

Collective phenomena emerge in various ways in complex systems. Assuming a network description, they can appear as cooperative patterns in the structure or the dynamics of interactions, as we just discussed in the previous two Chapters. On the other hand, cooperative phenomena may appear as collective states of nodes in a network. Such patterns may strongly depend on the underpinning structure or could even alter the network formation, this way leading to correlated patterns between the states and interactions of agents. In this Chapter, we are going to focus on the observations and modelling of such phenomena emerging \emph{on networks}.

During this discussion we are going to move along several dimensions characterising collective processes. First of all, one dimension may consider the coupling between the phenomena and the underlying network. If the system exhibits \emph{one-directional coupling}, only the evolving process is influenced by the (static/temporal/multiplex) structure, while the network evolves without being altered by the ongoing phenomena. Good examples are epidemic processes in case patients are not aware of their infection, thus they continue their interaction practise and propagate further the actual disease without intervention~\cite{barrat2008dynamical}. On the other hand, if the system is characterised by \emph{mutual coupling}, node states and interactions influence each other, which leads to adaptive systems with correlations between the phenomena and the network structure. This type of symmetry arguably lays behind the indistinguishability problem of social influence and homophily~\cite{mcpherson2001birds,aral2009distinguishing}, where one cannot decide whether two connected individuals in a social network are similar because they influenced each other after connection (social influence), or they became connected because they were similar at the outset (homophily). These effects commonly trouble the observation of social spreading phenomena, as we will discuss in the coming Chapter.

Another imaginary dimension may consider the ways of observing a system. While arguably all emerging phenomena in complex systems are the results of some dynamical process, sometimes data do not allow to follow their emergence via \emph{dynamic observations}, but to make only \emph{static observations} on their actual state. While in the former case a possible diachronic analysis may lead to deeper insight about the driving mechanisms (e.g. in case of epidemic spreading), in the latter case observations are limited to the detection of correlations in static informations.

Finally, we may distinguish between different modelling techniques we choose to explain the observed phenomena. One way involves statistical methods, which help us to identify various hidden dependencies. On the other way, we will discuss agent-based models of dynamical processes. Various such processes were proposed to model complex emerging behaviour, e.g. random walks for diffusion, sandpile models for self organised critical systems, or coupled oscillators for synchronisation processes, just to mention a few. Here we will mostly concentrate on spreading phenomena, thus we will limit our discussion to two generic modelling frameworks. On one hand, we will discuss models of \emph{simple contagion processes}, arguably describing epidemic like spreading~\cite{barrat2008dynamical}, while on the other hand we will also develop models of \emph{complex contagion processes}, commonly assumed behind social spreading phenomena~\cite{centola2007complex}.

The following Chapter is organised via these axes to explore some relation between social networks and ongoing social phenomena. First we will discuss empirical studies based on static or dynamic observations of collective behaviour, and subsequently we will summarise modelling results exploring the critical behaviour and dynamics of simple and complex contagion processes.

\section{Static observations of collective social phenomena}

The un-precedent availability and amount of digital data, recording traces of human behaviour, opened the gate for the quantitative analysis of various social phenomena. Due to these developments it recently became possible (a) to verify earlier social hypothesis, which were taken as granted in the scientific community, but has never been observed quantitatively, (b) to generalise earlier observations made in small-scale experiments usually on a non-representative groups of individuals, and (c) to observe new aspects of human behaviour to build novel hypothesises. Evidently, independently of the method, all such behavioural studies should consider the dynamical aspects of the observed phenomena, however, this is not always the case. Eventually, sometimes it is disregarded for simplicity, or the data records a horizontal view of a social system, involving a large population, but only capturing the state of the system at a given time point. 

Here we are going to discuss a set of studies~\cite{leo2016socioeconomic,leo2016correlations,abitbol2018socioeconomic,abitbol2018location,mocanu2015collective}, which take this static observation approach to identify correlated patterns of individual attributes such as socioeconomic status or language usage and their variability as the function of the social network structure and other factors. Our findings, on one hand, verify long lasting hypothesis about socioeconomic inequalities, social stratification and status homophily, and on the other hand, provide new observations of patterns in language variance and socioeconomic assortativity.

\subsection{Socioeconomic correlations in social-communication networks}

Socioeconomic imbalances, which universally characterise all modern societies~\cite{piketty2015capital, sernau2016social}, are partially induced by the uneven distribution of economic power between individuals. Such disparities are among the key forces behind the emergence of social inequalities~\cite{sernau2016social} and determinant of ones socioeconomic status. Although several competing hypothesis has been proposed for the definition and identification of socioeconomic~\cite{giddens1973class}, it is a common understanding that it is a combination of several factors. People who live in the same neighbourhood may belong to the same class, and may have similar levels of education, jobs, income, ethnic background, and may even share common political views, obtained similar social or cultural capital or reputation~\cite{bourdieu1984distinction}. These similarities together with homophily, i.e., the tendency that people build social ties with similar others \cite{mcpherson2001birds}, strongly influence the structure of social interactions and have indisputable consequences on the global social network as well. The coexistence of social classes and status homophily may lead to a strongly stratified social structure where people of the same social class tend to be better connected among each other, while connections between different classes are less frequent than one would expect from structural characteristics only~\cite{grusky2007theories,saunders2006social}.  Although this hypothesis has been drawn long time ago~\cite{grusky2007theories}, the empirical observation of spatial, socioeconomic, and structural correlations in large social systems has been difficult, as it requires simultaneous access to multimodal characters for a large number of individuals. Our aim in this study~\cite{leo2016socioeconomic} was to find evidence of social stratification through the analysis of a combined large-scale anonymised dataset that disclose simultaneously the social interactions, frequent locations, and the economic status of millions of individuals.

\subsubsection{Economic status indicators}

Our estimation of an individual's economic status is based on the measurement of consumption power. We use a dataset (see for detailed description as DS2 in Section~\ref{sec:datasets}), which contains the amount and type of daily debit/credit card purchases, monthly loan amounts, and some personal attributes such as age, gender, and zip code of billing address of $\sim 6$ million anonymised customers of a bank in the studied country over $8$ months. In addition, for a smaller subset of clients, the data provide the precise salary and total monthly income that we use for verification purposes as in~\cite{leo2016socioeconomic}.

By following the purchase history of each individual, we estimate their economic position from their average amount of debit card purchases. More precisely, for an individual $u$ who spent a total amount of $p_u(t)$ in month $t$, we estimate his/her average monthly purchase (AMP) as
\begin{equation}
P_u=\frac{\sum_{t\in T} p_u(t)}{|T|_u},
\end{equation}
where $|T|_u$ corresponds to the number of active months of the user (with at least one purchase). In order to verify this individual economic indicator we check its correlations with other indicators (not shown here), such as the salary and income and found significant and strong positive correlations between them (for definitions and results see~\cite{leo2016socioeconomic}). 
Further we completed an equivalent analysis (not shown here~\cite{leo2016socioeconomic}) for the average monthly debt (AMD) of people and found very similar distributions of the two seemingly independent quantities.

\begin{figure}[!ht]
\centering
\includegraphics[width=1.\textwidth]{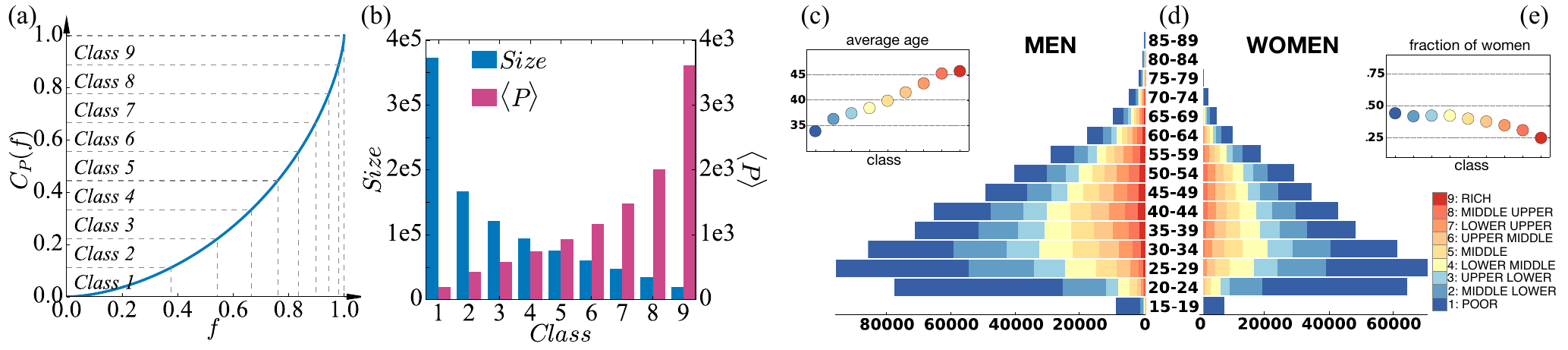}
\caption{\small (a) The cumulative average monthly purchase (AMP) function $C_P(f)$ (blue line) and the schematic demonstration of user partitions into 9 socioeconomic classes. Fraction of individuals belonging to a given class ($x$ axis) have the same sum of AMP $(\sum_u P_u)/n$ ($y$ axis) for each class. (b) Number of egos (blue), and the average AMP $\langle P \rangle$ per individual (pink) in different classes. (c) Average age of different classes. (d) Age pyramids for men and women with colours indicating the corresponding socioeconomic groups and with bars proportional to absolute numbers. (e) Fraction of women in different classes. This figure was prepared by Y. Leo and was published in~\cite{leo2016socioeconomic}.}
\label{fig:coll_SEC}
\end{figure}

The distribution of an individual economic indicator may disclose signs of socioeconomic imbalances on the population level. This hypothesis was first suggested by V. Pareto and later became widely known as the law named after him~\cite{pareto1927manual}. The present data provide a straightforward way to verify this hypothesis through the distribution of individual AMP. We measured the normalised cumulative function of AMP for $f$ fraction of people sorted by $P_u$ in an increasing order:
\begin{equation}
C_P(f)=\frac{1}{\sum_u P_u}\sum_{f} P_u
\end{equation}
We computed this distribution for the $6,002,192$ individuals assigned with AMP values (in USD \footnote{To assign purchase values in USD we used the daily average rate of the local currency on 2 March 2016.}). This function shows (see Fig.\ref{fig:coll_SEC}a blue line) that AMP is distributed with a large variance, i.e., indicating large economical imbalances just as suggested by the Pareto's law. A conventional way to quantify the variation of this distribution is provided by the Gini coefficient $G$ \cite{gastwirth1972estimation}, which characterises the deviation of the $C_P(f)$ function from a perfectly balanced situation, where wealth is evenly distributed among all individuals. In our case we found $G_P\approx 0.461$, which is relatively close to the World Bank reported value $G=0.481$ for the studied country~\footnote{World Bank - GINI index estimates data.worldbank.org/indicator/SI.POV.GINI (date of access: 1/2/2016).}, and corresponds to a Pareto index~\cite{souma2002physics} $\alpha=1.315$. This observation indicates a $0.73:0.27$ ratio characterising the uneven distribution of wealth, i.e., that the $27\%$ of people are responsible for the $73\%$ of total monthly purchases in the observed population. Note that we found similar imbalances in the distribution of AMD with $G_D \approx 0.627$ and $\alpha=1.140$, an observation which has never been reported before~\cite{leo2016socioeconomic}.

\subsubsection{Estimation of socioeconomic classes}
\label{sec:esSCL}

In order to investigate signatures of social stratification, we combine the bank transaction data with data disclosing the social connections between the bank's customers. To identify social ties, we use a mobile communication dataset, described as DS2 in Section~\ref{sec:datasets}, with a customer set that partially overlaps with the user set found in the bank data\footnote{For the combinations, filtering, network construction, and data matching policy of the datasets see~\cite{leo2016socioeconomic}.}.

Taking each individual in the selected social network, we assign each of them into one of $n=9$ socioeconomic classes based on their individual AMP values. This classification is defined by sorting individuals by their AMP, take the cumulative function $C_P(f)$ of AMP, and cut it in $n$ segments such that the sum of AMP in each class is equal to $(\sum_u P_u)/n$ (as shown in Fig.\ref{fig:coll_SEC}a). Our selection of nine distinct classes is based on the common three-stratum model \cite{brown2009concise,akhbar2010class}, which identifies three main social classes (lower, middle, and upper), and  three sub-classes for each of them~\cite{saunders2006social}. More importantly, this way of classification relies merely on individual economic estimators, $P_u$, and naturally partition individuals into classes with decreasing sizes, and increasing $\langle P \rangle$ per capita average AMP values for richer groups (for exact values see Fig.\ref{fig:coll_SEC}b). To explore the demographic structure of the classes we used data on the age and gender of customers. We draw the population pyramids for men and women in Fig.\ref{fig:coll_SEC}d with colour-bars indicating the number of people in a given social class at a given age. We found a positive correlation between social class and average age, suggesting that people in higher classes are also older on average (see Fig.\ref{fig:coll_SEC}c). In addition, our data verifies the presence of gender imbalance as the fraction of women varies from $0.45$ to $0.25$ by going from lower to upper socioeconomic classes (see Fig.\ref{fig:coll_SEC}e).

\begin{figure}[!ht]
\centering
\includegraphics[width=1.\textwidth]{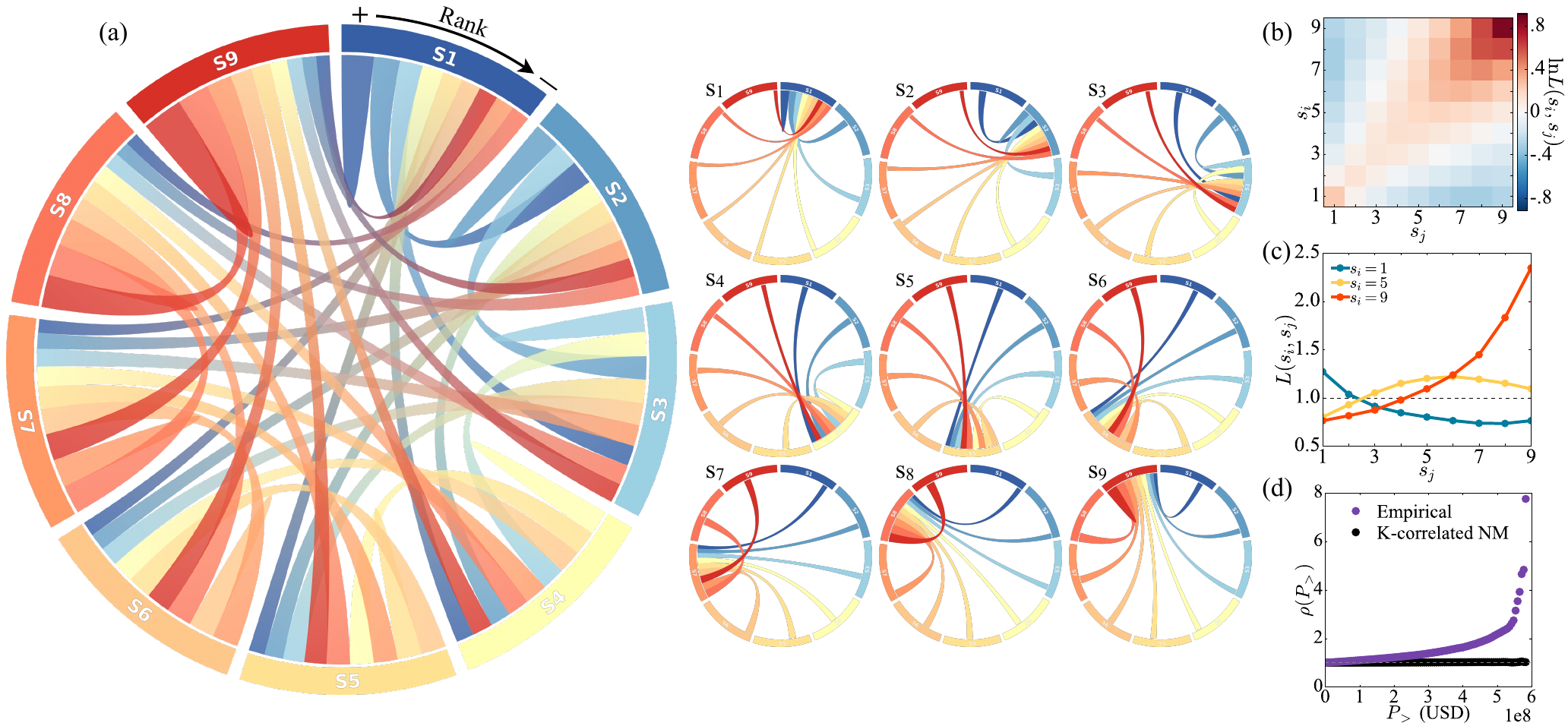}
\caption{\small Structural correlations in the socioeconomic network (a) Chord diagram of connectedness of socioeconomic classes $s_i$, where each segment represents a social class $s_i$ connected by chords with width proportional to the corresponding inter-class link fraction $\tilde{L}_{s_i}(s_j)$, and using gradient colours matched with opposite ends $s_j$. Note that the $\tilde{L}_{s_i}(s_j)=L(s_i,s_j)/\Sigma_{s_j}L(s_i,s_j)$ normalised fraction of $L(s_i,s_j)$ (in Eq.\ref{eq:Lsisj}) was introduced here to assign equal segments for each class for better visualisation. Chords for each class are sorted in decreasing width order in the direction shown above the main panel. (b) Matrix representation of $L(s_i,s_j)$ (for definition see Eq.\ref{eq:Lsisj}) with logarithmic colour scale. (c) The $L(s_i,s_j)$ function extracted for three selected classes ($1$ (blue), $5$ (yellow), and $9$ (red)). (d) ``Rich-club'' coefficient $\rho(P_>)$ (see Eq.\ref{eq:RCC}) based on the empirical (purple), and a degree-correlated null model (black) networks.  This figure was prepared by Y. Leo and was published in~\cite{leo2016socioeconomic}. 
}
\label{fig:coll_Strat}
\end{figure}

\subsubsection{Structural correlations and social stratification}
\label{sec:strat}

Using the above-defined socioeconomic classes and the social network structure, we turn to look for correlations in the inter-connected class structure. To highlight structural correlations, such as the probability of connectedness, we use a randomised reference system. It is defined as the corresponding configuration network model structure~\cite{newman2010networks} with $5\times |E|$ swaps per realisation. This degree-preserving randomisation keeps the number of links, individual degrees (and hence any degree-wealth correlations), individual economic indicators $P_u$, and the assigned class of people unchanged, but destroys any higher-order structural correlations in the social structure and consequently between socioeconomic layers as well. In each case, we present results averaged over the $100$ independent random realisations. Taking the original (resp. randomised) network we count the number of links $|E(s_i,s_j)|$ (resp. $|E_{rn}(s_i,s_j)|$) connecting people in different classes $s_i$ and $s_j$.  After repeating this procedure for each pair of classes in both networks, we take the fraction:
\begin{equation}
L(s_i,s_j)=\frac{|E(s_i,s_j)|}{|E_{rn}(s_i,s_j)|},
\label{eq:Lsisj}
\end{equation}
which gives us how many times more (or less) links are present between classes in the original structure as compared to the randomised one. Note that in the randomised structure the probability that two people from given classes are connected depends only on the number of social ties of the individuals and the size of the corresponding classes, but is independent of the effect of potential structural correlations. 

From the chord diagram visualisation of this measure in Fig.\ref{fig:coll_Strat}a, we can draw several conclusions. 
First, after sorting the chords of a given class $s_i$ in a decreasing $L(s_i,s_j)$ order, chords connecting a class to itself (self-links) appear always at top (or top 2nd) positions of the ranks. At the same time other top positions are always occupied by chords connecting to neighbouring social classes. These two observations (better visible in Fig.\ref{fig:coll_Strat}a insets) indicate strong effects of status homophily and the existence of stratified social structure where people from a given class are the most connected with similar others from their own or from neighbouring classes, while connections with individuals from remote classes are least frequent. A second conclusion can be drawn by looking at the sorting of links in the middle and lower upper classes ($S4-S8$). As demonstrated in the inset of Fig.\ref{fig:coll_Strat}a, people prefer to connect upward and tend to hold social ties with others from higher social classes rather than with people from lower classes.

These conclusions can be further verified by looking at other representations of the same measure. First we show a heat map matrix representation of Eq.\ref{eq:Lsisj} (see Fig.\ref{fig:coll_Strat}b), where $L(s_i,s_j)$ values are shown with logarithmic colour scales. This matrix has a strong diagonal component verifying that people of a given class are always better connected among themselves (red) and with others from neighbouring groups, while social ties with people from remote classes are largely underrepresented (blue) as compared to the expected value provided by the random reference model. This again indicates the presence of homophily and the stratified structure of the socioeconomic network. The upward-biased inter-class connectivity can also be concluded here from the increase of the red area around the diagonal by going towards richer classes. These conclusions are even more straightforward from Fig.\ref{fig:coll_Strat}c where the $L(s_i,s_j)$ is shown for three selected classes ($1$-poor, $5$-middle, and $9$-rich)~\footnote{Note, via an extensive correlation analysis we demonstrated (with results in~\cite{leo2016socioeconomic}) that the observed stratified structure is not induced by simultaneously present degree-degree and degree-wealth correlations, but they are a fundamental character of the network}. 


The above observations further suggest that the social structure may show assortative correlations in terms of socioeconomic status on the individual level. In other words, richer people may be better connected among themselves than one would expect them by chance and this way they form tightly connected ``rich clubs'' in the structure similar to the suggestion of  Mills \cite{mills2018power}. This can be actually verified by measuring the rich-club coefficient \cite{zhou2004rich, colizza2006detecting}, after we adjust its definition to our system as follows. We take the original social network structure, sort individuals by their AMP value $P_u$ and remove them in an increasing order from the network (together with their connected links). At the same time we keep track of the density of the remaining network defined as
\begin{equation}
\rho_{obs}(P_>)=\frac{2L_{P_>}}{N_{P_>}(N_{P_>}-1)}
\label{eq:phiP}
\end{equation}
where $L_{P_>}$ and $N_{P_>}$ are the number of links and nodes remaining in the network after removing nodes with $P_u$ smaller than a given value $P_>$. In our case, we consider $P_>$ as a cumulative quantity going from $0$ to $\sum_u P_u$ with values determined just like in case of $C_P(f)$ in Fig.\ref{fig:coll_SEC}a but now using $100$ segments. At the same time, we randomise the structure using a configuration network model and by removing nodes in the same order we calculate an equivalent measure $\rho_{rn}(P_{>})$ as defined in Eq.\ref{eq:phiP} but in the uncorrelated structure. For each randomisation process, we used the same parameters as earlier and calculated the average density $\langle {\rho}_{rn}\rangle (P_{>})$ of the networks over $100$ independent realisations. Using the two density functions we define the "rich-club" coefficient as
\begin{equation}
\rho(P_>)=\frac{\rho_{obs}(P_>)}{\langle {\rho}_{rn}\rangle(P_>)},
\label{eq:RCC}
\end{equation}
which indicates how many times the remaining network of richer people is denser connected than expected from the reference model. In our case (see Fig.\ref{fig:coll_Strat}d purple symbols) the rich-club coefficient increases monotonously with $P_>$ and grows rapidly once only the richer people remain in the network. At its maximum it shows that the richest people are $\sim 8$ times more connected in the original structure than in the uncorrelated case. This provides a direct evidence about the existence of tightly connected ``rich clubs'' \cite{mills2018power}, and the presence of strong assortative correlations in the social structure on the level of individuals in terms of their socioeconomic status. To exclude the possibility that the observed assortative patterns are induced by strong wealth-degree and degree-degree correlations, we completed an complementary analysis where we used reference models, where we kept node degrees, degree-degree, and degree-wealth correlations, but randomised the structure otherwise (for more on this analysis see~\cite{leo2016socioeconomic}). Results in Fig.\ref{fig:coll_Strat}d (black symbols) show that the obtained rich-club coefficient appears approximately as a constant function around $1$ thus the entangled effects of degree-degree and degree-wealth correlations cannot explain the emergence of ``rich-clubs'' observed in the empirical case. Further we analysed the actual dataset to identify correlations between commuting patterns and socioeconomic status of people~\cite{leo2016socioeconomic}. Without going in details, we found that people of the same socioeconomic class tend to live closer to each other as compared to people from other classes, and we found a positive correlation between their economic capacities and the typical distance they use to commute.

In a second set of works~\cite{leo2016correlations,leo2018correlations} we analysed this coupled dataset to understand how consumption patterns were correlated with socioeconomic status of people and the underlying social network, and how product are purchased together. To study these questions we considered the purchase history of each customer and analysed the $281$ merchant category codes (MCC) of products they bought. Note that we had no actual information what they purchased exactly, but only MCC information and the amount they spent on a given product. 
Through the analysis of spending distributions in different categories of people in given socioeconomic classes we showed, on one hand intuitive correlations characterising the system. We found that lower class people use dominantly cash and spend on necessities mostly as food, gas, and telecommunication, that richer people spend on extra and luxury services, while middle class people spend relatively the most on education. Further by analysing purchase similarities between connected people we found signatures of similar stratifications as earlier and showed that SES and social ties (influence) play deterministic roles in purchase behaviour. Finally, we showed that co-purchased products form a correlation networks with distinct community structure disclosing some hidden patterns of product category correlations.

\subsection{Socioeconomic dependencies of linguistic patterns in Twitter}
\label{sec:TwSESLang}

In a second set of studies~\cite{abitbol2018socioeconomic,abitbol2018location} we investigated a particularly different collective phenomena, written online language and its emergent variance as the function of arguably determinant factors. The data collection and the related studies were carried out within the SoSweet ANR project~\footnote{The participating teams were affiliated to the DAMTE team (Inria), ICAR lab (ENS Lyon), LIDILEM lab (Universit\'e Grenoble Alpes), and ALMANACH team (Inria). The project was supported by ANR (ANR-15-CE38-0011-03)}, which was a collaboration with three linguist groups and my team in order to investigate the sociolinguistics of the French Twitter space via a series of variational studies. 

Communication is highly variable and this variability contributes to language change and fulfils social functions. Analysing and modelling data from social media allows the high-resolution and long-term follow-up of large samples of speakers, whose social links and utterances are automatically collected. This empirical basis and long-standing collaboration between computer and social scientists could dramatically extend our understanding of the links between language variation, language change, and society.

Sociolinguistics has traditionally carried out research on the quantitative analysis of the so-called linguistic variables, i.e. points of the linguistic system which enable speakers to say the same thing in different ways~\cite{labov1972sociolinguistic}. 
Such variables have been described in many languages: variable pronunciation of -ing as [in] instead of [i\textipa{N}] in English (\emph{playing} pronounced \emph{playin'}); optional realisation of the first part of the French negation (\emph{je (ne) fume pas}, "I do not smoke"); optional realisation of the plural ending of verb in Brazilian Portuguese (eles disse(ram), "they said").  For decades, sociolinguistic studies have showed that hearing certain variants triggers social stereotypes~\cite{campbell2010new}. The so-called standard variants (\emph{e.g.} [i\textipa{N}], realisation of negative ne and plural -ram) are associated with social prestige, high education, professional ambition and effectiveness. They are more often produced in more formal situation. Non-standard variants are linked to social skills, solidarity and loyalty towards the local group, and they are produced more frequently in less formal situation.


In the coming set of works we analyse a large dataset of French tweets enriched with census sociodemographic information to look for the variation of two grammatical cues and an index of vocabulary size of users located in France. First we will discuss how these linguistic cues correlated with three features reflective of the socioeconomic status of the users, their most representative location and their daily periods of activity on Twitter. We will also observe whether connected people are more linguistically alike than disconnected ones. Further we will propose inventive ways to use linguistic information, inferred home locations, or occupational skills to predict the socioeconomic status of people.


\subsubsection{Data construction}

One of the main achievements of our study was the construction of a combined dataset for the analysis of sociolinguistic variables as a function of  socioeconomic status, geographic location, time, and the social network. Our first dataset consists of a large data corpus collected from the online news and social networking service, Twitter (for detailed description see DS3 in Section~\ref{sec:datasets}).

\paragraph{Linguistic data:}
To obtain meaningful linguistic data we preprocessed the incoming tweet stream in several ways. As our central question here deals with the variability of the language, retweets do not bring any additional information to our study, thus they were removed at the outset. Next, in order to facilitate the detection of the selected  linguistic markers we removed any URLs, emoticons, mentions of other users (denoted by the \texttt{@} symbol) and hashtags (denoted by the \texttt{\#} symbol) from each tweet. In addition we completed a last step of textual preprocessing by down-casing and stripping the punctuation out of the tweets body. POS-taggers such as MElt \cite{denis2012coupling} were also tested but they provided no significant improvement in the detection of the linguistic markers.

\paragraph{Network data:}
We used the collected tweets in another way to infer social relationships between users. Tweet messages may be direct interactions between users, who mention each other in the text by using the \texttt{@} symbol (\texttt{@username}). When one user $u$, mentions another user $v$, user $v$ will see the tweet posted by user $u$ directly in his / her feed and may tweet back. In our work we took direct mentions as proxies of social interactions and used them to identify social ties between pairs of users who at least once mutually mentioned each other during the observation. People who reciprocally mentioned each other express some mutual interest, which may be a stronger reflection of real social relationships as compared to the non-mutual cases~\cite{hours2016link}. This constraint provided us a network structure of $508,975$ users and $4,029,862$ undirected links.

\paragraph{Geolocated data:}
About $2\%$ of tweets included in our dataset contained some location information regarding either the customer's self-provided position or the place from which the tweet was posted. We considered only tweets which contained the exact GPS coordinates with resolution of $\sim 3$ meters of the location where the actual tweet was posted and excluded tweets with assigned places such as "Paris" or "France". Further, we discarded coordinates that appeared more than $500$ times throughout the whole GPS-tagged data. After this selection procedure we rounded up each tweet location to a $100$ meter precision.

To obtain a unique representative location of each user, we identified their most frequent location to be a proxy for the user's home location. This selection method provided us with $110,369$ geolocated users who are either detected as French speakers or assigned to be such by Twitter and all associated to specific 'home' GPS coordinates in France. For cross correlation of the geo-located population with census population data see~\cite{abitbol2018socioeconomic}.

\begin{figure}[!ht]
\centering
\includegraphics[width=1.\textwidth]{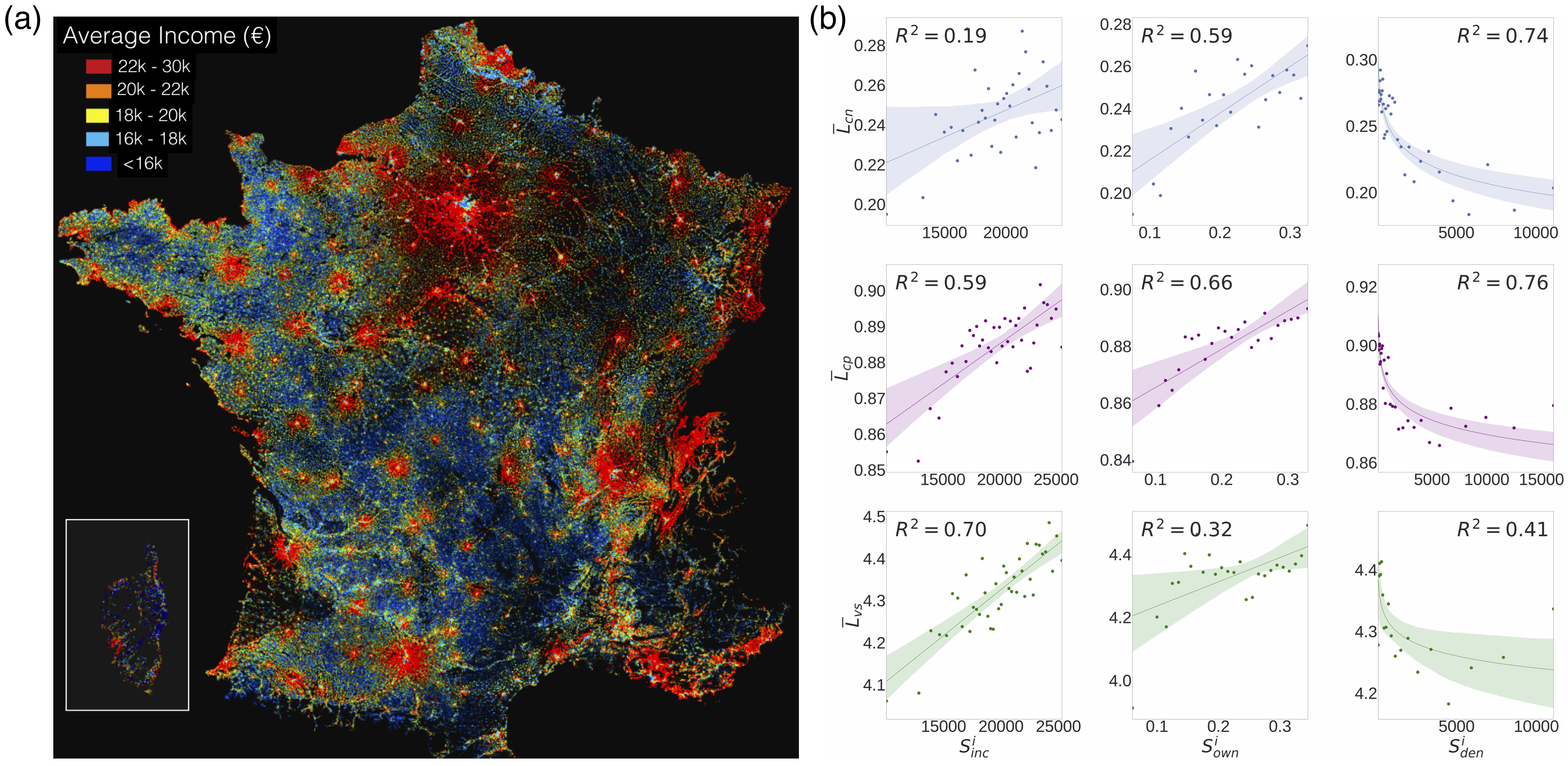}
\caption{\small Distributions and correlations of socioeconomic indicators and linguistic variables. (a) Spatial distribution of average income in France with $200m \times 200m$ resolution. (b) Pairwise correlations between three SES indicators and three linguistic markers. Columns correspond to SES indicators (resp. $S^i_\mathrm{inc}$, $S^i_\mathrm{own}$, $S^i_\mathrm{den}$), while rows correspond to linguistic variables (resp. $\overline{L}_{\mathrm{cn}}$,  $\overline{L}_{\mathrm{cp}}$ and $\overline{L}_\mathrm{vs}$). On each plot coloured symbols are binned data values and a linear regression curve are shown together with the $95$ percentile confidence interval and $R^2$ values. This figure was prepared by J. Levy Abitbol and was published in~\cite{abitbol2018socioeconomic}.}
\label{fig:TwLang1}
\end{figure}

\paragraph{Socioeconomic data:}
To obtain socioeconomic informations we used a dataset, which was released in December 2016 by the National Institute of Statistics and Economic Studies (INSEE) of France. This corpus~\cite{INSEEdata} contains a set of sociodemographic aggregated indicators, estimated from the 2010 tax return in France, for each 4 hectare ($200m \times 200m$) square patch across the whole French territory. We concentrated on three indicators for each patch $i$ as the $S^i_\mathrm{inc}$ average yearly income per capita (in euros), the $S^i_{\mathrm{own}}$ fraction of owners (not renters) of real estate, and the $S^i_\mathrm{den}$ density of population defined respectively as
\begin{equation}
S^i_\mathrm{inc}=\frac{{S}^i_{hh}}{{N}^i_{hh}}, \hspace{.15in} S^i_\mathrm{own}=\frac{N^i_\mathrm{own}}{N^i}, \hspace{.15in}\mbox{and}\hspace{.15in}  S^i_\mathrm{den}=\frac{N^i}{(200m)^2}.
\end{equation}
Here ${S}^i_{hh}$ and ${N}^i_{hh}$ assign respectively the cumulative income and total number of inhabitants of patch $i$, while $N^i_\mathrm{own}$ and $N^i$ are respectively the number of real estate owners  and the number of individuals living in patch $i$. As an illustration we show the spatial distribution of $S^i_\mathrm{inc}$ average income over the country in Fig.\ref{fig:TwLang1}a. For a cross correlation study between the different SES indicators see~\cite{abitbol2018socioeconomic}.

In order to assign socioeconomic status (SES) indicators to twitter users, we combined our Twitter data with the socioeconomic maps of INSEE by assigning each geolocated Twitter user to a patch closest to their estimated home location (within 1 km).   This way we obtained for all $110,369$ geolocated users their dynamical linguistic data, their egocentric social network as well as a set of SES indicators.

\subsubsection{Linguistic variables}

We identified the following three linguistic markers  to study across users from different socioeconomic backgrounds: 

\paragraph{Standard usage of negation:}
The basic form of negation in French includes two negative particles: \textit{ne} (no) before the verb and another particle after the verb that conveys more accurate meaning: \textit{pas} (not), \textit{jamais} (never), \textit{personne} (no one), \textit{rien} (nothing), etc. Due to this double construction, the first part of the negation (\textit{ne}) is optional in spoken French, but it is obligatory in standard writing. Sociolinguistic studies have previously observed the realisation of  \textit{ne} in corpora of recorded everyday spoken interactions. Although all the studies do not converge, a general trend is that \textit{ne} realisation is more frequent in speakers with higher socioeconomic status than in speakers with lower status~\cite{ashby2001nouveau,hansen2004ne}. We built upon this research to set out to detect both negation variants in the tweets using regular expressions.
We are namely interested in the rate of usage of the standard negation (featuring both negative particles) across users:
\begin{equation}
L^u_{\mathrm{cn}}=\frac{n^u_{\mathrm{cn}}}{n^u_{\mathrm{cn}}+n^u_{\mathrm{incn}}} \hspace{.2in} \mbox{and} \hspace{.2in} \overline{L}^{i}_{\mathrm{cn}}=\frac{\sum_{u\in i}L^u_{\mathrm{cn}}}{N_i},
\end{equation}
where $n^{u}_{\mathrm{cn}}$ and $n^{u}_{\mathrm{incn}}$ assign the number of correct negation and incorrect number of negation of user $u$, thus $L_{\mathrm{cn}}^u$ defines the rate of correct negation of a users and $\overline{L}_{\mathrm{cn}}^i$ its average over a selected $i$ group (like people living in a given place) of $N_i$ users.

\paragraph{Standard usage of plural ending of written words:}
In written French, adjectives and nouns are marked as being plural by generally adding the letters \textit{s} or \textit{x} at the end of the word. Because these endings are mute (without counterpart in spoken French), their omission is the most frequent spelling error in adults~\cite{billiez1994orthographe}. Moreover, studies showed correlations between standard spelling and social status of the writers, in preteens, teens and adults~\cite{totereau2013orthographe,billiez1994orthographe}. We then set to estimate the use of standard plural across users: 
\begin{equation}
L^u_{\mathrm{cp}}=\frac{n^u_{\mathrm{cp}}}{n^u_{\mathrm{cp}}+n^u_{\mathrm{incp}}} \hspace{.2in} \mbox{and} \hspace{.2in} \overline{L}^{i}_{\mathrm{cp}}=\frac{\sum_{u\in i}L^u_{\mathrm{cp}}}{N_i}
\end{equation}
where the notation follows as before ($\mathrm{cp}$ stands for correct plural and $\mathrm{incp}$ stands for incorrect plural).

\paragraph{Normalised vocabulary set size:}
A positive relationship between an adult's lexical diversity level and his or her socioeconomic status has been evidenced in the field of language acquisition. Specifically, converging results showed that the growth of child lexicon depends on the lexical diversity in the speech of the caretakers, which in turn is related to their socioeconomic status and their educational level~\cite{hoff2003specificity,huttenlocher2007varieties}. We thus proceeded to study the following metric: 
\begin{equation}
L^u_\mathrm{vs}=\frac{N^u_\mathrm{vs}}{N^u_{tw}} \hspace{.2in} \mbox{and} \hspace{.2in} \overline{L}^{i}_\mathrm{vs}=\frac{\sum_{u\in i}N^u_\mathrm{vs}}{N_i},
\end{equation}
where $N_vs^u$ assigns the total number of unique words used by user $u$ who tweeted $N_{tw}^u$ times during the observation period. As such $L_\mathrm{vs}^u$ gives the normalised vocabulary set size of a user $u$, while $\overline{L}_\mathrm{vs}^i$ defines its average for a population $i$.

\begin{figure}[!ht]
\centering
\includegraphics[width=1.\textwidth]{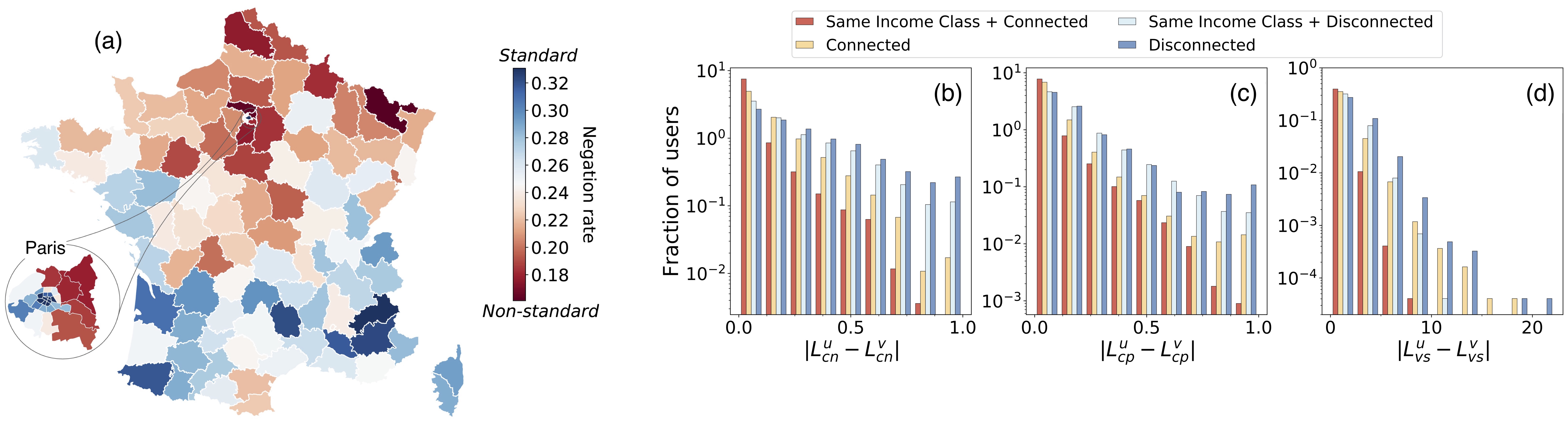}
\caption{\small Geographical and network variability of linguistic markers. (a) Variability of the rate of correct negation in France. Inset focuses on larger Paris. Plot depicts variability on the department level except the inset which is on the "arrondissements" level. (b-d) Distribution of the $|L^{u}_{*}-L^{v}_{*}|$ absolute difference of linguistic variables $*\in \{\mathrm{cn},\mathrm{cp},vs\}$ (resp. panels (b), (c), and (d)) of user pairs who were connected and from the same socioeconomic group (red), connected (yellow), disconnected and from the same socioeconomic group (light blue), disconnected randomly selected users (blue). This figure was prepared by J. Levy Abitbol and was published in~\cite{abitbol2018socioeconomic}.}
\label{fig:TwLang2}
\end{figure}

\subsubsection{Linguistic variations}

By measuring the defined linguistic variables in the Twitter timeline of users we are finally set to address the core questions of our study, whether the variation of language used online depend on the socioeconomic status of the users, on the location or time of usage, and on ones social network. To answer these questions we present here a multidimensional correlation study considering these factors one by one. 

\paragraph{Socioeconomic variation:}


To compute and visualise correlations between socioeconomic status and  linguistic variables we defined linear bins (in numbers varying from 20 to 50) for the socioeconomic indicators and computed the average of the given linguistic variables for people falling within the given bin. 

In \ref{fig:TwLang1}b we show the correlation plots of all nine pairs of SES indicators and linguistic variables together with the linear regression (or log regression for population density) curves, the corresponding $R^2$ values and the $95$ percentile confidence intervals. These results show that correlations between socioeconomic indicators and linguistic variables actually exist and suggest that people with lower SES may use more non-standard expressions (higher rates of incorrect negation and plural forms) have a smaller vocabulary set size than people with higher SES. Note that, although the observed variation of linguistic variables were limited, all the correlations were statistically significant ($p<10^{-2}$) with considerably high $R^2$ values~\cite{abitbol2018socioeconomic} ranging from $0.19$ (between $\overline{L}_{\mathrm{cn}}\sim S_\mathrm{inc}$) to $0.76$ (between $\overline{L}_{\mathrm{cp}}\sim S_\mathrm{den}$). For the rates of standard negation and plural terms the population density appeared to be the most determinant indicator with $R^2=0.74$ (and $0.76$ respectively), while for the vocabulary set size the average income provided the highest correlation (with $R^2=0.7$). Note that while these correlations only hold meaning at the population level but at the user level, the variability of individual language usage hinders the observation of the correlation values.

In addition, we performed a study to address the temporal variability of the linguistic patterns, which revealed an interesting pattern with higher standard usage in the morning and the contrary in the evening. However, we found that these temporal patterns are well explained by the varying population on Twitter in terms of socioeconomic status. Thus although these results are interesting, they do not indicate a new determinant factor for language variance, thus detailed results are not reported here (see~\cite{abitbol2018socioeconomic}).

\paragraph{Spatial variation:} Next we address the spatial variation of linguistic variables. Although officially a standard language is used over the whole country, geographic variations of language may exist due to regional variability resulting from remnants of local languages that have disappeared, uneven spatial distribution of socioeconomic potentials, or influence of neighbouring countries~\cite{kulkarni2016freshman,wieling2011quantitative}. For the observation of such variability, by using their representative locations, we assigned each user to one of the $97$ departments of France and computed the $\overline{L}^{i}_{\mathrm{cn}}$ (resp. $\overline{L}^{i}_{\mathrm{cp}}$) average rates of standard negation (resp. pluralisation) and the $\overline{L}^{i}_\mathrm{vs}$ average vocabulary set size for each "d\'epartement" $i$.

Results shown in Fig.\ref{fig:TwLang2}a revealed some surprising patterns, which appeared to be consistent for each linguistic variable (others than negation shown in~\cite{abitbol2018socioeconomic}). By considering latitudinal variability it appears that, overall, people living in the northern part of the country used a less standard language, i.e., negated and pluralised less standardly, and used a smaller number of words. On the other hand, people from the South used a language which is somewhat closer to the standard (in terms of analysed linguistic markers) and a more diverse vocabulary. On the other hand on shorter spatial scale, like in cities, socioeconomic differences determine language variance. As an example, in case of Paris it is evident from Fig.\ref{fig:TwLang2}a inset) that in the city centre and in the western part of the city (historically richer districts) people use more standard language, while the contrary is true for the eastern suburbs. Note that we also performed a multivariate regression analysis (not shown here), using the linguistic markers as target and considering as factors both location (in terms of latitude and longitude) and income as proxies of socioeconomic status. It confirmed that while location is a strong global determinant of language variability, socioeconomic variability may still be significant locally to determine standard language usage (just as we demonstrated in the case of Paris).

\paragraph{Network variation:} Finally we aimed to understand the effects of the social network on the variability of linguistic patterns. People in a social structure can be connected due to several reasons. Link creation mechanisms like focal or cyclic closure~\cite{kumpula2007emergence,laurent2015calls}, or preferential attachment~\cite{newman2001clustering} together with the effects of homophily~\cite{mcpherson2001birds} concerning age, gender, common interest or political opinion are all potentially driving the creation of social ties. We have already seen that status homophily between people of similar socioeconomic status are specifically important~\cite{leo2016socioeconomic} in driving social tie creation and to explain the stratified structure of society. By using our combined datasets, we aim here to identify the effects of status homophily and to distinguish them from other homophilic correlations and the effects of social influence inducing similarities among already connected people.

First, we took the geolocated Twitter users in France and partitioned them into nine socioeconomic classes using their inferred income $S_\mathrm{inc}^u$. Partitioning was done by the same method as in Section~\ref{sec:esSCL} but using the assigned income of users. As a result of a similar network analysis, we observed a stratified structure in the Twitter network, further strengthening the validity of our earlier observations.


In order to measure linguistic similarities between a pair of users $u$ and $v$, we simply computed the $|L^{u}_{*}-L^{v}_{*}|$ absolute difference of their corresponding individual linguistic variable $*\in \{\mathrm{cn},\mathrm{cp},vs\}$. This measure associates smaller values to more similar pairs of users. We computed the similarity distribution for pairs of people in four cases: for connected users from the same socioeconomic class; for disconnected randomly selected pairs of users from the same socioeconomic class; for connected users in the network; and randomly selected pairs of disconnected users in the network. Note that in each case the same number of user pairs were sampled from the network to obtain comparable averages. This number was naturally limited by the number of connected users in the smallest socioeconomic class, and were chosen to be $10,000$ in each cases. By comparing the distributions shown in Fig.\ref{fig:TwLang2} we see that (a) connected users (red and yellow bars) were the most similar in terms of any linguistic marker. This similarity was even greater when the considered tie was connecting people from the same socioeconomic group; (b) network effects can be quantified by comparing the most similar connected (red bar) and disconnected (light blue bar) users from the same socioeconomic group. Since the similarity between disconnected users here is mostly induced by status homophily, the difference of these two bars indicates additional effects that cannot be explained solely by status homophily. These additional similarities may rather be induced by other factors such as social influence, the physical proximity of users within a geographical area or other homophilic effects that were not accounted for; (c) Randomly selected pairs of users were more dissimilar than connected ones. We therefore can conclude that both the effects of network and status homophily mattered in terms of linguistic similarity between users of this social media platform.

\subsubsection{Inference of socioeconomic status of Twitter users}
\label{sec:TwSESinf}

Building on these determinant correlations of language variability, in a subsequent study~\cite{abitbol2018location} we addressed to inverse question and tried to infer socioeconomic status of users from their location and linguistic features. Beyond associating income to users from census data using their location, we proposed two other ways to solve this assignment problem and to build reliable data for training. In our second method, we considered users who shared their LinkedIn profile in their Twitter profile description and crawled their LinkedIn profile to obtain information about their occupation and professional skill set. After identifying the professions from skill sets and reported occupations for the obtained $4,140$ users, we used a salary classification table provided by INSEE~\cite{INSEEsalary}. After that, users were eventually assigned to one of two SES classes based on whether their salary was higher or lower than the average value of the income distribution. Finally, we designed a third method, where we took the inferred home locations of Twitter users and collected satellite and street view images from Google Maps about their habitual places. After filtering on residential areas using via a ResNet50 network using keras~\cite{chollet2017keras} pre-trained on ImageNet~\cite{deng2009imagenet}, we gave images of $1000$ location to architects who assigned SES scores to each sample of residential places.

To select features for the training we considered shallow features directly observable from the data (like profile informations, word counts, etc.), and features which were obtained via a pipeline of data processing methods to capture semantic user characteristics. On one hand, we took into account linguistic features by training a \emph{word2vec}~\cite{mikolov2013efficient} model to obtain a topic correlation matrix for the dataset, which we used then to calculate a topic distribution for each users. Further we assigned people with their lexical and semantic profile they displayed on Twitter, thus finally we arrived to a feature vector with $1117$ entries for each users. Using multiple methods to solve the inference problem, the XGBoost algorithm provided us the best performing solution and gave us the highest precision (around $0.798\pm0.015$ AUC score) in case we used occupation information from LinkedIn to assign SES of people (for further details see~\cite{abitbol2018location}).

\begin{center}
  $\ast$~$\ast$~$\ast$
\end{center}

Finally, note that I have some further contributions related to static analysis of collective social phenomena, which are not discussed in details here. In one paper~\cite{liao2017prepaid} we investigate the behavioural differences between mobile phone customers with prepaid and postpaid subscriptions. Our study reveals that (a) postpaid customers are more active in terms of service usage and (b) connections between customers of the same subscription type are much more frequent than those between customers of different subscription types. Based on these observations we provide methods to detect the subscription type of customers by using information about their personal call statistics, and also their egocentric networks simultaneously. We cast this classification problem as a problem of graph labelling, which can be solved by max-flow min-cut algorithms, and able to achieve a classification accuracy of $\sim 87\%$. Further, we infer the subscription type of customers of external operators via approximate methods by using node attributes, and a two-ways indirect inference method based on observed homophilic structural correlations.

In another study~\cite{mocanu2015collective} we address, on a sample of $2.3$ million individuals, how Facebook users consumed different information at the edge of political discussion and news during an Italian electoral competition. We categorise webpages according to their topics and the communities of interests they pertain to, in (a) alternative information sources (diffusing topics that are neglected by science and main stream media); (b) online political activism; and (c) main stream media. We show that attention patterns are similar despite the different qualitative nature of the information, meaning that unsubstantiated claims (mainly conspiracy theories) reverberate for as long as other (verifiable) information. Further in this study, we classify users according to their interaction patterns among the different topics and measure how they responded to the injection of $2788$ false information. Our analysis reveals that users, which are prominently interacting with conspiracists information sources are more prone to interact with intentional false claims.




\section{Dynamic observations of social spreading phenomena}
\label{sec:obssp}

In this Section we are going to summarise empirical studies reporting observations about the dynamics of social contagion phenomena. Social contagion evolves over networks of interconnected individuals, where links associated with social ties transferring influence between peers \cite{castellano2009statistical}. Several earlier studies aimed to identify the dominant mechanisms at play in such processes \cite{rogers2010diffusion, granovetter1978threshold,schelling1969models,axelrod1997dissemination}. One key element, termed behavioural threshold by Granovetter \cite{granovetter1978threshold}, is defined as \textit{``the number or proportion of others who must make one decision before a given actor does so''}. Following this idea, various network models have been introduced \cite{watts2002simple,valente1996social,watts2007influentials,melnik2013multi, karampourniotis2015impact} to understand threshold-driven spreading processes, commonly known as \textit{complex contagion} \cite{centola2007complex}. Although these models are related to a larger set of collective dynamics, they are particularly different from \textit{simple contagion} where the exposure of nodes is driven by independent contagion stimuli and one infected neighbour is always sufficient to expose a susceptible node \cite{bass1969new,barrat2008dynamical,hill2010infectious}. During the last ten years several studies contributed to the foundation of complex contagion  \cite{watts2002simple,backstrom2006group,gleeson2007seed,romero2011differences,bakshy2012role,singh2013threshold}, and in addition online experiments were carried out to provide empirical evidence about the effect of social influence \cite{centola2010spread,centola2011experimental}. Beyond the conventional threshold mechanism, the effect of homophily \cite{aral2009distinguishing,bakshy2012role,suri2011cooperation} and the role of external media influence \cite{toole2012modeling} were also investigated recently.

Most of my work on dynamical social phenomena are related to the analysis of social contagion processes, with observations in large online social systems like Skype or Twitter. As it follows, we will discuss empirical observations reported in~\cite{karsai2014complex,karsai2016local,hours2016link} and identify mechanisms, which arguably play central role in the adoption dynamics of online services or information. Subsequently, in Section~\ref{sec:ccpmodels}, we will incorporate the identified mechanisms into predictive dynamical models of complex contagion processes.

\subsection{Complex contagion process in spreading of online innovation}
\label{sec:ccpoi}


The propagation of innovations takes place in a social network \cite{bass1969new,toole2012modeling,onnela2010spontaneous,aral2009distinguishing} and is driven by the entanglement of individuals' decision-making processes \cite{qudrat2007complex} as well as by the influence of media and social interactions \cite{rogers2010diffusion}. Although the effects of network structure on contagion processes have recently been shown to be important \cite{barrat2008dynamical}, knowledge about the social network itself is rather limited since its structure and dynamics usually remain hidden. In this respect the digital age has opened up unprecedented opportunities, as online social networks and Voice over Internet Protocol services record detailed information of the connections and activities of their users. These services partially decode the underlying social structure by acting as proxies for the network of real social ties between individuals, and also provide accurate records of the users' adoption behaviour. In this way the different sources of influence on the decisions of an individual immersed in a perpetually changing environment of social interactions become traceable. 

In this project~\cite{karsai2014complex} we studied one of today's largest online communication services, the Skype network, with over $300$ million monthly connected users. Data covers the history of individuals that have adopted Skype from September 2003 until March 2011 (i.e. 2738 days), including registration events and contact network evolution for every registered user around the world. For our investigation we selected user accounts with an identified country of registration and considered only their mutually confirmed connections, both within the country and abroad. To receive the best estimation of node degrees in the underlying social network, we integrated the evolving Skype network for the whole available period and count the number of confirmed relationships per node (including international ties). The adoption dynamics of a given country can be directly observed by assigning times of adoption ($t_a$) and termination ($t_t$) to all the accounts. These are respectively defined as the dates of registration and last activity (as regards to any of the services) in Skype. Explicitly, we identified any account as terminated if its last activity happened earlier than one year prior to the end of the observation period. In this way we were able to build a complete adoption and termination history of Skype for 2373 days. 

\begin{figure}[!ht]
\centering
\includegraphics[width=.7\textwidth]{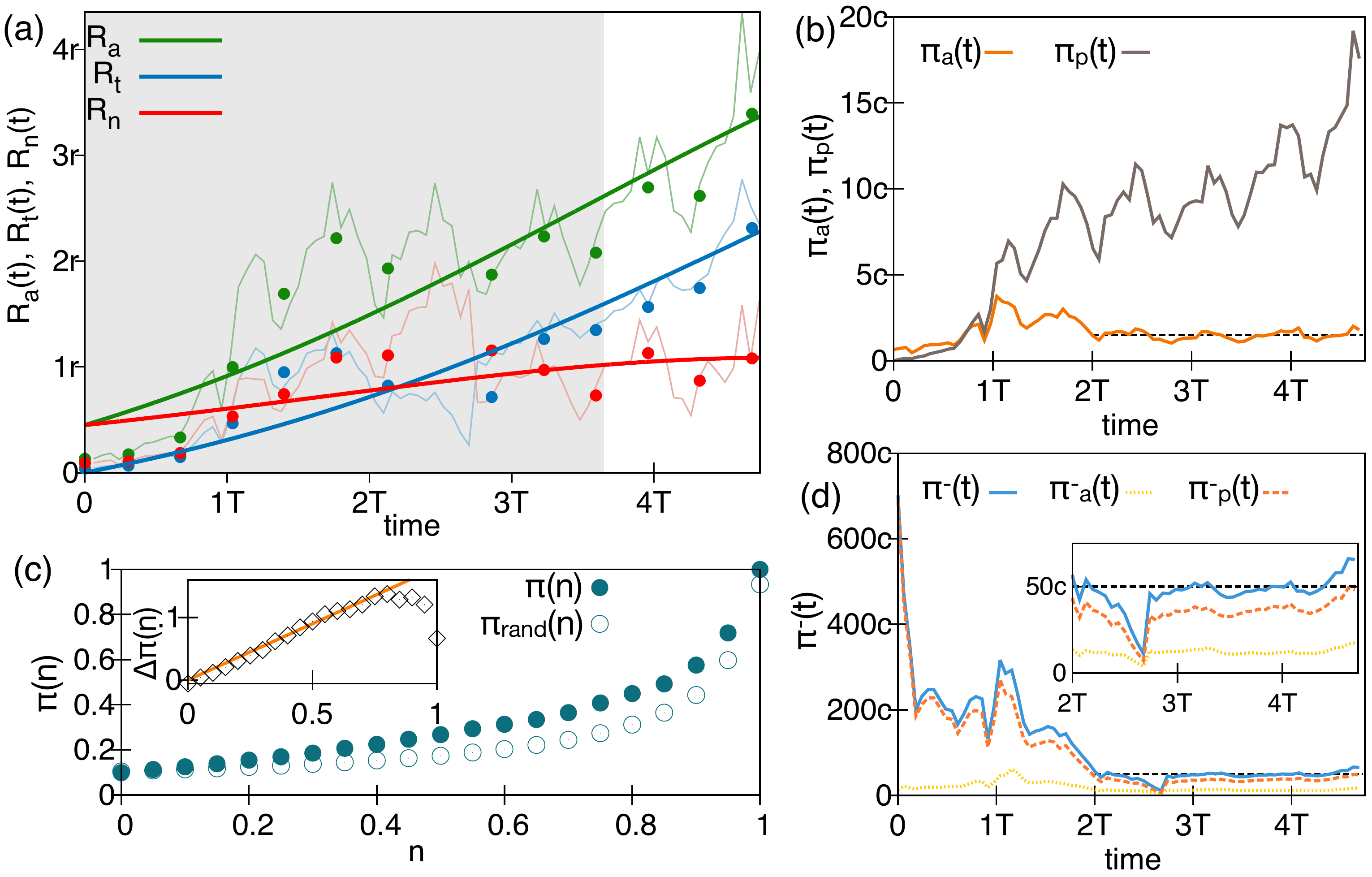}
\caption{\small Empirical rates and probabilities for Switzerland. (a) Thin curves denote empirical rates of adoption [$R_a(t)$], termination [$R_t(t)$], and net adoption [$R_n(t)$], while symbols are their corresponding binned values. A binned data point in $[2T,3T]$ has been removed due to systematic bias in $R_t(t)$ caused by a major software update during this period. A shaded (white) area indicates the training (predicted) period for the theoretical fit of our model, drawn as thick lines with the same colours as the empirical rates. (b) Probabilities of spontaneous [$\pi_a(t)$] and peer-pressure [$\pi_p(t)$] adoption per unit time. (c) Average conditional probability of adoption as a function of the fraction of adopting neighbours $n$, measured in the original data [$\pi(n)$, solid circles] and in the shuffled data corresponding to the null model [$p_{\mathrm{rand}}(n)$, open circles]. Inset shows the unbiased difference $\Delta \pi(n) = \pi(n) - \pi_{\mathrm{rand}}(n)$ (symbols) and a fitted linear function (continuous line). (d) Probabilities of overall termination [$\pi^-(t)$], and of spontaneous [$\pi_a^-(t)$] and peer-pressure [$\pi_p^-(t)$] termination per unit time. The inset depicts a zoom from time $2T$ onwards. $T$, $r$ and $c$ are arbitrary linear scaling constants, with time dimensions for $T$. Black lines in panels (b), (d) are fitted constants. This figure was published in~\cite{karsai2014complex}.}
\label{fig:Skype1Emp}
\end{figure}


\subsubsection{The adoption dynamics}
The spreading of the online service (in this case the adoption of the Skype free service) is determined by competing processes of adoption and termination, measured by the evolution of the corresponding rates $R_a(t)$ and $R_t(t)$ of all users that adopt or terminate the service in a given time window $\Delta t$ (Fig. \ref{fig:Skype1Emp}.a). These simple rate functions already disclose interesting features of the adoption dynamics, since their overall growth signals continuously accelerating processes of adoption and termination. Yet the actual time evolution of spreading service is better characterised by the net adoption rate $R_n(t)=R_a(t)-R_t(t)$.


\subsubsection{Mechanisms of adoption}
Opening a user account constitutes a single event in the decision-making process of an individual that is triggered by either spontaneous decisions, the influence of media or by the social environment \cite{watts2002simple,aral2009distinguishing}. On the other hand, users may terminate their accounts for several reasons including vanishing demand or dissatisfaction, by switching to another product permanently, or by simply abandoning the service with a chance of re-adoption (e.g. due to loss of password or intention for lower monitoring). An analysis of the evolving network structure around a given user can help us to detect some of these scenarios, by observing whether an ego adopted or terminated the product before any of its neighbours did; or else followed the decisions previously made by a fraction of them. In this way we can label the performed action as either spontaneous or driven by peer pressure.

To define the related measures we consider the underlying social network as static, meaning that its evolution requires a much larger temporal scale than the adoption process itself. This static structure is defined as the aggregated social network of Skype at the end of the recorded period, and provides a lower estimate for the total number of friends of each individual. Moreover, we assume that the maximum size of the static social network is the number $I$ of internet users in a given country at the end of the observation period (2011) \cite{Internet2013}, and thus define $I - N_a(t)$ as the population that has not yet adopted Skype at time $t$.

Under these assumptions, the probabilities per unit time that a user adopts either spontaneously or due to peer pressure are defined as,
\begin{equation}
\label{eq:empAdopt}
\pi_a(t)=\dfrac{\# ad(t+\Delta t|SF=0)}{I-N_a(t)}, \quad \text{and} \quad
\pi_p(t)=\dfrac{\# ad(t+\Delta t|SF\neq 0)}{I-N_a(t)},
\end{equation}
where $\# ad(t+\Delta t|SF=0)$ [$\# ad(t+\Delta t|SF\neq 0)$] is the number of users who adopt the service in a time window $\Delta t$, under the condition that their number of adopting neighbours at time $t$ is $SF=0$ ($SF \neq 0$). In a similar fashion, the probabilities per unit time that a user terminates the service either spontaneously or due to peer pressure are,
\begin{equation}
\label{eq:empTerm}
\pi_a^-(t) = \dfrac{\# tr(t+\Delta t,TF=0)}{N_a(t)}, \quad \text{and} \quad
\pi_p^-(t) = \dfrac{\# tr(t+\Delta t,TF\neq0)}{N_a(t)},
\end{equation}
where $TF$ stands for the number of neighbours of a user that have terminated usage up to time $t$ (for a discussion on the restrictions of these empirical quantities see~\cite{karsai2014complex}).

The data shows that after an initial, transient period, the rate of spontaneous adoption $\pi_a(t)$ (Fig. \ref{fig:Skype1Emp}b) and the rate of termination $p^-(t)=\pi_a^-(t)+\pi_p^-(t)$ (Fig. \ref{fig:Skype1Emp}d) become constant apart from small fluctuations. The same holds separately for the rates of spontaneous [$\pi_a^-(t)$] and peer-pressure [$\pi_p^-(t)$] termination. The time invariance of these rates is an obvious assumption for most biological epidemics, which, however, has never been empirically shown before in the case of social contagion phenomena, despite its wide use \cite{vespignani2012modelling,castellano2009statistical}. 

When the ego is not the first adopter among neighbours, the rate $\pi_p(t)$ of adoption via peer pressure is not constant but increases with time (Fig. \ref{fig:Skype1Emp}b). This is arguably due to social influence arising from the user's social circle. An appropriate way to quantify such effects is to measure the conditional probability $\pi(n)$ of adoption provided that a fraction $n$ of the ego's neighbours have adopted the product before as
\begin{equation}
\label{eq:pn}
\pi(n)=\dfrac{\# ad(n)}{N-\sum_{m=0}^{m<n}\# ad(m)}.
\end{equation}
Here the numerator counts the number of users with a fraction $n$ of adopter friends at the time of adoption, while the denominator is the number of people with a larger or equal fraction $m \geq n$, i.e. all individuals who had the chance to adopt Skype while having a fraction $n$ of adopter neighbours. We observe that the probability $\pi(n)$ is monotonically increasing (Fig. \ref{fig:Skype1Emp}c), an empirical finding in agreement with the assumptions of several threshold models for epidemic spreading and social dynamics \cite{watts2002simple,dodds2004universal,klimek2008opinion,takaguchi2013bursty}. However, since we cannot see the entire social network (only the part uncovered by the Skype graph), this probability is biased as $n \rightarrow 1$. To estimate such bias, we build a reference null model by shuffling the adoption times of all accounts and measuring the corresponding conditional probability $p_{\mathrm{rand}}(n)$ for this system. The shuffling procedure removes the effect of social influence but conserves the adoption rates and keeps the social structure unchanged. In other words, the reference probability is biased in the same way as the original measurement, but is not driven by social influence as all such correlations have been removed by the shuffling. Consequently, the difference $\Delta \pi(n)= \pi(n) - \pi_{\mathrm{rand}}(n)$ quantifies the effect of social influence in the adoption process (inset of Fig. \ref{fig:Skype1Emp}c): $\Delta \pi(n)$ increases approximately in a linear fashion with the fraction of adopting neighbours. This observation is in agreement with previous studies where a similar scaling of social influence has been recognised through small scale experiments \cite{centola2010spread}, data-driven observations \cite{bakshy2012role}, and modelling \cite{dodds2004universal,dodds2013limited}.

Thus based on these observations, for modelling purposes we can summarise that the behaviour of an agent can be characterised by four elementary processes: (a) \textit{Spontaneous adoption}, influenced by individual factors or external media independently of the social network. This is certainly the dominant mechanism for agents with no neighbours at the time of adoption. (b) \textit{Peer-pressure adoption}, an intrinsic social effect implemented here by making use of the observed linear scaling of the probability $\pi(n)$. (c) \textit{Temporary termination}, describing the case in which agents stop usage with a chance of re-adoption. (d) \textit{Permanent termination}, when users abandon the service altogether.

\subsection{Local cascades induced global contagion}
\label{sec:localcascades}

\begin{figure}[!ht]
\centering
\includegraphics[width=1.\textwidth]{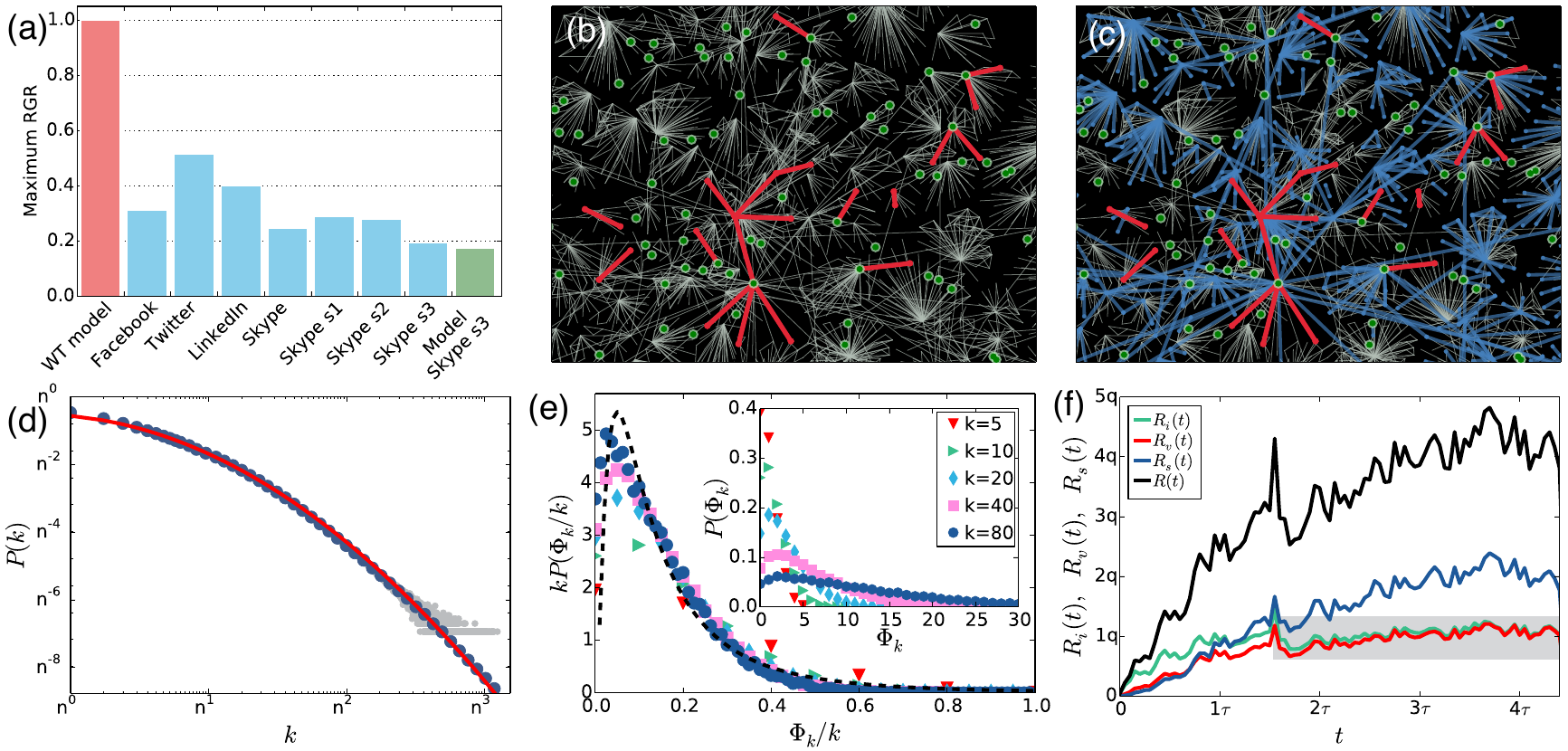}
\caption{\small Structure and dynamics of online service adoption. (a) Yearly maximum relative growth rate (RGR) of cumulative adoption~\cite{karsai2016local} for several online social-communication services \cite{SocialMedia}, including three Skype paid services (s1 - "subscription", s2  - "voicemail", and s3 - "buy credit"). The red bar corresponds to a rapid cascade of adoption suggested by the Watts threshold (WT) model, while the green bar is the model prediction for Skype s3. (b-c) Snowball sample of the Skype social network (grey links) with nodes and links coloured according to their adoption state: multiple innovators (green nodes), induced small vulnerable trees (red nodes and links), and the triggered connected stable cluster (blue nodes and links). (d) Degree distribution $P(k)$ of the Skype network (grey/blue circles for raw/binned data) on double log-scale with arbitrary base $n$. $P(k)$ is fitted by a lognormal distribution with parameters $\mu_D=1.2$ and $\sigma_D=1.39$, and average $z = 8.56$ (red line). (e) Distribution $P(\Phi_k)$ of integer thresholds $\Phi_k$ for several degree groups in Skype s3 (inset). By using $P(\Phi_k, k) = k P(\Phi_k/k)$, these curves collapse to a master curve approximated by a lognormal function (dashed line in main panel) with parameters $\mu_T=-2$ and $\sigma_T=1$, as constrained by the average threshold $w = 0.19$. (f) Adoption rate of innovators [$R_i(t)$], vulnerable nodes [$R_v(t)$], and stable nodes [$R_s(t)$], as well as net service adoption rate [$R(t)$]. Rates are measured with a 1-month time window, while $q$ and $\tau$ are arbitrary constants. The shaded area indicates the regime where innovators adopt approximately with constant rate. This figure was published in~\cite{karsai2016local}.}
\label{fig:Skype2Emp}
\end{figure}

Behavioural cascades are rare but potentially stupendous social spreading phenomena, where collective patterns of exposure emerge as a consequence of small initial perturbations~\cite{karsai2016local}. Some examples are the rapid emergence of political and grass-root movements \cite{gonzalez2011dynamics,borge2011structural,ellis2011information}, fast spreading of information \cite{dow2013anatomy,gruhl2004information,banos2013role, watts2007influentials,hale2013regime,leskovec2006patterns,leskovec2007dynamics,goel2012structure} or behavioural patterns \cite{fowler2010cooperative}, etc. The characterisation \cite{goel2012structure, borge2013cascading,gleeson2008cascades,brummitt2012suppressing,ghosh2011framework} and modelling \cite{watts2002simple, singh2013threshold,gleeson2007seed} of such processes have received plenty of attention and provide some basic understanding of the conditions and structure of empirical and synthetic cascades on various types of networks \cite{yaugan2012analysis, karimi2013threshold, backlund2014effects}. However, these studies commonly fail in addressing the temporal dynamics of the emerging cascades, which may vary considerably between different cases of social contagion. Moreover, they have not answered why real-world cascades can evolve through various dynamic pathways ranging from slow to rapid patterning, especially in systems where the threshold mechanisms play a role and social phenomena spread globally. Besides the case of rapid cascading mentioned above, an example of the other extreme is the propagation of products in social networks \cite{bass1969new}, where adoption evolves gradually even if it is driven by threshold mechanisms and may cover a large fraction of the total population \cite{karsai2014complex}. This behaviour characterises the adoption of online services such as Facebook, Twitter, LinkedIn and Skype (Fig.\ref{fig:Skype2Emp}a), since their yearly maximum relative growth of cumulative adoption \cite{white2010social} (which is the maximum of the yearly adoption rate normalised by the final observed adoption number of a given service) is lower than in the case of rapid cascades as suggested e.g. by the Watts threshold model.

To fill this gap in the modelling of social diffusion, we analyse and model real-world examples of social contagion phenomena. 
We follow the adoption dynamics of the Skype paid service ``buy credit'' for $89$ months since 2004, which evolves over the social network of one of the largest voice over internet providers in the world. Data includes the time of first payment of each user, an individual and conscious action that tracks adoption behaviour. In contrast to other empirical studies where incomplete knowledge about the underlying social network leads to unavoidable bias \cite{karsai2014complex}, in this study, we use the largest connected component of the aggregated free Skype service as the underlying structure, where nodes are Skype users and links confirmed contacts between them. This is a good approximation since it maps all connections in the Skype social network without sampling, and the paid service is only available for individuals already enrolled in the Skype network. Also note that the service adoption process evolves in a considerably faster time-scale than the underpinning social network. This way applying a time-scale separation, and considering the network to be static, may provide a good first approximation here. The underlying structure is an aggregate from September 2003 to November 2011 (i.e. over $99$ months) and contains roughly 4.4 billion links and 510 million registered users worldwide \cite{SkypeIPO}. The data is fully anonymised and considers only confirmed connections between users (for more on the data see DS4 in Section~\ref{sec:datasets}).


\subsubsection{Watts cascade conditions}
\label{sec:wattscond}

In his seminal paper about threshold dynamics, Watts \cite{watts2002simple} classified nodes into three categories based on the necessary social influence they needed for adoption. He assumed that each node has an individual threshold $\varphi \in [0, 1]$ drawn from a distribution $P(\varphi)$ with average $\phi = \langle \varphi \rangle$. This threshold determines the minimum fraction of exposed neighbours that triggers adoption and captures the resistance of an individual against engaging in spreading behaviour. He identified {\it innovator} nodes that spontaneously change state to $1$, thus starting the process. Such nodes have a trivial threshold $\varphi=0$. Then there are nodes with threshold $0 < \varphi \leq 1/k$, called {\it vulnerable}, which need one adopting neighbour before their own adoption. Finally, there are more resilient nodes with threshold $\varphi>1/k$, denoted as {\it stable}, referring to individuals in need of strong social influence to follow the actions of their acquaintances. Watts suggested~\cite{watts2002simple} that a small perturbations (like the spontaneous adoption of a single seed node) can trigger global cascading patterns, however, their emergence is subject to the so-called {\it cascade condition}: the innovator seed has to be linked to a percolating vulnerable cluster, which adopts immediately afterwards and further triggers a global cascade (i.e. a set of adopters larger than a fixed fraction of the finite network).  In addition, while models with more sophisticated social influence function can be introduced \cite{latane1981psychology, dodds2004universal,karsai2014complex} the original linear-threshold assumption proposed by Watts and Granovetter seems to be sufficient to interpret our observations.

\subsubsection{Heterogeneous thresholds and slow adoption dynamics}

Degree and threshold heterogeneities are indeed present in the social network of Skype. The degree distribution $P(k)$ is well approximated by a lognormal function
\begin{equation}
P(k) \propto k^{-1} e^{-(\ln k - \mu_D)^2/(2\sigma_D^2)}$ \hspace{.1in} \text{where} \hspace{.1in} $k \geq k_{\mathrm{min}}
\label{eq:SkypePK}
\end{equation}
with parameters $\mu_D=1.2$, $\sigma_D=1.39$ and $k_{\mathrm{min}}=1$ (Fig.~\ref{fig:Skype2Emp}d), giving an average degree $z = 8.56$. Moreover, at the time of adoption we can measure the threshold $\varphi=\Phi_k/k$ of a user by counting the number $\Phi_k$ of its neighbours who have adopted the service earlier. We then group users by degree and calculate the distribution $P(\Phi_k)$ of the integer threshold $\Phi_k$ \cite{gleeson2008cascades} (Fig.~\ref{fig:Skype2Emp}e). By using the scaling relation $P(\Phi_k, k) = k P(\Phi_k/k)$ all distributions collapse to a master curve well approximated by a lognormal function
\begin{equation}
P(\varphi) \propto \varphi^{-1} e^{-(\ln\varphi - \mu_T)^2/(2\sigma_T^2)},
\label{eq:SkypePhiK}
\end{equation}
with parameters $\mu_T=-2$ and $\sigma_T=1$ as constrained by the average threshold $w = 0.19$ (as explained in~\cite{karsai2016local}). These empirical observations, in addition to the broad degree distribution, provide quantitative evidence about the heterogeneous nature of adoption thresholds.

Since we know the complete structure of the online social network, as well as the first time of service usage for all adopters, we can follow the temporal evolution of the adoption dynamics. By counting the number of adopting neighbours of an ego, we identify innovators ($\Phi_k=0$), and vulnerable ($\Phi_k=1$) or stable ($\Phi_k>1$) nodes. The adoption rates for these categories behave rather differently from previous suggestions \cite{watts2002simple} (Fig.~\ref{fig:Skype2Emp}f). First, there is not only one seed but an increasing fraction of innovators in the system who, after an initial period, adopt approximately at a constant rate. Second, vulnerable nodes adopt approximately with the same rate as innovators suggesting a strong correlation between these types of adoption. This stationary behaviour is rather surprising as environmental effects, like competition or marketing campaigns, potentially influence the adoption dynamics. 
On the other hand, the overall adoption process accelerates due to the increasing rate of stable adoptions induced by social influence. At the same time a giant adoption cluster grows and percolates through the whole network. Despite of this expansion dynamics and connected structure of the service adoption cluster, the service reaches less than $6\%$ of the total number of active Skype users over a period of $7$ years \cite{SkypeIPO}. Therefore we ask whether one can refer to these adoption clusters as cascades. They are not triggered by a small perturbation but induced by several innovators; their evolution is not instantaneous but ranges through several years; and although they involve millions of individuals, they reach only a reduced fraction of the whole network. To answer this question, in Section~\ref{sec:ccpmodels} we will incorporate the above mentioned features into a dynamical threshold model \cite{ruan2015kinetics,karsai2016local} with a growing group of innovators and investigate their effects on the evolution of global social adoption. Note that although we cannot follow the direct pathways of social influence, we performed a null model study to demonstrate at the system level that social influence is present and dominates the contagion process, as compared to effects of homophily (for results see~\cite{karsai2016local}).

\begin{center}
  $\ast$~$\ast$~$\ast$
\end{center}

Finally note that I have some further contributions related to the dynamic analysis of collective social phenomena, not discussed in details here. In one study~\cite{hours2016link} we analysed a Twitter corpus (DS3 in Section~\ref{sec:datasets}) and quantified dynamic similarities between users by considering the evolving set of their common friends and the set of their commonly shared hashtags in order to predict the evolution of mention links among them. We showed that these similarity measures are correlated among connected people and that the combination of contextual and local structural features provides better predictions as compared to cases where they are considered separately.

In another work~\cite{alessandretti2016user} we studied multimodal transportation systems, with several coexisting services like bus, tram, and metro of several French municipal areas. Transportation systems can commonly be represented as time-resolved multilayer networks where the different transportation modes connecting the same set of nodes are associated to distinct network layers. In this study, we provided a novel user-based representation of public transportation systems, which combines representations, accounting for the presence of multiple lines and reducing the effect of spatial embeddedness, while considering the total travel time, its variability across the schedule, and taking into account the number of transfers necessary. After the adjustment of earlier techniques to the novel representation framework, using non-negative matrix factorisation, we identified hidden patterns of privileged connections linking places at a given distance with the fastest multimodal transportation connections. We also studied their efficiency as compared to the commuting flow.


\section{Modelling simple spreading phenomena}


A simplified mathematical description of spreading processes is proposed by compartment models, which commonly assume that at a given time a node can be in one of many mutually exclusive states and can be dynamically transferred between these states following some stochastic or deterministic rules. Based on the compartmental states and the transition rules we distinguish between \emph{simple}~\cite{barrat2008dynamical} and \emph{complex contagion processes}~\cite{romero2011differences,centola2007complex} modelling epidemic type of spreading in the former case, while social contagion in the latter. Discussion of my contributions to modelling simple~\cite{liu2014controlling,karsai2017control,tizzoni2015scaling,zhang2018link} and complex~\cite{karsai2014complex,ruan2015kinetics,karsai2016local,unicomb2018threshold} contagion will be the subject of this Section. Our central questions will be to understand the structural effects of temporal and static networks on spreading processes using synthetic and data-driven models, and to define methods to characterise the underlying network structure sampled by spreading processes.

\subsection{Simple contagion processes}
\label{sec:scp}

\begin{figure}[!ht]
\centering
\includegraphics[width=.8\textwidth]{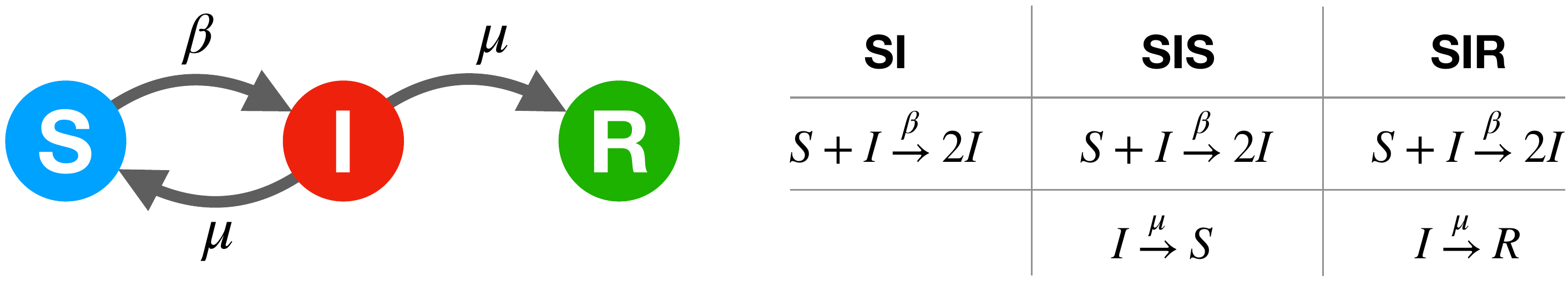}
\caption{\small Schematic summary of compartment models of epidemic spreading.}
\label{fig:SimplSpr}
\end{figure}

The spread of infectious diseases are common examples of simple contagion, which can be well described by compartment models of \emph{reaction-diffusion processes}~\cite{keeling2011modeling,barrat2008dynamical}. In such models one takes $N$ nodes, all of which can be in multiple but mutually exclusive states. One commonly used model family of simple spreading processes assumes that nodes can be susceptible (S), infected (I), or recovered/removed (R) and can transfer between these states in consecutive iteration steps via probabilistic rates. If in a given time step a node is susceptible and interacting with an infected other (reaction), in the next iteration step it can become infected (diffusion) with a rate $\beta$ (described by the spreading scheme $S+I\xrightarrow{\beta}2I$). Assuming only these two possible states and a single way of transition we obtain the definition of the well known Susceptible-Infected (SI) model. On the other hand, if we allow for an infected node to spontaneously transfer with a given rate $\mu$ into other states, we can get more complicated model definitions. In case an infected node can recover ($I\xrightarrow{\mu}R$) and reach a state from which it never get infected again, we arrive to the definition of the Susceptible-Infected-Recovered model. Otherwise, if it can transfer back to state $S$ ($I\xrightarrow{\mu}S$) and became susceptible again for infection, we arrive to the definition of the Susceptible-Infected-Susceptible (SIS) model. Schematic summary of states and transitions scheme of these models are summarised in Fig.~\ref{fig:SimplSpr}. Although the mathematical definition of such models are very similar, they display rather different critical behaviour when transitioning between phases of vanishing or global contagion~\cite{barrat2008dynamical}.

Compartment models of spreading processes were originally defined for unstructured populations assuming \emph{homogeneous mixing}, i.e., where every node interacts with any other nodes in each iteration. In this case, solutions for the dynamics of compartment sizes were proposed by the \emph{dynamical mean-field approach}. However, over the last decades this framework has been successfully extended for structured populations, where a network structure is used to code the possible interactions between nodes. Solution in this case was built on a \emph{degree decomposition method}~\cite{pastor2001epidemic}, which introduces classes of statistically equivalent nodes of degree $k$ in the description. This idea lead to the seminal result that degree heterogeneities decrease the critical point of epidemics leading to a vanishing infection threshold in scale-free networks with degree exponent $\gamma\leq 3$.

One common assumption in the majority of related works is to consider a \emph{time-scale separation} between the changes in network structures, $\tau_G$, and the contagion process $\tau_P$ (for a discussion on temporal scales see Section~\ref{sec:ts}). Indeed, spreading processes have been typically considered to take place in either static ($\tau_P\ll \tau_G$) or annealed ($\tau_P\gg \tau_G$) networks, where links are not evolving but present always or they evolve so rapidly that in turn they allow the infection to pass between nodes at any time. While these approximations can be used to study a range of processes such as the spreading of some diseases in contact networks or the propagation of energy in power grids they fail to describe many other phenomena in which the two timescales are comparable~\cite{karsai2011small,miritello2011dynamical,rocha2011simulated,morris1993telling,stehle2011simulation,kivela2012multiscale,karsai2014time,karsai2017control,ribeiro2013quantifying,scholtes2014causality,karimi2013threshold,valdano2015analytical}.

In these cases epidemics may strongly depend on the dynamics of the network, as infections can be transferred between nodes only at the time of their interactions. Spreading of ideas, memes, information and some type of diseases the diffusion processes can take place in such time-varying networks and their modelling needs to consider the interplay between the two simultaneously unfolding dynamics. These new directions land temporal networks in the focus of epidemic modelling~\cite{valdano2015analytical,liu2014controlling} and set a new direction in their investigations called \emph{temporal network epidemiology}~\cite{masuda2017temporal}.

Several other characters of networks have been considered for ever more realistic modelling of epidemic spreading including multi-layer description of social interactions and mobility patterns~\cite{granell2013dynamical,zhao2014immunization}, or meta-population networks~\cite{colizza2007reaction,colizza2007invasion,tizzoni2015scaling} to capture human mobility patterns at multiple scales. In the coming Section I am going to summarise some of my contributions in this direction, addressing how to effectively control epidemic spreading on temporal networks~\cite{liu2014controlling}. As another example for the usefulness of simple spreading processes, we will also introduce a new centrality measure based on deterministic SI processes to identify important links to impede epidemic outbreaks~\cite{zhang2018link}. Finally we will shortly summarise our other works addressing how epidemic spreading is influenced by effects of contact memory~\cite{karsai2014time} or emerging communities~\cite{laurent2015calls} in temporal networks and discuss how the super-linear scaling of contact densities with city sizes influences the outcome of meta-population models of spreading~\cite{tizzoni2015scaling}.

\subsection{Controlling contagion processes in time-varying networks}

Control strategies of contagion processes is a central question in network epidemiology and in information spreading~\cite{cohen2003efficient,pastor2002immunization}. The aim here is to identify the minimum set of nodes (or links), which removal would maximally decrease the probability of the emergence of global spreading on one hand, or to reach maximal involvement once initiate spreading from these sites. Spreading processes evolving on networks are influenced by a set of network features. In particular, heterogeneity observed in the distribution of networks' metrics, like the number of connections per node, degree, and the intensity of contacts, or weights have been shown to be critical. These quantities follow distributions characterised by heavy-tails, which imply the absence of characteristic scales and the presence of large fluctuations with respect to the average~\cite{vespignani2009predicting}. Other influencing factors identified are higher-order organisation of connectivity patterns associated to the presence of clusters/communities~\cite{newman2010networks,fortunato2010community}. 

The understanding of these properties and their effects on spreading phenomena has spurred the creation of strategies aimed at controlling or promoting diffusion processes. These can be classified in two main categories~\cite{wang2016statistical}. On one hand we can devise \emph{global} strategies that rely on the full knowledge of the network structure, while on the other hand we can define \emph{local} strategies, which relax this, often unrealistic assumption. In order to better understand the problem setting, let us imagine that we want to protect a network of computers against the spreading of malwares. The problem is to find a way to immunise a fraction $p$ of nodes to effectively protect the entire network. Each prescription for the selection of this fraction constitutes what we call a strategy. To this end, global strategies use centrality measures such as degree, k-core, betweenness and PageRank to rank the importance of each node ~\cite{barrat2008dynamical,newman2010networks,kitsak2010identification,pastor2002immunization}, while local strategies instead infer the role of nodes by local explorations and samples~\cite{cohen2003efficient}.

Here we investigate the effect of time-varying connectivity patterns on contagion control strategies by considering the specific class of activity driven network models~\cite{perra2012activity}. In particular, we consider the susceptible-infected-susceptible (SIS) model~\cite{keeling2011modeling} and derive analytically its critical immunisation threshold in case of three different control strategies. We also validate qualitatively the findings obtained in synthetic networks by studying the effect of each strategy in a large-scale mobile telephone call dataset.

\subsubsection{Controlling contagion processes in activity-driven networks}

A closed formula for the epidemic threshold of a SIS epidemic process unfolding on \emph{any} time-varying network has been derived~\cite{prakash2010virus}. In this approach the network is considered as a sequence of adjacency matrices $A_1, A_2, \ldots, A_T$ and has been shown that a disease cannot spread in the system if $\lambda_{\prod_i \mathbf{S_i}}<1$, where $\mathbf{S_i}=(1-\mu)\mathbf{I}+\alpha \mathbf{A_i}$, and $\alpha$ is the transmission rate per contact. In other words, the disease will die out if the largest eigenvalue of the system-matrix $\mathbf{S}=\prod_i \mathbf{S_i}$ is smaller than one. This result have been recently confirmed with a different approach~\cite{valdano2015analytical}. 

In case of activity-driven networks (for definition see Section~\ref{sec:tnet_adn}), a solution for the SIS model can be obtained by using the homogeneous mean-field theory (discussed for degree decomposition in Section~\ref{sec:scp}). However, here instead of degrees, we group nodes according to their activity assuming that nodes in the same class are statistically equivalent. At the mean-field level, the spreading process can be described by the number of infected individuals in the class of activity $a$ at time $t$, i.e.,  $I^{t}_{a}$. Following Ref.~\cite{perra2012activity}, the number of infected individuals of class $a$  at time $t + 1$ is given by:
\begin{equation}
\label{eq:pp1}
I^{t+1}_{a} = I^{t}_{a} -\mu I^{t}_{a}   +\alpha  m (N_{a}-I_{a}^{t})a   \int d a' \frac{I_{a'}^{t}}{N} +\alpha  m(N_{a}-I_{a}^{t})\int d a' \frac{I_{a'}^{t}a' }{N},
\end{equation}
where $N_a$ is the total number of individuals with activity rate $a$  (which is constant over time).  Each term in the Eq.~\ref{eq:pp1} has a clear physical interpretation. In fact, the number of infected nodes in the class $a$ at time $t+ 1$ is given by: the number of infected nodes in this class at time $t$ (first term), minus the number of nodes that recover and going back to the class $S_a$ (second term), plus the number of infected individuals generated when nodes in the class $S^t_a=N_a-I^t_a$ are active and connect with infected nodes in the other activity classes (third term), plus the number of infected nodes generated when nodes in the class $S^t_a$ are linked by active infected nodes in other activity classes. After finding the solution (not shown here~\cite{liu2014controlling}), and considering per capita spreading rate $\beta=\alpha \langle k\rangle$ we can write the threshold for the $SIS$ process, $\xi^{SIS}$, as:
\begin{equation}
\label{thre_SIS}
\frac{\beta}{\mu} \ge \xi^{SIS}\equiv \frac{2\langle a \rangle }{\langle a \rangle +\sqrt{\langle a^2 \rangle}}.
\end{equation}
In other words, the epidemic threshold is a function of the first and second moment of the activity distribution. Due to the co-evolution of the network structure and the spreading processes, the threshold is not depending on time-aggregated metrics such as the degree. It is defined by the interplay between the timescale of the contagion process and the convolution of the network timescales encoded in the moments of the activity distribution.

\begin{figure}[ht!]
\centering
\includegraphics[width=.9\textwidth]{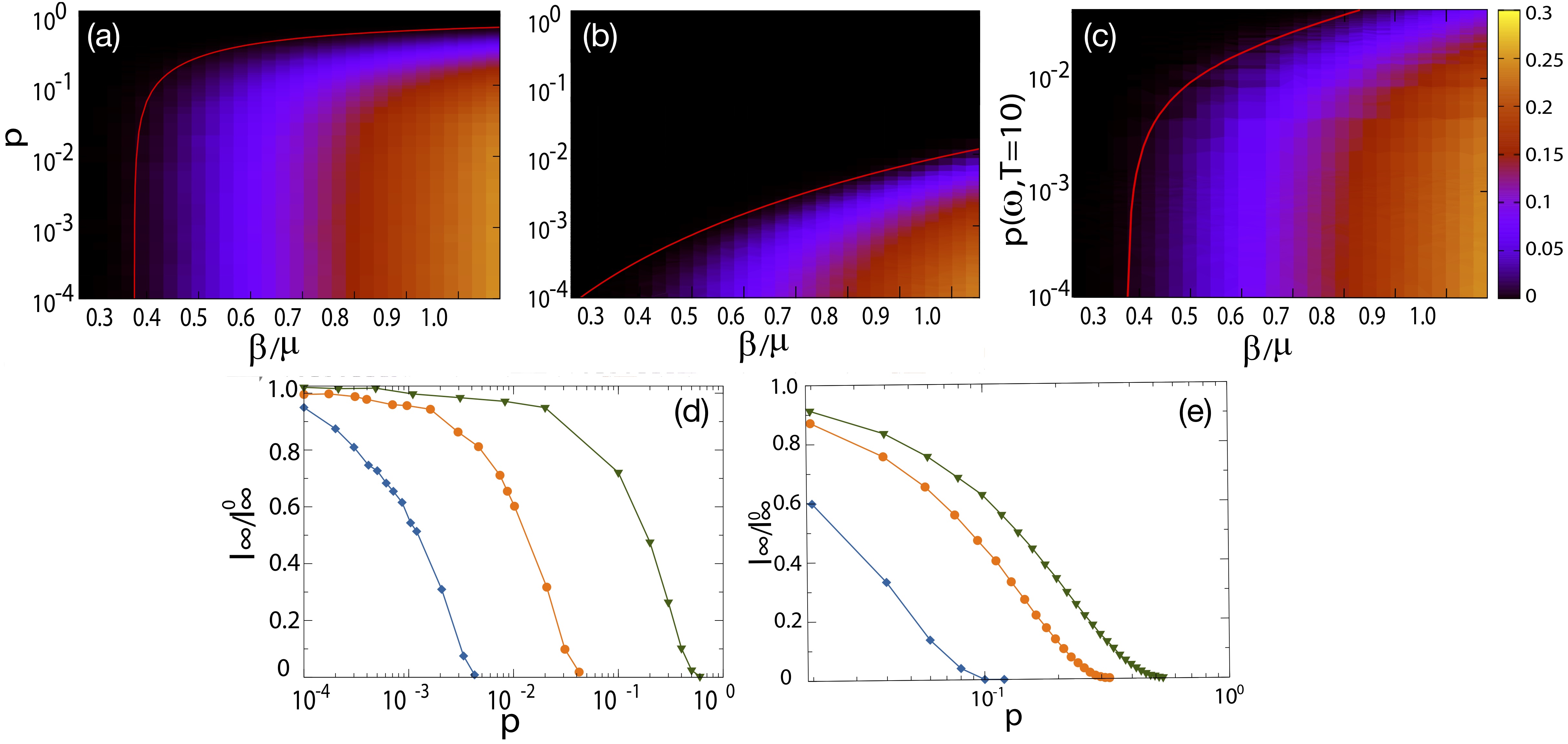} 
\caption{\small Panels a, b, and c show the phase space of an SIS process under random, targeted, and egocentric control strategy, respectively. Considering $N=10^4$, $m=3$, $\epsilon=10^{-3}$, activity distributed as $F(a)\sim a^{-2.2}$, we plot $I_\infty$ as a function of $\beta/\mu$ and $p$. Red curves represent the critical value $p_c$. Panel d shows the comparison of the stationary state of a SIS model with and without control strategy, $I_\infty^p/I^0_\infty$, as a function of $p$ when $\beta/\mu=0.81$. In green triangles, we consider the random strategy, in blue diamonds the targeted strategy, and in orange circles the egocentric strategy. Panel e shows the same results obtained on a real mobile call network. Every simulations was initiated with $1\%$ infected seed, executed $10^2$ times for $T=10^4$ step with $\beta/\mu=2.5$. Each step was integrated for $6\times10^2$ seconds, and periodic temporal boundary condition was applied. This figure was prepared by S. Liu and was published in~\cite{liu2014controlling}.}
\label{fig:control1}
\end{figure}

Let's now study different immunisation strategies. Following Ref.~\cite{liu2014controlling} and earlier works on controlling annealed and static networks~\cite{pastor2002immunization,cohen2003efficient}, we will consider three main strategies: random, global and local. In all the cases, we introduce a fraction $p$ of nodes as immunised. To account for this new class of removed nodes, we introduce a new compartment, $R$, in the classic SIS scheme. Thus, the Eq.~(\ref{eq:pp1}) becomes:
\begin{equation}
\label{eq:eq_1}
I^{t+1}_{a} = I^{t}_{a} -\mu I^{t}_{a}+ \alpha m (N_{a}-I_{a}^{t}-R_a^t)a \int d a'  \frac{I_{a'}^{t}}{N} + \alpha  m(N_{a}-I_{a}^{t}-R_a^t)\int d a'  \frac{I_{a'}^{t}a'}{N}.
\end{equation}
First, let us consider the random strategy (RS) in which a fraction $p$ of nodes is immunised with a uniform probability (see Fig.~\ref{fig:control1}a)~\cite{liu2014controlling}. In this case, the system of equations describing the dynamic process in activity-driven networks can be obtained by setting $R_a=p N_a$. The epidemic threshold condition changes as
\begin{equation}
\label{eq:thre_RI}
\frac{\beta}{\mu} \ge \xi^{RS} \equiv \frac{1}{1-p}\frac{2\langle a\rangle}{\langle a \rangle +\sqrt{ \langle a^2 \rangle}} =\frac{\xi^{SIS}}{1-p}.
\end{equation}
As expected, when a fraction $p$ of nodes is randomly immunised/removed, the epidemic threshold can be written as the threshold with no intervention, $\xi^{SIS}$, rescaled by the number of nodes still available to the spreading process. Another important quantity is the critical value of immunised/removed nodes, $p_c$, necessary to halt the contagion process. This quantity is a function of the network's structure and the specific features of the contagion process.  The explicit value of $p_c$ can be obtained by inverting Eq.~\ref{eq:thre_RI}. In Fig.~\ref{fig:control1}a, we plot $p_c$ as a function of $\beta/\mu$ keeping fixed the statistical properties underlying network. The phase space of the diffusion process is divided into two different regions separated by the red solid line that represents $p_c$ as derived by Eq.~\ref{eq:thre_RI}. In the region below the curve, the spreading process will take over, $p<p_c$, however, in the region above the curve, the fraction of removed/immunised nodes is enough to completely stop the diffusion process, $p \ge p_c$. To further assess the efficiency of the immunisation strategy in Fig.~\ref{fig:control1}d (green triangles), we plot, as a function of the density of removed/immunised nodes $p$, the ratio $I_{\infty}^p/I^0_{\infty}$ where $I^0_{\infty}$ is the asymptotic density of infected nodes when no-intervention is implemented. As shown clearly in the figure, the random strategy allows a reduction in the fraction of infected nodes just for large values of $p$.

In networks with heavy-tailed degree distributions,  targeting nodes with high degree centrality performs more efficiently than random strategies~\cite{barrat2008dynamical,cohen2010complex}. Analogously, in activity driven networks, effective strategies shall target high activity nodes. For this reason, we rank nodes in decreasing order of activity and immunising/removing the top ranking $pN$ nodes, and obtain the phase space shown in Fig.~\ref{fig:control1}b. This method is equivalent to fix a value $a_c$ so that any node with activity $a\ge a_c$ is immune to the contagion process. Also, for this scheme, it is possible to derive the analytic expression for the epidemic threshold~\cite{liu2014controlling}: 
\begin{equation}
\label{eq:thre_targ}
\frac{\beta}{\mu}\ge \xi^{TS}\equiv\frac{2 \langle a \rangle}{\langle a \rangle^{c}+\sqrt{(1-p) \langle a^2 \rangle^{c} }},
\end{equation}
where $ \xi^{TS}$ indicates the threshold for the targeted control strategy. In this case it is not possible to derive explicitly $p_c$, however, it can be easily evaluated numerically by solving the equation $\xi^{TS}-\beta/\mu=0$ for different values of $\beta/\mu$. In Fig.~\ref{fig:control1}b, we show $p_c$ (red line) as a function of $\beta/\mu$. From there it is evident that the immunising/removing a very small fraction of the most active nodes is enough to stop the contagion process, as also confirmed in Fig.~\ref{fig:control1}d (blue diamonds). The extreme efficiency of this strategy is due to the crucial role of high activity nodes in the spreading process. Immunising just the top 1\% of nodes is enough to halt the disease.

Unfortunately, the network-wide knowledge required to implement targeted control strategies is generally not available~\cite{cohen2003efficient}. In the case of evolving networks, this issue is even more pronounced as node's characterisation depends on how long it is possible to observe the network dynamics. To solve this problem we propose a local sampling strategy where first we select randomly a $w$ fraction of nodes act as ``probes". During an observation time $T$, we monitor their egocentric network generated by their interactions, which after we select randomly a node in the observed egocentric network of each probe to immunise/remove it. For the sake of comparison with the previous control strategies, we start the epidemic after a $p$ fraction of node have been immunised (for further notes on how to determine $p$ from $w$ see~\cite{liu2014controlling}). In case of this egocentric sampling scheme, after some analytical considerations, the epidemic threshold can be expressed as 
\begin{equation}
\label{thre_ES}
\frac{\beta}{\mu} \ge \xi^{ESS} \equiv \frac{2\langle a \rangle}{\Psi_1^T +\sqrt{ \Psi_0^T \Psi_2^T}},
\end{equation}
where we define  $\Psi_{n}^T=\int da \,\ a^{n}(1-P_a)^TF(a)$. This last integral is a function of the observing time window $T$, the probability of immunisation/removal of each class, and the activity distribution. We evaluate each $\Psi$ term through numerical integration with results shown in Fig.~\ref{fig:control1}c, which make clear that this strategy is much more efficient than the random one, although not as performant as the targeted scheme but with the advantage to rely only on local information. The efficiency of the ES strategy is due to the ability to reach active nodes by a local exploration done observing the systems for few time steps.

Real world time-varying networks add a number of complications to the simplified picture offered by activity driven networks. Indeed, they exhibit correlations among nodes, persistency of links, and burstiness of the activity pattern, as we have discussed it earlier. In order to see whether the above derived mean-field framework provides a good approximation in real world datasets we considered a mobile phone call network (a sample of DS1 in Section~\ref{sec:datasets}), consisting of $93,190$ connected phone users of a single city involved in almost five million calls over $120$ days. Remarkably, as reported in~\cite{liu2014controlling} and shown in Fig.\ref{fig:control1}e, we found a very good qualitative agreement between what is observed in real time-varying systems and the analytical results obtained in activity driven networks.

\subsection{Link transmission centrality in large-scale social networks}

Next we discuss another example of simple contagion processes to demonstrate how they can be used to design effective centrality measures in networks~\cite{zhang2018link}. Centrality measures of nodes or links generally rely on local and/or global structural information. Measures using local information, like the node degree or link overlap, are computed efficiently as they only require knowledge about the neighbours of a given node or link. On the other hand, these measures cannot provide information on which nodes or links play global roles in the network structure. On the contrary, centrality measures based on global information about the network structure, like betweenness and closeness  centrality~\cite{freeman1977set,bavelas1950communication}, Katz centrality~\cite{katz1953new}, k-shell index~\cite{bollobas1984graph,kitsak2010identification}, subgraph centrality~\cite{estrada2005subgraph} and induced centrality measures~\cite{everett2010induced} may better characterise the overall importance of a node or link. Unfortunately, although effective algorithms for approximating these quantities have recently been proposed~\cite{brandes2001faster,ercsey2010centrality}, estimating these measures in large scale networks is still computationally challenging.

While global centrality measures have been very successful in identifying structurally important nodes or links in networks, it has been argued~\cite{borgatti2005centrality} that they do not evidently identify nodes or links with a key role in dynamical processes. Other centrality metrics, which directly use dynamical processes to assign importance, such as PageRank \cite{brin1998anatomy}, eigenvector centrality \cite{leontief1941structure}, or accessibility \cite{travencolo2008accessibility},  were found to be more successful in this sense. However, these measures are based on random diffusion processes, and as such they do not fully capture the basic mechanisms behind contagion mediated spreading phenomena. In this Section we define a new link centrality measure, \textit{transmission centrality}, tailored to identify the role of nodes and links in controlling contagion phenomena. The transmission centrality measures the average number of nodes who are reached through each link during the spreading of a stochastic contagion process. This provides a direct measure of the centrality of the link in hindering or facilitating the contagion process. 

\subsubsection{Transmission centrality}

\begin{figure}[ht!]
	\begin{center}
		\includegraphics[width=.9\textwidth]{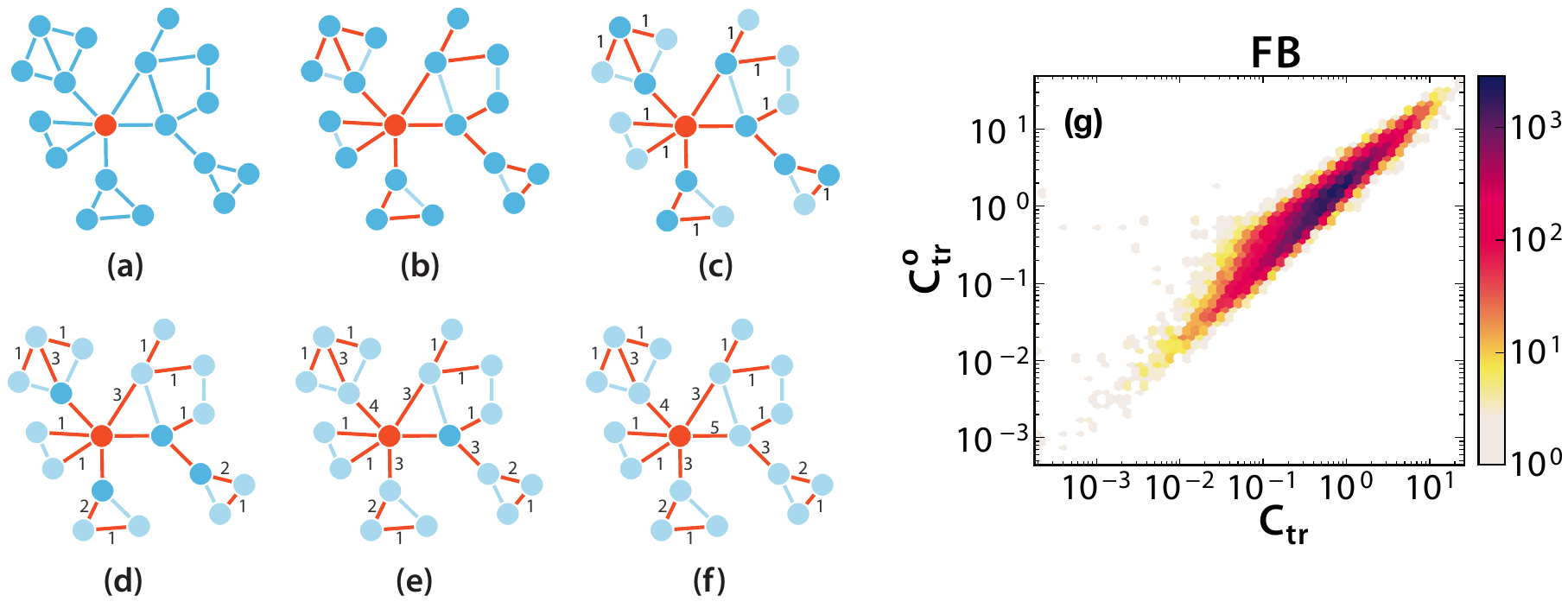}
	\end{center}
\caption{\small Calculation of \textit{transmission centrality} of links. (a) A network with a randomly selected seed node; (b) the branching tree rooted from the initial seed (root and edges in the tree are coloured in red); (c) for each leaf edge in the branching tree increase the counter by 1; (d)-(f) remove leafs and increase the counter of their ascendant by the counter of the removed leafs. (g) Correlation heat-map plot between the $C_{tr}^{o}$ exact and approximated $C_{tr}$ transmission centrality values of the FB network. Approximate measures were initiated from $5000$ seeds and unbiased in $d=3$ distance. This figure was prepared by Q. Zhang and was published in~\cite{zhang2018link}.}
\label{fig:riverbasin}
\end{figure}

Transmission centrality~\cite{zhang2018link} aims to measure for each link in a network its influence in disseminating some globally spreading information. More precisely it measures the number of nodes who received information during a diffusion process through a given link. Its definition intrinsically assumes a diffusion process to unfold on a network structure. In our definition we use the simplest possible information spreading process, the \textit{Susceptible-Infected (SI)} model (for definition see Section~\ref{sec:scp} and~\cite{barrat2008dynamical}), however this can be replaced by any other diffusion process. The SI process is controlled by the parameter $\beta$, which scales the speed of information/infection spreading. If it is $\beta=1$ (like in our case by default), it corresponds to the fastest possible information diffusion process determining the shortest diffusion routes between the seed and any other node in the network. In our case, we initiate the SI spreading process from a random seed$s$ and simulate the process with a modified breath-first-search algorithm~\cite{cormen2009introduction,zhang2018link}. Using this algorithmic solution one can record the branching tree $G_{BT}=(V_{BT},E_{BT})$ of the process by keeping track of the direct ascendant of each node from which it received the information. Exploiting the structure of the actual branching tree, \textit{transmission centrality} is formally defined as
\begin{equation}
C_{tr}(u,v) = \begin{cases}
  \max(|desc(u)|,|desc(v)|), & \text{if } (u,v)\in E_{BT}, \\
  0, & \text{otherwise}
\end{cases}
\end{equation}
where $|desc(i)|$ denotes the number of descendant nodes of node $i$ in the branching tree of the actual spreading.

The branching tree $G_{BT}$, which is a subgraph of $G$, encodes the diffusion paths that the information takes to reach the vertices of the network. Using its structure we can easily measure the actual $C_{tr}$ value of each link by performing a second step of calculation based on the \textit{river-basin algorithm}~\cite{rodriguez2001fractal}. In practice, taking the initial seed $s$ as the root of $G_{BT}$, and starting from the leafs of the branching tree, we can count the number of descendant nodes of each link, i.e., who received the information via the actual link. The algorithm is summarised in Alg.\ref{alg:riverbasin}, illustrated in Fig.\ref{fig:riverbasin} and works as follows:

\begin{algorithm}[h!]
\begin{algorithmic}[1]
\Require $G=(V,E)$ and $G_{BT}=(V_{BT},E_{BT})$
\Ensure $C_{tr}$ dictionary of \textit{transmission centrality} values
\State{$C_{tr}=dict()$}
\For{$(u,v) \in G.E$}
\State{$C_{tr}((u,v))=0$   \hspace{.4in} // set counter to zero for each link}
\EndFor
\While{$G_{BT}.E_{BT} \neq \emptyset $}{
\For{$ v \in G_{BT}.V_{BT}$}
\If{$k_v==1$}
		\State{$p=asc(v)$	\hspace{.4in} // parent node of v}
		\State{$gp=asc(p)$	\hspace{.4in} // grandparent node of v}
	  	\State{$C_{tr}((v,p)) = C_{tr}((v,p))+1$}
	  	\State{$C_{tr}((p,gp)) = C_{tr}((p,gp))+C_{tr}((v,p))$}
	    \State{$G_{BT}.E_{BT} \longleftarrow G_{BT}.E_{BT} - \{ (v,p) \}$}
	    \State{$G_{BT}.V_{BT} \longleftarrow G_{BT}.V_{BT} - \{ v \}$}
\EndIf
\EndFor
}\EndWhile
\end{algorithmic}
\caption{\small Transmission centrality}
\label{alg:riverbasin}
\end{algorithm}

First we define a dictionary $C_{tr}$, which associates a counter to each link $(i,j)\in G.E$, that we set to zero initially (lines 1-3 in Alg.\ref{alg:riverbasin}). Then we do the following for every node $v\in G_{BT}.V_{BT}$, which appears with degree $k_v=1$ in $G_{BT}$:
\begin{enumerate}[label=(\roman*)]
\item Increase by one the counter $C_{tr}((v,p))$ of the (leaf) edge $e_f=(v,p)\in G_{BT}.E_{BT}$, which connects $v$ to its parent node $p=asc_{BT}(v)$ in $G_{BT}.V_{BT}$ (line 9 in Alg.\ref{alg:riverbasin}).
\item Increase by $C_{tr}((v,p))$ the counter $C_{tr}((p,gp))$ of its ascendant edge $asc_{BT}(e_f)=(p,gp)$, where $gp=asc(p)$ is the grandparent node of $v$ in $G_{BT}.V_{BT}$ (line 10 in Alg.\ref{alg:riverbasin}).
\item Remove $v$ from $G_{BT}.V_{BT}$ and $e_f$ from $G_{BT}.E_{BT}$ (line 11 and 12 in Alg.\ref{alg:riverbasin}). The final transmission transmission centrality value of the actual link $e_f=(v,p)$ is stored in $C_{tr}((v,p))$.
\end{enumerate}
By repeating (a)-(c) for each appearing leaf edge we assign a non-zero value for each link in the branching tree as it is demonstrated in Fig.\ref{fig:riverbasin}.c-f.

The transmission centrality of a link can take values between 0 (for links, which are not in the branching tree) and $(N-1)$ (e.g. in the case the seed is propagating information via a single link). Its actual value depends on the choice of the seed node and on the structure of the branching tree determined by the diffusion process. In this way centrality values of the same link may vary from one realisation to another. To eliminate the effects of such fluctuations, the final definition of transmission centrality of links is taken as the average centrality value for each link computed over processes initiated from every node in the network (for a algorithmic definition see~\cite{zhang2018link}). Note that from now on $C_{tr}$ always assigns an average quantity if not stated otherwise.

\subsubsection{Heuristic calculation of transmission centrality}

One iteration to measure $C_{tr}$ performs with $\mathcal{O}(|E|)$ time complexity, thus in the case when we initiate its calculation from every node $v\in V$, its overall complexity is $\mathcal{O}(|V||E|)$. It is however possible to define a heuristic estimate of transmission centrality at a considerably small computational cost. As the branching trees of different realisations may largely overlap, a relatively small number of independent realisations, initiated from a reduced set of randomly selected seeds, could provide a good approximation.

Link transmission centrality initiated from a single node provides a locally biased measure as it assigns higher values to links closer to the actual seed. This bias is averaged out if we initiate the spreading process from every node in the network, but in case of a limited number of seeds it has some residual effects. One way to eliminate this residual bias is by assigning zero centrality values to links connecting nodes closer than a distance $d$ to the actual seed. The best value of $d$ depends on the network; however this can be estimated via parameter scanning, as demonstrated in~\cite{zhang2018link}. 

To illustrate the computation of the heuristic estimate, we use a Facebook network with $20,244$ nodes and $70,132$ edges~\cite{zhang2018link} and compute the average transmission centrality for each link via the exact method by initiating an SI process from each node, and the heuristic method where we initiate processes from $5000$ random seeds (i.e. $\sim 25\%$ of all nodes). Further we eliminate biases in distance $d=3$ around each seed. In Fig.~\ref{fig:riverbasin}g we present a  heat-map plot about the correlation between the exact (assigned as $C^{o}_{tr}$ here) and the approximated (assigned as $C_{tr}$) centrality values of each link. It is evident that there is a strong correlation between these values, quantified by an $r=0.96$ ($p<10^{-6}$) Pearson correlation coefficient. Consequently, this unbiased sampling method can provide very close approximations to the exact transmission centrality values, while considerably reducing the computational cost ($\sim 75\%$ in this case).

Due to its advantage to locate central link laying on multiple shortest paths, transmission centrality is very efficient in identifying weak links transmitting information between communities, and which in turn control dynamical processes. In~\cite{zhang2018link} we performed an extensive study to find the best combined tie strength measure, to propose the best link control strategy to impede SIR epidemic spreading. As a result we found that ordering links first by their overlap, and then ordering links having the same overlap values by their transmission centrality provides an efficient method to identify the weakest ties. It is demonstrated there that in some cases the control of only the $30\%$ of the identified weakest links may lead to $90\%$ of reduction of infection size.

Transmission centrality can be generalised in various ways. First, it can be easily defined as a \textit{node centrality metric} by counting for each node the number of their descendant nodes in the branching tree. Moreover it can be extended for \textit{directed and/or weighted networks} by restricting the SI process to respect the direction of links during spreading or by scaling the transmission rate with the normalised weight of links. In addition, for an SI process one can explore central links in the case when the process does not diffuse along the shortest paths. By taking $\beta<1$, longer spreading paths become plausible allowing the inference of links, which are central in any scenario. Transmission centrality can be easily defined for \textit{temporal networks}~\cite{holme2012temporal} as well. 
Finally, note that transmission centrality as a path based measure is not equivalent but closely related to betweenness centrality. However, while betweenness centrality considers all shortest paths between every pairs of nodes, transmission centrality takes only a single one from the potentially many others. This is especially true when $\beta=1$, when the SI process is fully deterministic inducing somewhat different but closely matching values with the corresponding betweenness measure, on a considerably lower computational cost.

\begin{center}
  $\ast$~$\ast$~$\ast$
\end{center}

In addition to the above studies me and colleagues published several other contributions addressing simple contagion processes on networks. In one line of research~\cite{tizzoni2015scaling} we studied the dynamics of reaction-diffusion processes (in particular the spreading of SIR processes) on heterogeneous meta-population networks where interaction rates scale with subpopulation sizes. We presented a new empirical evidence, based on the analysis of the interactions of $13$ million users on Twitter, about the scaling of human interactions with population size. We found that they scale super-linearly with an exponent $\gamma$ ranging between 1.11 and 1.21, as observed in recent studies based on mobile phone data. We integrated these observations into a reaction-diffusion meta-population framework and provided an explicit analytical expression for the global invasion threshold. Interestingly, we found that the super-linear scaling of human contacts facilitate the spreading dynamics. This behaviour is enhanced by increasing heterogeneities in the mobility flows coupling the subpopulations. Our results show that the scaling properties of human interactions can significantly affect dynamical processes mediated by human contacts such as the spread of diseases, ideas and behavioural patterns.

In some other works, we studied simple contagion processes evolving on \emph{temporal networks}. In one set of works~\cite{karsai2011small,kivela2012multiscale} we focused on data-driven simulations of information diffusion on random reference models of temporal networks (as already discussed in Section~\ref{sec:tnet_rrm}) to identify relevant structural and temporal correlations which influence the dynamics of the process. In this case we used deterministic SI spreading (with $\beta=1$) as a stereotypic model for information diffusion and showed that bursty temporal heterogeneities in contact dynamics and Granovetterian weight-topology correlations are mostly responsible for the observed slow spreading dynamics. In another line of works, we studied the effects of contact memory and communities on rumour and information spreading processes evolving on synthetic temporal networks simulated by the \emph{activity-driven network} model. In one case, as we discussed in Section~\ref{sec:ADNmemory} and in~\cite{karsai2014time}, we incorporated a simple statistical law that characterises memory in the temporal evolution of users' egocentric networks. We encoded this mechanism in a reinforcement process defining a time-varying network model that exhibits the emergence of strong and weak ties. On this network we studied the effects of time-varying and heterogeneous interactions on the classic rumour spreading model in both synthetic, and real-world networks. The model used here was the Delay-Kendall model~\cite{daley1964epidemics}, which is very similar to an SIR spreading, except that transition to a removed (stifler) state is not spontaneous but due to some interaction with other infected or removed nodes. We observe that strong ties severely inhibit information diffusion by confining the spreading process among agents with recurrent communication patterns. This provided the counterintuitive evidence that strong ties may have a negative role in the spreading of information across networks.




\section{Modelling complex spreading phenomena}
\label{sec:ccpmodels}

There are remarkable analogies between the social contagion of information, behavioural patterns or innovation and some physical or epidemic spreading processes, where global phenomena emerge through the diffusion of microscopic states \cite{barrat2008dynamical,easley2010networks,goffman1964generalization,jackson2008social}. All evolve in networks with nodes characterised by relevant state variables, and links that represent direct interactions between nodes. In biological systems epidemics are driven by binary interactions that lead to the emergence of {\it simple contagion} phenomena \cite{barrat2008dynamical}, as we discussed before in Section~\ref{sec:scp}. On the other hand, social diffusion processes are usually characterised by {\it complex contagion} mechanisms, where node states are determined by comparing individual thresholds with all neighbour states \cite{easley2010networks,karsai2014complex,watts2002simple,wejnert2002integrating, centola2007complex}. This property, capturing the effect of peer pressure and commonly assumed in social spreading phenomena \cite{granovetter1983threshold,centola2010spread}, has consequences on the dynamics and the final outcome of the social contagion process. Moreover, the theoretical approach to these systems has much in common \cite{barrat2008dynamical,watts2002simple,gleeson2007seed}, which greatly helps us to understand their behaviour. However, as we have already denoted in Section~\ref{sec:obssp}, real world observations of social spreading processes suggest significantly different picture as drawn from their modelling. In the following I am going to present a series of studies we published to introduce a framework based on the original model design of Watts~\cite{watts2002simple} and the formalism known as the \emph{approximate mean-field equations} introduced by Gleeson~\cite{gleeson2011high,gleeson2013binary}. Our aim with these contributions~\cite{karsai2014complex,ruan2015kinetics,karsai2016local,unicomb2018threshold,karsai2019multi} was to extend and re-define a threshold driven modelling framework, to better capture the dynamics of social contagion and its dependences on structural heterogeneities.

\subsection{Complex contagion processes}

Models employing threshold mechanisms mostly focus on cascading phenomena where, under some circumstances, a macroscopic fraction of nodes in the network is converted rapidly due to microscopic perturbations. This approach is motivated by earlier social theories \cite{granovetter1983threshold, schelling1969models} and has been implemented by Watts in an elegant model of cascading behaviour \cite{watts2002simple}. Watts showed that a global cascade (occupying a macroscopic fraction of the network and induced by local perturbations) can occur due to the interplay between network structure and individual thresholds (as briefly discussed in Section~\ref{sec:wattscond}). He further identified the phase with a non-zero probability of global cascades in the space $(\phi, z)$ of the average threshold $\phi$ of nodes and the average degree $z$ of the network.

\begin{figure}[!ht]
\centering
\includegraphics[width=.8\textwidth]{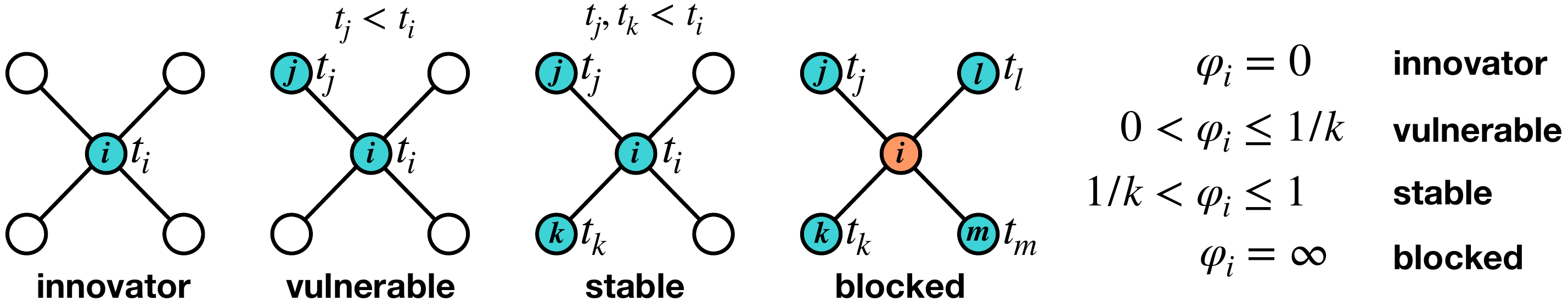}
\caption{\small Schematic summary of (non-)adopter types and threshold driven transitions in complex contagion processes. Colours assign states of nodes as white-susceptible, blue-adopter and orange-blocked.}
\label{fig:CpxSpr}
\end{figure}

Watts' threshold model \cite{watts2002simple} is defined on networks where nodes are associated to a state and can observe the state of their network neighbours at any time during the process. All nodes are initially in a susceptible state $0$, except for a single adopter seed in state $1$. The process evolves as each node with degree $k$ changes its state from $0$ to $1$ if a fraction $\varphi$ of its neighbours have adopted before (as demonstrated in Fig.~\ref{fig:CpxSpr}). Since nodes cannot change their state after exposure, the system evolves towards a state where no further adoptions are possible. The emergence of a global cascade depends on the degree distribution $P(k)$ of the network, the distribution $P({\varphi})$ of individual thresholds, and the initial seed. As we have already discussed in Section~\ref{sec:wattscond}, the condition for a global cascade is the existence of a percolating component of {\it vulnerable} nodes with thresholds $0 < \varphi \leq 1/k$ (who need one adopting neighbour before exposure) connected to the innovator seed with threshold $\varphi=0$. This percolating vulnerable tree is quickly converted after adoption of the seed and may trigger further adoption of {\it stable} nodes with thresholds $\varphi > 1/k$ (who need more than one adopting neighbour to adopt). Assuming an Erd\H{o}s-R\'{e}nyi (ER) random network \cite{erdos1960evolution} and a single adopter seed, there is a phase boundary in $(\phi, z)$-space encompassing a regime where global cascades occur. The properties of this cascading regime have been investigated for the case of heterogeneous thresholds, different network topologies \cite{watts2002simple,gleeson2007seed}, and variable seed size \cite{gleeson2008cascades,singh2013threshold}.

While the relevance of the Watts' model is indisputable \cite{watts2002simple,gleeson2007seed,watts2007influentials,gleeson2013binary,gleeson2011high,gleeson2008cascades,yaugan2012analysis,singh2013threshold},
its limitations become clear from real social spreading data. The Watts model focuses on the (instantaneous) emergence of global cascades triggered by single local perturbations, while there are empirical examples where threshold mechanisms do play a role yet global adoption phenomena emerge through other scenarios (for examples see Fig.~\ref{fig:Skype2Emp}a). Contrary, the Watts criterion for macroscopic adoption is purely deterministic, coded in the network structure, threshold distribution and perturbation site -- it does not concern time, which is clearly a feature of empirical stochastic processes of adoption spreading. To resolve this shortcoming, we extend the conventional Watts model with empirically motivated mechanisms (some already reported in Section~\ref{sec:localcascades}) to see whether they are eligible to explain the observed dynamical scenarios of social contagion phenomena.

Empirical studies support the intuition that some individuals in the social network may refuse to adopt technological innovations for various reasons -- due to another favourite product, aversion towards a firm, or some criticism on principle~\cite{karsai2016local}. Such individuals will never be exposed, irrespective of the state of their neighbours \cite{yildiz2013binary}, thus their threshold can be regarded as $\varphi=\infty$ (as demonstrated in Fig.\ref{fig:CpxSpr}). To consider this behavioural pattern we introduce a third state of nodes who are blocked (immunised) to participate in the spreading process. We block the adoption of a fraction $r$ of randomly selected nodes in the network, who in turn do count when their neighbours consider the decision to adopt, and thus will make it harder for neighbours to fulfil the threshold criterion. While blocked nodes hinder the spreading process, there are reasons other than social influence that could motivate individuals to adopt a social pattern, like external influence from mass media. This {\it spontaneous} adoption has been studied theoretically by introducing a given density of adopters at the outset of the Watts model~\cite{singh2013threshold}. However, spontaneous adopters may get active at any time during a real social contagion. Thus we include a stochastic dynamics where a susceptible node may become adopter with rate $p$ at any time, irrespective of the status of its neighbours. This assumption is based on our observations reported in Section~\ref{sec:localcascades} and~\cite{karsai2016local}, where after an initial period, innovators adopted a service with an approximately constant rate.

Considering both extensions, we have a threshold-driven dynamics with three node states: blocked, susceptible and adopter (Fig.\ref{fig:CpxSpr}). At the outset, all nodes are susceptible except for a fraction $r$ that remains blocked. At each time step of the simulation, a randomly selected susceptible node $i$ adopts spontaneously with probability $p$, otherwise it adopts if at least a fraction $\varphi$ of its neighbours has already adopted. If $r = 0$ and $p > 0$ all nodes will eventually adopt (Fig.\ref{fig:local}c), following a kinetics reminiscent of the approach to a unique ground state in a physics system. On the other hand, if we introduce quenched randomness and stochastic perturbations ($r,p > 0$), our model allows various temporal regimes and a transition from rapid to slow spreading dynamics.


\subsection{Dynamical threshold model with immune nodes}
\label{sec:dmodel}

Our threshold model can be studied analytically by extending the framework of \emph{approximate master equations} (AMEs) for monotone binary-state dynamics developed by Gleeson~\cite{gleeson2013binary,gleeson2011high,gleeson2008cascades}, where the transition rate between susceptible and adoption states only depends on the number $m$ of neighbours that have already adopted. We describe a node by the property vector $\mathbf{k} = (k, c)$, where $k = k_0, k_1, \ldots k_{M-1}$ is its degree and $c = 0, 1, \ldots, M$ its type, i.e. $c = 0$ is the type of the fraction $r$ of immune nodes, while $c \neq 0$ is the type of all non-immune nodes that have threshold $\varphi_c$. In this way, $P(\varphi)$ is substituted by the discrete distribution of types $P(c)$ (for $c > 0$). The integer $M$ is the maximum number of degrees (or non-zero types) considered in the AME framework, which can be increased to improve the accuracy of the analytical approximation at the expense of speed in its numerical computation.

We characterise the static social network by the extended distribution $P({\mathbf{k}})$, where $P({\mathbf{k}}) = r P(k)$ for $c = 0$ and $P({\mathbf{k}}) = (1 - r) P(k) P(c)$ for $c > 0$. Non-immune and susceptible nodes with property vector $\mathbf{k}$ adopt spontaneously with a constant rate $p$, otherwise they adopt only if a fraction $\varphi_c$ of their $k$ neighbours has adopted before. These rules are condensed into the probability $F_{\mathbf{k}, m} dt$ that a node will adopt within a small time interval $dt$, given that $m$ of its neighbours are already adopters,
\begin{equation}
\label{eq:thresRule}
F_{\mathbf{k}, m} =
\begin{cases}
p_r & \text{if} \quad m < k \varphi_c \\
1 & \text{if} \quad m \geq k \varphi_c
\end{cases}, \quad \forall m \quad \text{and} \quad  k, c \neq 0,
\end{equation}
with $F_{(k,0),m} = 0$ $\forall k, m$ and $F_{(0,c),0} = p_r$ $\forall c \neq 0$ (for immune and isolated nodes, respectively). The rescaled rate $p_r = p / (1 - r)$ (with $p_r = 1$ for $p > 1 - r$) is necessary if we wish to obtain a rate $p$ of innovators for early times of the dynamics, regardless of the value of $r$.

The dynamics of adoption is well described by an AME for the fraction $s_{\mathbf{k}, m}(t)$ of $\mathbf{k}$-nodes that are susceptible at time $t$ and have $m=0,\ldots,k$ adopting neighbours~\cite{porter2016dynamical,gleeson2013binary,gleeson2011high},
\begin{equation}
\label{eq:AMEsThres}
\dot{s}_{\mathbf{k}, m} = -F_{\mathbf{k}, m} s_{\mathbf{k}, m} -\beta_s (k - m) s_{\mathbf{k}, m} + \beta_s (k - m + 1) s_{\mathbf{k}, m-1},
\end{equation}
where
\begin{equation}
\label{eq:rateBs}
\beta_s(t) = \frac{\sum_{\mathbf{k}} P({\mathbf{k}}) \sum_m (k - m) F_{\mathbf{k}, m} s_{\mathbf{k}, m}(t)}{\sum_{\mathbf{k}} P({\mathbf{k}}) \sum_m (k - m) s_{\mathbf{k}, m}(t)},
\end{equation}
and the sum is over all the degrees and types, i.e. $\sum_{\mathbf{k}} \bullet = \sum_k \sum_c \bullet$. To reduce the dimensionality of Eq.~(\ref{eq:AMEsThres}), we consider the ansatz
\begin{equation}
\label{eq:AMEansatz}
s_{\mathbf{k}, m}(t) = B_{k, m} [\nu(t)] e^{-p_r t}
\quad \text{for} \quad   m < k\varphi_c \quad   \text{and} \quad  c \neq 0,
\end{equation}
with $\nu(t)$ the probability that a randomly-chosen neighbour of a susceptible node is an adopter.

Introducing the ansatz of Eq.~(\ref{eq:AMEansatz}) into the AME system of Eq.~(\ref{eq:AMEsThres}) leads to the condition $\dot{\nu} = \beta_s (1 - \nu)$. With some algebra, the AMEs for our dynamical threshold model are reduced to the pair of ordinary differential equations
\begin{subequations}
\label{eq:reducedAMEs}
\begin{align}
\dot{\rho} &= h(\nu, t) - \rho, \\
\dot{\nu} &= g(\nu, t) - \nu,
\end{align}
\end{subequations}
where $\rho(t) =  1 - \sum_{\mathbf{k}} P({\mathbf{k}}) \sum_m s_{\mathbf{k}, m}(t)$ is the fraction of adopters in the network, and the initial conditions are $\rho(0) = \nu(0) = 0$. Here,
\begin{equation}
\label{eq:hTerm}
h = (1 - r) \Big[ f_t + (1 - f_t) \sum_{\mathbf{k} | c \neq 0} P(k) P(c) \sum_{m \geq k\varphi_c} B_{k, m}(\nu) \Big],
\end{equation}
and
\begin{equation}
\label{eq:gTerm}
g = (1 - r) \Big[ f_t + (1 - f_t) \sum_{\mathbf{k} | c \neq 0} \frac{k}{z} P(k) P(c) \sum_{m \geq k\varphi_c} B_{k-1, m}(\nu) \Big],
\end{equation}
where $f_t = 1 - (1 - p_r) e^{-p_r t}$, and $B_{k, m}(\nu) = \binom{k}{m} \nu^m (1 - \nu)^{k - m}$ is the binomial distribution. The fraction of adopters $\rho$ is then obtained by solving Eq.~(\ref{eq:reducedAMEs}) numerically. Since the susceptible nodes adopt spontaneously with rate $p$, the fraction of innovators $\rho_0(t)$ in the network is given by
\begin{equation}
\label{eq:innovFrac}
\rho_0(t) = p_r \int_0^t [1 - r - \rho(\tau)] d\tau.
\end{equation}

\begin{figure}[h!]
\centering
\includegraphics[width=.65\textwidth]{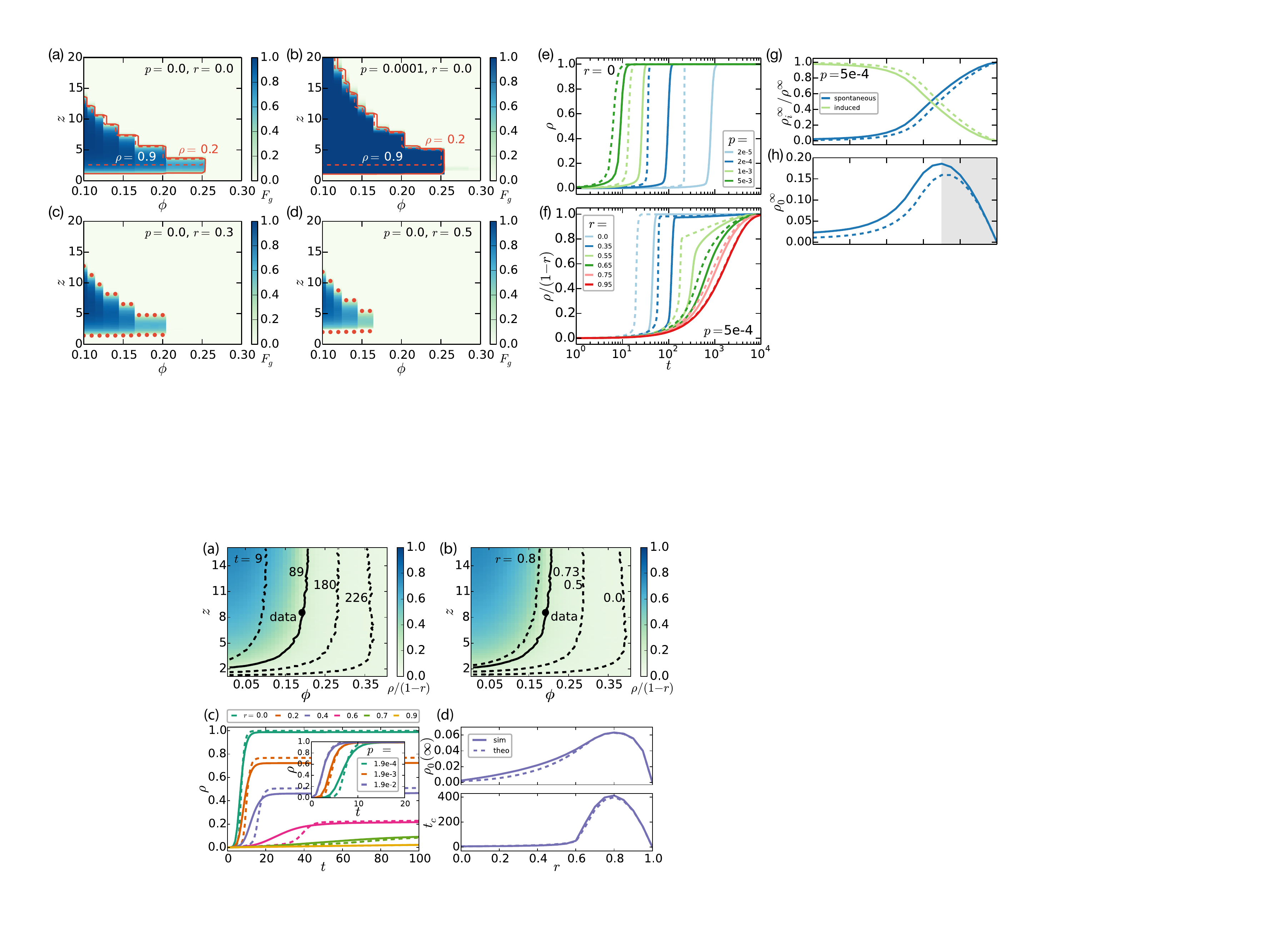}
\caption{\small (a-b) Surface plot of the normalised fraction of adopters $\rho / (1 - r)$ in $(\phi, z)$-space, for $r = 0.73$ and $t = 89$. Contour lines signal the parameter values for which $20\%$ of non-immune nodes have adopted, for fixed $r$ and varying time (a), and for fixed time and varying $r$ (b). The continuous contour line and dot indicate parameter values of the last observation of Skype s3. A regime of maximal adoption ($\rho \approx 1 - r$) grows as time goes by, and shrinks for larger $r$. (c) Time series of the fraction of adopters $\rho$ for fixed $p = 0.00019$ and varying $r$ (main), and for fixed $r = 0$ and varying $p$ (inset). These curves are well approximated by the solution of Eq.~(\ref{eq:reducedAMEs}) for $k_0 = 3$, $k_{M-1} = 150$ and $M = 25$ (dashed lines). The dynamics is clearly faster for larger $p$ values. As $r$ increases, the system enters a regime where the dynamics is slowed down and adopters are mostly innovators. (d) Final fraction of innovators $\rho_{0,\infty}$ and the time $t_c$ when $50\%$ of non-immune nodes have adopted as a function of $r$, both simulated and theoretical. The crossover to a regime of slow adoption is characterised by a maximal fraction of innovators and time $t_c$. Unless otherwise stated, $p=0.00019$ and we use $N=10^4$, $\mu_D=1.09$, $\sigma_D=1.39$, $k_{min}=1$, $\mu_T=-2$, and $\sigma_T=1$ to obtain $z = 8.56$ and $\phi = 0.19$ as in Skype s3. The difference in $\mu_D$ between data and model is due to finite-size effects. Numerical results are averaged over $10^2$ (a-b) and $10^3$ (c-d) realisations. This figure was prepared by G. I\~{n}iguez and was published in~\cite{karsai2016local}.}
\label{fig:local}
\end{figure}

We may also implement the model numerically via a Monte Carlo simulation in a network of size $N$, with a log-normal degree distribution and a log-normal threshold distribution as observed empirically in the case of Skype (see Section~\ref{sec:localcascades}). Hence, we can explore the behaviour of the fractions of adopters and innovators in the network, $\rho$ and $\rho_0$, as a function of $z$, $\phi$, $p$ and $r$, both in the numerical simulation and in the theoretical approximation given by Eqs.~(\ref{eq:reducedAMEs}) and~(\ref{eq:innovFrac}). For $p > 0$ some nodes adopt spontaneously as time passes by, leading to a frozen state characterised by the final fraction of adopters $\rho(\infty) = 1 - r$. However, the time needed to reach such a state depends heavily on the distribution of degrees and thresholds, as indicated by a region of large adoption ($\rho \approx 1 - r$) that grows in $(\phi, z)$-space with time (contour lines in Fig.~\ref{fig:local}a). If we fix the time in the dynamics and vary the fraction of immune nodes instead, this region shrinks as $r$ increases (contour lines in Fig.~\ref{fig:local}b). In other words, the set of networks (defined by their average degree and threshold) that allow the spread of adoption is larger at later times in the dynamics, or when the fraction of immune nodes is small. When both $t$ and $r$ are fixed, the normalised fraction of adopters $\rho / (1 - r)$ gradually decreases for less connected networks with larger thresholds (surface plot in Fig.~\ref{fig:local}a and b).

Both numerical simulations and analytical approximations show how the dynamics of spreading changes by introducing immune individuals in the social network. For $r \approx 0$, the adoption cascade appears sooner for larger $p$, since this parameter regulates how quickly we reach the critical fraction of innovators necessary to trigger a cascade of fast adoption throughout all susceptible nodes (Fig.~\ref{fig:local}c, inset). Yet as we increase $r$ above a critical value $r_c$ (and thus introduce random quenching), the system enters a regime where rapid cascades disappear and adoption is slowed down, since stable nodes have more immune neighbours and it is difficult to fulfil their threshold condition. The crossover between these fast and slow regimes is gradual, as seen in the shape of $\rho$ for increasing $r$ (Fig.~\ref{fig:local}c, main panel). We may identify $r_c$ in various ways: by the maximum in both the final fraction of innovators $\rho_{0,\infty} = \rho_0(\infty)$ and the critical time $t_c$ when $\rho = (1-r)/2$ (Fig.~\ref{fig:local}d), or as the $r$ value where the inflection point in $\rho$ disappears. These measures indicate $r_c \approx 0.8$ for parameter values calibrated with Skype data. All global properties of the dynamics (like the functional dependence of $\rho$ and $\rho_0$) are very well approximated by the solution of Eqs.~(\ref{eq:reducedAMEs}) and~(\ref{eq:innovFrac}) (dashed lines in Fig.~\ref{fig:local}c and d). Indeed, the AME framework is able to capture the shape of the $\rho$ time series, the crossover between regimes of fast and slow adoption, as well as the maximum in $\rho_{0,\infty}$ and $t_c$.

Thus with our new model extensions we have shown, that outside of the cascading regime of the Watts model, there is possibility of global contagion mediated by spontaneous adopters. However, the speed of spreading depends strongly on the density of blocked or immune nodes. For a small fraction $r$ of blocked nodes, few spontaneous adopters enable the formation of large clusters by initiating cascades. For large $r$, spreading slows down as it is dominated by spontaneous adopters as suggested by the empirical observations in Section~\ref{sec:localcascades}. This way, our intrinsically dynamic model shows a novel percolation transition of induced clusters and is able to describe various scenarios of real social contagion as well as the crossover between them.

\subsubsection{Model verification via data-driven simulations}

After we defined our dynamical threshold model, let's return back to our empirical observations on the spreading of Skype services reported in Section~\ref{sec:localcascades}. As demonstrated above, our model provides insight on the role of innovators and immune nodes in controlling the speed of the adoption process. However, in empirical datasets information about the fraction of non-adopters is usually not available, which makes it difficult to predict the future dynamics of service adoption. Here we use our modelling framework to perform data-driven simulations for two reasons: (a) to estimate the fraction $r$ of immune nodes in the real system; and (b) to validate our modelling as compared to real data.

\begin{figure}[h!]
\centering
\includegraphics[width=.8\textwidth]{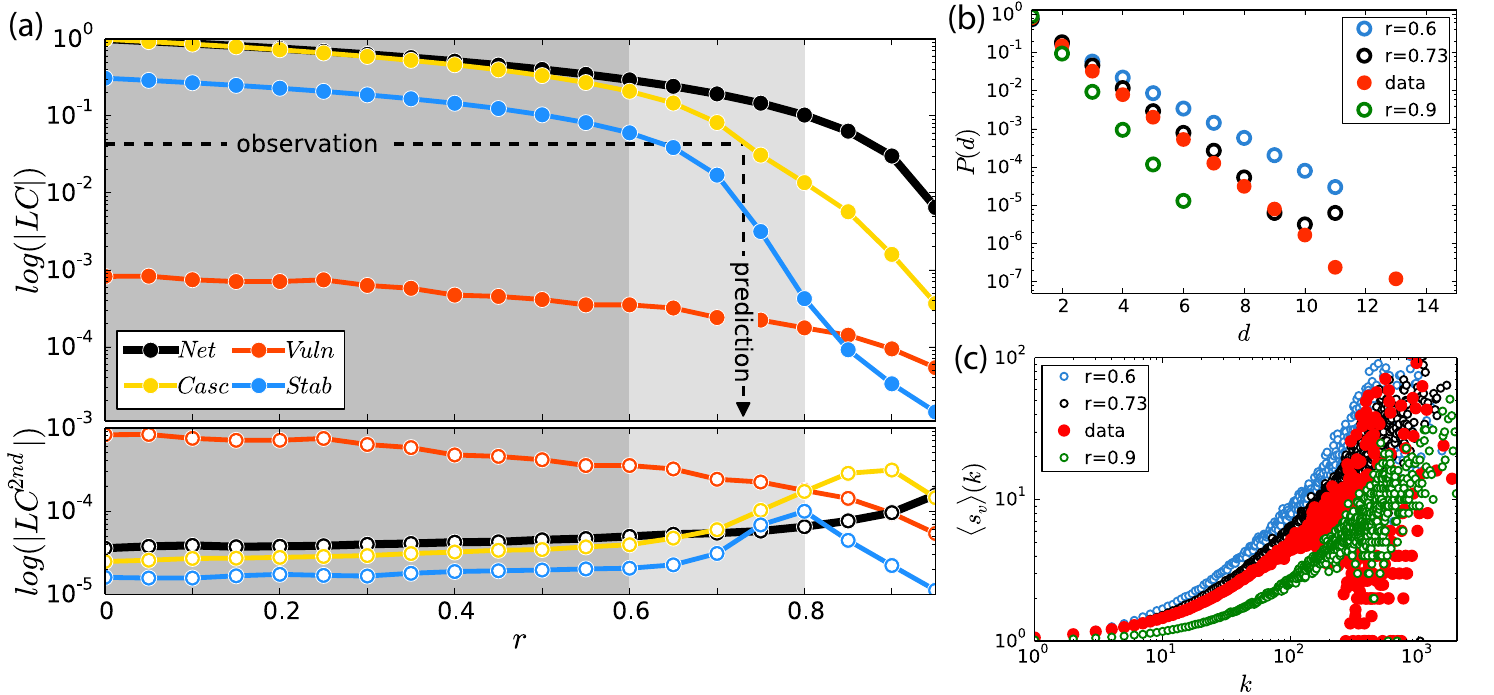}
\caption{\small (a) Average size of the largest ($LC$) and 2nd largest ($LC^{2nd}$) components of the model network (`Net'), adoption network (`Casc'), stable network (`Stab'), and induced vulnerable trees (`Vuln') as a function of $r$. Dashed lines show the observed relative size of the real $LC$ of the adopter network in $2011$ and the predicted $r$ value. (b) Distribution $P(d)$ of depths of induced vulnerable trees in both data and model for several $r$ values, showing a good fit with the data for $r=0.73$. The difference in the tail is due to finite-size effects. (c) Correlation $\langle s_v \rangle (k)$ between innovator degree and average size of vulnerable trees in both data and model with the same $r$ values as in (b). Calculations were carried out on networks with size $N=10^6$ and were averaged over $10^2$ realisations. This figure was published in~\cite{karsai2016local}.}
\label{fig:kineticsverif}
\end{figure}

To set up our data-driven simulations we use the Skype data to directly determine all model parameters, apart from the fraction $r$ of immune nodes. As we already discussed in Section~\ref{sec:localcascades}, the best approximation of the degree distribution of the real network is a log-normal function (Eq.~\ref{eq:SkypePK}) with parameters $\mu_D=1.2$, $\sigma_D=1.39$, minimum degree $k_{min} = 1$ and average degree $z = 8.56$. To account for finite-size effects in the model results for low $N$, we decrease $\mu_D$ slightly to obtain the same value of $z$ as in the real network. We also observe in Fig.~\ref{fig:Skype2Emp}e that the threshold distribution of each degree group collapses into a master curve after normalisation by using the scaling relation $P(\Phi_k,k)=k P (\Phi_k/k)$. This master curve can be well-approximated by the log-normal distribution shown in Eq.~\ref{eq:SkypePhiK}, with parameters $\mu_T=-2$ and $\sigma_T=1$ as determined by the empirical average threshold $\phi = 0.19$ and standard deviation $0.233$. We estimate a rate of innovators $p = 0.00019$ by fitting a constant function to $R_i(t)$ for $t > 2\tau$ (Fig.~\ref{fig:Skype2Emp}f). The fit to $p$ also matches the time-scale of a Monte Carlo iteration in the model to 1 month. To model the observed dynamics and explore the effect of immune nodes, we use a configuration-model network~\cite{newman2010networks} with log-normal degree and threshold distributions and $p$ as the constant rate of innovators, all determined from the empirical data. Model results in Fig.~\ref{fig:kineticsverif} are averaged over $100$ networks of size $N=10^5$ ($10^6$) after $T=89$ iterations, matching the length of the observation period in Skype.

As a function of $r$, the underlying and adoption networks pass through three percolation-type phase transitions. First, the appearance of immune nodes (for increasing $r$) can be considered as a removal process of nodes available for adoption from the underlying network structure. After the appearance of a critical fraction of immune nodes, $r_c^{net}$, the effective network structure available for adoption will be fragmented and will consist of small components only, limiting the size of the largest adoption cluster possible. Second, $r$ also controls the size of the emergent adoption cascades evolving on top of the network structure. While for small $r$ the adoption network is connected into a large component, for larger $r$ cascades cannot evolve since there are not enough nodes to fulfil the threshold condition of susceptible stable nodes, even if the underlying network is still connected. The transition point between these two phases of the adoption network is located at $r_c^{casc}\leq r_c^{net}$, limited from above by the critical point $r_c^{net}$. Finally, we observe from the empirical data and model results that the adoption network is held together by a large connected component of stable nodes. Consequently, for increasing $r$ the stable adoption network goes through a percolation transition as well, with a critical point $r_c^{stab}\leq r_c^{casc}\leq r_c^{net}$.

To characterise these percolation phase transitions we compute the average size of the largest ($LC$) and second largest ($LC^{2nd}$) connected components (Fig.~\ref{fig:kineticsverif}). We measure these quantities for the underlying network, and for the stable, vulnerable and global adoption networks, as a function of the fraction of immune nodes $r$. After $T=89$ iterations (matching the length of the real observation period), we identify three regimes of the dynamics: if $0<r<0.6$ (dark-shaded area) the spreading process is very rapid and evolves as a global cascade, which reaches most of the nodes of the shrinking susceptible network in a few iteration steps. About $10\%$ of adopters are connected in a percolating stable cluster, while vulnerable components remain very small in accordance with empirical observations. In the crossover regime $0.6<r<0.8$ (light-shaded area), the adoption process slows down considerably (Fig.~\ref{fig:kineticsverif}, upper panel), as stable adoptions become less likely due to the quenching effect of immune nodes. The adoption process becomes the slowest at $r_c^{stab}=0.8$ when the percolating stable cluster falls apart, as demonstrated by a peak in the corresponding $LC^{2nd}$ curve in Fig.~\ref{fig:kineticsverif} (diamonds in lower panel). Finally, around $r_c^{casc}=0.9$ the adoption network becomes fragmented and no global cascade takes place. Since the underlying network has a broad degree distribution, it is robust against random node removal processes~\cite{newman2010networks}. That is why its critical percolation point $r_c^{net}$ appears after $95\%$ or more nodes are immune. Note that similar calculations for another service have been presented~\cite{karsai2016local} with qualitatively the same results, but with the crossover regime shifted towards larger $r$ due to different parameter values of the model process.

We can use these calculations to estimate the only unknown parameter, namely the fraction $r$ of immune nodes in Skype, by matching the relative size of the largest component ($LC_{Net}$) between real and model adoption networks at time $T$. Empirically, this value is the relative size $s_a^{LC}\simeq 0.043$ (for more details see~\cite{karsai2016local}). Matching this relative size with the simulation results (see the observation line in Fig.~\ref{fig:kineticsverif}a upper panel), we find that it corresponds to $r^{emp} = 0.73$ (prediction line in Fig.~\ref{fig:kineticsverif}a), suggesting that the real adoption process lies in the crossover regime. In other words, large adoption cascades could potentially evolve in Skype but with reduced speed, as $73\%$ of users might not be interested in adopting a service within the network.

To test the validity of the predicted $r^{emp}$ value we perform three different calculations. First we measure the maximum relative growth rate of cumulative adoptions and find a good match between model and data (see Skype s3 and Model Skype s3 in Fig.~\ref{fig:Skype2Emp}a). In other words, the model correctly estimates the speed of the adoption process. Second, we measure the distribution $P(d)$ of the depths of induced vulnerable trees (Fig.~\ref{fig:kineticsverif}b). Vulnerable trees evolve with a shallow structure in the empirical and model processes. After measuring the distribution $P(d)$ for various $r$ values below, above and at $r^{emp}$, we find that the distribution corresponding to the predicted $r^{emp}$ value fits the best with the empirical data. Finally, in order to verify earlier theoretical suggestions~\cite{singh2013threshold}, we look at the correlation $\langle s_v \rangle (k)$ between the degree of innovators and the average size of vulnerable trees induced by them (Fig.~\ref{fig:kineticsverif}c). Similar to the distribution $P(d)$, we perform this measurement on the real data and in the model for $r=0.6$ and $0.9$, as well as for the predicted value $r_{emp}=0.73$. We find a strong positive correlation in the data, explained partially by degree heterogeneities in the underlying social network, but surprisingly well emulated by the model as well. More importantly, although this quantity appears to scale with $r$, the estimated $r$ value fits the empirical data remarkably well, thus validating our estimation method for $r$ based on a matching of relative component sizes.

\subsection{Threshold driven contagion on weighted networks}

Weighted networks provide meaningful representations of the architecture of a large number of complex systems where 
weights capture the strength of interactions between connected entities. This way, weights can help to differentiate between links of varying importance, influence, and role, which may play crucial roles in dynamical processes evolving on networks. 
In threshold models on networks, links are usually considered unweighted, such that the stimuli or influence arriving from each neighbour contributes equally to reaching the behavioural threshold. Although this assumption simplifies their modelling, it does not lead to an accurate representation of real world dynamics. For example, in neural systems synaptic connections have weights that quantify the strength of incoming stimuli, and contribute unequally in bringing neurons to an excited state, as recognised recently in models of neural population dynamics~\cite{iyer2013influence}. In social systems link weights are associated with tie strengths that quantify the social influence that individuals have on their peers. Measurement of tie strength is a long standing challenge, but it is generally accepted that social ties are not equal, as some are more influential than others on one's decision making. Surprisingly, apart from some recent studies~\cite{kempe2003maximizing,hurd2013watts,cox2016spread}, weights have been commonly overlooked in models of threshold driven phenomena. In the next Section, we summarise our work on threshold driven contagion on weighted networks. This study is a natural extension of our previously discussed dynamical threshold model, but it sheds the light on more exotic, not yet observed dynamical behaviour. Results have been published originally in~\cite{unicomb2018threshold}, while here we present the essence of our findings only.

\vspace{.2in}

To study threshold driven dynamical processes over weighted networks we build on the model defined in the previous Section and in~\cite{ruan2015kinetics,karsai2016local}. Following its standard formulation~\cite{watts2002simple,singh2013threshold,ruan2015kinetics,karsai2016local}, we define a monotone binary-state dynamics over a weighted, undirected network where edge weights $w > 0$ are continuous variables drawn from the distribution $P(w)$. The edge weight $w_{ij}$ represents the capacity of connected nodes $i$ and $j$ to influence each other. Accordingly, the node strength $q_k(i) = \sum_{j=1}^k w_{ij}$ is the total influence what node $i$ receives from its $k$ neighbours. However, influence may vary from neighbour to neighbour. We implement this idea by defining the partial strength $q_m(i) = \sum_{j=1}^m w_{ij}$ associated with the influence of the $m$ infected neighbours on node $i$ (where $0 \leq m \leq k$). If the condition $q_m \geq \varphi q_k$ is fulfilled, node $i$ becomes infected and remains so indefinitely. For simplicity we assume that all nodes have the same threshold $\phi=\varphi$, just as in many other studies~\cite{watts2002simple,singh2013threshold}.

Similar as earlier, we build our analytical description on Gleeson's approximate master equation formalism for stochastic binary-state dynamics~\cite{porter2016dynamical,gleeson2013binary,gleeson2008cascades,gleeson2011high}. In order to extend this formalism to weighted networks, we discretise $P(w)$ and assume only $n$ possible weight types $w_j$, such that all distinct weights in the network are contained in the weight vector $\mathbf{w} = (w_1, \ldots, w_n)^{\mathrm{T}}$. Then, a node in class $(k, m)$ has $k_j$ links with weight $w_j$ and $m_j = 0, \ldots, k_j$ infected neighbours across these links, such that $k = \sum_{j=1}^n k_j$ and $m = \sum_{j=1}^n m_j$. Furthermore, we can define a degree vector $\mathbf{k} = (k_1, \ldots, k_n)^{\mathrm{T}}$ and a partial degree vector $\mathbf{m} = (m_1, \ldots, m_n)^{\mathrm{T}}$, generalising the strength and partial strength to $q_{\mathbf{k}} = \mathbf{k} \cdot \mathbf{w}$ and $q_{\mathbf{m}} = \mathbf{m} \cdot \mathbf{w}$, respectively. Nodes in class $(\mathbf{k}, \mathbf{m})$ have identical strengths and partial strengths, and follow the same pair of rate equations for the fraction $s_{\mathbf{k}, \mathbf{m}} (t)$ (resp. $i_{\mathbf{k}, \mathbf{m}} (t)$) of $\mathbf{k}$-nodes that are susceptible (resp. infected) at time $t$ and have partial degree vector $\mathbf{m}$.
 
In our threshold driven model, a susceptible node can become infected in two ways, either spontaneously with rate $p$, or if its weighted threshold $\phi$ is reached. As such, the infection rate of susceptible nodes in class $(\mathbf{k}, \mathbf{m})$ is
\begin{equation}
\label{eq:thresRule}
F_{\mathbf{k}, \mathbf{m}}\mathbf{k} =
\begin{cases}
p & \quad q_{\mathbf{m}} < \phi q_{\mathbf{k}} \\
1 & \quad q_{\mathbf{m}} \geq \phi q_{\mathbf{k}}
\end{cases}, \quad k > 0,
\end{equation}
with $F_{\mathbf{0}, \mathbf{0}} = p$. The stepwise nature of $F_{\mathbf{k}, \mathbf{m}}\mathbf{k}$ allows us to map the rate equations for $s_{\mathbf{k}, \mathbf{m}}$ and $i_{\mathbf{k}, \mathbf{m}}$ to a reduced-dimension system, as has been done earlier in Eq.\ref{eq:reducedAMEs} for unweighted complex contagion~\cite{ruan2015kinetics,karsai2016local}. Namely, if we consider as aggregated variables the density $\rho(t)$ of infected nodes and the probability $\nu_j(t)$ that a randomly chosen neighbour (across a $j$-type edge) of a susceptible node is infected, then the description of the dynamics can be reduced to the system of $n + 1$ equations
\begin{subequations}
\label{eq:reducedAMEs}
\begin{align}
\dot{\nu}_j &= g_j(\boldsymbol{\nu}, t) - \nu_j, \\
\dot{\rho} &= h(\boldsymbol{\nu}, t) - \rho,
\end{align}
\end{subequations}
where $\boldsymbol{\nu} = (\nu_1, \ldots, \nu_n)^{\mathrm{T}}$ is the vector of probabilities $\nu_j$ for all weight types, and $g_j(\boldsymbol{\nu}, t)$ and $h(\boldsymbol{\nu}, t)$ are functions of binomial terms.

\subsection*{Regular networks with bimodal weights}

To study the dynamics of our model we first consider a simple structure, the configuration-model $k$-regular network, with $k=7$. Edge weights are sampled from a bimodal distribution with $n=2$ values, denoted strong ($w_1$) and weak ($w_2$). The weight distribution is characterised by its average $\mu$, standard deviation $\sigma \geq 0$, and the fraction $\delta$ of links that are strong. Thus, weights take the values $w_1 = \mu + \sigma \sqrt{ (1 - \delta) / \delta }$ and $w_2 = \mu - \sigma \sqrt{ \delta / (1 - \delta) }$. The parameter $\delta$ contributes to the skewness of $P(w)$, initially fixed to the symmetric case $\delta = 0.5$. The parameter $\sigma$ interpolates weight heterogeneity between the homogeneous case of an unweighted network ($\sigma=0$), and the most heterogeneous case of a diluted network ($\sigma = \mu \sqrt{ (1 - \delta) / \delta }$), where only strong links have influence and the weak are functionally absent. After fixing the spontaneous infection rate $p$ and skewness $\delta$, our model has only two parameters, $\sigma$ and $\phi$. To characterise the speed of dynamics we introduce the quantity $t_a$, the time when infection density reaches a set value ($\rho=0.75$), called the absolute time of cascade emergence. We measure $t_a$ via numerical simulations of the $(\sigma, \phi)$-parameter space, which shows unexpected dependencies on both parameters. On one hand, for fixed $\sigma$ and increasing $\phi$ the dynamics slows down, since nodes with higher thresholds require more infected neighbours to become infected. On the other, for fixed $\phi$ the dynamics depends \emph{non-monotonously} on $\sigma$, where cascades may evolve either faster or slower as we increase weight heterogeneity, relative to the unweighted case ($\sigma=0$).

\begin{figure*}[t]
 \centering
  \includegraphics[width=1.0\textwidth]{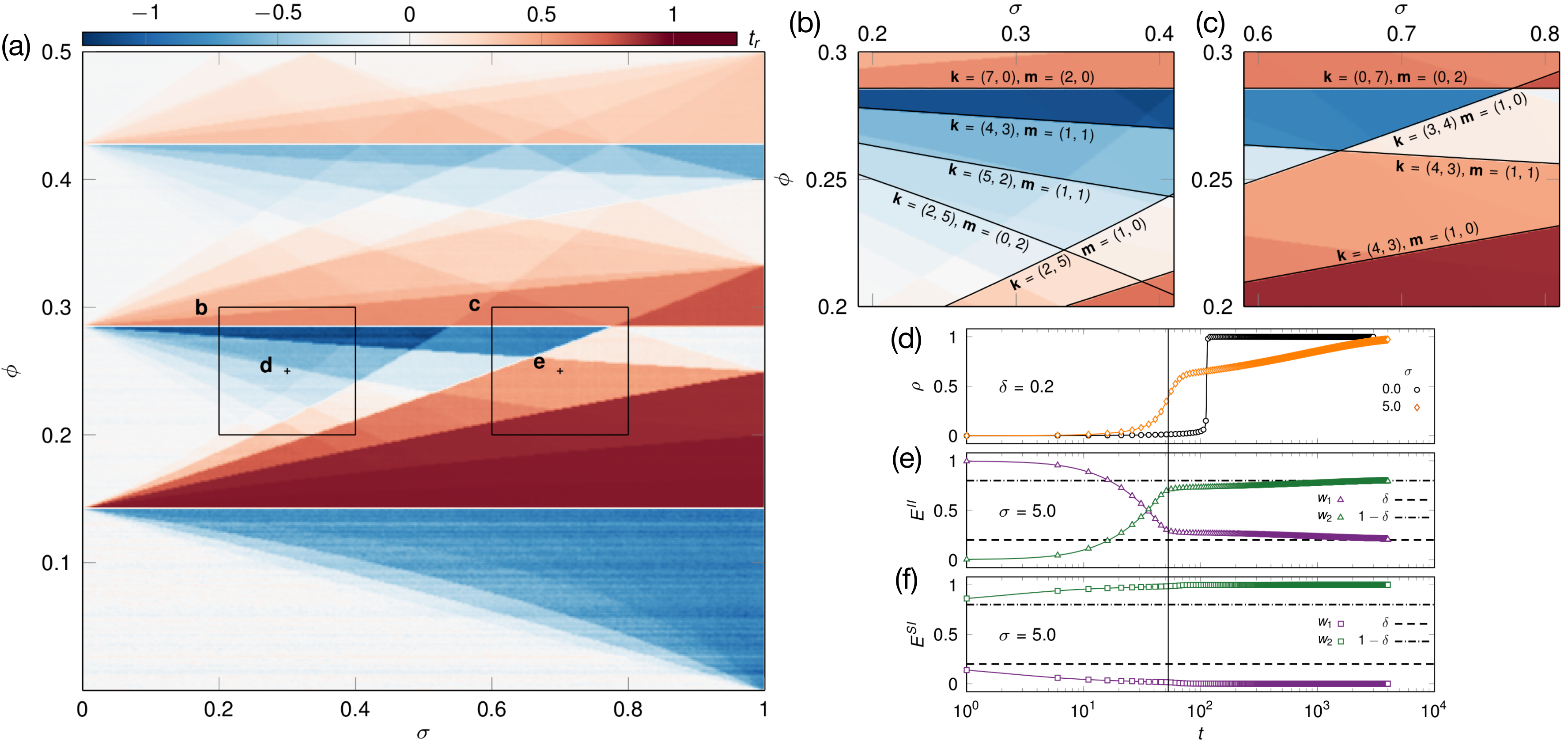}
  \caption{\small \small Relative time of threshold driven cascades on weighted networks. (a) Relative time $t_r$ of cascade emergence on $(\sigma, \phi)$-parameter space, simulated over $k$-regular regular networks ($k=7$) with $\mu = 1$, $\delta = 0.5$, $p = 2\times 10^{-4}$, $N = 10^4$ and averaged over 25 realisations. (b-c) Selected regions of parameter space in (a), where $t_r$ is instead calculated from the numerical solution of the AME systems in Eq.~\ref{eq:reducedAMEs}. Boundaries are obtained from a combinatorial argument~\cite{unicomb2018threshold} for various $( \mathbf{k}, \mathbf{m} )$ classes. (d-f) Effect of skewed weight distributions on cascade   evolution. \text{(d)} Infection density $\rho (t)$ on $k$-regular networks ($k=7$) and a bimodal weight distribution with $\mu = 3$ and $\delta=0.2$, both for unweighted ($\sigma = 0$) and heterogeneous ($\sigma > 0$) cases. \text{(d-e)} Fractions of strong ($w_1$) and weak ($w_2$) links connecting two infected nodes in the bulk of the infected component [$E^{II} (t)$, b] and susceptible and infected nodes on its surface [$E^{SI} (t)$, c] in the heterogeneous spreading scenario of (a). Simulations (symbols) are averaged over 25 realisations with $p=2 \times 10^{-4}$ and $N=10^4$, and compared with the corresponding AME solutions (lines). Dashed lines are the expected fractions of weak and strong links as determined by $\delta$, and the vertical line shows the inflection point of $\rho$ in the heterogeneous case of (a), which coincides with a turning point of $E^{II}$ in (b). This figure was prepared by S. Unicomb and was published in~\cite{unicomb2018threshold}.}
    \label{fig:wcasc1}
\end{figure*}

To better characterise the dynamics, first we concentrate on the $\sigma$ dependency by calculating $t_r =[ t_a(0,\phi) - t_a(\sigma,\phi) ] / t_a(0,\phi)$, the time of cascade emergence relative to the unweighted case with the same $\phi$ value. (Fig.~\ref{fig:wcasc1}a). The relative time $t_{r}$ will be positive if the weighted process evolves faster than the unweighted case, zero if they evolve at the same speed, and negative if slower than the unweighted case. The $(\sigma, \phi)$-parameter space for $t_r$ is highly structured and driven by competing effects of key $(\textbf{k},\textbf{m})$ classes, which either reduce or enhance the speed of the spreading process as compared to the unweighted case. We also explore the corresponding numerical solution of the AME systems in Eq.~\ref{eq:reducedAMEs}, as well as an independent combinatorial solution~\cite{unicomb2018threshold} for the boundaries between regions of low and high cascade speed (Fig.~\ref{fig:wcasc1}b-c). Both the AME and combinatorial solutions perfectly recover the parameter space obtained by simulations. To further explore how weight heterogeneities produce slow or fast cascades, we partition the system according to the number $m$ of infected neighbours required for infection, and measure the aggregated infection rate $F_{k,m}(t) = \sum_{\mathbf{k}, \mathbf{m}} P({\mathbf{k}}) F_{\mathbf{k}, \mathbf{m}} s_{\mathbf{k}, \mathbf{m}}(t) / \sum_{\mathbf{k}, \mathbf{m}} P({\mathbf{k}}) s_{\mathbf{k}, \mathbf{m}}(t)$ and other determinant quantities in several spreading scenarios (Fig.~\ref{fig:wcasc1}d-e).

In the neutral scenario, all $(\mathbf{k}, \mathbf{m})$ classes of the weighted network share the same dynamics as the corresponding $(k,m)$ class in an unweighted network, so $F_{k,m} = p$ or $1$ and weights have no impact on contagion, meaning $t_r = 0$. In a decelerative scenario like $\phi=0.25$ and $\sigma=0.3$ (Fig.~\ref{fig:wcasc1}d), $F_{k, m}$ for any $m$ is equal to its unweighted counterpart, except for the $m = 2$ class. Here, the adoption rate is $1$ in the unweighted case but strongly suppressed in the weighted case, thus decreasing the overall spreading speed. For an accelerative scenario, like $\phi=0.25$ and $\sigma=0.7$, competing effects from several $(\mathbf{k}, \mathbf{m})$ classes combine to determine the overall dynamics (Fig.~\ref{fig:wcasc1}e). The rate $F_{k, m}$ for $m = 2, \dots, 4$ is lower than 1 which is a decelerative effect (as in the previous case), but the rate $F_{k, 1}$, which is equal to $p$ in the unweighted case, is significantly larger than $p$ here. Since at the early stages of contagion the number of nodes in class $m=1$ is larger than in any other class with $m > 1$, spreading evolves rapidly to an early cascade. It should be noted that competition between the accelerative and decelerative effects of the weight distribution is one of the defining characteristics of threshold driven contagion on weighted networks. It is this competition that leads to the interference patterns evident in Fig. \ref{fig:wcasc1}a.


Up until now we have considered the symmetric case $\delta = 0.5$ with equal numbers of strong and weak links. However, by skewing the weight distribution we observe an additional effect of weight heterogeneities on the spreading behaviour. When $\delta = 0.2$ the extent of the cascade decreases for large $\sigma$ with respect to the unweighted case (Fig.~\ref{fig:wcasc1}d). In this case, despite their sparsity, strong links again drive the contagion, but are soon exhausted causing spreading to slow down and continue via spontaneous or infrequent threshold driven infections over weak ties (Fig.~\ref{fig:wcasc1}e). Indeed, strong links dominate the bulk of the infected component, but disappear quickly from its surface (Fig.~\ref{fig:wcasc1}f). These so-called {\it partial cascades}, which do not infect the whole system through the cascade, are associated with skewness and a sufficiently large standard deviation in the weight distribution and are reminiscent of the slow spreading caused by immune nodes, as well as low connectivity networks in unweighted complex contagion~\cite{ruan2015kinetics,karsai2016local,watts2002simple}. Overall, we identify non-monotonous spreading behaviour and partial cascades as the main consequences of weight heterogeneities in threshold driven contagion.

\vspace{.2in}

Although regular networks and bimodal weights are useful in characterising the qualitative impact of weights on contagion, they are rather unrealistic since real complex networks commonly appear with broad degree and weight distributions~\cite{barrat2004architecture}. For this reason we further explored (not shown here~\cite{unicomb2018threshold}) the effects of heterogeneous degree and weight distributions on the threshold driven contagion. Interestingly, it turned out that the observed non-monotonous dependency of cascade time is robust against such heterogeneities, and were found even in case of data-driven simulations, where the contagion was iterated on real weighted networks.

To summarise, in this study we explored weighted networks with increasing complexity, from configuration-model structures with bimodal or lognormal weights, to real world networks with broad degree and weight distributions. We showed that threshold driven contagion depends non-monotonously on weight heterogeneity, creating slow or fast cascades relative to the equivalent unweighted spreading process. Via numerical simulations, master equations and combinatorial arguments, we found that this effect is the result of competing configurations of degree, weight, and infected neighbours that slow down or speed up contagion. We also observed that an imbalance in the amount of large and small weights leads to partial cascades, and smoother temporal patterns of spreading than those in unweighted networks. By analysing a range of degree and weight configurations, we show that these features are systemic and thus may drive a variety of real world contagion phenomena.

\begin{center}
  $\ast$~$\ast$~$\ast$
\end{center}

As the continuation of this line of research we carried out two other projects. In one direction, under publication, we address threshold models on multiplex networks and observe a new type of parameter space indicating re-entrant phase transitions as the function of the number of layers and the average degree of the network. On another direction (in progress) we address threshold models on temporal networks and extend the solution provided by AME formalism for this case. Finally, in a separate work~\cite{karsai2016local}, we proposed another model of complex contagion phenomena, which build on an SIR model scheme, but relying on the observations reported in Section~\ref{sec:ccpoi}. It extends the model with a linearly increasing influence of adoption with the fraction of adopting neighbours.

\section{Conclusion}

This Chapter focused on my contributions on the observation and modelling of collective social phenomena. After a brief introduction I summarised selected studies on static and dynamic observations of collective behaviour, together with statistical and mechanistic models built on the obtained observations. I payed special focus on spreading processes, by walking through studies considering simple and complex contagion phenomena. My aim in this Chapter was to give an overview about my motivations and main contributions to the understanding of collective social behaviour. This way I decided to exclude several papers from the detailed discussion, but mentioning them in short summaries in the end of each Section. Moreover, many of my actual works address open challenges in the area of collective social phenomena, which will be discussed as my potential future research directions in the coming last Chapter of my Thesis.





\biblio



\chapter{Discussion}
\label{ch:disc}

\section{Perspectives}
The potential future of my research and my field can be discussed at different time-scales. Here I am going to describe some immediate and midterm directions related to my own research, while I will also synthesise my view on the longterm future potentials of my associated fields.

\subsection{Future directions of my research}

With my collaborators we have continued several research lines summarised in this Thesis, while we also initiated new spin-off directions based on recent ideas and developments. Next I shortly describe the most promising angles.

\paragraph{Data analysis and experiments:} Together with some socio-linguist colleagues we recently initiated a large-scale social experiment, where we collect social and verbal interactions of children between age in $3$ and $6$. Our experiment employs RFID sensors, which are capable of recording face-to-face interactions and speech at the same time, and is accompanied with frequent questionaries about the socioeconomic background and linguistic development of the participants. This project will continue over 3 years by recording one week of data every months with the involvement of $~110$ children. Our scientific goal is to understand the emergent consensus on the usage of standard patterns of French language as the function of the social network, time, linguistic and socioeconomic background.

In another line of research we continue to analyse large-scale datasets recording the social networks and sociodemographic, location, and linguistic characters of individuals using Twitter, mobile communication, and bank credit datasets. Our overall goal is to better understand the emergence of socioeconomic inequalities, limited social mobility, linguistic variance, and to infer average socioeconomic status at any level of granularity.

\paragraph{Higher order correlations in networks:}
The study on weighted event graph representation of temporal networks we reported in Section~\ref{sec:wDAGs} is only a first step in this direction. As a next step we are working on its extension to explore higher-order temporal-structural patterns, to more precisely model spreading processes on temporal networks, to identify large-scale temporal motifs, to define computationally cheap methods to calculate temporal network centralities, and to use this way of representation in systems where temporal interactions are important like in neural networks, transportation, or communication networks.

Rich datasets describing the actions and interactions of individuals provide outstanding sources of information and knowledge and fuel a wide spectrum of data-driven numerical simulations of dynamical processes. Data alone, however, even in huge amounts, do not easily transform into knowledge or predictive models. The rich and diverse information they contain raises crucial challenges concerning their analysis, representation and interpretation, the extraction of meaningful structures, and their integration into data-driven models. In a recent project we aim to build a methodological framework to reduce networked data complexity, while preserving its richness, by working at intermediate scales (“mesoscales”). Our objective is to reach a theoretical understanding and representation of rich and complex networked datasets for their use in predictive data-driven models.

In other exploratory projects we develop unsupervised learning approaches to build network embeddings (a) for static networks to infer correlations/patterns between node features and the mesoscopic structure of networks, and (b) for temporal networks to find correlations between events and infected populations induced by a spreading initiated from the actual event.

\paragraph{Modelling collective phenomena:}
In terms of social contagion phenomena one of my oldest personal puzzle is related to the phenomenological differences between simple and complex contagion processes. At the observation level these two processes appear very similar in real settings thus it is difficult to identify which of them is driving the actual spreading process. As both simple and complex mechanisms are arguably present simultaneously for a single social contagion process, beyond the identification of one or the other, it is important to decide which one plays a dominant role during the adoption of a single individual. Solving this problem would (a) close a long lasting debate about the dominant mechanisms driving social contagion and (b) would bring us closer to understand the role of social influence during decision making, which would contribute to another historic open question about the role of influence vs. homophily in network formation.

\paragraph{Statistical learning for data-driven modelling:}
In a new line of research I participate in a project, which aims to develop distributed and adaptive machine learning methods for the optimal parametrisation of generative models of graphs and complex systems. Our aim is to design distributed optimisation-based learning models for adaptive and distributed model inference in high dimensions. Our real challenge is to simultaneously consider individual features of humans and information provided by their relationships to infer latent correlations and to identify behavioural mechanisms to contribute to a better modelling of emergent social phenomena.

\subsection{Perspectives of my associated fields}
Computational social science builds on decades of knowledge accumulated in conventional social sciences, but as a separate field it is relatively young~\cite{lazer2009life}. First of all, its innovation is methodological as it relies on large and automatically collected behavioural data and novel experimental opportunities provided by the digital age. This new era just started and delivers new datasets, collection methods and challenges in a rapid pace. These developments fuel quantitative studies of social systems to identify mechanisms and emergent patterns in individual and social behaviour, with the promise of better understanding and predictability of human behaviour.

All three research domains I discussed in this Thesis have some relevance in these developments, however, they are at different phases of their developments. The observation of bursty human behaviour is relatively recent~\cite{barabasi2005origin}, but was followed by a very intense period of investigations~\cite{karsai2018bursty}, which brought us to the point that its first-order characterisation is well developed. New questions in this direction are related to higher-order temporal and structural correlations~\cite{scholtes2014causality,lambiotte2018understanding,kivela2017mapping}, which would bring us closer to understand the possible explanations and various consequences of such behavioural patterns. Temporal networks is a field which was also fuelled by the emergence of recently collected datasets recording time-varying interactions of entities on a high temporal resolution. This field has reached its first milestones~\cite{holme2012temporal,holme2015modern} but still offers several open theoretical and methodological challenges. Most interesting questions to answer are related to a more formal foundation of the representation, analytical tools and computability of temporal networks, in addition to the development of mathematically treatable network models. In addition, coupled dynamics of networks and processes are also very much un-understood. Due to the broad relevance of temporal networks in several fields, in my opinion, this field will remain in the centre of network science for a long period. Finally, the observation and modelling of the emergence of collective phenomena is a historic challenge, with some recent developments, which introduce ever more realistic data-driven prediction of collective processes. The observation of global information diffusion and formation of collective opinion, the precise data-driven simulations of epidemic processes, or the observation and prediction of the mobility of millions are all promising advances which will carry this field on to explore and explain more complex phenomena.

\section{Conclusions}

While writing this Thesis I had two main goals in mind. On one hand, I wanted to write a concise summary of my most interesting scientific contributions, and on the other hand, I aimed to provide and up-to-date view and perspectives about my field. Due to my training and scientific interests, this Thesis emerged as an interdisciplinary work discussing topics and methods associated to computer science, physics, statistics, applied mathematics and computational social science. As a consequence, it is not easy to read with an approach from a single discipline. To help the reader on this end, beyond the technical in-depth discussion of the actual methodologies and findings, I was trying to design a line of description, which is accessible for a broader audience. Despite the diverse subjects of my works, I needed to identify a common ground bringing them all together under the hood of a single thesis. As an outcome, I entitled my thesis as \emph{computational human dynamics}, which precisely captured the common denominator of all my contributions. In accordance, I consciously built all my reasoning and motivations on examples borrowed from social phenomena. On the other, some of my methodological work may be relevant in other fields as neural science or system biology, what I emphasised on the run.

\vspace{.2in}

I started my dissertation with an introduction to position the reader on the landscape of my field, open up actual challenges, and to put in perspective my contributions. To purposefully ground the terminology used throughout the whole dissertation, I briefly introduced the reader to the basic concepts of complex networks and to the characteristics and theories of social networks. To avoid redundant descriptions and confusing cross-referencing, I described various datasets, which had been used on multiple occasions in the reported studies. Subsequently, I summarised my contributions in three scientific domains. First, I concentrated on the bursty, heterogeneous temporal characters of individual human dynamics. Recently written a book on this topic, I had the advantage to easily provide an overall view on this field, and to introduce my contributions about the observations, characterisation and modelling of \emph{bursty phenomena}. Next, I concentrated on my works related to \emph{temporal networks}. This Chapter synthesises my publications I co-authored over the last years on various novel methods for the representation, characterisation, and modelling of time-varying structures. Our contributions range from system level characterisation and random reference models to the detection of higher-order correlations and the development of activity-driven generative network models of temporal networks. Finally, as a third area of my contributions, I discussed my work on the data-driven observations and modelling of \emph{collective social phenomena}. First I summarised studies on the static and dynamical observations of emergent patterns of collective processes such as language variance, socio-economic inequalities, or macroscopic and microscopic patterns of social contagion phenomena. Partially building on these observations, I subsequently discussed the modelling of simple and contagion phenomena, to understand how to control them on temporal structure, how to predict their dynamical emergence on society-large networks with various heterogeneous characters.

\vspace{.2in}

This Thesis was written for the purpose of obtaining the degree of habilitation of directing research. However, it was meant for something more than being simply a summary of my contributions. I wrote it as a research statement and a milestone (for myself) to summarise my past achievements, to clarify my actual position, and to help building plans for my future research. I hope that some others will also find this synthesis useful and motivating to find new ideas to solve and to explore paths towards a more comprehensive understanding the world around.

\biblio





\cleardoublepage

\phantomsection

\addcontentsline{toc}{chapter}{Bibliography}

\bibliographystyle{plain}



\cleardoublepage
\phantomsection
\setlength{\columnsep}{0.75cm}
\addcontentsline{toc}{chapter}{\textcolor{cyan}{Index}}
\printindex


\end{document}